\ifdefined\aurorasatmlclassloaded\else
\documentclass[conference]{IEEEtran}
\fi

\ifdefined\pdfinfoomitdate\pdfinfoomitdate=1\fi
\ifdefined\pdftrailerid\pdftrailerid{}\fi
\ifdefined\pdfsuppressptexinfo\pdfsuppressptexinfo=-1\fi

\usepackage{cite}

\usepackage{tcolorbox}
\usepackage{float}
\usepackage{enumitem}
\usepackage{mdframed}
\usepackage{xcolor}
\usepackage{tikz} % Figure 1 (figs/fig1_cade_tikz.tex)
\usetikzlibrary{arrows.meta, shapes.geometric, calc}
\usepackage{colortbl}
\usepackage{multirow}
\usepackage{booktabs}
\usepackage{amsmath}
\usepackage{amssymb}
\usepackage[final,markup=default,authormarkup=none,commandnameprefix=always]{changes}

\usepackage{algorithm}
\usepackage{algpseudocode}

\usepackage[hidelinks]{hyperref}
\usepackage{bm}
\usepackage{graphicx}

\usepackage{soul}
\definecolor{changemarkcolor}{RGB}{255, 215, 0}

\definecolor{newtextcolor}{RGB}{255,224,178}
\newmdenv[backgroundcolor=newtextcolor,linewidth=0pt,innerleftmargin=3pt,innerrightmargin=3pt,innertopmargin=3pt,innerbottommargin=3pt,skipabove=4pt,skipbelow=4pt,leftmargin=0pt,rightmargin=0pt]{newtext}
\definecolor{reliablecolor}{RGB}{184, 222, 162}
\definecolor{stablecolor}{RGB}{180, 190, 255}
\definecolor{reliablecolortable}{RGB}{230, 245, 220}
\definecolor{stablecolortable}{RGB}{235, 240, 250}
\definecolor{baselinecolor}{RGB}{245, 245, 245}
\definecolor{abstentioncolor}{RGB}{173, 216, 230}
\definecolor{darkbluebar}{RGB}{0, 90, 160}
\definecolor{lightgrey}{RGB}{245, 245, 245}

\newtcbox{\highlightterm}[2][]{on line, box align=base,
  colback=#2!50, colframe=#2!50,
  boxrule=0pt, arc=3pt, auto outer arc,
  left=2pt, right=2pt, top=-1pt, bottom=-1pt,
  #1}

\newtcbox{\highlighttermnobreaks}{%
  tcbox raise base,
  colback=stablecolor,
  colframe=stablecolor,
  boxrule=0pt,
  arc=3pt,
  left=2pt, right=2pt, top=0pt, bottom=0pt}

\sethlcolor{stablecolor!60}

\newtcolorbox[auto counter]{protocolbox}[2][]{%
  colback=gray!5,
  colframe=gray!65,
  title={\textbf{Protocol~\thetcbcounter: #2}},
  sharp corners,
  fonttitle=\normalsize\bfseries,
  left=2mm, right=2mm, top=0.5mm, bottom=0.5mm,
  boxsep=2pt, before skip=4pt, after skip=4pt,#1}

\newenvironment{obs}{%
    \vspace{0.3em}
    \begin{mdframed}[
        topline=false, rightline=false, bottomline=false, leftline=true,
        linecolor=darkbluebar, linewidth=3pt,
        backgroundcolor=lightgrey,
        innerleftmargin=10pt, innerrightmargin=10pt,
        innertopmargin=3pt, innerbottommargin=3pt, skipabove=2pt, skipbelow=2pt
    ]
}{%
    \end{mdframed}
    \vspace{0.3em}
}

\newcommand{\sigrej}{\ensuremath{\sigma[\text{Rej}]^*}}

\newcommand{\budgetmonthlymath}{B_M}  % monthly annotation budget (standardized from B_{M_i})
\newcommand{\cutoffrejmath}{\rho}
\newcommand{\datainitmath}{\mathcal{D}_0}

\newcommand{\codename}{\textsc{Aurora}}

\begin{document}
% =======================================================

\title{On the Reliability and Stability of Selective Methods in Malware Classification Tasks}

% De-anonymized version: author names and affiliations.
% Superscript numbers mark affiliations; * marks equal contribution
% (explained in a footnote after \maketitle).
\author{%
\IEEEauthorblockN{%
Alexander Herzog\textsuperscript{1,2,*}\quad
Aliai Eusebi\textsuperscript{1,*}\quad
Lorenzo Cavallaro\textsuperscript{1,3}}
\IEEEauthorblockA{\small
\textsuperscript{1}University College London\quad
\textsuperscript{2}CORE64\quad
\textsuperscript{3}Mohamed bin Zayed University of Artificial Intelligence}
}
\hypersetup{pdfauthor={Alexander Herzog, Aliai Eusebi, Lorenzo Cavallaro},
  pdftitle={On the Reliability and Stability of Selective Methods in Malware Classification Tasks}}

\maketitle
{\renewcommand{\thefootnote}{*}\footnotetext{Equal contribution.}}
\begin{abstract}
Drift-adaptive malware classifiers report strong performance metrics, yet accuracy alone says little about whether confidence estimates remain reliable under distribution shift -- a critical requirement for active learning, selective classification, and operational trust. Standard evaluation paradigms neglect whether confidence estimates are trustworthy or stable over time, leaving a systematic blind spot in how we assess production-ready classifiers. We propose \codename{}, a framework that evaluates malware classifiers along two axes: reliability and stability. \codename{} verifies whether a model’s confidence profile is well-calibrated and stable over time, exposing failure modes that point-in-time accuracy metrics conceal. Miscalibrated confidence erodes operator trust, misdirects annotation budgets toward non-informative samples, and allows high-error instances to escape selective classification filters. \codename{} is further complemented by a set of metrics designed to go beyond point-in-time performance, striving towards a more holistic assessment of operational stability throughout temporal evaluation periods. By applying \codename{}, we expose the gap between the reported performance of state-of-the-art frameworks and their stability and reliability over time, opening promising avenues for future research.

\end{abstract}


% =======================================================
\section{Introduction}
% TODO-SUBMISSION (carried over from removed todobox): add link to GitHub repo (needs anonymous URL).

Research contributions to malware classification under distribution drifts\footnote{Without any loss of generality, we use distribution shift, distribution drift, concept shift and concept drift} interchangeably throughout. have primarily centered on predictive accuracy, often treating trustworthiness as an implicit byproduct of performance. In Strathern's well-known paraphrase of Goodhart's law, ``when a measure becomes a target, it ceases to be a good measure''~\cite{goodhart1975monetary,strathern1997improving}. 

Malware classifiers experience performance degradation over time due to concept drift, caused by adversarial adaptation, the evolution of benign software, or fragile feature spaces. This violates the i.i.d. assumption upon which manually engineered or learned representations depend. Therefore, existing state-of-the-art (SOTA) drift-adaptation solutions aim to identify observations at test time that are likely to have drifted, enabling targeted mitigation. Each solution implements its own distinct detection criteria. CADE~\cite{CADE} calculates the minimum class-centroid distance normalized by Median Absolute Deviation (MAD); HCC~\cite{HCC} quantifies embedding misalignment via averaged pairwise contrastive losses between test samples and nearest training neighbors; Transcendent~\cite{barbero2022transcendent} applies conformal prediction theory to evaluate sample nonconformity relative to calibration data of the predicted class. Whether framed as confidence, nonconformity, or OOD-ness, these approaches all converge on the same aim: to attempt to produce a reliable ranking over incoming observations.

This ranking is then used in two closely related ways: (i) in selective classification, where the model abstains from predicting on drifted observations~\cite{barbero2022transcendent}, or ii) in active learning, where observations are ranked by their ``out-of-distribution (OOD) score'' and routed to human labeling for model retraining~\cite{CADE,HCC}.\footnote{Conceptually, rejected samples in selective classification are natural candidates for active learning.} In active learning, selection follows either a threshold-based rule that annotates all samples exceeding a score threshold~\cite{CADE} or a budget-based rule that annotates a fixed number of top-ranked samples~\cite{HCC}. Selective classification prioritizes classification reliability at the cost of coverage (the proportion of accepted predictions), while active learning considers drift signals as informative for model adaptation.

Existing solutions are, however, predominantly assessed using standard  performance metrics, with improvements in F1, FPR, or FNR treated as sufficient evidence of trustworthiness under realistic conditions~\cite{pendlebury2019tesseract}. Crucially, offline metrics presuppose access to ground truth labels -- an assumption that collapses once the model is deployed. Once released, the only signal available to assess decision quality is the model’s own confidence. A trustworthy system must therefore ensure that high-confidence predictions are likely correct and that low-confidence (or high OOD-score) ones are routed for further inspection. This requirement becomes most fragile under distribution shift. Modern neural networks are already known to be severely miscalibrated~\cite{on_the_calibration_of_modern_neural_networks} -- meaning their confidence fails to reflect true correctness -- and they frequently assign high-confidence scores to incorrect OOD predictions~\cite{ovadia2019can}. Whether the goal is abstention or adaptation, the system’s behavior in deployment is therefore conditioned not by raw in-lab accuracy, but by the quality and stability of its confidence-based ranking.

The centrality of confidence-based ranking for selective classification has recently been established on firm theoretical grounds. 
Rabanser and Papernot~\cite{rabanser2025does} formalize the selective-classification gap, i.e., the discrepancy between a model's selective performance and that of an oracle that perfectly ranks predictions by correctness, and show that, among its contributing factors, only \emph{ranking error} governs which points enter the acceptance set at a given coverage; standard calibration cannot reduce it, since it preserves score ordering, and distribution shift further destabilizes the ranking even when accuracy and calibration are controlled. This places the quality and stability of confidence-induced ranking at the core of operational reliability and directly motivates our work: we measure ranking quality under sustained temporal drift, moving beyond their focus on isolated shift events.

Despite the theoretical centrality of ranking quality, confidence scoring functions are rarely subjected to direct assessment in practice. Evaluations focus on downstream performance metrics and implicitly assume that strong F1, FPR, or FNR validates the underlying confidence function. This approach fails to differentiate whether improvements result from better ranking of predictions or from a robust model compensating for suboptimal confidence estimates. The assumption becomes particularly questionable under distribution shift, where confidence functions are most likely to mis-rank errors (refer to Figure~\ref{fig:example_cade_bad} for a motivating illustration). 

We propose \codename{}, a framework to evaluate malware classifiers based on their confidence quality and operational resilience. \codename{} subjects the confidence profile of a given model to verification using the excess Area Under the Risk--Coverage Curve (E-AURC) to measure the quality of its confidence ranking~\cite{aurc}. We define a model’s ``confidence profile'' as the behavior of both (i) its OOD scoring function and (ii) its vanilla confidence function. In this way, we evaluate both the reliability of a classifier’s OOD scoring function and its predictive uncertainty. E-AURC quantifies how effectively confidence scores rank instances according to their actual probability of misclassification. Intuitively, a prediction with a higher OOD score or predictive uncertainty should be more likely to be incorrect. Therefore, we expect performance to be monotonically non-decreasing when observations are rejected in order of increasing confidence (least confident first). This shows the model correctly quantifies both \emph{what} samples fall outside its training distribution and \emph{when} it is likely to make mistakes. If not, we are allowing error-prone instances to remain undetected, relying on systems that \emph{do not know what they do not know}.

\codename{} is further complemented by a set of metrics designed to move beyond point-in-time performance, enabling the assessment of operational stability across temporal evaluation periods. We focus on a deployment simulation, an abstention-based protocol that closely reflects real operating conditions. We set a monthly rejection quota $\cutoffrejmath$ equal to the monthly annotation budget $\budgetmonthlymath$, using the fixed monthly budgets of drift-adaptive Android malware detection~\cite{HCC,mcfadden2026drmd,haque2025citadel}, so that the abstention workload is expressed in the same analyst-capacity unit as labeling; the rejected and labeled samples, nevertheless, remain separate sets. The threshold is re-calibrated at the start of every month to this quota on the unlabeled scores of the six preceding months, adapting to the evolving score-distribution. Unlike Transcendent~\cite{barbero2022transcendent}, which selects class-conditional thresholds on a labeled calibration set to maximize F1 on the accepted set under a rejection-rate constraint, this calibration uses no labels (\S\ref{subsec:operational_assessment}). Under these constraints, a reliable confidence function should lead to stable and decision-relevant behavior: accepted samples ought to be correct with high probability, while rejected ones should concentrate the residual risk. By evaluating performance in this setting, we assess not only conventional accuracy but also the practical usefulness of the confidence ranking itself over time.

Our primary benchmark is Hypercube~\cite{chow2025breaking}, the state-of-the-art (SOTA) dataset for evaluating Android malware classifiers under drift. Unlike earlier benchmarks, it is curated to current guidance for sound evaluation, with market publication timestamps, single-market provenance of goodware and malware, and representative sampling at a realistic 9:1 ratio~\cite{pendlebury2019tesseract}. We compare all methods on Hypercube VTT2 (2021--2024). To test whether our findings depend on the benchmark, we additionally evaluate dataset-specific configurations of a subset of the methods on the AndroZoo, APIGraph and Transcendent datasets.  The ordering of confidence quality largely carries over across datasets, whereas the monthly F1 decline is severe on Hypercube, mild on Transcendent, and absent on APIGraph and AndroZoo. While we evaluate \codename{} on Android malware classification only, the methodology and its metrics apply to any classifier that exposes a confidence or OOD score over a temporally ordered dataset and selective classification or active learning.%

In summary, our contributions are as follows:
\begin{itemize}[leftmargin=10pt]
    \item We propose \codename{} (\S\ref{sec:evaluationframework}), an evaluation framework that extends beyond headline performance metrics such as F1, FNR, and FPR along two axes. For \emph{reliability} (\S\ref{subsec:reliability}), it evaluates the confidence ranking with risk--coverage curves and E-AURC, error capture at a fixed labeling budget (Cap$_{\budgetmonthlymath}$), and accepted-set F1 under abstention (F1$^{*}$, $\Delta$F1$^{*}$). For \emph{temporal stability} (\S\ref{subsec:stability}), it measures the rate of performance decay (Sen[F1]), the trend of the confidence ranking itself (Sen[AUROC$_f$]), and adherence to the rejection quota via the deployed threshold (Sen[$q^{*}$], $\bar q^{*}$).
    \item We formalize an operational perspective where we treat models and their associated confidence functions as selective classifiers whose rejection threshold is calibrated on unlabeled past scores to target a monthly rejection quota. Within this setting, we evaluate not only performance under rejection but also adherence to rejection quotas under drift and silent failures, i.e., high-confidence errors that remain accepted without review (\S\ref{subsec:operational_assessment}, \S\ref{riskcoverage}).
    \item We benchmark ten Android malware classifiers, updated monthly by uncertainty-based active learning, and four Transcendent conformal configurations, for a total of  sixteen confidence functions, on Hypercube VTT2~\cite{chow2025breaking} over 30 evaluation months through December 2024, with five seeds and monthly label budgets $\budgetmonthlymath\in\{50,100,200,400\}$ (\S\ref{sec:results}, Appendix~\ref{appendix:vtt2_budgets}). At every budget, headline performance and reliability diverge: the method with the highest F1 is never the one with the most reliable confidence ranking (lowest E-AURC), and abstention at the calibrated threshold lowers accepted-set F1 for most classifiers; all four Transcendent configurations and HCC-PL gain at every budget. No method is temporally stable: the Theil--Sen slope of monthly F1 is negative for every method at every budget and in every seed. We additionally evaluate dataset-specific configurations of a subset of these methods (DeepDrebin, Drebin, HCC, CADE and Transcendent) on AndroZoo, APIGraph~\cite{HCC} and Transcendent~\cite{barbero2022transcendent} datasets in the same metric set: the confidence ordering largely carries over, while the monthly F1 decline is severe on Hypercube.
\end{itemize}

\begin{figure}[!t]
    \centering
    % ============================================================================
% Figure 1 -- FINAL HYBRID: "Horizontal compact" (Proposal C) layout
%             + "Confidence field" (Proposal B) greyscale encoding
%
% CADE's distance-based OOD assumption (A) vs. observed reality (B)
%
%   Layout/skeleton from Proposal C: side-by-side borderless cards,
%   integrated panel titles, shared one-line marker legend.
%   Fill/encoding from Proposal B: greyscale radial confidence field
%   (dark = high confidence near centroids, fading to the card tint at the
%   decision boundary), subtle iso-confidence contours, and a small
%   "Model confidence: low -> high" colorbar in the legend.
%
%   Panel A: misclassification triangles in the LIGHT (low-confidence) zone
%            near the boundary + dashed distance-to-centroid arrow.
%   Panel B: CADE's score min_c |d_c - m_c| / MAD_c is lowest (confidence
%            highest) at the TYPICAL (median) training distance m_c, so the
%            high-confidence region is an annulus (dark ring at radius m_c,
%            light at the centroid itself and far outside); triangles sit on
%            that dark ring.
%
% Usage:  \input{figs/fig1_cade_hybrid} inside a figure environment.
% Natural size: ~8.6cm x ~4.8cm (fits IEEE single column without scaling).
% Requires in preamble:
%   \usepackage{tikz}
%   \usetikzlibrary{arrows.meta, shapes.geometric, calc}
% ============================================================================
\begingroup%
% --- palette (colorblind-safe markers, greyscale field) ---------------------
\definecolor{phBenign}{HTML}{00796B}% teal (benign)
\definecolor{phMal}{HTML}{D32F2F}% warm red (malware)
\colorlet{phBenignDark}{phBenign!75!black}%
\colorlet{phMalDark}{phMal!75!black}%
\colorlet{phFieldDark}{black!42}% field intensity reused from Proposal B
\colorlet{phCard}{black!4}% flat card tint (Proposal C)
\tikzset{%
  % observation: filled dot, thin white rim so it pops on the dark field (B)
  phobs/.style={circle, fill=#1, draw=white, line width=0.5pt,%
                inner sep=0pt, minimum size=5.6pt},%
  % misclassification: filled triangle, white halo (crisp on dark and light)
  phmiss/.style={regular polygon, regular polygon sides=3, fill=#1,%
                 draw=white, line width=0.8pt,%
                 inner sep=0pt, minimum size=9.5pt},%
  % class-centroid cross (pic), white under-stroke for contrast
  pics/phcross/.style={code={%
    \draw[line width=3.4pt, line cap=round, white]%
      (-3.6pt,-3.6pt) -- (3.6pt,3.6pt) (-3.6pt,3.6pt) -- (3.6pt,-3.6pt);%
    \draw[line width=2.4pt, line cap=round, #1]%
      (-3.1pt,-3.1pt) -- (3.1pt,3.1pt) (-3.1pt,3.1pt) -- (3.1pt,-3.1pt);}},%
  % flat panel card (no border) -- Proposal C
  phcard/.style={fill=phCard, rounded corners=3pt},%
  % decision boundary
  phbound/.style={draw=black!55, line width=0.8pt,%
                  dash pattern=on 4pt off 2.8pt},%
  % iso-confidence contour -- Proposal B
  phiso/.style={draw=black!28, line width=0.4pt},%
  % distance-to-centroid arrow
  phdist/.style={draw=black!75, line width=0.9pt,%
                 dash pattern=on 3pt off 2.2pt,%
                 -{Triangle[length=5.5pt, width=4.6pt]}},%
  % panel title -- Proposal C
  phtitle/.style={anchor=south west, font=\scriptsize\bfseries, inner sep=0pt}%
}%
%
% --- shared per-panel scenery (card interior: 4.15 x 3.35) ------------------
% Greyscale confidence field replaces Proposal C's color blobs:
% radial shading, dark at the class centroids, fading to the card tint at
% the decision boundary; subtle iso-confidence contours (Proposal B,
% radii rescaled by ~0.52 to the compact panel).
\newcommand{\phScenery}{%
  \path[phcard] (0,0) rectangle (4.15,3.35);%
  \begin{scope}%
    \clip[rounded corners=3pt] (0,0) rectangle (4.15,3.35);%
    \shade[inner color=phFieldDark, outer color=phCard]%
      (0.95,1.70) circle (1.20);%
    \shade[inner color=phFieldDark, outer color=phCard]%
      (3.18,1.78) circle (1.27);%
    \foreach \r in {0.40,0.76,1.10}{\path[phiso] (0.95,1.70) circle (\r);}%
    \foreach \r in {0.42,0.81,1.17}{\path[phiso] (3.18,1.78) circle (\r);}%
  \end{scope}%
  \path[phbound] (2.075,0.10) -- (2.075,3.25);%
}%
%
% --- panel-B scenery: annular confidence field ------------------------------
% Radial profile: card tint at the centroid -> dark on the ring at the median
% training distance m_c (radius ~0.62 / 0.66) -> card tint at the field edge.
% (TikZ maps 25bp of the shading to ~the circle radius.)
\pgfdeclareradialshading{phringB}{\pgfpoint{0bp}{0bp}}{%
  color(0bp)=(phCard); color(2bp)=(phCard);%
  color(12.5bp)=(phFieldDark);%
  color(23bp)=(phCard); color(50bp)=(phCard)}%
\newcommand{\phSceneryRing}{%
  \path[phcard] (0,0) rectangle (4.15,3.35);%
  \begin{scope}%
    \clip[rounded corners=3pt] (0,0) rectangle (4.15,3.35);%
    \shade[shading=phringB] (0.95,1.70) circle (1.20);%
    \shade[shading=phringB] (3.18,1.78) circle (1.27);%
    \foreach \r in {0.30,0.92}{\path[phiso] (0.95,1.70) circle (\r);}%
    \foreach \r in {0.32,0.98}{\path[phiso] (3.18,1.78) circle (\r);}%
    % median-distance ring m_c (typical training distance)
    \path[draw=black!60, line width=0.5pt, dash pattern=on 1.2pt off 1.2pt]%
      (0.95,1.70) circle (0.62) (3.18,1.78) circle (0.66);%
  \end{scope}%
  \path[phbound] (2.075,0.10) -- (2.075,3.25);%
}%
\begin{tikzpicture}[x=1cm, y=1cm]%
% ===========================================================================
% PANEL A (left): CADE's assumption -- errors in the LIGHT low-confidence
% zone, hugging the boundary, far from both centroids
% ===========================================================================
\begin{scope}[shift={(0,0)}]%
  \node[phtitle] at (0.02,3.46) {A --- CADE's assumption};%
  \phScenery%
  % --- benign (teal) observations ---
  \node[phobs=phBenign] at (0.62,1.50) {};%
  \node[phobs=phBenign] at (0.80,2.02) {};%
  \node[phobs=phBenign] at (1.26,1.42) {};%
  \node[phobs=phBenign] at (1.46,1.78) {};%
  \node[phobs=phBenign] at (1.18,2.50) {};%
  % --- malware (red) observations ---
  \node[phobs=phMal] at (2.80,2.02) {};%
  \node[phobs=phMal] at (3.62,1.58) {};%
  \node[phobs=phMal] at (3.05,1.30) {};%
  \node[phobs=phMal] at (3.46,2.30) {};%
  \node[phobs=phMal] at (2.72,0.95) {};%
  % --- centroids ---
  \pic at (0.95,1.70) {phcross=phBenignDark};%
  \pic at (3.18,1.78) {phcross=phMalDark};%
  % --- misclassifications: LIGHT zone, hugging the boundary ---
  \node[phmiss=phMal]    at (1.74,2.58) {};%
  \node[phmiss=phMal]    at (1.88,2.98) {};%
  \node[phmiss=phBenign] at (2.28,3.00) {};%
  \node[phmiss=phBenign] at (2.42,2.60) {};%
  \node[phmiss=phMal]    at (1.72,0.74) {};%
  \node[phmiss=phMal]    at (1.88,0.38) {};%
  \node[phmiss=phBenign] at (2.28,0.36) {};%
  \node[phmiss=phBenign] at (2.44,0.72) {};%
  % --- distance-to-centroid arrow: benign centroid -> boundary errors ---
  \path[phdist] (1.10,1.86) -- (1.58,2.40);%
\end{scope}%
% ===========================================================================
% PANEL B (right): observed -- errors on the DARK high-confidence ring at
% the typical (median) training distance, not at the centroids
% ===========================================================================
\begin{scope}[shift={(4.45,0)}]%
  \node[phtitle] at (0.02,3.46) {B --- Observed};%
  \phSceneryRing%
  % --- benign (teal) observations: spread around the typical distance ---
  \node[phobs=phBenign] at (0.38,1.42) {};%
  \node[phobs=phBenign] at (0.62,2.30) {};%
  \node[phobs=phBenign] at (0.82,0.92) {};%
  \node[phobs=phBenign] at (1.30,2.55) {};%
  \node[phobs=phBenign] at (0.22,2.02) {};%
  % --- malware (red) observations ---
  \node[phobs=phMal] at (3.78,2.18) {};%
  \node[phobs=phMal] at (3.88,1.50) {};%
  \node[phobs=phMal] at (3.34,1.02) {};%
  \node[phobs=phMal] at (3.02,2.62) {};%
  \node[phobs=phMal] at (3.60,2.78) {};%
  % --- centroids (sit in the LIGHT hole: atypically close = low confidence) ---
  \pic at (0.95,1.70) {phcross=phBenignDark};%
  \pic at (3.18,1.78) {phcross=phMalDark};%
  % --- misclassifications: ON the dark ring at the typical distance m_c ---
  \node[phmiss=phMal]    at (1.49,2.01) {};%
  \node[phmiss=phMal]    at (1.46,1.35) {};%
  \node[phmiss=phBenign] at (2.61,2.11) {};%
  \node[phmiss=phBenign] at (2.60,1.45) {};%
\end{scope}%
% ===========================================================================
% LEGEND: one compact marker line (Proposal C) + confidence colorbar (B)
% ===========================================================================
\begin{scope}[shift={(0,-0.42)}]%
  \node[phobs=black!75] at (0.20,0) {};%
  \node[anchor=west, font=\scriptsize] at (0.38,0) {Observation};%
  \node[phmiss=black!75] at (2.34,0.02) {};%
  \node[anchor=west, font=\scriptsize] at (2.54,0) {Misclassified};%
  \pic at (4.62,0) {phcross=black!75};%
  \node[anchor=west, font=\scriptsize] at (4.84,0) {Centroid};%
  \path[phbound] (6.62,0) -- (7.12,0);%
  \node[anchor=west, font=\scriptsize] at (7.20,0) {Boundary};%
  % --- confidence colorbar row ---
  \node[anchor=east, font=\scriptsize] at (3.92,-0.50) {Model confidence:};%
  \node[anchor=east, font=\tiny, text=black!70] at (4.36,-0.50) {low};%
  \shade[left color=phCard, right color=phFieldDark, draw=black!55,%
         line width=0.4pt] (4.44,-0.59) rectangle (6.24,-0.41);%
  \node[anchor=west, font=\tiny, text=black!70] at (6.32,-0.50) {high};%
\end{scope}%
\end{tikzpicture}%
\endgroup%
    \caption{\small CADE~\cite{CADE} assumes errors lie far from class centroids (case {\setlength{\fboxsep}{1pt}\fbox{\textbf{A}}}). Hypercube VTT2 risk--coverage evaluation (Fig.~\ref{fig:eaurc_vtt2}a) shows the \textbf{opposite}: errors concentrate at low OOD scores (distances typical of the training group), where confidence is high~\cite{ovadia2019can} (case {\setlength{\fboxsep}{1pt}\fbox{\textbf{B}}}; see also Fig.~\ref{fig:risk_coverage_tau50}).}
    \label{fig:example_cade_bad}
\end{figure}

\section{Motivation and Related Work}
\label{sec:background}
Existing Android malware detectors either optimize for stationary conditions~\cite{drebin,DeepDrebin,mariconti2016mamadroid}, harden against adversarial manipulation~\cite{demontis2019secureml,li2021ramda}, or aim for robustness to temporal drift~\cite{apigraph,HCC,CADE,barbero2022transcendent,xu2019droidevolver,narayanan2017context,zheng2026tif,mcfadden2026drmd,haque2025citadel,qian2025lamdcontextdrivenandroidmalware}.  
Traditional approaches rely on static or dynamic analysis to derive behavioral proxies: Drebin~\cite{drebin} encodes eight statically extracted feature sets (permissions, intents, API calls, etc.) as sparse multi-hot vectors; MaMaDroid~\cite{mariconti2016mamadroid} models package transitions via Markov chains; and APIGraph~\cite{apigraph} embeds relational API graphs to expose functional clusters.  
Because hand-crafted features suffer from inherent semantic limitations and struggle to capture the novel patterns introduced by emerging threats, recent work shifts to learned representations, often realized with deep neural networks~\cite{DeepDrebin,saxe2017expose,kolosnjaji2016deep,hardy2016dl4md,azmoodeh2018robust}.

Despite advances in representation learning~\cite{zheng2026tif}, detectors still need periodic retraining to track distribution shift~\cite{pendlebury2019tesseract,barbero2022transcendent,HCC}. This is expensive: labels are scarce, yet threat volume is huge. In 2025, Kaspersky alone blocked over 14 million attacks involving mobile malware, adware, or unwanted software, and observed $815{,}735$ new malicious installation packages~\cite{kaspersky2026mobile}; manual analysis of each sample is costly, so analysts can review only a small fraction of them~\cite{miller2016reviewer}. To stem performance decay under such constraints, SOTA systems integrate selective classification and active learning to maintain accuracy with minimal human effort.

Selective classification -- also called classification with rejection, learning with rejection, or misclassification detection --  allows a model to abstain when its prediction is unreliable~\cite{zhang2023survey,hendrickx2024machine,chow1970optimum,hendrycks2016baseline,corbiere2021confidence,zhu2022rethinking,moon2020confidence,galil2023can, linusson2018classification}. The objective is to minimize the \emph{selective risk}, expected misclassification cost, while maximizing \emph{coverage}, the share of accepted predictions~\cite{geifman2017selective}.  Practically, this is realized via a confidence estimator whose threshold acts as a ``dial'': tightening it lowers the error rate at the expense of accepting fewer samples, and vice versa.

Malware frameworks such as Transcendent~\cite{barbero2022transcendent} embed this mechanism: a conformal evaluator scores each prediction by its credibility, the p-value of the predicted class with respect to labeled calibration data, and rejects predictions whose credibility falls below class-conditional thresholds fixed before deployment (\S\ref{subsec:operational_assessment}). MalCertain~\cite{li2024malcertain} instead uses the uncertainty of a deep detector's predictions to identify and correct likely misclassifications. Building on the same principle, production security scanners already quarantine files whose risk exceeds a preset threshold, deferring their release until an analyst can vet them~\cite{trend_quarantine,norton_mobile}. Our deployment simulation adopts this quarantine-first stance (\S\ref{subsec:operational_assessment}).

Continual learning (CL) incrementally adapts a model to a non-stationary data stream and is often termed incremental or lifelong learning~\cite{wang2024comprehensive}. Classical CL assumes labels are plentiful throughout training -- a condition seldom met. Active continual learning addresses this gap by using active learning to query only the most informative points~\cite{vu2023activecl,HCC}. The dominant query rule is uncertainty sampling: select inputs for which the model is most unsure, thereby sharpening the decision boundary. A standard baseline measures this uncertainty with the classifier’s posterior, treating samples with $p \approx 0.5$ (binary case) as lying in the ``uncertainty region'' where confidence is lowest. Existing active-learning studies for malware detection each introduce bespoke uncertainty scores. CADE flags samples whose distance to class centroids (normalized by median absolute deviation) exceeds a threshold~\cite{CADE}; HCC computes a pseudo-loss from contrastive pairings with training neighbors~\cite{HCC}. More recent systems couple acquisition with the learner: DRMD~\cite{mcfadden2026drmd} learns classification and rejection jointly with deep reinforcement learning, and CITADEL~\cite{haque2025citadel} combines semi-supervised learning with active sample selection. 

Confidence-based selection relies on the premise that a model’s confidence scores must correlate with true predictive risk.  In a system with a reliable confidence ranking, samples given high OOD scores or large predictive uncertainty should coincide with misclassifications, signaling that the model recognizes when its data move beyond its training distribution.  \textbf{The reliability of selective classification and active learning therefore depends on confidence estimates that rank errors correctly}. When this ranking degrades -- an outcome especially likely under distribution shift~\cite{ovadia2019can} -- two problems arise. Selective classification may pass high-confidence errors while discarding easy cases, and active learning may waste a limited annotation budget on  uncertain yet uninformative samples, overlooking those that the model truly finds difficult to score (Figure~\ref{fig:example_cade_bad}).

In sum, the field has traversed a chain of corrections. Evaluations based on k-fold cross-validation inflate scores under temporal drift by mixing past and future samples. Temporal evaluation, established by TESSERACT~\cite{pendlebury2019tesseract}, prescribes \emph{when} we measure, and recent dataset-curation guidance~\cite{chow2025breaking} \emph{what} we measure on. Selective classification and active learning then emerged as the operational answer to the drift that temporal evaluation exposes. Yet the confidence machinery on which both rest is rarely verified -- and when it fails silently, analyst hours are wasted on wrongly rejected samples while wrongly accepted errors create a false sense of security. \codename{} closes this link by quantifying how effectively a classifier’s confidence or OOD score orders inputs by their actual error probability and by validating this alignment under simulated temporal drift and resource constraints.

% \subsection{Research Questions}
% \label{subsec:rq}

% We formulate two research questions to assess SOTA Android malware classifiers under realistic deployment conditions:

% \textbf{RQ1}: How \highlightterm{reliablecolor}{reliable} are the OOD scores of SOTA Android malware classifiers under distribution shift?

% \textbf{RQ2}: How \highlightterm{stablecolor}{temporally stable} are SOTA Android malware classifiers with respect to baseline performance and under selective classification conditions? 

\section{\codename{}}
\label{sec:evaluationframework}
As mentioned in the introduction, ranking error, or the misalignment between confidence scores and true correctness, is the primary failure mode of selective classification, which cannot be recovered by post-hoc calibration since it rarely alters the model's underlying score ranking~\cite{rabanser2025does}.
In our framework, we therefore introduce two additional dimensions for evaluation. Reliability, whether the confidence correctly ranks errors, and temporal stability, whether this ranking and the classifier's overall behavior remain consistent as the data distribution evolves over time. We frame two research questions for assessing SOTA Android malware classifiers under realistic deployment conditions:

\vspace{1mm}

\noindent \textbullet\ \textbf{RQ1}: How \highlightterm{reliablecolor}{reliable} are the confidence scores of representative Android malware classifiers under distribution shift?

\noindent \textbullet\ \textbf{RQ2}: How \highlightterm{stablecolor}{temporally stable} are representative Android malware classifiers with respect to detection performance, confidence ranking and referral workload? 

\vspace{1mm}

\noindent We first benchmark \highlightterm{reliablecolor}{reliability}, the degree to which a model's confidence function $\kappa$ orders predictions by their probability of error; we use ``confidence quality'' in this ranking sense, not as expected calibration error (ECE)-style calibration. A reliable $\kappa$ assigns lower confidence to incorrect than to correct predictions, which both sample prioritization and selective classification rest on. We measure this at three depths of the ranking: E-AURC integrates the risk--coverage curve over every coverage (\S\ref{riskcoverage}); Cap$_{\budgetmonthlymath}$ is the share of a month's errors among the $\budgetmonthlymath$ samples of the month's labeling batch (\S\ref{subsub:cap400}); and F1$^{*}$, reported alongside $\Delta$F1$^{*}$, is the F1 score on predictions accepted by a threshold fixed before observing that month's data (\S\ref{subsub:f1star}).

\noindent Secondly, we assess \highlightterm{stablecolor}{temporal stability}, that is the robustness of a model’s predictive performance and rejection behavior over time under distribution shift. Each stability metric is the Theil--Sen slope of a monthly series. Sen[F1] tracks detection performance without abstention (\S\ref{subsub:sen_f1}); Sen[AUROC$_f$] tracks whether the confidence function keeps separating errors from correct predictions, with its first- and last-six-month levels as companion (\S\ref{subsub:sen_auroc}); and Sen[$q^{*}$] tracks the drift of the realized referral rate $q^{*}$ from the quota's share $\rho/n_i$ of monthly arrivals, with its mean $\bar q^{*}$ as companion (\S\ref{subsub:rejection_stability}). Section~\ref{subsec:operational_assessment} details the deployment simulation underlying all experiments.

% First, we benchmark \highlightterm{reliablecolor}{reliability} by measuring how well each confidence function separates correct from incorrect predictions: the Risk--Coverage (RC) curve and its scalar AURC provide this ranking-based view, while the benefit fraction (BF*) reveals how often abstention translates that ranking into tangible F1 gains (\S\ref{riskcoverage}, \S\ref{subsub:benefit_frequency}).

\paragraph{\textbf{Non-Stationary Evaluation Setting}} Our evaluation framework is explicitly designed for non-stationary malware classification, where concept drift continuously shifts the data distribution~\cite{transcend}. Standard machine learning evaluation assumptions, such as i.i.d.\ samples and stable decision boundaries, do not hold in this setting. As established by prior work~\cite{pendlebury2019tesseract}, temporal evaluation with strict train-test temporal ordering is essential for the realistic assessment of malware classifiers. Throughout, we denote by $\datainitmath$ the fully-labeled bootstrap set (first 12 months of data), by $M_i$ the batch of samples arriving in calendar month $i$ of the test stream ($i=1,\dots,N$), and by $\budgetmonthlymath \in \{50,100,200,400\}$ the monthly annotation budget, the number of samples per month selected for labeling via the model’s confidence function $\kappa$. The rejection quota of online selective classification is denoted as $\rho$. These symbols underpin the temporal active learning protocol described in detail in \S\ref{subsec:training-setup}.

\paragraph{\textbf{Threat Model}} We inherit the threat model from selective classification literature, where the primary challenge is natural distribution shift. In Android malware detection, this arises from malware evolution, benign app updates, and fragile feature spaces that violate i.i.d. assumptions. Under stationary distributions, existing methods achieve strong performance; our evaluation examines whether selective methods maintain their core operational properties under natural distribution shifts: reliable confidence ranking, consistent rejection behavior, and temporal stability. We assume realistic deployment constraints, including limited annotation budgets, no ground-truth labels post-deployment and confidence functions as the sole signal for operational decisions.

%\subsection{Operational Assessment via Selective Classification}
\subsection{Online Assessment via a Deployment Simulation}
\label{subsec:operational_assessment}

SOTA approaches escalate the most uncertain apps
(typically the top~$\budgetmonthlymath$ by an uncertainty proxy) to analysts~\cite{CADE,HCC}.
Instead of first issuing a prediction \emph{and then} flagging those
cases for review, we abstain \emph{immediately} on that same fraction in a \emph{deployment simulation} -- adopting the quarantine-first stance of production scanners discussed in \S\ref{sec:background}, where a calibrated rejection quota turns the confidence score into an operational ``risk dial'' in the spirit of Chow's optimal reject rule~\cite{chow1970optimum}.

Let $\rho$ be the desired rejection quota (e.g.\ $\rho\!=\!400$
apps/month) and let $S_{M_i}$ denote the stored uncertainty scores
for the $i$-th calendar month $M_i$ in the temporally ordered
test stream $\mathcal{D}_{\text{test}}=\{M_1,\dots,M_N\}$.
Note that $\rho$ is a per-month \emph{count} of rejected samples, whereas the
coverage $c$ used in risk--coverage analysis is a \emph{fraction} of retained
samples.

For our comparison we re-calibrate the threshold at the start of every month on the unlabeled scores of the six preceding months, starting from January--June 2022 on the Hypercube dataset~\cite{chow2025breaking}, and keep it fixed for that month only. With uncertainty $U$, the entries of $S_{M_i}$, which order samples inversely to $\kappa$, we reject when $U>t_i$. The threshold targets $\rho$ rejections per month on average over its window and is not a hard monthly capacity limit; drift within the month still moves the realized workload. The rejection experiment is post hoc and does not affect the active-learning loop. Appendix~\ref{appendix:threshold_protocol} compares this rolling calibration with a frozen threshold and with a growing pool, Algorithm~\ref{alg:posthoc_rej_routine} with all preceding months, and shows that it keeps the realized referrals closest to the quota.

\begin{protocolbox}[label={prot:posthoc_abstention}]{Rolling-calibration deployment simulation}
\begin{enumerate}[leftmargin=*,label=\bfseries P\arabic*.]
\item At the start of month $M_i$, set $t_i$ so that the six preceding months would have yielded $\rho$ rejections per month on average: the $6\rho$-th largest of their uncertainty scores.
\item Reject the samples of $M_i$ with $U>t_i$ and evaluate predictions on the accepted samples.
\item Record the realized rejection count and accepted-set performance; add the scores of $M_i$ to the window and drop its oldest month.
\end{enumerate}
\end{protocolbox}

Algorithm~\ref{alg:posthoc_rej_routine} (Appendix~\ref{appendix:posthoc_rej_simulation}) states this protocol in pseudocode.

\paragraph{\textbf{Workload-based calibration}}
\label{par:not_transcendent}

Transcendent~\cite{barbero2022transcendent} selects class-specific rejection thresholds using labeled calibration data; its experiments maximize F1 on accepted samples subject to a rejection rate below 15\%. \codename{} instead calibrates with respect to a monthly acquisition budget $\budgetmonthlymath$ in accordance with recent work on active learning~\cite{HCC,mcfadden2026drmd,haque2025citadel,chen2026hydra,han2026alpha}. In our protocol, new labels are available only post hoc for the $\budgetmonthlymath$ samples acquired monthly by uncertainty sampling. These selectively acquired samples do not provide a representative basis for estimating accepted-set F1. Antivirus-derived labels can change over time~\cite{zhu2020labeldynamics,miller2016reviewer}; retrospective labels for unreviewed samples are reserved for evaluation (\S\ref{subsec:datasets}).

We calibrate the threshold from past scores without their labels, following the principle of coverage calibration on unlabeled data~\cite{geifman2019selectivenet}. The realized workload can depart from its target under shift, as Transcendent also reports~\cite{barbero2022transcendent}; we measure this departure in \S\ref{subsec:answer_rq2}.

Setting $\rho=\budgetmonthlymath$ uses the labeling budget as a workload reference. Acquisition determines retraining; rejection is applied post hoc through a threshold re-calibrated each month based on preceding scores and does not affect the active-learning loop. The acquired and rejected sets need not coincide, and referrals may exceed or fall below $\rho$. For fixed data, predictions, and scores, E-AURC does not depend on $\rho$. The Transcendent rows in Table~\ref{tab:main-vtt2-b400} therefore evaluate conformal credibility as an uncertainty-measure but replace its label-dependent threshold search with our workload-based calibration.

\subsection{Assessing \texorpdfstring{\highlightterm{reliablecolor}{Reliability}}{Reliability}}
\label{subsec:reliability}

\subsubsection{\normalfont\textbf{Risk--Coverage Curves and E-AURC}}
\label{riskcoverage}

F1 measures detection performance but does not show whether confidence distinguishes correct from incorrect predictions. We therefore evaluate confidence through risk--coverage (RC) curves, which relate the error rate among accepted predictions to the fraction accepted~\cite{elyaniv2010foundations,geifman2017selective}. For a monthly evaluation set $X$ of $n$ samples, let $e_x=\mathbf{1}\{\hat{y}_x\neq y_x\}$ for each $x\in X$, with higher values of $\kappa$ indicating greater confidence. A threshold $h$ retains $X_h=\{x\in X:\kappa(x)\geq h\}$, giving
\begin{equation*}
  c_h=\frac{|X_h|}{n},\qquad
  R(c_h)=\frac{1}{|X_h|}\sum_{x\in X_h}e_x,
\end{equation*}
for $X_h\neq\varnothing$. Lower risk at a given coverage means fewer errors among the retained predictions (Figure~\ref{fig:eaurc_vtt2}(a)).

We rank predictions separately within each month, without pooling scores across time. For tied scores, we use the group's average error at each position in the area calculation. This is equivalent to averaging AURC over all within-group orderings. Denoting the resulting loss at confidence rank $k$ by $\tilde e_k$, we compute
\begin{equation*}
  \mathrm{AURC}=\frac{1}{n}\sum_{j=1}^{n}\frac{1}{j}\sum_{k=1}^{j}\tilde e_k.
\end{equation*}
Thus, AURC averages selective risk uniformly over the coverages $1/n,\ldots,1$, rather than evaluating a single operating point~\cite{franc2023optimalreject}.

We report the excess area, $\text{E-AURC}=\text{AURC}-\text{AURC}_{\mathrm{oracle}}$, where the oracle ranks all correct predictions before all errors~\cite{aurc}; tables, figures and text give it $\times$100. This measures excess area relative to the best ordering of the same predictions, but does not eliminate dependence on the classifier's error rate~\cite{galil2023can,cattelan2024naurc}. We therefore interpret E-AURC alongside full-coverage F1, rather than as an accuracy-independent measure of ranking quality.

An error near the top of the confidence ranking contributes to retained-set risk at many coverage levels. Specifically, rank $k$ has weight $\frac{1}{n}\sum_{j=k}^{n}\frac{1}{j}$ in AURC, with rank $1$ denoting the most confident prediction~\cite{traub2024overcoming}. High-confidence errors therefore receive greater weight: they remain accepted as lower-confidence predictions are rejected. Such errors can remain \emph{silent to the rejection mechanism}: the prediction is wrong, but its confidence does not trigger referral.

\paragraph{\textbf{Link to Model Calibration}} Confidence calibration asks whether numerical confidence agrees with the frequency of correct predictions. ECE estimates this discrepancy by averaging the absolute confidence--accuracy gap within confidence bins, weighted by bin-size~\cite{naeini2015obtaining}. Good calibration does not, however, guarantee a useful rejection ranking: assigning every prediction the evaluation set's accuracy empirically yields ECE$=0$ but does not distinguish errors~\cite{ding2020revisiting}. Conversely, a strictly increasing transformation of confidence preserves the RC curve while potentially changing calibration. We therefore use RC-based evaluation to assess rejection, rather than treating ECE as a substitute. This evaluation requires an ordered confidence score and not a calibrated probability of correctness, and applies across model types~\cite{geifman2017selective}. 

\paragraph{\textbf{Choice of summary metrics}} AURC has precedent in evaluating selective classification under distribution shift~\cite{liang2024selectiveshift,heng2026knowwhen}. Transcend and Transcendent report accepted-set performance and rejection rates, but not AURC~\cite{transcend,barbero2022transcendent}. Failure AUROC (area under the ROC curve) evaluates uncertainty as a detector of prediction errors, with errors treated as positives~\cite{hendrycks2016baseline,galil2023can}. It summarizes pairwise discrimination without reporting error frequency; equal AUROC values can also conceal different concentrations of high-confidence errors. We use E-AURC as our primary excess-risk summary to emphasize these errors and failure AUROC to track monthly error-discrimination performance (\S\ref{subsec:stability}). 
We also considered AUGRC, which integrates generalized risk $cR(c)$ rather than selective risk $R(c)$~\cite{traub2024overcoming}. Generalized risk measures accepted errors as a fraction of all inputs, whereas selective risk measures them among accepted inputs. AUGRC therefore assigns less weight to low-coverage selective risk than AURC, which is the region \codename{} targets; its ranking agrees closely with E-AURC (Spearman $r_s=0.91$), and we list it among the candidate metrics not adopted as headline metrics in Appendix~\ref{appendix:augrc}.

\subsubsection{\normalfont\textbf{Error Capture at a Fixed Labeling Budget (Cap$_{\budgetmonthlymath}$)}}
\label{subsub:cap400}

E-AURC integrates ranking over every coverage-point, but a drift-adaptive detector makes its labeling decision at a fixed quota: at the end of each month, the arrivals are scored, ranked, and exactly $\budgetmonthlymath$ of them are handed to analysts. Hence, as a second reliability lens, we introduce Cap$_{\budgetmonthlymath}$: the share of each month's misclassifications that fall among its $\budgetmonthlymath$ most uncertain samples. Ties are handled as for E-AURC, and the random reference is $\budgetmonthlymath/n_i$, the batch's share of the $n_i$ arrivals of month $M_i$. Cap$_{\budgetmonthlymath}$ and its counterpart, the share of that batch's samples that are misclassified, mark the recall and precision of failure detection at a fixed labeling budget.
In this line, Kivlichan et al.~\cite{kivlichan2021moderator} introduce a review-efficiency measure for a fixed review-budget, and in the security literature, CADE evaluates its ranking of drifting samples by the precision--recall curve over the number of inspected samples~\cite{CADE}.
The malware literature states analyst capacity as a labeling budget~\cite{miller2016reviewer,pendlebury2019tesseract}. All methods in Table~\ref{tab:main-vtt2-b400} acquire $\budgetmonthlymath=400$ labels per month under the convention of \S\ref{subsec:operational_assessment}. We therefore report Cap$_{\budgetmonthlymath}$ at $\budgetmonthlymath=400$ (Cap$_{400}$), the budget of Table~\ref{tab:main-vtt2-b400}.
Cap$_{400}$ is computed as a post-hoc construction on the recorded scores of each month, which coincides with the batch that the active-learning loop actually labeled, only for rows whose evaluated confidence function is also used for acquisition. For example, HCC's pseudo-loss row (HCC-PL) is acquired by MSP, the ICE rows by CCE credibility, and CADE-MSP by its native OOD score (\S\ref{subsec:confidencefunctions}). It is also not the rejected set of the deployment simulation (\S\ref{subsub:f1star}). Because Cap$_{\budgetmonthlymath}$ fixes the batch size by construction and needs no calibration set, it isolates within-month ranking at the quota from threshold transfer.

\subsubsection{\normalfont\textbf{Accepted-Set F1 (F1$^{*}$, $\Delta$F1$^{*}$)}}
\label{subsub:f1star}
\label{subsub:aurc_f1}
\label{subsub:benefit_frequency}

As a third reliability lens, we evaluate the detector's decision to serve or withhold each prediction. Under Protocol~\ref{prot:posthoc_abstention}, we recalibrate $t_i$ using the preceding six months to target $\rho=\budgetmonthlymath=400$ referrals per month, then hold it fixed throughout the upcoming month. In contrast to Cap$_{\budgetmonthlymath}$, which measures the error recall of the month-end batch, F1$^{*}_i$ measures the F1 score on the accepted samples of month $i$, and F1$^{*}$ is its equal-weight mean over the evaluation months within a seed. Months with fewer than 20 accepted samples are excluded from F1$^{*}$ and $\Delta$F1$^{*}$ (on VTT2, 31 of 150 seed-months for CADE-MSP and 1 of 150 for RAMDA). Besides, we report $\Delta$F1$^{*}=$ F1$^{*}-$F1 as the effect of abstention in percentage points (pp), which is positive when rejection removes more errors than correct predictions.
Our F1$^{*}$ measure is inspired by Transcend and Transcendent~\cite{transcend,barbero2022transcendent}, who evaluate the performance of the accepted set. Similarly, TESSERACT counts the rejected volume as quarantine cost~\cite{pendlebury2019tesseract}, and DRMD reports kept AUT (area under time) with the mean monthly rejection rate~\cite{mcfadden2026drmd}. Condessa et al.\ define non-rejected accuracy as the working-point measure of a classifier with a reject option~\cite{condessa2017rejectionmeasures}; Traub et al.\ likewise reserve fixed working-point measures such as selective risk (for 0/1 loss, one minus non-rejected accuracy) for applied evaluation~\cite{traub2024overcoming}.
Two properties follow from a threshold fixed before the month is seen and distinguish F1$^{*}$ from E-AURC and Cap$_{400}$. First, it is the only reliability metric whose operating point is determined at the beginning of the month, measuring the transfer of the preceding months' score distribution as well as the ranking. Fixing the operating point in retrospect instead, as F1 on the samples outside the month's $\rho$ most uncertain, lowers F1$^{*}$ by 1.8\,pp on average and preserves the ranking (Spearman $r_s=0.97$); the transfer itself is reported in Appendix~\ref{appendix:threshold_protocol}. The realized referral rate $q^{*}$ measures how much the transfer deviated from the quota's share $\rho/n_i$ of monthly arrivals (refer to \S\ref{subsub:rejection_stability}). Second, F1$^{*}$ can fall below F1 when rejection removes correct predictions. Because F1 ignores a removed true negative, rewards a removed error, and penalizes a removed true positive, $\Delta$F1$^{*}$ is governed by which predicted class the confidence function $\kappa$ marks as uncertain: a score that flags borderline detections trades detections for errors, whereas a score that flags only predicted-benign samples cannot lose a detection. Cap$_{\budgetmonthlymath}$ counts errors of both classes alike and does not see this cost, so the two metrics rank methods differently (Spearman $|r_s|<0.4$ across Table~\ref{tab:main-vtt2-b400}), which is why we report both. A negative $\Delta$F1$^{*}$ therefore indicates a confidence function that holds back detections faster than it removes errors.

\subsection{Assessing \texorpdfstring{\highlightterm{stablecolor}{Stability}}{Stability}}
\label{subsec:stability}

\subsubsection{\normalfont\textbf{Performance Trend (Sen[F1])}}
\label{subsub:sen_f1}
\label{subsub:sigma_f1}
\label{subsub:mann_kendall}

As a central measure for stability, we ask whether detection performance decays temporally and at what rate. Time decay is a central question of temporal evaluation in malware detection: TESSERACT's area under time~\cite{pendlebury2019tesseract} and its extension to Windows PE and PDF malware~\cite{kan2025tesseractextended} average monthly performance, APIGraph reports model aging~\cite{apigraph}, DroidEvolver shows a decline per year~\cite{xu2019droidevolver}, and Chow et al.\ trace the steepest monthly drops back to their causes~\cite{chow2023forensics}. We report the rate of decline as a metric: Sen[F1] is the Theil--Sen slope of the monthly F1 series, the median of the slopes over all pairs of months~\cite{sen1968slope}, in percentage points per month. A value of $-1$ means the method loses one point of F1 per month over the window on average. The estimator is the magnitude companion of the Mann--Kendall statistic $\tau$~\cite{mann1945nonparametric}: $\tau$ reports only whether changes are predominantly negative, whereas the slope reports their magnitude. Both statistics rank the methods almost identically (Spearman $r_s=0.93$; Appendix~\ref{appendix:retired_metrics}). 
Since the Theil--Sen slope is computed as the median of pairwise slopes, it is robust to outliers (e.g., single ``bad months'') and does not make any assumptions about the distribution.

\subsubsection{\normalfont\textbf{Ranking Stability (Sen[AUROC$_f$])}}
\label{subsub:sen_auroc}

\begin{comment}
The reliability metrics describe the confidence ranking; the second stability question is whether that ranking itself holds up as the stream drifts. Ovadia et al.\ show on benchmark shifts that uncertainty quality degrades with shift intensity even where accuracy is still acceptable~\cite{ovadia2019can}, and RQ2 asks whether a confidence ranking remains consistent as the distribution evolves; no drift study in malware detection has reported the trend of a ranking metric over deployment time. We track the failure AUROC~\cite{hendrycks2016baseline}, the probability that a misclassified sample receives higher uncertainty than a correctly classified one within the month, because it is the pure ranking axis (\S\ref{riskcoverage}): it does not depend on the error rate, so its trend separates a decaying ranking from a decaying classifier, which Sen[F1] already records. Sen[AUROC$_f$] is the Theil--Sen slope of the monthly failure AUROC within a seed, in units of $10^{-3}$ per month, with the same robustness and reading as Sen[F1]; the companion column reports the mean failure AUROC over the first and the last six evaluation months, so that the slope can be read against the level it starts from and ends at, with 0.5 as the random reference. A negative slope means that the confidence function is losing its ability to place errors above correct predictions as the data shifts, an erosion that neither F1 nor a within-month reliability metric alone would show, and that bears directly on every mechanism that spends a review budget by rank.
\end{comment}

As a second stability question, we ask how the confidence ranking of the confidence functions evolves with drift. Ovadia et al.\ show that uncertainty quality degrades with drift intensity even where accuracy is still acceptable~\cite{ovadia2019can}. In RQ2 we deliberately ask whether a confidence ranking remains consistent as the distribution evolves. To this end we track the failure AUROC~\cite{hendrycks2016baseline}, the probability that a misclassified sample receives higher uncertainty than a correctly classified one within the month. As a pure ranking measure (\S\ref{riskcoverage}), it does not depend on the error rate and its trend separates a decaying ranking from a decaying classifier (as per our Sen[F1] metric). Sen[AUROC$_f$] is the Theil--Sen slope of the monthly failure AUROC, in units of $10^{-3}$ per month. Additionally, we report the average failure AUROC over the first and last six evaluation months to contrast early from later periods in the dataset, where a value of 0.5 denotes random chance.
A negative slope indicates that the confidence function is losing its ability to rank errors above correct predictions over time. This erosion would neither be reported by baseline F1 nor any within-month reliability metric alone.

% ============================================================================
% ============================================================================
\subsubsection{\normalfont\textbf{Workload Tracking (Sen[$q^{*}$], $\bar q^{*}$)}}
\label{subsub:rejection_stability}

Thirdly, we ask whether the deployed rejection-threshold adheres to its calibrated target $\rho$. Transcendent observes that calibrated rejection rates are surpassed under drift~\cite{barbero2022transcendent}, TESSERACT counts the rejected volume as a quarantine cost~\cite{pendlebury2019tesseract}, and DRMD reports the mean monthly rejection rate alongside kept performance~\cite{mcfadden2026drmd}. Similarly, adaptive procedures for conformal prediction~\cite{gibbs2021adaptive,angelopoulos2023pid} control for the realized mis-coverage of a threshold set on past data via trend-statistics and deviation from the target-rate.

We report workload as a referral rate because monthly arrivals vary from 5.6k to 1.7k over the test period. Let $q_i^{*}$ denote the share of month $i$ referred under threshold $t_i$ (Protocol~\ref{prot:posthoc_abstention}). The fixed count quota $\rho$ corresponds to a target rate of $\rho/n_i$, where $n_i$ is that month's arrival count. As arrivals decline, this target rate rises for all confidence functions. We therefore estimate trends in the excess $q_i^{*}-\rho/n_i$, rather than in the raw referral rate.

We report Sen[$q^{*}$], the Theil--Sen slope of this excess, in percentage points per month. Positive and negative slopes indicate upward and downward trends relative to the quota, respectively. A zero slope indicates no estimated linear trend in the discrepancy. To assess average adherence, we also report the mean realized referral rate, $\bar q^{*}$, alongside the quota's average share of arrivals at 11.7\%. Count-based measures of referral bias and dispersion, which also depend on arrival volume, are reported in Appendix~\ref{appendix:retired_metrics}.

\subsection{Interpreting the Metrics}
\label{subsec:no_single_metric}

Consider a classifier deployed to triage incoming Android applications, referring its most uncertain predictions for analyst review. \codename{} replays this deployment retrospectively (Protocol~\ref{prot:posthoc_abstention}): the model is updated only with the $\budgetmonthlymath$ labels acquired each month, while the labels of all other samples are reserved for evaluation (\S\ref{subsec:operational_assessment}). Each metric exposes a failure of this deployment that the others can miss. \textbf{E-AURC} exposes a confidence function that fails to rank errors ahead of correct predictions across coverage levels. \textbf{Cap$_{\budgetmonthlymath}$} exposes a ranking that fails at the labeling budget, where few of a month's errors fall among its $\budgetmonthlymath$ most uncertain samples; this set coincides with the acquired batch when the evaluated score also drives acquisition. \textbf{F1$^{*}$} and \textbf{$\Delta$F1$^{*}$} expose abstention at the quota-calibrated threshold that lowers the F1 of accepted predictions. \textbf{Sen[F1]} exposes a model that is strong on average but decays steadily under drift, and \textbf{Sen[AUROC$_f$]} a confidence function that gradually loses its ability to separate errors from correct predictions. These metrics require labels and are assessed retrospectively. Only \textbf{Sen[$q^{*}$]} and $\boldsymbol{\bar q^{*}}$ are label-free: they expose a deployed threshold that departs from its quota, which the operator experiences directly as unpredictable analyst workload. A method that performs well on one criterion may fail on another (\S\ref{sec:results}). The choice therefore depends on which failure a deployment can least afford: with a fixed analyst capacity, an operator should first check $\bar q^{*}$ and Sen[$q^{*}$]; if accepted predictions drive decisions, F1$^{*}$ and $\Delta$F1$^{*}$; and if retraining is infrequent, Sen[F1] and Sen[AUROC$_f$].

% ============================================================================

\section{Experimental Setup}
\label{sec:experimentalsetup}
\subsection{Expanded VTT2 Evaluation}
\label{subsec:datasets}
Our main comparison uses Hypercube VTT2~\cite{chow2025breaking,hypercubeRepo} (the extended\_2024 release): 224,965 Android applications spanning 2021--2024, with 67,953 fully labeled applications from 2021 for initial training and 157,012 applications in the January 2022--December 2024 deployment stream. VTT2 (VirusTotal threshold 2) labels malware at two or more VirusTotal detections and excludes the ambiguous one-detection group. We reserve January--June 2022 for rejection-threshold calibration and report July 2022--December 2024 (30 months). The first comparison fixes $\budgetmonthlymath=400$ and uses five seeds.  

\subsection{Reference Frameworks and Monthly Updates}
\label{subsec:referenceframeworks}
\label{subsec:training-setup}
We evaluate Drebin~\cite{drebin}, DeepDrebin~\cite{DeepDrebin}, HCC~\cite{HCC}, CADE~\cite{CADE}, Random Forest~\cite{breiman2001random}, SecSVM~\cite{demontis2019secureml}, DRMD~\cite{mcfadden2026drmd}, RAMDA~\cite{li2021ramda}, TIF~\cite{zheng2026tif}, and MalCertain~\cite{li2024malcertain}, using the existing VTT2 benchmark implementations and their common 10,000-feature Drebin representation. 
Each model predicts the current month before acquiring $\budgetmonthlymath=400$ labels via uncertainty-sampling for subsequent updates. Drebin, SecSVM and Random Forest refit on the cumulative labeled pool. DeepDrebin, HCC, RAMDA and TIF update using at most the most recent 5,000 labeled observations, with method-specific optimizers; DRMD keeps its weights and retrains on the full cumulative labeled pool (Appendix~\ref{appendix:hp_vtt2}). DeepDrebin uses 10 SGD epochs, and HCC uses 40 Adam epochs per update. MalCertain cold-refits its components on the cumulative pool with an internal temporal 80/20 training/correction split. Model configurations are fixed to the released benchmark settings; the selected hyperparameters on VTT2 data are listed in Appendix~\ref{appendix:hp_vtt2}.

The HCC rows use the benchmark adaptation with common uncertainty acquisition, distinct from the original family-supervised, pseudo-loss-acquisition protocol. Their paired MSP and pseudo-loss (PL) evaluations, HCC-MSP and HCC-PL, share monthly checkpoints, predictions and MSP acquisition histories. HCC-PL uses only the pre-month labeled reference pool, with 1,023 neighbors, margin 10 and cross-entropy weight 100 in deterministic evaluation mode. CADE~\cite{CADE} uses pinned HCC reference components, cold-fitting its contrastive autoencoder and binary MLP on the cumulative labeled pool. Its two rows, CADE-OOD and CADE-MSP, share predictions and acquisition by MAD-normalized centroid-distance OOD scores.

We additionally evaluate Transcendent~\cite{barbero2022transcendent} with Drebin and DeepDrebin as base classifiers. Five persistent, stratified folds separate fitting from conformal calibration data: the inductive evaluator (ICE) uses the first fold, the cross-conformal evaluator (CCE) all five, with majority-vote predictions and median class p-values. Both share acquisition by CCE credibility. Drebin refits on cumulative proper-training data; DeepDrebin warm-updates on at most the latest 5,000 examples. Thresholds follow \codename{}'s label-free quota calibration (\S\ref{subsec:operational_assessment}), not Transcendent's labeled threshold search. Acquisition thus differs from the baselines.

\subsection{Confidence Functions and Metric Replay}
\label{subsec:confidencefunctions}
Baseline neural rows use maximum softmax probability (MSP); when logits are available, uncertainty is computed as $2/(1+\exp(|z_1-z_0|))$ to avoid loss of ordering from rounded probabilities. Drebin and SecSVM rows use the negative absolute decision margin, so larger values denote greater uncertainty. Random Forest uses its class probability. RAMDA uses its classifier's confidence, not reconstruction error. MalCertain uses base-model MSP with its corrected predictions; its mirrored output score is not interpreted as a calibrated corrected probability. Transcendent rows use one minus conformal credibility.

All metrics are recomputed from the recorded predictions and uncertainty scores after active learning. Every quantity in Table~\ref{tab:main-vtt2-b400} is computed per month within a seed, summarized over the 30 evaluation months within the seed, and then averaged over the five seeds. Our quantities F1$^{*}$, $\Delta$F1$^{*}$ and $q^{*}$ use the threshold of Protocol~\ref{prot:posthoc_abstention}: the monthly loop starts in January 2022, and the first six months are scored but serve only to calibrate the first threshold. From July 2022 on, the threshold $t_i$ of each month is the $6\rho$-th largest uncertainty score of the six preceding months, so that those months would have yielded $\rho=400$ referrals per month on average, and the window advances by one month after each evaluation. We selected a rolling window for threshold calibration because it produces the lowest error with respect to the rejection quota $\rho$ among the protocols compared in Appendix~\ref{appendix:threshold_protocol}.

\subsection{Reproducibility and Compute}
\label{subsec:computational}
Every result in this paper can be recomputed from the recorded predictions and confidence scores, without retraining any model. Before a metric is computed, the replay confirms that all methods and seeds were evaluated on the same monthly cohorts, with identical sample indices and labels. For the baselines, it checks all 45 source files against their manifest hashes. For Transcendent, it checks 360 monthly files (two base classifiers, five seeds and 36 months), each holding both the ICE and the CCE outputs. Each of the 180 monthly HCC-PL score files is paired with a verified HCC checkpoint and the reference pool available before that month, and the 180 monthly CADE files hold both OOD and MSP scores, with complete trajectory manifests and checks on the order of acquisition. For each table, we release the input hashes, metric source, calibration/test split and per-seed rejection summaries.

\section{Results}
\label{sec:results}
% Generated by generate_satml_vtt2_frozen_table.py; do not edit numbers by hand.
% Layout, colors, headers and number formatting reuse the original Aurora table.
\providecolor{tabbaseband}{RGB}{240,240,240}
\providecolor{tabreliband}{RGB}{230,245,220}
\providecolor{tabstabband}{RGB}{235,240,250}
\providecolor{worstcolor}{RGB}{176,32,32}
\providecolor{subheadcolor}{RGB}{246,246,246}
\providecolor{basetext}{RGB}{90,90,90}
\providecolor{relitext}{RGB}{46,125,50}
\providecolor{stabtext}{RGB}{40,53,147}
\providecommand{\codename}{\textsc{Aurora}}
% equal widths of the two sub-columns under a grouped metric (widest unit label or value)
\ifdefined\frzwtmp\else\newlength{\frzwtmp}\fi
\ifdefined\frzwA\else\newlength{\frzwA}\fi\setlength{\frzwA}{0pt}
\settowidth{\frzwtmp}{\footnotesize 31.1}\ifdim\frzwtmp>\frzwA\setlength{\frzwA}{\frzwtmp}\fi
\settowidth{\frzwtmp}{\footnotesize {\color{basetext}$-$7.4}}\ifdim\frzwtmp>\frzwA\setlength{\frzwA}{\frzwtmp}\fi
\settowidth{\frzwtmp}{\footnotesize 42.5}\ifdim\frzwtmp>\frzwA\setlength{\frzwA}{\frzwtmp}\fi
\settowidth{\frzwtmp}{\footnotesize {\color{basetext}$+$4.2}}\ifdim\frzwtmp>\frzwA\setlength{\frzwA}{\frzwtmp}\fi
\settowidth{\frzwtmp}{\footnotesize \textcolor{worstcolor}{\itshape 2.7}}\ifdim\frzwtmp>\frzwA\setlength{\frzwA}{\frzwtmp}\fi
\settowidth{\frzwtmp}{\footnotesize {\color{basetext}$-$35.5}}\ifdim\frzwtmp>\frzwA\setlength{\frzwA}{\frzwtmp}\fi
\settowidth{\frzwtmp}{\footnotesize 43.2}\ifdim\frzwtmp>\frzwA\setlength{\frzwA}{\frzwtmp}\fi
\settowidth{\frzwtmp}{\footnotesize {\color{basetext}$+$4.1}}\ifdim\frzwtmp>\frzwA\setlength{\frzwA}{\frzwtmp}\fi
\settowidth{\frzwtmp}{\footnotesize 25.8}\ifdim\frzwtmp>\frzwA\setlength{\frzwA}{\frzwtmp}\fi
\settowidth{\frzwtmp}{\footnotesize {\color{basetext}$-$13.3}}\ifdim\frzwtmp>\frzwA\setlength{\frzwA}{\frzwtmp}\fi
\settowidth{\frzwtmp}{\footnotesize 23.1}\ifdim\frzwtmp>\frzwA\setlength{\frzwA}{\frzwtmp}\fi
\settowidth{\frzwtmp}{\footnotesize {\color{basetext}$-$7.3}}\ifdim\frzwtmp>\frzwA\setlength{\frzwA}{\frzwtmp}\fi
\settowidth{\frzwtmp}{\footnotesize 24.7}\ifdim\frzwtmp>\frzwA\setlength{\frzwA}{\frzwtmp}\fi
\settowidth{\frzwtmp}{\footnotesize {\color{basetext}$-$13.0}}\ifdim\frzwtmp>\frzwA\setlength{\frzwA}{\frzwtmp}\fi
\settowidth{\frzwtmp}{\footnotesize 22.5}\ifdim\frzwtmp>\frzwA\setlength{\frzwA}{\frzwtmp}\fi
\settowidth{\frzwtmp}{\footnotesize {\color{basetext}$-$4.3}}\ifdim\frzwtmp>\frzwA\setlength{\frzwA}{\frzwtmp}\fi
\settowidth{\frzwtmp}{\footnotesize 43.6}\ifdim\frzwtmp>\frzwA\setlength{\frzwA}{\frzwtmp}\fi
\settowidth{\frzwtmp}{\footnotesize {\color{basetext}$+$0.1}}\ifdim\frzwtmp>\frzwA\setlength{\frzwA}{\frzwtmp}\fi
\settowidth{\frzwtmp}{\footnotesize 34.2}\ifdim\frzwtmp>\frzwA\setlength{\frzwA}{\frzwtmp}\fi
\settowidth{\frzwtmp}{\footnotesize 20.2}\ifdim\frzwtmp>\frzwA\setlength{\frzwA}{\frzwtmp}\fi
\settowidth{\frzwtmp}{\footnotesize {\color{basetext}$-$18.1}}\ifdim\frzwtmp>\frzwA\setlength{\frzwA}{\frzwtmp}\fi
\settowidth{\frzwtmp}{\footnotesize {\color{basetext}$-$5.0}}\ifdim\frzwtmp>\frzwA\setlength{\frzwA}{\frzwtmp}\fi
\settowidth{\frzwtmp}{\footnotesize 37.5}\ifdim\frzwtmp>\frzwA\setlength{\frzwA}{\frzwtmp}\fi
\settowidth{\frzwtmp}{\footnotesize {\color{basetext}$+$6.0}}\ifdim\frzwtmp>\frzwA\setlength{\frzwA}{\frzwtmp}\fi
\settowidth{\frzwtmp}{\footnotesize 38.2}\ifdim\frzwtmp>\frzwA\setlength{\frzwA}{\frzwtmp}\fi
\settowidth{\frzwtmp}{\footnotesize {\color{basetext}$+$6.1}}\ifdim\frzwtmp>\frzwA\setlength{\frzwA}{\frzwtmp}\fi
\settowidth{\frzwtmp}{\footnotesize 44.2}\ifdim\frzwtmp>\frzwA\setlength{\frzwA}{\frzwtmp}\fi
\settowidth{\frzwtmp}{\footnotesize {\color{basetext}$+$6.4}}\ifdim\frzwtmp>\frzwA\setlength{\frzwA}{\frzwtmp}\fi
\settowidth{\frzwtmp}{\footnotesize \textbf{45.4}}\ifdim\frzwtmp>\frzwA\setlength{\frzwA}{\frzwtmp}\fi
\settowidth{\frzwtmp}{\footnotesize {\color{basetext}$+$7.0}}\ifdim\frzwtmp>\frzwA\setlength{\frzwA}{\frzwtmp}\fi
\settowidth{\frzwtmp}{\footnotesize {\scriptsize \%}}\ifdim\frzwtmp>\frzwA\setlength{\frzwA}{\frzwtmp}\fi
\settowidth{\frzwtmp}{\footnotesize {\scriptsize\color{basetext} $\Delta$F1$^{*}$ (pp)}}\ifdim\frzwtmp>\frzwA\setlength{\frzwA}{\frzwtmp}\fi
\ifdefined\frzwB\else\newlength{\frzwB}\fi\setlength{\frzwB}{0pt}
\settowidth{\frzwtmp}{\footnotesize $-$3.6}\ifdim\frzwtmp>\frzwB\setlength{\frzwB}{\frzwtmp}\fi
\settowidth{\frzwtmp}{\footnotesize {\color{basetext}0.72$\to$0.65}}\ifdim\frzwtmp>\frzwB\setlength{\frzwB}{\frzwtmp}\fi
\settowidth{\frzwtmp}{\footnotesize $-$1.4}\ifdim\frzwtmp>\frzwB\setlength{\frzwB}{\frzwtmp}\fi
\settowidth{\frzwtmp}{\footnotesize {\color{basetext}0.56$\to$0.51}}\ifdim\frzwtmp>\frzwB\setlength{\frzwB}{\frzwtmp}\fi
\settowidth{\frzwtmp}{\footnotesize \textbf{$+$1.8}}\ifdim\frzwtmp>\frzwB\setlength{\frzwB}{\frzwtmp}\fi
\settowidth{\frzwtmp}{\footnotesize {\color{basetext}0.54$\to$0.59}}\ifdim\frzwtmp>\frzwB\setlength{\frzwB}{\frzwtmp}\fi
\settowidth{\frzwtmp}{\footnotesize \textcolor{worstcolor}{\itshape $-$5.8}}\ifdim\frzwtmp>\frzwB\setlength{\frzwB}{\frzwtmp}\fi
\settowidth{\frzwtmp}{\footnotesize {\color{basetext}0.62$\to$0.48}}\ifdim\frzwtmp>\frzwB\setlength{\frzwB}{\frzwtmp}\fi
\settowidth{\frzwtmp}{\footnotesize $+$0.4}\ifdim\frzwtmp>\frzwB\setlength{\frzwB}{\frzwtmp}\fi
\settowidth{\frzwtmp}{\footnotesize {\color{basetext}0.64$\to$0.65}}\ifdim\frzwtmp>\frzwB\setlength{\frzwB}{\frzwtmp}\fi
\settowidth{\frzwtmp}{\footnotesize $-$0.9}\ifdim\frzwtmp>\frzwB\setlength{\frzwB}{\frzwtmp}\fi
\settowidth{\frzwtmp}{\footnotesize {\color{basetext}0.67$\to$0.65}}\ifdim\frzwtmp>\frzwB\setlength{\frzwB}{\frzwtmp}\fi
\settowidth{\frzwtmp}{\footnotesize $-$4.2}\ifdim\frzwtmp>\frzwB\setlength{\frzwB}{\frzwtmp}\fi
\settowidth{\frzwtmp}{\footnotesize {\color{basetext}0.72$\to$0.64}}\ifdim\frzwtmp>\frzwB\setlength{\frzwB}{\frzwtmp}\fi
\settowidth{\frzwtmp}{\footnotesize $-$1.3}\ifdim\frzwtmp>\frzwB\setlength{\frzwB}{\frzwtmp}\fi
\settowidth{\frzwtmp}{\footnotesize {\color{basetext}0.64$\to$0.62}}\ifdim\frzwtmp>\frzwB\setlength{\frzwB}{\frzwtmp}\fi
\settowidth{\frzwtmp}{\footnotesize $-$5.1}\ifdim\frzwtmp>\frzwB\setlength{\frzwB}{\frzwtmp}\fi
\settowidth{\frzwtmp}{\footnotesize {\color{basetext}0.74$\to$0.62}}\ifdim\frzwtmp>\frzwB\setlength{\frzwB}{\frzwtmp}\fi
\settowidth{\frzwtmp}{\footnotesize $-$0.4}\ifdim\frzwtmp>\frzwB\setlength{\frzwB}{\frzwtmp}\fi
\settowidth{\frzwtmp}{\footnotesize {\color{basetext}0.62$\to$0.62}}\ifdim\frzwtmp>\frzwB\setlength{\frzwB}{\frzwtmp}\fi
\settowidth{\frzwtmp}{\footnotesize $+$0.5}\ifdim\frzwtmp>\frzwB\setlength{\frzwB}{\frzwtmp}\fi
\settowidth{\frzwtmp}{\footnotesize {\color{basetext}0.62$\to$0.63}}\ifdim\frzwtmp>\frzwB\setlength{\frzwB}{\frzwtmp}\fi
\settowidth{\frzwtmp}{\footnotesize $-$1.7}\ifdim\frzwtmp>\frzwB\setlength{\frzwB}{\frzwtmp}\fi
\settowidth{\frzwtmp}{\footnotesize {\color{basetext}0.67$\to$0.64}}\ifdim\frzwtmp>\frzwB\setlength{\frzwB}{\frzwtmp}\fi
\settowidth{\frzwtmp}{\footnotesize {\color{basetext}0.66$\to$0.63}}\ifdim\frzwtmp>\frzwB\setlength{\frzwB}{\frzwtmp}\fi
\settowidth{\frzwtmp}{\footnotesize $-$1.8}\ifdim\frzwtmp>\frzwB\setlength{\frzwB}{\frzwtmp}\fi
\settowidth{\frzwtmp}{\footnotesize $-$5.6}\ifdim\frzwtmp>\frzwB\setlength{\frzwB}{\frzwtmp}\fi
\settowidth{\frzwtmp}{\footnotesize {\color{basetext}0.71$\to$0.59}}\ifdim\frzwtmp>\frzwB\setlength{\frzwB}{\frzwtmp}\fi
\settowidth{\frzwtmp}{\footnotesize {\color{basetext}0.71$\to$0.61}}\ifdim\frzwtmp>\frzwB\setlength{\frzwB}{\frzwtmp}\fi
\settowidth{\frzwtmp}{\footnotesize {\scriptsize $10^{-3}$/month}}\ifdim\frzwtmp>\frzwB\setlength{\frzwB}{\frzwtmp}\fi
\settowidth{\frzwtmp}{\footnotesize {\scriptsize\color{basetext} first$\to$last}}\ifdim\frzwtmp>\frzwB\setlength{\frzwB}{\frzwtmp}\fi
\ifdefined\frzwC\else\newlength{\frzwC}\fi\setlength{\frzwC}{0pt}
\settowidth{\frzwtmp}{\footnotesize $+$0.01}\ifdim\frzwtmp>\frzwC\setlength{\frzwC}{\frzwtmp}\fi
\settowidth{\frzwtmp}{\footnotesize {\color{basetext}9.9}}\ifdim\frzwtmp>\frzwC\setlength{\frzwC}{\frzwtmp}\fi
\settowidth{\frzwtmp}{\footnotesize $-$0.24}\ifdim\frzwtmp>\frzwC\setlength{\frzwC}{\frzwtmp}\fi
\settowidth{\frzwtmp}{\footnotesize {\color{basetext}12.7}}\ifdim\frzwtmp>\frzwC\setlength{\frzwC}{\frzwtmp}\fi
\settowidth{\frzwtmp}{\footnotesize $-$0.48}\ifdim\frzwtmp>\frzwC\setlength{\frzwC}{\frzwtmp}\fi
\settowidth{\frzwtmp}{\footnotesize {\color{basetext}23.6}}\ifdim\frzwtmp>\frzwC\setlength{\frzwC}{\frzwtmp}\fi
\settowidth{\frzwtmp}{\footnotesize $+$0.23}\ifdim\frzwtmp>\frzwC\setlength{\frzwC}{\frzwtmp}\fi
\settowidth{\frzwtmp}{\footnotesize {\color{basetext}16.2}}\ifdim\frzwtmp>\frzwC\setlength{\frzwC}{\frzwtmp}\fi
\settowidth{\frzwtmp}{\footnotesize $-$0.04}\ifdim\frzwtmp>\frzwC\setlength{\frzwC}{\frzwtmp}\fi
\settowidth{\frzwtmp}{\footnotesize {\color{basetext}10.2}}\ifdim\frzwtmp>\frzwC\setlength{\frzwC}{\frzwtmp}\fi
\settowidth{\frzwtmp}{\footnotesize $-$0.05}\ifdim\frzwtmp>\frzwC\setlength{\frzwC}{\frzwtmp}\fi
\settowidth{\frzwtmp}{\footnotesize {\color{basetext}10.1}}\ifdim\frzwtmp>\frzwC\setlength{\frzwC}{\frzwtmp}\fi
\settowidth{\frzwtmp}{\footnotesize $-$0.11}\ifdim\frzwtmp>\frzwC\setlength{\frzwC}{\frzwtmp}\fi
\settowidth{\frzwtmp}{\footnotesize {\color{basetext}10.9}}\ifdim\frzwtmp>\frzwC\setlength{\frzwC}{\frzwtmp}\fi
\settowidth{\frzwtmp}{\footnotesize {\color{basetext}9.6}}\ifdim\frzwtmp>\frzwC\setlength{\frzwC}{\frzwtmp}\fi
\settowidth{\frzwtmp}{\footnotesize {\color{basetext}10.3}}\ifdim\frzwtmp>\frzwC\setlength{\frzwC}{\frzwtmp}\fi
\settowidth{\frzwtmp}{\footnotesize $-$0.21}\ifdim\frzwtmp>\frzwC\setlength{\frzwC}{\frzwtmp}\fi
\settowidth{\frzwtmp}{\footnotesize {\color{basetext}11.2}}\ifdim\frzwtmp>\frzwC\setlength{\frzwC}{\frzwtmp}\fi
\settowidth{\frzwtmp}{\footnotesize $-$0.12}\ifdim\frzwtmp>\frzwC\setlength{\frzwC}{\frzwtmp}\fi
\settowidth{\frzwtmp}{\footnotesize {\color{basetext}8.9}}\ifdim\frzwtmp>\frzwC\setlength{\frzwC}{\frzwtmp}\fi
\settowidth{\frzwtmp}{\footnotesize $-$0.18}\ifdim\frzwtmp>\frzwC\setlength{\frzwC}{\frzwtmp}\fi
\settowidth{\frzwtmp}{\footnotesize {\color{basetext}9.1}}\ifdim\frzwtmp>\frzwC\setlength{\frzwC}{\frzwtmp}\fi
\settowidth{\frzwtmp}{\footnotesize $-$0.15}\ifdim\frzwtmp>\frzwC\setlength{\frzwC}{\frzwtmp}\fi
\settowidth{\frzwtmp}{\footnotesize $-$0.16}\ifdim\frzwtmp>\frzwC\setlength{\frzwC}{\frzwtmp}\fi
\settowidth{\frzwtmp}{\footnotesize $-$0.07}\ifdim\frzwtmp>\frzwC\setlength{\frzwC}{\frzwtmp}\fi
\settowidth{\frzwtmp}{\footnotesize {\color{basetext}11.4}}\ifdim\frzwtmp>\frzwC\setlength{\frzwC}{\frzwtmp}\fi
\settowidth{\frzwtmp}{\footnotesize {\scriptsize pp/month}}\ifdim\frzwtmp>\frzwC\setlength{\frzwC}{\frzwtmp}\fi
\settowidth{\frzwtmp}{\footnotesize {\scriptsize\color{basetext} $\bar{q}^{*}$ (\%)}}\ifdim\frzwtmp>\frzwC\setlength{\frzwC}{\frzwtmp}\fi
\begin{table*}[!t]
  \centering
  \caption{\codename{} evaluation on \textbf{Hypercube VTT2}~\cite{chow2025breaking} at $B_M=400$. {\color{basetext}Gray} sub-columns are companion values; \textbf{bold}=best, {\color{worstcolor}\itshape dark-red italics}=worst per ranked column.}
  \label{tab:main-vtt2-b400}
  \footnotesize
  \setlength{\tabcolsep}{4.2pt}
  \renewcommand{\arraystretch}{0.97}
  \begin{tabular}{@{}cl@{\hspace{7pt}}c@{\hspace{7pt}}ccw{c}{\frzwA}w{c}{\frzwA}@{\hspace{7pt}}cw{c}{\frzwB}w{c}{\frzwB}w{c}{\frzwC}w{c}{\frzwC}}
    \toprule
     &  & \multicolumn{1}{>{\columncolor{tabbaseband}[\tabcolsep][\tabcolsep]}c}{Baseline} & \multicolumn{4}{>{\columncolor{tabreliband}[\tabcolsep][\tabcolsep]}c}{Reliability} & \multicolumn{5}{>{\columncolor{tabstabband}[\tabcolsep][\tabcolsep]}c}{Stability} \\
    $B_M$ & Method & \cellcolor{tabbaseband}F1 $\uparrow$ & \cellcolor{tabreliband}E-AURC $\downarrow$ & \cellcolor{tabreliband}Cap$_{400}$ $\uparrow$ & \multicolumn{2}{>{\columncolor{tabreliband}[\tabcolsep][\tabcolsep]}c}{\makebox[0pt][c]{F1$^{*}$ $\uparrow$}\makebox[0pt][c]{\smash{\raisebox{-2.6pt}{\rule{\dimexpr 2\frzwA+4\tabcolsep-5pt\relax}{0.4pt}}}}} & \cellcolor{tabstabband}Sen[F1] $\uparrow$ & \multicolumn{2}{>{\columncolor{tabstabband}[\tabcolsep][\tabcolsep]}c}{\makebox[0pt][c]{Sen[AUROC$_f$] $\uparrow$}\makebox[0pt][c]{\smash{\raisebox{-2.6pt}{\rule{\dimexpr 2\frzwB+4\tabcolsep-5pt\relax}{0.4pt}}}}} & \multicolumn{2}{>{\columncolor{tabstabband}[\tabcolsep][\tabcolsep]}c}{\makebox[0pt][c]{Sen[$q^{*}$]}\makebox[0pt][c]{\smash{\raisebox{-2.6pt}{\rule{\dimexpr 2\frzwC+4\tabcolsep-5pt\relax}{0.4pt}}}}} \\
     &  & \cellcolor{tabbaseband}{\scriptsize \%} & \cellcolor{tabreliband}{\scriptsize $\times$100} & \cellcolor{tabreliband}{\scriptsize \% of errors} & \cellcolor{tabreliband}{\scriptsize \%} & \cellcolor{tabreliband}{\scriptsize\color{basetext} $\Delta$F1$^{*}$ (pp)} & \cellcolor{tabstabband}{\scriptsize pp/month} & \cellcolor{tabstabband}{\scriptsize $10^{-3}$/month} & \cellcolor{tabstabband}{\scriptsize\color{basetext} first$\to$last} & \cellcolor{tabstabband}{\scriptsize pp/month} & \cellcolor{tabstabband}{\scriptsize\color{basetext} $\bar{q}^{*}$ (\%)} \\
    \midrule
    \multirow{17}{*}{400} & DeepDrebin & 38.5 & \textbf{4.53} & 30.3 & 31.1 & {\color{basetext}$-$7.4} & $-$1.23 & $-$3.6 & {\color{basetext}0.72$\to$0.65} & $+$0.01 & {\color{basetext}9.9} \\
     & CADE-OOD & 38.3 & \textcolor{worstcolor}{\itshape 8.20} & 18.0 & 42.5 & {\color{basetext}$+$4.2} & \textbf{$-$0.81} & $-$1.4 & {\color{basetext}0.56$\to$0.51} & $-$0.24 & {\color{basetext}12.7} \\
     & CADE-MSP & 38.3 & 7.04 & 21.4 & \textcolor{worstcolor}{\itshape 2.7} & {\color{basetext}$-$35.5} & \textbf{$-$0.81} & \textbf{$+$1.8} & {\color{basetext}0.54$\to$0.59} & $-$0.48 & {\color{basetext}23.6} \\
     & HCC-PL & 39.1 & 7.12 & \textcolor{worstcolor}{\itshape 17.2} & 43.2 & {\color{basetext}$+$4.1} & $-$1.04 & \textcolor{worstcolor}{\itshape $-$5.8} & {\color{basetext}0.62$\to$0.48} & $+$0.23 & {\color{basetext}16.2} \\
     & HCC-MSP & 39.1 & 5.36 & 27.4 & 25.8 & {\color{basetext}$-$13.3} & $-$1.04 & $+$0.4 & {\color{basetext}0.64$\to$0.65} & $-$0.04 & {\color{basetext}10.2} \\
     & Drebin & 30.4 & 5.18 & 29.7 & 23.1 & {\color{basetext}$-$7.3} & $-$1.32 & $-$0.9 & {\color{basetext}0.67$\to$0.65} & $-$0.05 & {\color{basetext}10.1} \\
     & Random Forest & 37.8 & 4.99 & 29.6 & 24.7 & {\color{basetext}$-$13.0} & $-$0.82 & $-$4.2 & {\color{basetext}0.72$\to$0.64} & $-$0.11 & {\color{basetext}10.9} \\
     & SecSVM & \textcolor{worstcolor}{\itshape 26.8} & 5.61 & 27.4 & 22.5 & {\color{basetext}$-$4.3} & \textcolor{worstcolor}{\itshape $-$1.61} & $-$1.3 & {\color{basetext}0.64$\to$0.62} & $-$0.05 & {\color{basetext}9.6} \\
     & DRMD & \textbf{43.5} & 5.41 & \textbf{30.6} & 43.6 & {\color{basetext}$+$0.1} & $-$0.93 & $-$5.1 & {\color{basetext}0.74$\to$0.62} & $-$0.05 & {\color{basetext}10.3} \\
     & RAMDA & 41.5 & 6.15 & 30.1 & 34.2 & {\color{basetext}$-$7.3} & $-$0.89 & $-$0.4 & {\color{basetext}0.62$\to$0.62} & $-$0.21 & {\color{basetext}11.2} \\
     & TIF & 38.3 & 5.68 & 27.5 & 20.2 & {\color{basetext}$-$18.1} & $-$1.20 & $+$0.5 & {\color{basetext}0.62$\to$0.63} & $-$0.12 & {\color{basetext}8.9} \\
     & MalCertain & 29.7 & 6.35 & 28.4 & 24.7 & {\color{basetext}$-$5.0} & $-$1.33 & $-$1.7 & {\color{basetext}0.67$\to$0.64} & $-$0.18 & {\color{basetext}9.1} \\
    \addlinespace[2pt]
     & \multicolumn{11}{>{\columncolor{subheadcolor}[\tabcolsep][\tabcolsep]}l}{\itshape Transcendent: inductive (ICE) and cross-conformal (CCE)} \\
     & ICE (Drebin) & 31.5 & 5.24 & 27.2 & 37.5 & {\color{basetext}$+$6.0} & $-$1.34 & $-$1.7 & {\color{basetext}0.66$\to$0.63} & $-$0.15 & {\color{basetext}9.9} \\
     & CCE (Drebin) & 32.2 & 5.17 & 27.2 & 38.2 & {\color{basetext}$+$6.1} & $-$1.33 & $-$1.8 & {\color{basetext}0.66$\to$0.63} & $-$0.16 & {\color{basetext}9.9} \\
     & ICE (DeepDrebin) & 37.7 & 4.87 & 25.3 & 44.2 & {\color{basetext}$+$6.4} & $-$1.23 & $-$5.6 & {\color{basetext}0.71$\to$0.59} & $-$0.07 & {\color{basetext}11.2} \\
     & CCE (DeepDrebin) & 38.4 & 4.70 & 26.7 & \textbf{45.4} & {\color{basetext}$+$7.0} & $-$1.33 & $-$5.1 & {\color{basetext}0.71$\to$0.61} & $-$0.04 & {\color{basetext}11.4} \\
    \bottomrule
  \end{tabular}
\end{table*}

Table~\ref{tab:main-vtt2-b400} compares ten classifiers, including paired HCC-MSP and HCC-PL scoring, the paired CADE-OOD and CADE-MSP evaluations, and four Transcendent configurations on Hypercube VTT2 at $\budgetmonthlymath=400$. All sixteen score rows contain five complete seeds; their 30 reported months extend through December 2024. Additionally, we report results for a subset of the baselines on the legacy Androzoo, APIGraph and Transcendent datasets in the Appendix~\ref{tab:legacy-azoo}--\ref{tab:legacy-trans}

Table~\ref{tab:main-vtt2-b400} reports the baseline metric F1 and three metrics per dimension, each computed per month within seed, summarized over the 30 evaluation months within  a seed and then averaged over the five seeds. The reliability dimension reads each confidence function at three operating points: E-AURC integrates the ranking of the whole month; Cap$_{400}$ scores the top of the ranking that the acquisition protocol labels each month, as the share of the month's misclassifications among its 400 most uncertain samples (random reference 11.7\% on average over months); F1$^{*}$ and $\Delta$F1$^{*}$ score the accepted set under a threshold re-calibrated at the start of each month to 400 referrals/month on the six preceding months (Appendix~\ref{appendix:threshold_protocol}). The stability dimension reports the Sen slope of monthly F1, the failure AUROC over the first and the last six evaluation months with its Sen slope, and the Sen slope and mean of the realized referral rate $q^{*}$ under that threshold; the quota corresponds to 11.7\% of the month's arrivals on average.

\subsection{Answering RQ1: How Reliable Are the Confidence Scores?}
\label{subsec:answer_rq1}
DRMD has the highest mean F1 among the evaluated rows (43.5\%), followed by RAMDA (41.5\%) and the HCC in two variants (39.1\%). DeepDrebin has a lower F1 (38.5\%) but the lowest E-AURC (4.53), illustrating that the strongest detector and the strongest error ranking need not coincide. On these two axes the Transcendent rows stay close to their base classifiers: CCE (DeepDrebin) has the second-lowest E-AURC (4.70) at 38.4\% F1, and CCE (Drebin) matches Drebin's E-AURC (5.17 vs.~5.18) with higher F1 (32.2\% vs.~30.4\%). They follow a different acquisition history from their bases, however, and diverge from them on Cap$_{400}$ and F1$^{*}$, discussed below. RAMDA combines the second-highest F1 with E-AURC 6.15; MalCertain reports E-AURC 6.35 with its corrected predictions and base-model confidence. 

\begin{figure*}[t]
\centering
\begin{minipage}[t]{0.49\textwidth}\centering
\includegraphics[width=\linewidth]{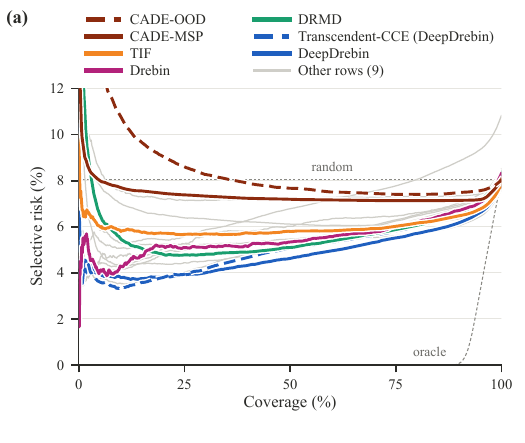}
\end{minipage}\hfill
\begin{minipage}[t]{0.49\textwidth}\centering
\includegraphics[width=\linewidth]{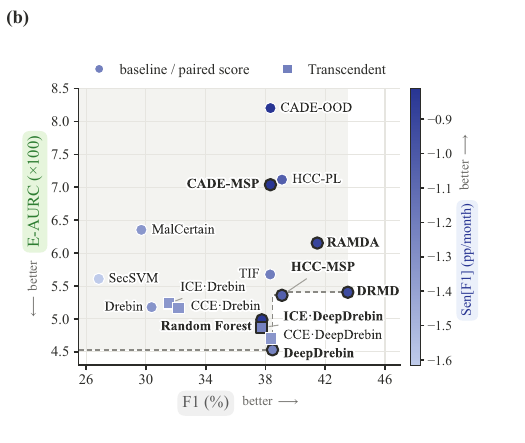}
\end{minipage}
\caption{\small Hypercube VTT2 at $\budgetmonthlymath=400$, all sixteen rows of Table~\ref{tab:main-vtt2-b400}. \textbf{(a)} Month-averaged risk--coverage curves, with seven confidence functions highlighted and nine in gray. The dashed line is the pointwise median of the rows' oracles; each row's E-AURC is its area above its own oracle. The random level is 8.1\% (median across rows of the mean monthly error rate), and the axis is truncated at 12\%. \textbf{(b)} F1 (equal-month mean) against E-AURC; right and down are better. Marker color shows Sen[F1] (darker means a shallower decline) and squares mark Transcendent rows. The dashed staircase is the F1$\times$E-AURC frontier; bold labels mark rows that no other row dominates in F1, E-AURC and Sen[F1] together (unrounded values).}
\label{fig:eaurc_vtt2}
\end{figure*}

Figure~\ref{fig:eaurc_vtt2}(a) resolves these areas along the coverage range. DeepDrebin has the lowest selective risk over most of it (3.9\% at 25\% and 4.6\% at 50\% coverage). CCE (DeepDrebin) lies below it at most coverages under 20\% (3.4\% vs.\ 3.8\% at 10\%) but above it thereafter (5.1\% vs.\ 4.6\% at 50\%), which accounts for its higher E-AURC. DRMD's most confident predictions are less reliable than its average prediction: its selective risk at 1\% coverage (13.0\%) is higher than at full coverage and above the 8.1\% random level, before falling to 5.5\% at 10\%. The two CADE scores share predictions and differ only in their ranking: with the centroid-distance score, selective risk is 16.8\% at 1\% coverage and stays above the 8.1\% random level until about 35\% coverage, compared to 9.5\% at 1\% for MSP. Figure~\ref{fig:eaurc_vtt2}(b) places the sixteen rows in the F1$\times$E-AURC plane. At the displayed precision, DeepDrebin, HCC-MSP, and DRMD are the only rows that no other row beats on both axes (dashed staircase). The color scale shows that the shallowest F1 declines (both CADE rows, $-$0.81; Random Forest, $-$0.82; RAMDA, $-$0.89\,pp/month) belong to rows off this frontier; with Sen[F1] as a third criterion (on unrounded values), CADE-MSP, Random Forest, RAMDA and ICE (DeepDrebin) also become non-dominated (bold labels).
Cap$_{400}$ and F1$^{*}$ provide yet another ordering. Every row places more of its errors among its 400 most uncertain samples than random selection would (Cap$_{400}$ of 17.2--30.6\%, against 11.7\%), but none captures more than a third. DRMD has the highest Cap$_{400}$ (30.6\%, narrowly ahead of DeepDrebin's 30.3\%): a larger share of its monthly errors falls among its 400 most uncertain samples than any other row, even though its most confident predictions are unreliable. Under the threshold re-calibrated to 400 referrals per month, CCE (DeepDrebin) attains the highest F1$^{*}$ (45.4\%, compared to 31.1\% for DeepDrebin), and DRMD keeps its baseline F1 almost unchanged (43.6\%, $\Delta$F1$^{*}=+0.1$\,pp). Holding HCC predictions and acquisition histories fixed, Pseudo-Loss worsens E-AURC (5.36 to 7.12) and Cap$_{400}$ (27.4\% to 17.2\%) yet raises F1$^{*}$ from 25.8\% to 43.2\%.

Nine of the sixteen rows have a negative $\Delta$F1$^{*}$: their confidence functions hold back detections faster than they remove errors. With identical predictions, the score decides the sign: HCC-MSP loses 13.3\,pp while HCC-PL gains 4.1\,pp, and CADE-MSP loses 35.5\,pp (on the 119 of 150 seed-months with at least 20 accepted samples) while CADE-OOD gains 4.2\,pp. Because F1$^{*}$ also reflects how the threshold transfers from the preceding months, these pairs differ in referral rate as well as in which samples they refer (mean realized referral rates $\bar q^{*}$ of 10.2\% vs.~16.2\% for HCC-MSP and HCC-PL, and 23.6\% vs.~12.7\% for CADE-MSP and CADE-OOD). The four Transcendent rows gain most ($+6.0$ to $+7.0$\,pp), whereas DeepDrebin, despite the lowest E-AURC, loses 7.4\,pp. Cap$_{400}$ does not anticipate these differences, since it counts errors of both predicted classes alike. A good error ranking within a month, therefore, does not guarantee that a quota-calibrated threshold improves the samples it accepts. F1$^{*}$ concerns accepted samples; it is not an estimate of end-to-end detection after analyst review. Thus, error-ranking quality, class-sensitive performance at the rejection quota, and baseline detection capture different properties.

Overall, on VTT2 no confidence function captures more than a third of a month's errors among its 400 most uncertain samples on average, and reliability does not track detection: F1 is only weakly rank-correlated with E-AURC, Cap$_{400}$ and $\Delta$F1$^{*}$. Table~\ref{tab:cand-budgets} shows that this divergence between detection and error ranking does not depend on the budget: at every VTT2 budget, DRMD has the highest F1 and DeepDrebin or CCE (DeepDrebin) the lowest E-AURC, RAMDA captures the largest share of errors at the three lower budgets, and abstention lowers accepted-set F1 for nine or ten rows (Appendix~\ref{appendix:vtt2_budgets}; the lower budgets come from a separate campaign, so differences across budgets are not strictly paired). Table~\ref{tab:cand-datasets} summarizes the additional datasets, on which five of the methods were run. Their horizons and training protocols also differ from VTT2. Among the five methods, the divergence holds on AndroZoo, where HCC leads F1 and CCE (Drebin) E-AURC at every budget, whereas on APIGraph and Transcendent DeepDrebin or CCE (DeepDrebin) leads both (Appendix~\ref{appendix:prior_hypercube}, Tables~\ref{tab:legacy-azoo}--\ref{tab:legacy-trans}).

\begin{table}[!t]
\centering
\caption{Leaders per budget on Hypercube VTT2; full results in
Table~\ref{tab:main-vtt2-b400} and Table~\ref{tab:vtt2-lower-budgets}
(Appendix~\ref{appendix:vtt2_budgets}). DD = DeepDrebin.}
\label{tab:cand-budgets}
\footnotesize
\setlength{\tabcolsep}{3pt}
\resizebox{\columnwidth}{!}{%
\begin{tabular}{@{}rllllc@{}}
\toprule
& \multicolumn{1}{>{\columncolor{tabbaseband}}c}{Baseline} & \multicolumn{3}{>{\columncolor{tabreliband}}c}{Reliability} & \multicolumn{1}{>{\columncolor{tabstabband}}c}{Stability} \\
$B_M$ & \cellcolor{tabbaseband}F1 $\uparrow$ & \cellcolor{tabreliband}E-AURC $\downarrow$ & \cellcolor{tabreliband}Cap$_{B_M}$ $\uparrow$ & \cellcolor{tabreliband}$\Delta$F1$^{*}{<}0$ & \cellcolor{tabstabband}Sen[F1] \\
 & \cellcolor{tabbaseband}{\scriptsize highest, \%} & \cellcolor{tabreliband}{\scriptsize lowest, $\times$100} & \cellcolor{tabreliband}{\scriptsize highest, \%} & \cellcolor{tabreliband}{\scriptsize rows} & \cellcolor{tabstabband}{\scriptsize range, pp/month} \\
\midrule
50  & DRMD 33.1 & CCE (DD) 4.95     & RAMDA 7.2  & 10/16 & $-1.55$ to $-1.15$ \\
100 & DRMD 36.0 & DD, CCE (DD) 4.85 & RAMDA 12.0 & 10/16 & $-1.56$ to $-1.13$ \\
200 & DRMD 40.2 & DD 4.69           & RAMDA 19.2 & 10/16 & $-1.58$ to $-1.01$ \\
400 & DRMD 43.5 & DD 4.53           & DRMD 30.6  & 9/16  & $-1.61$ to $-0.81$ \\
\bottomrule
\end{tabular}}
\end{table}

\begin{table}[!t]
\centering
\caption{Leaders per budget on the additional datasets, for the five methods run there; full results in Appendix~\ref{appendix:prior_hypercube} (Tables~\ref{tab:legacy-azoo}--\ref{tab:legacy-trans}); DD = DeepDrebin.}
\label{tab:cand-datasets}
\footnotesize
\setlength{\tabcolsep}{3pt}
\resizebox{\columnwidth}{!}{%
\begin{tabular}{@{}lrlll@{}}
\toprule
& & \multicolumn{1}{>{\columncolor{tabbaseband}}c}{Baseline} & \multicolumn{1}{>{\columncolor{tabreliband}}c}{Reliability} & \multicolumn{1}{>{\columncolor{tabstabband}}c}{Stability} \\
Dataset & $B_M$ & \cellcolor{tabbaseband}F1 $\uparrow$ & \cellcolor{tabreliband}E-AURC $\downarrow$ & \cellcolor{tabstabband}Sen[F1] \\
 & & \cellcolor{tabbaseband}{\scriptsize highest, \%} & \cellcolor{tabreliband}{\scriptsize lowest, $\times$100} & \cellcolor{tabstabband}{\scriptsize range, pp/month} \\
\midrule
\multirow{4}{*}{AndroZoo} & 50  & HCC 75.1 & CCE (Drebin) 1.24 & $-0.66$ to $+1.81$ \\
                          & 100 & HCC 76.9 & CCE (Drebin) 0.93 & $+0.11$ to $+1.17$ \\
                          & 200 & HCC 78.3 & CCE (Drebin) 0.82 & $+0.65$ to $+1.56$ \\
                          & 400 & HCC 84.5 & CCE (Drebin) 0.63 & $+0.67$ to $+1.87$ \\
\midrule
\multirow{4}{*}{APIGraph} & 50  & DD 89.7 & DD 0.61 & $+0.03$ to $+0.12$ \\
                          & 100 & DD 91.2 & DD 0.54 & $+0.03$ to $+0.10$ \\
                          & 200 & DD 92.2 & DD 0.46 & $+0.03$ to $+0.08$ \\
                          & 400 & CCE (DD) 92.9 & DD 0.40 & $-0.01$ to $+0.08$ \\
\midrule
\multirow{4}{*}{Transcendent} & 50  & CCE (DD) 83.2 & CCE (DD) 1.13 & $-0.46$ to $+0.07$ \\
                              & 100 & CCE (DD) 85.8 & CCE (DD) 1.05 & $-0.55$ to $+0.14$ \\
                              & 200 & CCE (DD) 87.5 & DD 0.91 & $-0.34$ to $+0.11$ \\
                              & 400 & DD 89.9 & DD 0.66 & $-0.12$ to $+0.01$ \\
\bottomrule
\end{tabular}}
\end{table}

\begin{obs}
\textbf{RQ1 Takeaways:}
\textbf{(1)} Strong detection does not imply reliable error ranking: on Hypercube, the best detector and the best error ranker are different methods.
\textbf{(2)} Even the best confidence functions place only a minority of each month's errors among their most uncertain samples.
\textbf{(3)} Some methods are most often wrong where they are most confident, so these errors escape rejection.
\textbf{(4)} Abstention can \emph{hurt}: for most methods, rejecting uncertain samples lowers F1 on the samples that are kept, and with the same classifier the choice of confidence score can turn this loss into a gain.
\end{obs}

\subsection{Answering RQ2: How Stable Are Performance, Ranking and Workload Over Time?}
\label{subsec:answer_rq2}
On Hypercube, performance does not persist. Every evaluated method loses F1 over the 30 evaluation months, with Sen[F1] ranging from $-1.61$\,pp/month (SecSVM) to $-0.81$\,pp/month (both CADE rows). The decline is consistent across seeds: all 80 per-seed slopes lie between $-1.61$ and $-0.67$\,pp/month. Error ranking deteriorates as well, and 13 of the 16 rows have a negative Sen[AUROC$_f$] (12 of them in all five seeds), and HCC-PL's failure AUROC falls from 0.62 over the first six months to 0.48 over the last six, below the random level of 0.5. Only CADE-MSP ($+1.8\times10^{-3}$/month), TIF ($+0.5$) and HCC-MSP ($+0.4$) avoid a decline, and CADE-MSP does so from a near-random start (0.54). The results for methods with alternative uncertainty functions show that, with identical predictions, the score determines whether error ranking persists: HCC-MSP holds its failure AUROC (0.64 to 0.65) while HCC-PL loses it, and CADE-MSP improves while CADE-OOD declines ($-1.4\times10^{-3}$/month, ending at 0.51). 

Under the threshold re-calibrated at the start of each month to 400 referrals on the six preceding months, 13 rows have a mean realized referral rate ($\bar q^{*}$) below the quota's average share of monthly arrivals ($\bar q^{*}$ of 8.9--11.4\%, against 11.7\%), whereas CADE-OOD (12.7\%), HCC-PL (16.2\%) and CADE-MSP (23.6\%) exceed it. For 12 of the 13 under-referring rows, the gap widens over time (negative Sen[$q^{*}$], down to $-0.21$\,pp/month for RAMDA), where only DeepDrebin shows no trend ($+0.01$). Among the over-referring rows, HCC-PL drifts further above the quota ($+0.23$), whereas the excess of CADE-OOD and CADE-MSP declines ($-0.24$ and $-0.48$ at its steepest). A threshold calibrated on the score-history therefore does not retain the  calibrated workload under drift, with high variance across confidence functions.

CADE isolates the effect of scoring on the same predictions and acquisition history: MSP refers to 23.6\% of arrivals on average, about twice the quota, compared to 12.7\% for the centroid-distance score. HCC shows the same dependence on the score: PL over-refers (16.2\%) with a rising excess ($+0.23$\,pp/month), whereas MSP stays below the quota (10.2\%) with a slight negative trend ($-0.04$). Excessive rejections indicate that a score rates new months as more uncertain compared to the calibration window; indicating that the threshold transfers poorly.

The four Transcendent configurations lose F1 at similar rates compared to their model bases (Sen[F1] of $-1.34/-1.33$\,pp/month for ICE/CCE (Drebin) against $-1.32$ for Drebin, and $-1.23/-1.33$ for ICE/CCE (DeepDrebin) against $-1.23$ for DeepDrebin), but their error ranking degrades faster: Sen[AUROC$_f$] is $-1.7/-1.8\times10^{-3}$/month against $-0.9$ for Drebin, and $-5.6/-5.1$ against $-3.6$ for DeepDrebin. All four reject below the quota on average ($\bar q^{*}$ of 9.9\% for both Drebin variants, 11.2\% and 11.4\% for the DeepDrebin variants), and this effect increases at a faster rate compared to their bases (Sen[$q^{*}$] of $-0.15/-0.16$ against $-0.05$ for Drebin, and $-0.07/-0.04$ against $+0.01$ for DeepDrebin). 

Within these five methods, the decline is specific to Hypercube (Table~\ref{tab:cand-datasets}): every VTT2 row loses 0.81 to 1.61\,pp of F1 per month at every budget (Appendix~\ref{appendix:vtt2_budgets}), whereas the slopes lie between $-0.55$ and $+0.14$\,pp/month on Transcendent and within 0.12\,pp/month of zero on APIGraph, and are positive for almost every row on AndroZoo (Tables~\ref{tab:legacy-azoo}--\ref{tab:legacy-trans}). Even where F1 remains high, the confidence ranking need not: on AndroZoo at $\budgetmonthlymath=50$, DeepDrebin's F1 rises by $+1.81$\,pp/month while its failure AUROC falls from 0.76 to 0.46.

\begin{obs}
\textbf{RQ2 Takeaways:}
\textbf{(1)} On Hypercube, every method loses detection performance over time, but this is not universal: on some of the legacy datasets, performance stays flat or improves.
\textbf{(2)} Confidence rankings decay too but can hide: F1 may hold or even rise while the ranking degrades.
\textbf{(3)} With the same classifier, the choice of confidence score can decide whether the ranking and the analyst workload stay stable over time.
\textbf{(4)} Conformal wrappers add no stability: their F1 declines like that of their base classifiers, while their ranking and workload tracking degrade faster.
\textbf{(5)} Re-calibrating the threshold on past scores does not keep the workload on target: most methods increasingly under-reject while a few exceed it.
\end{obs}

\subsection{Scope of the Expanded Comparison}
\label{subsec:pareto_synthesis}
This first expansion holds the monthly acquisition budget at 400. Appendix~\ref{appendix:vtt2_budgets} reports the 50/100/200 budget grid. No universal winner follows from the reported summaries: DRMD leads F1 (43.5\%) and Cap$_{400}$ (30.6\%); DeepDrebin leads E-AURC (4.53) and has the flattest workload trend (Sen[$q^{*}$] $=+0.01$\,pp/month), yet loses 7.4\,pp of F1 on accepted samples; CCE (DeepDrebin) has the highest F1$^{*}$ (45.4\%) at a referral rate close to the quota (11.4\% vs.~11.7\%), but its failure AUROC declines faster than DeepDrebin's ($-5.1$ vs.~$-3.6\times10^{-3}$/month); and the CADE rows lose F1 most slowly ($-0.81$\,pp/month), while CADE-OOD has the highest E-AURC (8.20). The frontier in Figure~\ref{fig:eaurc_vtt2}(b) is descriptive.

\section{Limitations}
The main comparison covers one dataset version at $\budgetmonthlymath=400$, with five seeds and method-specific update procedures. In particular, the paired HCC-MSP and HCC-PL rows use benchmark training and acquisition, while CADE uses a pinned HCC reference adaptation. The lower-budget grid (Appendix~\ref{appendix:vtt2_budgets}) is incomplete for MalCertain and CADE and comes from a separate campaign, so budget effects are not strictly paired. The AndroZoo, APIGraph and Transcendent results use different datasets and protocols and do not fill these missing VTT2 experiments.

Every quantity in Table~\ref{tab:main-vtt2-b400} is a per-seed summary under a threshold re-calibrated monthly on the unlabeled score-history. These metrics describe different aspects of performance; they do not establish statistical superiority or guarantee a hard analyst-capacity limit. Confidence-ranked acquisition, monthly updates and immediate, correct annotations simplify operational deployments. Accepted-set F1 also does not measure the outcome of analyst review. Generalization beyond Android malware in Drebin feature space remains to be evaluated.

\section{Conclusion}
\codename{} evaluates confidence reliability and temporal stability alongside detection performance. Expanding the comparison on Hypercube VTT2 shows distinct trade-offs: the highest F1, the lowest excess risk--coverage area and the closest average referral rate occur in different methods. Across the reported 30-month window, all evaluated methods show declining F1 trends, and quota-calibrated abstention often reduces F1 on accepted samples. The expanded comparison therefore reinforces the need to evaluate detection, confidence ranking and operational rejection behavior together, while keeping dataset versions and training protocols explicit.

% =======================================================
% SaTML excludes these disclosure sections from the body-text page limit.
% Keep Open Science followed by LLM usage considerations before references.
\section{Ethical Considerations}
\label{sec:ethics}
% NOTE (page budget, restructure 2026-06-12): phrasing tightened; original in
% .backup-pre-restructure-0900.
This work evaluates existing malware classifiers using publicly available datasets (APIGraph, AndroZoo, Transcendent, Hypercube) that have been widely used in prior peer-reviewed research~\cite{HCC,CADE,apigraph,barbero2022transcendent,chow2025breaking}. All datasets consist of aggregated application metadata and extracted features rather than raw user data and contain no personally identifiable information. Our research is purely evaluative: we propose no new attacks, identify no vulnerabilities in deployed systems (so coordinated disclosure does not apply), and our framework cannot be repurposed to evade detection, as it assesses classifier reliability rather than exposing exploitable weaknesses. The evaluation we advocate can, if adopted, reduce harm by exposing confidence functions that produce confident misclassifications before they are deployed. We see no direct ethical concerns arising from this work.%

\section{Open Science}
\label{appendix:open_science}

% NOTE (page budget, restructure 2026-06-12): Artifacts-Overview itemize folded
% into prose; phrasing tightened; original in .backup-pre-restructure-0900.
This section enumerates all artifacts required to reproduce and evaluate the contributions of this paper, in accordance with the IEEE SaTML open science policy.%

\subsection{Code and Scripts}
%For the expanded VTT2 comparison, the manuscript project includes the full-precision table, input-file hashes, metric-source hashes and validation summaries. The complete VTT2 benchmark artifacts and HCC scoring additions will be packaged for the revised release. The previously released archive described below covers the earlier evaluation and should not be taken as a complete release of the new VTT2 comparison.

The released artifacts comprise the \codename{} evaluation framework, all experiment scripts and model implementations (DeepDrebin, CADE, HCC, Drebin, Random Forest, SecSVM, DRMD, RAMDA, TIF, MalCertain, and Transcendent), configuration files (App.~\ref{hyperparameters_configurations}), analysis notebooks, and raw result files with per-month predictions -- anonymous, standalone files to be released under an open-source license, with documentation for reproducing all experiments and figures. During double-blind review, the code is available at: \url{https://anonymous.4open.science/r/eval-artifact-2027-2BCA/README.md}.%

The accompanying manuscript files include the complete five-seed lower-budget tables, per-seed values and source hashes. These completed lower-budget results postdate the linked evaluation-artifact snapshot; their full raw outputs and checkpoints are retained in a separate preservation archive.

\subsection{Datasets}

\textbf{APIGraph and AndroZoo:} We use the preprocessed datasets from Chen et al.~\cite{HCC}, which extract Drebin-style features from Android APKs. These datasets are available at: \url{https://drive.google.com/file/d/1O0upEcTolGyyvasCPkZFY86FNclk29XO/view?usp=drive_link}.%

\noindent \textbf{Transcendent:} We use the dataset from Barbero et al.~\cite{barbero2022transcendent}, available at: \url{https://github.com/s2labres/transcendent-release}.%

\noindent \textbf{Hypercube:} We use the dataset of Chow et al.~\cite{chow2025breaking}, released with their paper.%

\section{LLM usage considerations}
\label{sec:ai_disclosure}
This work used generative AI tools for assistance with code development, data analysis, literature review, and drafting and editing the manuscript. All AI-generated content was reviewed and validated by the authors, who take full responsibility for the accuracy and integrity of the final submission.

% =======================================================
\bibliographystyle{IEEEtran}
\bibliography{sample-base}

% =======================================================
\appendices
\section{Hypercube VTT2 at Lower Monthly Budgets}
\label{appendix:vtt2_budgets}

Table~\ref{tab:vtt2-lower-budgets} repeats the comparison of Table~\ref{tab:main-vtt2-b400} at $\budgetmonthlymath\in\{50,100,200\}$ with the same sixteen score rows, metrics, rolling threshold protocol and aggregation. The Cap batch size and the rejection quota both equal the budget ($\rho=\budgetmonthlymath$), so Cap$_{\budgetmonthlymath}$ scores the $\budgetmonthlymath$ most uncertain samples of each month and the threshold is re-calibrated to $\budgetmonthlymath$ referrals per month. Model configurations are those of the $\budgetmonthlymath=400$ runs (Appendix~\ref{hyperparameters_configurations}); only the number of acquired labels changes. The lower-budget runs share initial models and code with each other but were
run independently of Table~\ref{tab:main-vtt2-b400}, so differences relative to $B_M = 400$ are not strictly paired budget effects.

The grid is complete: all sixteen score rows have five complete seeds (IDs 1--5) at each budget. Every seed contains all 36 monthly outputs; the 30 evaluation months run from July 2022 through December 2024. Partial trajectories are not used.

F1 rises across the three lower budgets for every row except MalCertain, whose five-seed means are 30.3, 29.9 and 30.1\%. DRMD has the highest F1 at every budget (33.1, 36.0 and 40.2\% at 50, 100 and 200, against 43.5\% at 400). The lowest E-AURC belongs to DeepDrebin or CCE~(DeepDrebin) at every budget (4.95 to 4.69, against 4.53 at 400), as in Table~\ref{tab:main-vtt2-b400}. RAMDA captures the largest share of errors in its $\budgetmonthlymath$ most uncertain samples at each lower budget (7.2, 12.0 and 19.2\%, against random references of 1.5, 2.9 and 5.8\%). HCC-PL captures the smallest share at 50 and 100 (tied with ICE (DeepDrebin) at 50), and CADE-OOD the smallest at 200 (11.2\%). Abstention raises accepted-set F1 for the four Transcendent rows at every budget ($+0.7$ to $+3.7$\,pp), for HCC-PL ($+0.2$ to $+1.4$\,pp) and for CADE-OOD; it lowers it for the other ten rows at every lower budget. Every row declines over the window, by 1.01 to 1.58\,pp of F1 per month. Under the rolling protocol, CADE-MSP at $\budgetmonthlymath=200$ misses its calibration target by 27.2\% on average over seeds and rolling windows; seeds 1, 2, 3 and 5 exceed the 25\% guard, and its quota-dependent cells are flagged and unranked. No other lower-budget row exceeds the guard. These checks are specific to the rolling protocol; the separate frozen-threshold replay retains its calibration warnings. The same rolling check applied to Table~\ref{tab:main-vtt2-b400} flags CADE-MSP, whose rolling windows miss the 400-referral target by 49\% on average. Figure~\ref{fig:overview_rejection} shows the monthly trajectories at $\budgetmonthlymath=50$ under a fixed quota of 400 referrals.

% Generated by generate_satml_vtt2_budget_tables.py; do not edit numbers by hand.
% Layout, metrics and threshold protocol are those of Table I (generate_satml_vtt2_frozen_table.py).
\providecolor{tabbaseband}{RGB}{240,240,240}
\providecolor{tabreliband}{RGB}{230,245,220}
\providecolor{tabstabband}{RGB}{235,240,250}
\providecolor{worstcolor}{RGB}{176,32,32}
\providecolor{subheadcolor}{RGB}{246,246,246}
\providecolor{basetext}{RGB}{90,90,90}
\providecolor{relitext}{RGB}{46,125,50}
\providecolor{stabtext}{RGB}{40,53,147}
\providecommand{\codename}{\textsc{Aurora}}
\ifdefined\frzwtmp\else\newlength{\frzwtmp}\fi
\ifdefined\bgwA\else\newlength{\bgwA}\fi\setlength{\bgwA}{0pt}
\settowidth{\frzwtmp}{\footnotesize 28.3}\ifdim\frzwtmp>\bgwA\setlength{\bgwA}{\frzwtmp}\fi
\settowidth{\frzwtmp}{\footnotesize {\color{basetext}$-$1.5}}\ifdim\frzwtmp>\bgwA\setlength{\bgwA}{\frzwtmp}\fi
\settowidth{\frzwtmp}{\footnotesize 28.2}\ifdim\frzwtmp>\bgwA\setlength{\bgwA}{\frzwtmp}\fi
\settowidth{\frzwtmp}{\footnotesize {\color{basetext}$+$1.3}}\ifdim\frzwtmp>\bgwA\setlength{\bgwA}{\frzwtmp}\fi
\settowidth{\frzwtmp}{\footnotesize \textcolor{worstcolor}{\itshape 18.9}}\ifdim\frzwtmp>\bgwA\setlength{\bgwA}{\frzwtmp}\fi
\settowidth{\frzwtmp}{\footnotesize {\color{basetext}$-$8.2}}\ifdim\frzwtmp>\bgwA\setlength{\bgwA}{\frzwtmp}\fi
\settowidth{\frzwtmp}{\footnotesize 28.1}\ifdim\frzwtmp>\bgwA\setlength{\bgwA}{\frzwtmp}\fi
\settowidth{\frzwtmp}{\footnotesize {\color{basetext}$+$0.2}}\ifdim\frzwtmp>\bgwA\setlength{\bgwA}{\frzwtmp}\fi
\settowidth{\frzwtmp}{\footnotesize 26.9}\ifdim\frzwtmp>\bgwA\setlength{\bgwA}{\frzwtmp}\fi
\settowidth{\frzwtmp}{\footnotesize {\color{basetext}$-$1.0}}\ifdim\frzwtmp>\bgwA\setlength{\bgwA}{\frzwtmp}\fi
\settowidth{\frzwtmp}{\footnotesize 21.4}\ifdim\frzwtmp>\bgwA\setlength{\bgwA}{\frzwtmp}\fi
\settowidth{\frzwtmp}{\footnotesize {\color{basetext}$-$2.8}}\ifdim\frzwtmp>\bgwA\setlength{\bgwA}{\frzwtmp}\fi
\settowidth{\frzwtmp}{\footnotesize 21.8}\ifdim\frzwtmp>\bgwA\setlength{\bgwA}{\frzwtmp}\fi
\settowidth{\frzwtmp}{\footnotesize 21.2}\ifdim\frzwtmp>\bgwA\setlength{\bgwA}{\frzwtmp}\fi
\settowidth{\frzwtmp}{\footnotesize {\color{basetext}$-$2.3}}\ifdim\frzwtmp>\bgwA\setlength{\bgwA}{\frzwtmp}\fi
\settowidth{\frzwtmp}{\footnotesize \textbf{32.3}}\ifdim\frzwtmp>\bgwA\setlength{\bgwA}{\frzwtmp}\fi
\settowidth{\frzwtmp}{\footnotesize {\color{basetext}$-$0.8}}\ifdim\frzwtmp>\bgwA\setlength{\bgwA}{\frzwtmp}\fi
\settowidth{\frzwtmp}{\footnotesize 29.9}\ifdim\frzwtmp>\bgwA\setlength{\bgwA}{\frzwtmp}\fi
\settowidth{\frzwtmp}{\footnotesize {\color{basetext}$-$1.1}}\ifdim\frzwtmp>\bgwA\setlength{\bgwA}{\frzwtmp}\fi
\settowidth{\frzwtmp}{\footnotesize 28.9}\ifdim\frzwtmp>\bgwA\setlength{\bgwA}{\frzwtmp}\fi
\settowidth{\frzwtmp}{\footnotesize {\color{basetext}$-$1.6}}\ifdim\frzwtmp>\bgwA\setlength{\bgwA}{\frzwtmp}\fi
\settowidth{\frzwtmp}{\footnotesize {\color{basetext}$-$1.4}}\ifdim\frzwtmp>\bgwA\setlength{\bgwA}{\frzwtmp}\fi
\settowidth{\frzwtmp}{\footnotesize 25.9}\ifdim\frzwtmp>\bgwA\setlength{\bgwA}{\frzwtmp}\fi
\settowidth{\frzwtmp}{\footnotesize {\color{basetext}$+$0.7}}\ifdim\frzwtmp>\bgwA\setlength{\bgwA}{\frzwtmp}\fi
\settowidth{\frzwtmp}{\footnotesize 26.2}\ifdim\frzwtmp>\bgwA\setlength{\bgwA}{\frzwtmp}\fi
\settowidth{\frzwtmp}{\footnotesize {\color{basetext}$+$0.8}}\ifdim\frzwtmp>\bgwA\setlength{\bgwA}{\frzwtmp}\fi
\settowidth{\frzwtmp}{\footnotesize 31.1}\ifdim\frzwtmp>\bgwA\setlength{\bgwA}{\frzwtmp}\fi
\settowidth{\frzwtmp}{\footnotesize 32.2}\ifdim\frzwtmp>\bgwA\setlength{\bgwA}{\frzwtmp}\fi
\settowidth{\frzwtmp}{\footnotesize 29.8}\ifdim\frzwtmp>\bgwA\setlength{\bgwA}{\frzwtmp}\fi
\settowidth{\frzwtmp}{\footnotesize {\color{basetext}$-$3.1}}\ifdim\frzwtmp>\bgwA\setlength{\bgwA}{\frzwtmp}\fi
\settowidth{\frzwtmp}{\footnotesize 33.4}\ifdim\frzwtmp>\bgwA\setlength{\bgwA}{\frzwtmp}\fi
\settowidth{\frzwtmp}{\footnotesize {\color{basetext}$+$2.2}}\ifdim\frzwtmp>\bgwA\setlength{\bgwA}{\frzwtmp}\fi
\settowidth{\frzwtmp}{\footnotesize \textcolor{worstcolor}{\itshape 16.6}}\ifdim\frzwtmp>\bgwA\setlength{\bgwA}{\frzwtmp}\fi
\settowidth{\frzwtmp}{\footnotesize {\color{basetext}$-$15.2}}\ifdim\frzwtmp>\bgwA\setlength{\bgwA}{\frzwtmp}\fi
\settowidth{\frzwtmp}{\footnotesize 30.6}\ifdim\frzwtmp>\bgwA\setlength{\bgwA}{\frzwtmp}\fi
\settowidth{\frzwtmp}{\footnotesize {\color{basetext}$+$0.4}}\ifdim\frzwtmp>\bgwA\setlength{\bgwA}{\frzwtmp}\fi
\settowidth{\frzwtmp}{\footnotesize 27.4}\ifdim\frzwtmp>\bgwA\setlength{\bgwA}{\frzwtmp}\fi
\settowidth{\frzwtmp}{\footnotesize 20.7}\ifdim\frzwtmp>\bgwA\setlength{\bgwA}{\frzwtmp}\fi
\settowidth{\frzwtmp}{\footnotesize {\color{basetext}$-$4.6}}\ifdim\frzwtmp>\bgwA\setlength{\bgwA}{\frzwtmp}\fi
\settowidth{\frzwtmp}{\footnotesize 22.7}\ifdim\frzwtmp>\bgwA\setlength{\bgwA}{\frzwtmp}\fi
\settowidth{\frzwtmp}{\footnotesize {\color{basetext}$-$5.3}}\ifdim\frzwtmp>\bgwA\setlength{\bgwA}{\frzwtmp}\fi
\settowidth{\frzwtmp}{\footnotesize 21.3}\ifdim\frzwtmp>\bgwA\setlength{\bgwA}{\frzwtmp}\fi
\settowidth{\frzwtmp}{\footnotesize {\color{basetext}$-$3.0}}\ifdim\frzwtmp>\bgwA\setlength{\bgwA}{\frzwtmp}\fi
\settowidth{\frzwtmp}{\footnotesize \textbf{35.3}}\ifdim\frzwtmp>\bgwA\setlength{\bgwA}{\frzwtmp}\fi
\settowidth{\frzwtmp}{\footnotesize {\color{basetext}$-$0.7}}\ifdim\frzwtmp>\bgwA\setlength{\bgwA}{\frzwtmp}\fi
\settowidth{\frzwtmp}{\footnotesize 30.5}\ifdim\frzwtmp>\bgwA\setlength{\bgwA}{\frzwtmp}\fi
\settowidth{\frzwtmp}{\footnotesize 30.1}\ifdim\frzwtmp>\bgwA\setlength{\bgwA}{\frzwtmp}\fi
\settowidth{\frzwtmp}{\footnotesize {\color{basetext}$-$3.5}}\ifdim\frzwtmp>\bgwA\setlength{\bgwA}{\frzwtmp}\fi
\settowidth{\frzwtmp}{\footnotesize 27.5}\ifdim\frzwtmp>\bgwA\setlength{\bgwA}{\frzwtmp}\fi
\settowidth{\frzwtmp}{\footnotesize {\color{basetext}$-$2.4}}\ifdim\frzwtmp>\bgwA\setlength{\bgwA}{\frzwtmp}\fi
\settowidth{\frzwtmp}{\footnotesize 29.0}\ifdim\frzwtmp>\bgwA\setlength{\bgwA}{\frzwtmp}\fi
\settowidth{\frzwtmp}{\footnotesize {\color{basetext}$+$1.5}}\ifdim\frzwtmp>\bgwA\setlength{\bgwA}{\frzwtmp}\fi
\settowidth{\frzwtmp}{\footnotesize 29.2}\ifdim\frzwtmp>\bgwA\setlength{\bgwA}{\frzwtmp}\fi
\settowidth{\frzwtmp}{\footnotesize 34.3}\ifdim\frzwtmp>\bgwA\setlength{\bgwA}{\frzwtmp}\fi
\settowidth{\frzwtmp}{\footnotesize {\color{basetext}$+$1.6}}\ifdim\frzwtmp>\bgwA\setlength{\bgwA}{\frzwtmp}\fi
\settowidth{\frzwtmp}{\footnotesize 35.1}\ifdim\frzwtmp>\bgwA\setlength{\bgwA}{\frzwtmp}\fi
\settowidth{\frzwtmp}{\footnotesize {\color{basetext}$+$1.8}}\ifdim\frzwtmp>\bgwA\setlength{\bgwA}{\frzwtmp}\fi
\settowidth{\frzwtmp}{\footnotesize {\color{basetext}$-$4.4}}\ifdim\frzwtmp>\bgwA\setlength{\bgwA}{\frzwtmp}\fi
\settowidth{\frzwtmp}{\footnotesize 37.0}\ifdim\frzwtmp>\bgwA\setlength{\bgwA}{\frzwtmp}\fi
\settowidth{\frzwtmp}{\footnotesize {\color{basetext}$+$2.6}}\ifdim\frzwtmp>\bgwA\setlength{\bgwA}{\frzwtmp}\fi
\settowidth{\frzwtmp}{\footnotesize 7.3\rlap{$^{\dagger}$}}\ifdim\frzwtmp>\bgwA\setlength{\bgwA}{\frzwtmp}\fi
\settowidth{\frzwtmp}{\footnotesize {\color{basetext}$-$27.7}\rlap{$^{\dagger}$}}\ifdim\frzwtmp>\bgwA\setlength{\bgwA}{\frzwtmp}\fi
\settowidth{\frzwtmp}{\footnotesize 36.7}\ifdim\frzwtmp>\bgwA\setlength{\bgwA}{\frzwtmp}\fi
\settowidth{\frzwtmp}{\footnotesize {\color{basetext}$+$1.4}}\ifdim\frzwtmp>\bgwA\setlength{\bgwA}{\frzwtmp}\fi
\settowidth{\frzwtmp}{\footnotesize 29.5}\ifdim\frzwtmp>\bgwA\setlength{\bgwA}{\frzwtmp}\fi
\settowidth{\frzwtmp}{\footnotesize {\color{basetext}$-$5.8}}\ifdim\frzwtmp>\bgwA\setlength{\bgwA}{\frzwtmp}\fi
\settowidth{\frzwtmp}{\footnotesize 22.8}\ifdim\frzwtmp>\bgwA\setlength{\bgwA}{\frzwtmp}\fi
\settowidth{\frzwtmp}{\footnotesize 24.2}\ifdim\frzwtmp>\bgwA\setlength{\bgwA}{\frzwtmp}\fi
\settowidth{\frzwtmp}{\footnotesize {\color{basetext}$-$8.7}}\ifdim\frzwtmp>\bgwA\setlength{\bgwA}{\frzwtmp}\fi
\settowidth{\frzwtmp}{\footnotesize \textcolor{worstcolor}{\itshape 21.9}}\ifdim\frzwtmp>\bgwA\setlength{\bgwA}{\frzwtmp}\fi
\settowidth{\frzwtmp}{\footnotesize {\color{basetext}$-$3.7}}\ifdim\frzwtmp>\bgwA\setlength{\bgwA}{\frzwtmp}\fi
\settowidth{\frzwtmp}{\footnotesize \textbf{39.6}}\ifdim\frzwtmp>\bgwA\setlength{\bgwA}{\frzwtmp}\fi
\settowidth{\frzwtmp}{\footnotesize {\color{basetext}$-$0.6}}\ifdim\frzwtmp>\bgwA\setlength{\bgwA}{\frzwtmp}\fi
\settowidth{\frzwtmp}{\footnotesize 31.0}\ifdim\frzwtmp>\bgwA\setlength{\bgwA}{\frzwtmp}\fi
\settowidth{\frzwtmp}{\footnotesize {\color{basetext}$-$5.5}}\ifdim\frzwtmp>\bgwA\setlength{\bgwA}{\frzwtmp}\fi
\settowidth{\frzwtmp}{\footnotesize {\color{basetext}$-$10.7}}\ifdim\frzwtmp>\bgwA\setlength{\bgwA}{\frzwtmp}\fi
\settowidth{\frzwtmp}{\footnotesize 26.4}\ifdim\frzwtmp>\bgwA\setlength{\bgwA}{\frzwtmp}\fi
\settowidth{\frzwtmp}{\footnotesize 33.3}\ifdim\frzwtmp>\bgwA\setlength{\bgwA}{\frzwtmp}\fi
\settowidth{\frzwtmp}{\footnotesize {\color{basetext}$+$3.2}}\ifdim\frzwtmp>\bgwA\setlength{\bgwA}{\frzwtmp}\fi
\settowidth{\frzwtmp}{\footnotesize 33.8}\ifdim\frzwtmp>\bgwA\setlength{\bgwA}{\frzwtmp}\fi
\settowidth{\frzwtmp}{\footnotesize {\color{basetext}$+$3.3}}\ifdim\frzwtmp>\bgwA\setlength{\bgwA}{\frzwtmp}\fi
\settowidth{\frzwtmp}{\footnotesize 38.1}\ifdim\frzwtmp>\bgwA\setlength{\bgwA}{\frzwtmp}\fi
\settowidth{\frzwtmp}{\footnotesize {\color{basetext}$+$3.4}}\ifdim\frzwtmp>\bgwA\setlength{\bgwA}{\frzwtmp}\fi
\settowidth{\frzwtmp}{\footnotesize 39.2}\ifdim\frzwtmp>\bgwA\setlength{\bgwA}{\frzwtmp}\fi
\settowidth{\frzwtmp}{\footnotesize {\color{basetext}$+$3.7}}\ifdim\frzwtmp>\bgwA\setlength{\bgwA}{\frzwtmp}\fi
\settowidth{\frzwtmp}{\footnotesize {\scriptsize \%}}\ifdim\frzwtmp>\bgwA\setlength{\bgwA}{\frzwtmp}\fi
\settowidth{\frzwtmp}{\footnotesize {\scriptsize\color{basetext} $\Delta$F1$^{*}$ (pp)}}\ifdim\frzwtmp>\bgwA\setlength{\bgwA}{\frzwtmp}\fi
\ifdefined\bgwB\else\newlength{\bgwB}\fi\setlength{\bgwB}{0pt}
\settowidth{\frzwtmp}{\footnotesize $-$4.7}\ifdim\frzwtmp>\bgwB\setlength{\bgwB}{\frzwtmp}\fi
\settowidth{\frzwtmp}{\footnotesize {\color{basetext}0.73$\to$0.62}}\ifdim\frzwtmp>\bgwB\setlength{\bgwB}{\frzwtmp}\fi
\settowidth{\frzwtmp}{\footnotesize $-$0.6}\ifdim\frzwtmp>\bgwB\setlength{\bgwB}{\frzwtmp}\fi
\settowidth{\frzwtmp}{\footnotesize {\color{basetext}0.55$\to$0.52}}\ifdim\frzwtmp>\bgwB\setlength{\bgwB}{\frzwtmp}\fi
\settowidth{\frzwtmp}{\footnotesize \textbf{$+$0.4}}\ifdim\frzwtmp>\bgwB\setlength{\bgwB}{\frzwtmp}\fi
\settowidth{\frzwtmp}{\footnotesize {\color{basetext}0.53$\to$0.54}}\ifdim\frzwtmp>\bgwB\setlength{\bgwB}{\frzwtmp}\fi
\settowidth{\frzwtmp}{\footnotesize {\color{basetext}0.64$\to$0.53}}\ifdim\frzwtmp>\bgwB\setlength{\bgwB}{\frzwtmp}\fi
\settowidth{\frzwtmp}{\footnotesize $-$3.0}\ifdim\frzwtmp>\bgwB\setlength{\bgwB}{\frzwtmp}\fi
\settowidth{\frzwtmp}{\footnotesize {\color{basetext}0.67$\to$0.58}}\ifdim\frzwtmp>\bgwB\setlength{\bgwB}{\frzwtmp}\fi
\settowidth{\frzwtmp}{\footnotesize $-$1.8}\ifdim\frzwtmp>\bgwB\setlength{\bgwB}{\frzwtmp}\fi
\settowidth{\frzwtmp}{\footnotesize {\color{basetext}0.68$\to$0.61}}\ifdim\frzwtmp>\bgwB\setlength{\bgwB}{\frzwtmp}\fi
\settowidth{\frzwtmp}{\footnotesize \textcolor{worstcolor}{\itshape $-$7.3}}\ifdim\frzwtmp>\bgwB\setlength{\bgwB}{\frzwtmp}\fi
\settowidth{\frzwtmp}{\footnotesize {\color{basetext}0.73$\to$0.56}}\ifdim\frzwtmp>\bgwB\setlength{\bgwB}{\frzwtmp}\fi
\settowidth{\frzwtmp}{\footnotesize $-$2.8}\ifdim\frzwtmp>\bgwB\setlength{\bgwB}{\frzwtmp}\fi
\settowidth{\frzwtmp}{\footnotesize {\color{basetext}0.66$\to$0.60}}\ifdim\frzwtmp>\bgwB\setlength{\bgwB}{\frzwtmp}\fi
\settowidth{\frzwtmp}{\footnotesize $-$5.6}\ifdim\frzwtmp>\bgwB\setlength{\bgwB}{\frzwtmp}\fi
\settowidth{\frzwtmp}{\footnotesize {\color{basetext}0.72$\to$0.60}}\ifdim\frzwtmp>\bgwB\setlength{\bgwB}{\frzwtmp}\fi
\settowidth{\frzwtmp}{\footnotesize $-$3.6}\ifdim\frzwtmp>\bgwB\setlength{\bgwB}{\frzwtmp}\fi
\settowidth{\frzwtmp}{\footnotesize {\color{basetext}0.64$\to$0.56}}\ifdim\frzwtmp>\bgwB\setlength{\bgwB}{\frzwtmp}\fi
\settowidth{\frzwtmp}{\footnotesize $-$1.1}\ifdim\frzwtmp>\bgwB\setlength{\bgwB}{\frzwtmp}\fi
\settowidth{\frzwtmp}{\footnotesize {\color{basetext}0.60$\to$0.60}}\ifdim\frzwtmp>\bgwB\setlength{\bgwB}{\frzwtmp}\fi
\settowidth{\frzwtmp}{\footnotesize $-$1.6}\ifdim\frzwtmp>\bgwB\setlength{\bgwB}{\frzwtmp}\fi
\settowidth{\frzwtmp}{\footnotesize {\color{basetext}0.67$\to$0.64}}\ifdim\frzwtmp>\bgwB\setlength{\bgwB}{\frzwtmp}\fi
\settowidth{\frzwtmp}{\footnotesize $-$0.7}\ifdim\frzwtmp>\bgwB\setlength{\bgwB}{\frzwtmp}\fi
\settowidth{\frzwtmp}{\footnotesize {\color{basetext}0.65$\to$0.61}}\ifdim\frzwtmp>\bgwB\setlength{\bgwB}{\frzwtmp}\fi
\settowidth{\frzwtmp}{\footnotesize $-$0.9}\ifdim\frzwtmp>\bgwB\setlength{\bgwB}{\frzwtmp}\fi
\settowidth{\frzwtmp}{\footnotesize {\color{basetext}0.66$\to$0.62}}\ifdim\frzwtmp>\bgwB\setlength{\bgwB}{\frzwtmp}\fi
\settowidth{\frzwtmp}{\footnotesize $-$5.0}\ifdim\frzwtmp>\bgwB\setlength{\bgwB}{\frzwtmp}\fi
\settowidth{\frzwtmp}{\footnotesize {\color{basetext}0.74$\to$0.62}}\ifdim\frzwtmp>\bgwB\setlength{\bgwB}{\frzwtmp}\fi
\settowidth{\frzwtmp}{\footnotesize $-$4.5}\ifdim\frzwtmp>\bgwB\setlength{\bgwB}{\frzwtmp}\fi
\settowidth{\frzwtmp}{\footnotesize {\color{basetext}0.73$\to$0.63}}\ifdim\frzwtmp>\bgwB\setlength{\bgwB}{\frzwtmp}\fi
\settowidth{\frzwtmp}{\footnotesize $-$1.2}\ifdim\frzwtmp>\bgwB\setlength{\bgwB}{\frzwtmp}\fi
\settowidth{\frzwtmp}{\footnotesize {\color{basetext}0.56$\to$0.52}}\ifdim\frzwtmp>\bgwB\setlength{\bgwB}{\frzwtmp}\fi
\settowidth{\frzwtmp}{\footnotesize \textbf{$+$0.6}}\ifdim\frzwtmp>\bgwB\setlength{\bgwB}{\frzwtmp}\fi
\settowidth{\frzwtmp}{\footnotesize {\color{basetext}0.53$\to$0.55}}\ifdim\frzwtmp>\bgwB\setlength{\bgwB}{\frzwtmp}\fi
\settowidth{\frzwtmp}{\footnotesize $-$4.9}\ifdim\frzwtmp>\bgwB\setlength{\bgwB}{\frzwtmp}\fi
\settowidth{\frzwtmp}{\footnotesize $-$2.0}\ifdim\frzwtmp>\bgwB\setlength{\bgwB}{\frzwtmp}\fi
\settowidth{\frzwtmp}{\footnotesize {\color{basetext}0.65$\to$0.60}}\ifdim\frzwtmp>\bgwB\setlength{\bgwB}{\frzwtmp}\fi
\settowidth{\frzwtmp}{\footnotesize $-$0.2}\ifdim\frzwtmp>\bgwB\setlength{\bgwB}{\frzwtmp}\fi
\settowidth{\frzwtmp}{\footnotesize {\color{basetext}0.66$\to$0.64}}\ifdim\frzwtmp>\bgwB\setlength{\bgwB}{\frzwtmp}\fi
\settowidth{\frzwtmp}{\footnotesize \textcolor{worstcolor}{\itshape $-$6.8}}\ifdim\frzwtmp>\bgwB\setlength{\bgwB}{\frzwtmp}\fi
\settowidth{\frzwtmp}{\footnotesize {\color{basetext}0.72$\to$0.57}}\ifdim\frzwtmp>\bgwB\setlength{\bgwB}{\frzwtmp}\fi
\settowidth{\frzwtmp}{\footnotesize $-$2.1}\ifdim\frzwtmp>\bgwB\setlength{\bgwB}{\frzwtmp}\fi
\settowidth{\frzwtmp}{\footnotesize {\color{basetext}0.64$\to$0.61}}\ifdim\frzwtmp>\bgwB\setlength{\bgwB}{\frzwtmp}\fi
\settowidth{\frzwtmp}{\footnotesize {\color{basetext}0.72$\to$0.62}}\ifdim\frzwtmp>\bgwB\setlength{\bgwB}{\frzwtmp}\fi
\settowidth{\frzwtmp}{\footnotesize {\color{basetext}0.62$\to$0.60}}\ifdim\frzwtmp>\bgwB\setlength{\bgwB}{\frzwtmp}\fi
\settowidth{\frzwtmp}{\footnotesize $-$0.3}\ifdim\frzwtmp>\bgwB\setlength{\bgwB}{\frzwtmp}\fi
\settowidth{\frzwtmp}{\footnotesize {\color{basetext}0.60$\to$0.62}}\ifdim\frzwtmp>\bgwB\setlength{\bgwB}{\frzwtmp}\fi
\settowidth{\frzwtmp}{\footnotesize $-$1.0}\ifdim\frzwtmp>\bgwB\setlength{\bgwB}{\frzwtmp}\fi
\settowidth{\frzwtmp}{\footnotesize {\color{basetext}0.67$\to$0.65}}\ifdim\frzwtmp>\bgwB\setlength{\bgwB}{\frzwtmp}\fi
\settowidth{\frzwtmp}{\footnotesize $-$1.4}\ifdim\frzwtmp>\bgwB\setlength{\bgwB}{\frzwtmp}\fi
\settowidth{\frzwtmp}{\footnotesize {\color{basetext}0.64$\to$0.62}}\ifdim\frzwtmp>\bgwB\setlength{\bgwB}{\frzwtmp}\fi
\settowidth{\frzwtmp}{\footnotesize $-$1.7}\ifdim\frzwtmp>\bgwB\setlength{\bgwB}{\frzwtmp}\fi
\settowidth{\frzwtmp}{\footnotesize $-$5.2}\ifdim\frzwtmp>\bgwB\setlength{\bgwB}{\frzwtmp}\fi
\settowidth{\frzwtmp}{\footnotesize $-$3.5}\ifdim\frzwtmp>\bgwB\setlength{\bgwB}{\frzwtmp}\fi
\settowidth{\frzwtmp}{\footnotesize {\color{basetext}0.72$\to$0.65}}\ifdim\frzwtmp>\bgwB\setlength{\bgwB}{\frzwtmp}\fi
\settowidth{\frzwtmp}{\footnotesize {\color{basetext}0.57$\to$0.52}}\ifdim\frzwtmp>\bgwB\setlength{\bgwB}{\frzwtmp}\fi
\settowidth{\frzwtmp}{\footnotesize \textbf{$+$1.7}}\ifdim\frzwtmp>\bgwB\setlength{\bgwB}{\frzwtmp}\fi
\settowidth{\frzwtmp}{\footnotesize {\color{basetext}0.53$\to$0.57}}\ifdim\frzwtmp>\bgwB\setlength{\bgwB}{\frzwtmp}\fi
\settowidth{\frzwtmp}{\footnotesize {\color{basetext}0.63$\to$0.54}}\ifdim\frzwtmp>\bgwB\setlength{\bgwB}{\frzwtmp}\fi
\settowidth{\frzwtmp}{\footnotesize $+$0.1}\ifdim\frzwtmp>\bgwB\setlength{\bgwB}{\frzwtmp}\fi
\settowidth{\frzwtmp}{\footnotesize {\color{basetext}0.63$\to$0.63}}\ifdim\frzwtmp>\bgwB\setlength{\bgwB}{\frzwtmp}\fi
\settowidth{\frzwtmp}{\footnotesize $-$0.1}\ifdim\frzwtmp>\bgwB\setlength{\bgwB}{\frzwtmp}\fi
\settowidth{\frzwtmp}{\footnotesize {\color{basetext}0.65$\to$0.64}}\ifdim\frzwtmp>\bgwB\setlength{\bgwB}{\frzwtmp}\fi
\settowidth{\frzwtmp}{\footnotesize \textcolor{worstcolor}{\itshape $-$6.2}}\ifdim\frzwtmp>\bgwB\setlength{\bgwB}{\frzwtmp}\fi
\settowidth{\frzwtmp}{\footnotesize {\color{basetext}0.72$\to$0.58}}\ifdim\frzwtmp>\bgwB\setlength{\bgwB}{\frzwtmp}\fi
\settowidth{\frzwtmp}{\footnotesize {\color{basetext}0.73$\to$0.61}}\ifdim\frzwtmp>\bgwB\setlength{\bgwB}{\frzwtmp}\fi
\settowidth{\frzwtmp}{\footnotesize $-$0.4}\ifdim\frzwtmp>\bgwB\setlength{\bgwB}{\frzwtmp}\fi
\settowidth{\frzwtmp}{\footnotesize {\color{basetext}0.62$\to$0.62}}\ifdim\frzwtmp>\bgwB\setlength{\bgwB}{\frzwtmp}\fi
\settowidth{\frzwtmp}{\footnotesize {\color{basetext}0.65$\to$0.66}}\ifdim\frzwtmp>\bgwB\setlength{\bgwB}{\frzwtmp}\fi
\settowidth{\frzwtmp}{\footnotesize $-$1.5}\ifdim\frzwtmp>\bgwB\setlength{\bgwB}{\frzwtmp}\fi
\settowidth{\frzwtmp}{\footnotesize {\color{basetext}0.65$\to$0.62}}\ifdim\frzwtmp>\bgwB\setlength{\bgwB}{\frzwtmp}\fi
\settowidth{\frzwtmp}{\footnotesize {\color{basetext}0.65$\to$0.63}}\ifdim\frzwtmp>\bgwB\setlength{\bgwB}{\frzwtmp}\fi
\settowidth{\frzwtmp}{\footnotesize $-$4.6}\ifdim\frzwtmp>\bgwB\setlength{\bgwB}{\frzwtmp}\fi
\settowidth{\frzwtmp}{\footnotesize {\color{basetext}0.71$\to$0.62}}\ifdim\frzwtmp>\bgwB\setlength{\bgwB}{\frzwtmp}\fi
\settowidth{\frzwtmp}{\footnotesize {\scriptsize $10^{-3}$/month}}\ifdim\frzwtmp>\bgwB\setlength{\bgwB}{\frzwtmp}\fi
\settowidth{\frzwtmp}{\footnotesize {\scriptsize\color{basetext} first$\to$last}}\ifdim\frzwtmp>\bgwB\setlength{\bgwB}{\frzwtmp}\fi
\ifdefined\bgwC\else\newlength{\bgwC}\fi\setlength{\bgwC}{0pt}
\settowidth{\frzwtmp}{\footnotesize $-$0.01}\ifdim\frzwtmp>\bgwC\setlength{\bgwC}{\frzwtmp}\fi
\settowidth{\frzwtmp}{\footnotesize {\color{basetext}1.1}}\ifdim\frzwtmp>\bgwC\setlength{\bgwC}{\frzwtmp}\fi
\settowidth{\frzwtmp}{\footnotesize $-$0.04}\ifdim\frzwtmp>\bgwC\setlength{\bgwC}{\frzwtmp}\fi
\settowidth{\frzwtmp}{\footnotesize {\color{basetext}4.0}}\ifdim\frzwtmp>\bgwC\setlength{\bgwC}{\frzwtmp}\fi
\settowidth{\frzwtmp}{\footnotesize {\color{basetext}2.3}}\ifdim\frzwtmp>\bgwC\setlength{\bgwC}{\frzwtmp}\fi
\settowidth{\frzwtmp}{\footnotesize {\color{basetext}1.2}}\ifdim\frzwtmp>\bgwC\setlength{\bgwC}{\frzwtmp}\fi
\settowidth{\frzwtmp}{\footnotesize $-$0.03}\ifdim\frzwtmp>\bgwC\setlength{\bgwC}{\frzwtmp}\fi
\settowidth{\frzwtmp}{\footnotesize $-$0.02}\ifdim\frzwtmp>\bgwC\setlength{\bgwC}{\frzwtmp}\fi
\settowidth{\frzwtmp}{\footnotesize $+$0.01}\ifdim\frzwtmp>\bgwC\setlength{\bgwC}{\frzwtmp}\fi
\settowidth{\frzwtmp}{\footnotesize {\color{basetext}1.0}}\ifdim\frzwtmp>\bgwC\setlength{\bgwC}{\frzwtmp}\fi
\settowidth{\frzwtmp}{\footnotesize 0.00}\ifdim\frzwtmp>\bgwC\setlength{\bgwC}{\frzwtmp}\fi
\settowidth{\frzwtmp}{\footnotesize {\color{basetext}1.3}}\ifdim\frzwtmp>\bgwC\setlength{\bgwC}{\frzwtmp}\fi
\settowidth{\frzwtmp}{\footnotesize {\color{basetext}2.2}}\ifdim\frzwtmp>\bgwC\setlength{\bgwC}{\frzwtmp}\fi
\settowidth{\frzwtmp}{\footnotesize $-$0.12}\ifdim\frzwtmp>\bgwC\setlength{\bgwC}{\frzwtmp}\fi
\settowidth{\frzwtmp}{\footnotesize {\color{basetext}6.3}}\ifdim\frzwtmp>\bgwC\setlength{\bgwC}{\frzwtmp}\fi
\settowidth{\frzwtmp}{\footnotesize $-$0.05}\ifdim\frzwtmp>\bgwC\setlength{\bgwC}{\frzwtmp}\fi
\settowidth{\frzwtmp}{\footnotesize {\color{basetext}5.6}}\ifdim\frzwtmp>\bgwC\setlength{\bgwC}{\frzwtmp}\fi
\settowidth{\frzwtmp}{\footnotesize {\color{basetext}2.6}}\ifdim\frzwtmp>\bgwC\setlength{\bgwC}{\frzwtmp}\fi
\settowidth{\frzwtmp}{\footnotesize {\color{basetext}2.5}}\ifdim\frzwtmp>\bgwC\setlength{\bgwC}{\frzwtmp}\fi
\settowidth{\frzwtmp}{\footnotesize {\color{basetext}2.0}}\ifdim\frzwtmp>\bgwC\setlength{\bgwC}{\frzwtmp}\fi
\settowidth{\frzwtmp}{\footnotesize {\color{basetext}2.8}}\ifdim\frzwtmp>\bgwC\setlength{\bgwC}{\frzwtmp}\fi
\settowidth{\frzwtmp}{\footnotesize {\color{basetext}2.4}}\ifdim\frzwtmp>\bgwC\setlength{\bgwC}{\frzwtmp}\fi
\settowidth{\frzwtmp}{\footnotesize {\color{basetext}2.7}}\ifdim\frzwtmp>\bgwC\setlength{\bgwC}{\frzwtmp}\fi
\settowidth{\frzwtmp}{\footnotesize $+$0.03}\ifdim\frzwtmp>\bgwC\setlength{\bgwC}{\frzwtmp}\fi
\settowidth{\frzwtmp}{\footnotesize {\color{basetext}4.8}}\ifdim\frzwtmp>\bgwC\setlength{\bgwC}{\frzwtmp}\fi
\settowidth{\frzwtmp}{\footnotesize $-$0.22}\ifdim\frzwtmp>\bgwC\setlength{\bgwC}{\frzwtmp}\fi
\settowidth{\frzwtmp}{\footnotesize {\color{basetext}7.6}}\ifdim\frzwtmp>\bgwC\setlength{\bgwC}{\frzwtmp}\fi
\settowidth{\frzwtmp}{\footnotesize $-$0.20\rlap{$^{\dagger}$}}\ifdim\frzwtmp>\bgwC\setlength{\bgwC}{\frzwtmp}\fi
\settowidth{\frzwtmp}{\footnotesize {\color{basetext}13.0}\rlap{$^{\dagger}$}}\ifdim\frzwtmp>\bgwC\setlength{\bgwC}{\frzwtmp}\fi
\settowidth{\frzwtmp}{\footnotesize {\color{basetext}5.7}}\ifdim\frzwtmp>\bgwC\setlength{\bgwC}{\frzwtmp}\fi
\settowidth{\frzwtmp}{\footnotesize {\color{basetext}5.1}}\ifdim\frzwtmp>\bgwC\setlength{\bgwC}{\frzwtmp}\fi
\settowidth{\frzwtmp}{\footnotesize {\color{basetext}4.9}}\ifdim\frzwtmp>\bgwC\setlength{\bgwC}{\frzwtmp}\fi
\settowidth{\frzwtmp}{\footnotesize $-$0.10}\ifdim\frzwtmp>\bgwC\setlength{\bgwC}{\frzwtmp}\fi
\settowidth{\frzwtmp}{\footnotesize {\color{basetext}5.2}}\ifdim\frzwtmp>\bgwC\setlength{\bgwC}{\frzwtmp}\fi
\settowidth{\frzwtmp}{\footnotesize $-$0.06}\ifdim\frzwtmp>\bgwC\setlength{\bgwC}{\frzwtmp}\fi
\settowidth{\frzwtmp}{\footnotesize {\color{basetext}4.4}}\ifdim\frzwtmp>\bgwC\setlength{\bgwC}{\frzwtmp}\fi
\settowidth{\frzwtmp}{\footnotesize {\color{basetext}5.0}}\ifdim\frzwtmp>\bgwC\setlength{\bgwC}{\frzwtmp}\fi
\settowidth{\frzwtmp}{\footnotesize $-$0.08}\ifdim\frzwtmp>\bgwC\setlength{\bgwC}{\frzwtmp}\fi
\settowidth{\frzwtmp}{\footnotesize {\color{basetext}5.8}}\ifdim\frzwtmp>\bgwC\setlength{\bgwC}{\frzwtmp}\fi
\settowidth{\frzwtmp}{\footnotesize {\scriptsize pp/month}}\ifdim\frzwtmp>\bgwC\setlength{\bgwC}{\frzwtmp}\fi
\settowidth{\frzwtmp}{\footnotesize {\scriptsize\color{basetext} $\bar{q}^{*}$ (\%)}}\ifdim\frzwtmp>\bgwC\setlength{\bgwC}{\frzwtmp}\fi
\begin{table*}[p]
  \centering
  \caption{\codename{} evaluation on \textbf{Hypercube VTT2} at the lower monthly budgets $B_M\in\{50,100,200\}$, in the layout and metric set of Table~\ref{tab:main-vtt2-b400} (full initial 2021 training set, calibration Jan--Jun 2022, evaluation Jul 2022--Dec 2024). The Cap batch size and the rejection quota equal $B_M$ ($\rho=B_M$). Values are computed per month within seed, summarized over the 30 months and averaged over five complete seeds (IDs 1--5) for every row. \textbf{Bold}=best, {\color{worstcolor}\itshape dark-red italics}=worst displayed value per ranked column and $B_M$ block.}
  \label{tab:vtt2-lower-budgets}
  \footnotesize
  \setlength{\tabcolsep}{4.2pt}
  \renewcommand{\arraystretch}{0.95}
  \begin{tabular}{@{}cl@{\hspace{7pt}}c@{\hspace{7pt}}ccw{c}{\bgwA}w{c}{\bgwA}@{\hspace{7pt}}cw{c}{\bgwB}w{c}{\bgwB}w{c}{\bgwC}w{c}{\bgwC}}
    \toprule
     &  & \multicolumn{1}{>{\columncolor{tabbaseband}[\tabcolsep][\tabcolsep]}c}{Baseline} & \multicolumn{4}{>{\columncolor{tabreliband}[\tabcolsep][\tabcolsep]}c}{Reliability} & \multicolumn{5}{>{\columncolor{tabstabband}[\tabcolsep][\tabcolsep]}c}{Stability} \\
    $B_M$ & Method & \cellcolor{tabbaseband}F1 $\uparrow$ & \cellcolor{tabreliband}E-AURC $\downarrow$ & \cellcolor{tabreliband}Cap$_{B_M}$ $\uparrow$ & \multicolumn{2}{>{\columncolor{tabreliband}[\tabcolsep][\tabcolsep]}c}{\makebox[0pt][c]{F1$^{*}$ $\uparrow$}\makebox[0pt][c]{\smash{\raisebox{-2.6pt}{\rule{\dimexpr 2\bgwA+4\tabcolsep-5pt\relax}{0.4pt}}}}} & \cellcolor{tabstabband}Sen[F1] $\uparrow$ & \multicolumn{2}{>{\columncolor{tabstabband}[\tabcolsep][\tabcolsep]}c}{\makebox[0pt][c]{Sen[AUROC$_f$] $\uparrow$}\makebox[0pt][c]{\smash{\raisebox{-2.6pt}{\rule{\dimexpr 2\bgwB+4\tabcolsep-5pt\relax}{0.4pt}}}}} & \multicolumn{2}{>{\columncolor{tabstabband}[\tabcolsep][\tabcolsep]}c}{\makebox[0pt][c]{Sen[$q^{*}$]}\makebox[0pt][c]{\smash{\raisebox{-2.6pt}{\rule{\dimexpr 2\bgwC+4\tabcolsep-5pt\relax}{0.4pt}}}}} \\
     &  & \cellcolor{tabbaseband}{\scriptsize \%} & \cellcolor{tabreliband}{\scriptsize $\times$100} & \cellcolor{tabreliband}{\scriptsize \% of errors} & \cellcolor{tabreliband}{\scriptsize \%} & \cellcolor{tabreliband}{\scriptsize\color{basetext} $\Delta$F1$^{*}$ (pp)} & \cellcolor{tabstabband}{\scriptsize pp/month} & \cellcolor{tabstabband}{\scriptsize $10^{-3}$/month} & \cellcolor{tabstabband}{\scriptsize\color{basetext} first$\to$last} & \cellcolor{tabstabband}{\scriptsize pp/month} & \cellcolor{tabstabband}{\scriptsize\color{basetext} $\bar{q}^{*}$ (\%)} \\
    \midrule
    \multirow{17}{*}{50} & DeepDrebin & 29.9 & 5.07 & 6.5 & 28.3 & {\color{basetext}$-$1.5} & $-$1.38 & $-$4.7 & {\color{basetext}0.73$\to$0.62} & $-$0.01 & {\color{basetext}1.1} \\
     & CADE-OOD$^{c}$ & 26.9 & \textcolor{worstcolor}{\itshape 8.27} & 4.5 & 28.2 & {\color{basetext}$+$1.3} & $-$1.28 & $-$0.6 & {\color{basetext}0.55$\to$0.52} & $-$0.04 & {\color{basetext}4.0} \\
     & CADE-MSP$^{c}$ & 26.9 & 7.84 & 6.0 & \textcolor{worstcolor}{\itshape 18.9} & {\color{basetext}$-$8.2} & $-$1.28 & \textbf{$+$0.4} & {\color{basetext}0.53$\to$0.54} & $-$0.01 & {\color{basetext}2.3} \\
     & HCC-PL$^{a}$ & 27.9 & 6.56 & \textcolor{worstcolor}{\itshape 4.3} & 28.1 & {\color{basetext}$+$0.2} & $-$1.33 & $-$4.7 & {\color{basetext}0.64$\to$0.53} & $-$0.01 & {\color{basetext}1.2} \\
     & HCC-MSP$^{a}$ & 27.9 & 6.88 & 5.9 & 26.9 & {\color{basetext}$-$1.0} & $-$1.33 & $-$3.0 & {\color{basetext}0.67$\to$0.58} & $-$0.03 & {\color{basetext}1.2} \\
     & Drebin & 24.2 & 5.77 & 6.0 & 21.4 & {\color{basetext}$-$2.8} & $-$1.21 & $-$1.8 & {\color{basetext}0.68$\to$0.61} & $-$0.02 & {\color{basetext}1.1} \\
     & Random Forest & 24.7 & 5.62 & 6.0 & 21.8 & {\color{basetext}$-$2.8} & $-$1.16 & \textcolor{worstcolor}{\itshape $-$7.3} & {\color{basetext}0.73$\to$0.56} & $+$0.01 & {\color{basetext}1.2} \\
     & SecSVM & \textcolor{worstcolor}{\itshape 23.5} & 5.79 & 5.5 & 21.2 & {\color{basetext}$-$2.3} & \textcolor{worstcolor}{\itshape $-$1.55} & $-$2.8 & {\color{basetext}0.66$\to$0.60} & $-$0.03 & {\color{basetext}1.0} \\
     & DRMD & \textbf{33.1} & 5.70 & 6.2 & \textbf{32.3} & {\color{basetext}$-$0.8} & $-$1.39 & $-$5.6 & {\color{basetext}0.72$\to$0.60} & $-$0.01 & {\color{basetext}1.2} \\
     & RAMDA & 31.0 & 7.66 & \textbf{7.2} & 29.9 & {\color{basetext}$-$1.1} & $-$1.33 & $-$3.6 & {\color{basetext}0.64$\to$0.56} & $-$0.02 & {\color{basetext}1.2} \\
     & TIF & 30.5 & 6.55 & 6.4 & 28.9 & {\color{basetext}$-$1.6} & $-$1.46 & $-$1.1 & {\color{basetext}0.60$\to$0.60} & 0.00 & {\color{basetext}1.3} \\
     & MalCertain & 30.3 & 7.60 & 6.7 & 28.9 & {\color{basetext}$-$1.4} & \textbf{$-$1.15} & $-$1.6 & {\color{basetext}0.67$\to$0.64} & $-$0.02 & {\color{basetext}1.0} \\
    \addlinespace[2pt]
     & \multicolumn{11}{>{\columncolor{subheadcolor}[\tabcolsep][\tabcolsep]}l}{\itshape Transcendent: inductive (ICE) and cross-conformal (CCE)$^{b}$} \\
     & ICE (Drebin) & 25.2 & 6.01 & 4.8 & 25.9 & {\color{basetext}$+$0.7} & $-$1.17 & $-$0.7 & {\color{basetext}0.65$\to$0.61} & $-$0.02 & {\color{basetext}1.1} \\
     & CCE (Drebin) & 25.4 & 5.84 & 5.1 & 26.2 & {\color{basetext}$+$0.8} & $-$1.17 & $-$0.9 & {\color{basetext}0.66$\to$0.62} & $-$0.02 & {\color{basetext}1.1} \\
     & ICE (DeepDrebin) & 30.3 & 5.14 & \textcolor{worstcolor}{\itshape 4.3} & 31.1 & {\color{basetext}$+$0.7} & $-$1.31 & $-$5.0 & {\color{basetext}0.74$\to$0.62} & $-$0.01 & {\color{basetext}1.2} \\
     & CCE (DeepDrebin) & 31.4 & \textbf{4.95} & 4.7 & 32.2 & {\color{basetext}$+$0.8} & $-$1.45 & $-$4.7 & {\color{basetext}0.74$\to$0.62} & $-$0.01 & {\color{basetext}1.2} \\
    \midrule
    \multirow{17}{*}{100} & DeepDrebin & 32.9 & \textbf{4.85} & 11.3 & 29.8 & {\color{basetext}$-$3.1} & $-$1.30 & $-$4.5 & {\color{basetext}0.73$\to$0.63} & 0.00 & {\color{basetext}2.2} \\
     & CADE-OOD$^{c}$ & 31.2 & \textcolor{worstcolor}{\itshape 7.93} & 7.7 & 33.4 & {\color{basetext}$+$2.2} & $-$1.15 & $-$1.2 & {\color{basetext}0.56$\to$0.52} & $-$0.12 & {\color{basetext}6.3} \\
     & CADE-MSP$^{c}$ & 31.2 & 7.53 & 8.7 & \textcolor{worstcolor}{\itshape 16.6} & {\color{basetext}$-$15.2} & $-$1.15 & \textbf{$+$0.6} & {\color{basetext}0.53$\to$0.55} & $-$0.05 & {\color{basetext}5.6} \\
     & HCC-PL$^{a}$ & 30.2 & 6.40 & \textcolor{worstcolor}{\itshape 7.3} & 30.6 & {\color{basetext}$+$0.4} & $-$1.32 & $-$4.9 & {\color{basetext}0.64$\to$0.53} & 0.00 & {\color{basetext}2.6} \\
     & HCC-MSP$^{a}$ & 30.2 & 6.42 & 10.2 & 27.4 & {\color{basetext}$-$2.8} & $-$1.32 & $-$2.0 & {\color{basetext}0.65$\to$0.60} & $-$0.03 & {\color{basetext}2.6} \\
     & Drebin & 25.3 & 5.65 & 10.8 & 20.7 & {\color{basetext}$-$4.6} & $-$1.30 & $-$0.2 & {\color{basetext}0.66$\to$0.64} & $-$0.04 & {\color{basetext}2.3} \\
     & Random Forest & 28.0 & 5.70 & 10.6 & 22.7 & {\color{basetext}$-$5.3} & \textbf{$-$1.13} & \textcolor{worstcolor}{\itshape $-$6.8} & {\color{basetext}0.72$\to$0.57} & $-$0.02 & {\color{basetext}2.5} \\
     & SecSVM & \textcolor{worstcolor}{\itshape 24.4} & 5.73 & 10.0 & 21.3 & {\color{basetext}$-$3.0} & \textcolor{worstcolor}{\itshape $-$1.56} & $-$2.1 & {\color{basetext}0.64$\to$0.61} & $-$0.04 & {\color{basetext}2.0} \\
     & DRMD & \textbf{36.0} & 5.50 & 11.0 & \textbf{35.3} & {\color{basetext}$-$0.7} & $-$1.20 & $-$4.9 & {\color{basetext}0.72$\to$0.62} & $-$0.01 & {\color{basetext}2.5} \\
     & RAMDA & 33.3 & 7.15 & \textbf{12.0} & 30.5 & {\color{basetext}$-$2.8} & $-$1.24 & $-$1.2 & {\color{basetext}0.62$\to$0.60} & $-$0.02 & {\color{basetext}2.8} \\
     & TIF & 33.6 & 6.17 & 11.5 & 30.1 & {\color{basetext}$-$3.5} & $-$1.41 & $-$0.3 & {\color{basetext}0.60$\to$0.62} & $-$0.01 & {\color{basetext}2.4} \\
     & MalCertain & 29.9 & 7.15 & 11.3 & 27.5 & {\color{basetext}$-$2.4} & $-$1.25 & $-$1.0 & {\color{basetext}0.67$\to$0.65} & $-$0.04 & {\color{basetext}2.0} \\
    \addlinespace[2pt]
     & \multicolumn{11}{>{\columncolor{subheadcolor}[\tabcolsep][\tabcolsep]}l}{\itshape Transcendent: inductive (ICE) and cross-conformal (CCE)$^{b}$} \\
     & ICE (Drebin) & 27.5 & 5.68 & 9.2 & 29.0 & {\color{basetext}$+$1.5} & $-$1.28 & $-$1.4 & {\color{basetext}0.64$\to$0.62} & $-$0.03 & {\color{basetext}2.3} \\
     & CCE (Drebin) & 27.8 & 5.56 & 9.2 & 29.2 & {\color{basetext}$+$1.5} & $-$1.28 & $-$1.7 & {\color{basetext}0.66$\to$0.62} & $-$0.03 & {\color{basetext}2.3} \\
     & ICE (DeepDrebin) & 32.7 & 4.95 & 8.3 & 34.3 & {\color{basetext}$+$1.6} & $-$1.18 & $-$5.2 & {\color{basetext}0.74$\to$0.62} & $-$0.02 & {\color{basetext}2.7} \\
     & CCE (DeepDrebin) & 33.4 & \textbf{4.85} & 8.9 & 35.1 & {\color{basetext}$+$1.8} & $-$1.22 & $-$4.5 & {\color{basetext}0.72$\to$0.62} & 0.00 & {\color{basetext}2.7} \\
    \midrule
    \multirow{17}{*}{200} & DeepDrebin & 35.4 & \textbf{4.69} & 19.0 & 31.1 & {\color{basetext}$-$4.4} & $-$1.39 & $-$3.5 & {\color{basetext}0.72$\to$0.65} & $+$0.03 & {\color{basetext}4.8} \\
     & CADE-OOD$^{c}$ & 34.4 & \textcolor{worstcolor}{\itshape 7.77} & \textcolor{worstcolor}{\itshape 11.2} & 37.0 & {\color{basetext}$+$2.6} & $-$1.07 & $-$1.7 & {\color{basetext}0.57$\to$0.52} & $-$0.22 & {\color{basetext}7.6} \\
     & CADE-MSP$^{c}$ & 34.4 & 7.45 & 12.7 & 7.3\rlap{$^{\dagger}$} & {\color{basetext}$-$27.7}\rlap{$^{\dagger}$} & $-$1.07 & \textbf{$+$1.7} & {\color{basetext}0.53$\to$0.57} & $-$0.20\rlap{$^{\dagger}$} & {\color{basetext}13.0}\rlap{$^{\dagger}$} \\
     & HCC-PL$^{a}$ & 35.3 & 6.32 & 12.9 & 36.7 & {\color{basetext}$+$1.4} & $-$1.14 & $-$3.6 & {\color{basetext}0.63$\to$0.54} & $+$0.01 & {\color{basetext}5.7} \\
     & HCC-MSP$^{a}$ & 35.3 & 5.95 & 18.2 & 29.5 & {\color{basetext}$-$5.8} & $-$1.14 & $+$0.1 & {\color{basetext}0.63$\to$0.63} & $-$0.02 & {\color{basetext}5.1} \\
     & Drebin & 28.7 & 5.43 & 18.1 & 22.8 & {\color{basetext}$-$5.8} & $-$1.43 & $-$0.1 & {\color{basetext}0.65$\to$0.64} & $-$0.02 & {\color{basetext}4.9} \\
     & Random Forest & 32.8 & 5.51 & 17.9 & 24.2 & {\color{basetext}$-$8.7} & \textbf{$-$1.01} & \textcolor{worstcolor}{\itshape $-$6.2} & {\color{basetext}0.72$\to$0.58} & $-$0.10 & {\color{basetext}5.2} \\
     & SecSVM & \textcolor{worstcolor}{\itshape 25.6} & 5.64 & 17.7 & \textcolor{worstcolor}{\itshape 21.9} & {\color{basetext}$-$3.7} & \textcolor{worstcolor}{\itshape $-$1.58} & $-$1.8 & {\color{basetext}0.64$\to$0.62} & $-$0.06 & {\color{basetext}4.4} \\
     & DRMD & \textbf{40.2} & 5.53 & 18.6 & \textbf{39.6} & {\color{basetext}$-$0.6} & $-$1.19 & $-$5.2 & {\color{basetext}0.73$\to$0.61} & $-$0.02 & {\color{basetext}5.2} \\
     & RAMDA & 36.5 & 6.80 & \textbf{19.2} & 31.0 & {\color{basetext}$-$5.5} & $-$1.20 & $-$0.4 & {\color{basetext}0.60$\to$0.60} & $-$0.02 & {\color{basetext}5.0} \\
     & TIF & 34.9 & 5.94 & 17.6 & 24.2 & {\color{basetext}$-$10.7} & $-$1.40 & $-$0.4 & {\color{basetext}0.62$\to$0.62} & $+$0.01 & {\color{basetext}5.1} \\
     & MalCertain & 30.1 & 6.64 & 18.5 & 26.4 & {\color{basetext}$-$3.7} & $-$1.32 & $-$0.6 & {\color{basetext}0.65$\to$0.66} & $-$0.06 & {\color{basetext}4.4} \\
    \addlinespace[2pt]
     & \multicolumn{11}{>{\columncolor{subheadcolor}[\tabcolsep][\tabcolsep]}l}{\itshape Transcendent: inductive (ICE) and cross-conformal (CCE)$^{b}$} \\
     & ICE (Drebin) & 30.1 & 5.40 & 16.9 & 33.3 & {\color{basetext}$+$3.2} & $-$1.34 & $-$1.5 & {\color{basetext}0.65$\to$0.62} & $-$0.08 & {\color{basetext}4.9} \\
     & CCE (Drebin) & 30.5 & 5.33 & 17.1 & 33.8 & {\color{basetext}$+$3.3} & $-$1.33 & $-$1.6 & {\color{basetext}0.65$\to$0.63} & $-$0.08 & {\color{basetext}4.9} \\
     & ICE (DeepDrebin) & 34.7 & 4.92 & 15.1 & 38.1 & {\color{basetext}$+$3.4} & $-$1.30 & $-$5.2 & {\color{basetext}0.73$\to$0.62} & 0.00 & {\color{basetext}5.7} \\
     & CCE (DeepDrebin) & 35.5 & 4.77 & 16.1 & 39.2 & {\color{basetext}$+$3.7} & $-$1.45 & $-$4.6 & {\color{basetext}0.71$\to$0.62} & $+$0.03 & {\color{basetext}5.8} \\
    \bottomrule
  \end{tabular}
  \par\smallskip
  {\scriptsize\raggedright Cap$_{B_M}$: share of the month's misclassifications among its $B_M$ most uncertain samples (random reference 1.5\%, 2.9\%, 5.8\% at $B_M=50,100,200$). F1$^{*}$, $q^{*}$: under a threshold re-calibrated each month to $B_M$ referrals/month on the six preceding months. Lower-budget runs come from a separate campaign and are not paired with the Table~\ref{tab:main-vtt2-b400} trajectories; differences across budgets are not strictly paired budget effects. $^{a}$HCC: benchmark training with common uncertainty acquisition; MSP and pseudo-loss (PL) share checkpoints, predictions and MSP acquisition. $^{b}$Transcendent: ICE one fixed fold, CCE five folds; both share CCE-credibility acquisition. $^{c}$CADE: pinned HCC reference adaptation; OOD and MSP share predictions and OOD acquisition. $^{\dagger}$Tied scores make the rolling window miss its referral target by more than 25\% on average (CADE-MSP at $B_M=200$ (seeds 1, 2, 3, 5)); quota-dependent cells are unranked. Months with fewer than 20 accepted samples have no F1$^{*}$ (CADE-MSP at $B_M=50$: 2 of 150 seed-months; CADE-MSP at $B_M=100$: 6 of 150 seed-months; CADE-MSP at $B_M=200$: 16 of 150 seed-months).\par}
\end{table*}

\begin{figure}[!t]
\centering
\includegraphics[width=\columnwidth]{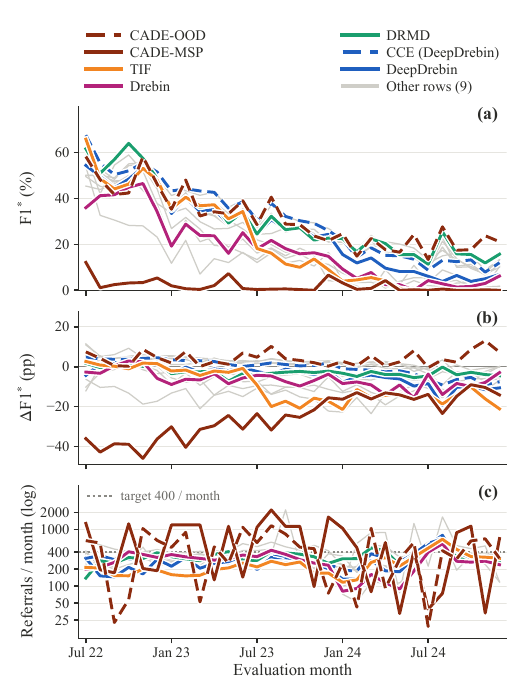}
\caption{Hypercube VTT2 at $\budgetmonthlymath=50$ with a fixed rejection quota $\rho=400$ under the rolling calibration of \S\ref{subsec:operational_assessment}: (a) accepted-set F1$^{*}$, (b) $\Delta$F1$^{*}$ and (c) realized referrals per month (log scale), averaged over all five seeds of every row in Table~\ref{tab:vtt2-lower-budgets}. Highlighted rows as in Fig.~\ref{fig:eaurc_vtt2}(a). Most rows refer fewer than 400 samples per month (median over rows about 330); CADE-MSP (mean 706, monthly range 33--2,216), CADE-OOD (485) and RAMDA (484) exceed the quota.}
\label{fig:overview_rejection}
\end{figure}

\section{Additional Datasets: AndroZoo, APIGraph and Transcendent}
\label{appendix:prior_hypercube}
\label{appendix:legacy_results}

Tables~\ref{tab:legacy-azoo}--\ref{tab:legacy-trans} report the AndroZoo and APIGraph datasets~\cite{HCC} and the Transcendent dataset~\cite{barbero2022transcendent} in the metric set, threshold protocol and layout of Table~\ref{tab:main-vtt2-b400}. No model was retrained for these tables; all metrics are computed from the stored predictions and scores. As in Table~\ref{tab:main-vtt2-b400}, the first six test months of each dataset calibrate and the remaining months evaluate, the threshold is re-calibrated each month on the six preceding months, and every quantity is computed per seed before averaging; the Cap batch size and the rejection quota $\rho$ equal $\budgetmonthlymath$.

These datasets differ from the VTT2 comparison in horizon and training protocol, and their numbers are not combined with Tables~\ref{tab:main-vtt2-b400} and~\ref{tab:vtt2-lower-budgets}. HCC is family-supervised with warm updates and CADE uses a cold-refitted neural head, not the VTT2 benchmark adaptations; DeepDrebin and Drebin use cold monthly refits (Appendix~\ref{hyperparameters_configurations}). The Transcendent ICE/CCE rows come from a separate conformal grid (seeds 0--4) with shared CCE-credibility acquisition. Random Forest, SecSVM, DRMD, RAMDA, TIF and MalCertain were not run on these datasets.

Three readings carry over from the VTT2 comparison. First, the lowest E-AURC belongs to DeepDrebin or CCE~(DeepDrebin) on APIGraph and Transcendent at every budget, and to CCE~(Drebin) on AndroZoo, while a CADE score has the highest E-AURC on every dataset and budget except AndroZoo at $\budgetmonthlymath\in\{200,400\}$, consistent with Figure~\ref{fig:example_cade_bad}. Second, the highest F1 and the lowest E-AURC again fall to different methods on AndroZoo, where HCC-PL leads F1 at every budget. Third, the decline is steepest on Hypercube: on the Transcendent dataset 27 of 40 rows have a negative Sen[F1] (down to $-0.55$\,pp/month), whereas the APIGraph slopes lie within 0.12\,pp/month of zero and most AndroZoo slopes are positive. Abstention helps more on these datasets than on VTT2: DeepDrebin and the four Transcendent rows have a positive $\Delta$F1$^{*}$ on every dataset and budget.

% Generated by generate_satml_vtt2_budget_tables.py; do not edit numbers by hand.
% Layout, metrics and threshold protocol are those of Table I (generate_satml_vtt2_frozen_table.py).
\providecolor{tabbaseband}{RGB}{240,240,240}
\providecolor{tabreliband}{RGB}{230,245,220}
\providecolor{tabstabband}{RGB}{235,240,250}
\providecolor{worstcolor}{RGB}{176,32,32}
\providecolor{subheadcolor}{RGB}{246,246,246}
\providecolor{basetext}{RGB}{90,90,90}
\providecolor{relitext}{RGB}{46,125,50}
\providecolor{stabtext}{RGB}{40,53,147}
\providecommand{\codename}{\textsc{Aurora}}
\ifdefined\frzwtmp\else\newlength{\frzwtmp}\fi
\ifdefined\lgwAA\else\newlength{\lgwAA}\fi\setlength{\lgwAA}{0pt}
\settowidth{\frzwtmp}{\footnotesize \textcolor{worstcolor}{\itshape 55.2}}\ifdim\frzwtmp>\lgwAA\setlength{\lgwAA}{\frzwtmp}\fi
\settowidth{\frzwtmp}{\footnotesize {\color{basetext}$+$1.0}}\ifdim\frzwtmp>\lgwAA\setlength{\lgwAA}{\frzwtmp}\fi
\settowidth{\frzwtmp}{\footnotesize 62.4\rlap{$^{\dagger}$}}\ifdim\frzwtmp>\lgwAA\setlength{\lgwAA}{\frzwtmp}\fi
\settowidth{\frzwtmp}{\footnotesize {\color{basetext}0.0}\rlap{$^{\dagger}$}}\ifdim\frzwtmp>\lgwAA\setlength{\lgwAA}{\frzwtmp}\fi
\settowidth{\frzwtmp}{\footnotesize 60.7}\ifdim\frzwtmp>\lgwAA\setlength{\lgwAA}{\frzwtmp}\fi
\settowidth{\frzwtmp}{\footnotesize {\color{basetext}$-$1.7}}\ifdim\frzwtmp>\lgwAA\setlength{\lgwAA}{\frzwtmp}\fi
\settowidth{\frzwtmp}{\footnotesize \textbf{76.5}}\ifdim\frzwtmp>\lgwAA\setlength{\lgwAA}{\frzwtmp}\fi
\settowidth{\frzwtmp}{\footnotesize {\color{basetext}$+$1.3}}\ifdim\frzwtmp>\lgwAA\setlength{\lgwAA}{\frzwtmp}\fi
\settowidth{\frzwtmp}{\footnotesize 76.1}\ifdim\frzwtmp>\lgwAA\setlength{\lgwAA}{\frzwtmp}\fi
\settowidth{\frzwtmp}{\footnotesize 61.3}\ifdim\frzwtmp>\lgwAA\setlength{\lgwAA}{\frzwtmp}\fi
\settowidth{\frzwtmp}{\footnotesize {\color{basetext}$-$1.3}}\ifdim\frzwtmp>\lgwAA\setlength{\lgwAA}{\frzwtmp}\fi
\settowidth{\frzwtmp}{\footnotesize 68.6\rlap{$^{\dagger}$}}\ifdim\frzwtmp>\lgwAA\setlength{\lgwAA}{\frzwtmp}\fi
\settowidth{\frzwtmp}{\footnotesize {\color{basetext}$+$5.5}\rlap{$^{\dagger}$}}\ifdim\frzwtmp>\lgwAA\setlength{\lgwAA}{\frzwtmp}\fi
\settowidth{\frzwtmp}{\footnotesize 67.7}\ifdim\frzwtmp>\lgwAA\setlength{\lgwAA}{\frzwtmp}\fi
\settowidth{\frzwtmp}{\footnotesize {\color{basetext}$+$4.8}}\ifdim\frzwtmp>\lgwAA\setlength{\lgwAA}{\frzwtmp}\fi
\settowidth{\frzwtmp}{\footnotesize 56.4}\ifdim\frzwtmp>\lgwAA\setlength{\lgwAA}{\frzwtmp}\fi
\settowidth{\frzwtmp}{\footnotesize {\color{basetext}$+$1.9}}\ifdim\frzwtmp>\lgwAA\setlength{\lgwAA}{\frzwtmp}\fi
\settowidth{\frzwtmp}{\footnotesize 55.6}\ifdim\frzwtmp>\lgwAA\setlength{\lgwAA}{\frzwtmp}\fi
\settowidth{\frzwtmp}{\footnotesize {\color{basetext}$+$1.6}}\ifdim\frzwtmp>\lgwAA\setlength{\lgwAA}{\frzwtmp}\fi
\settowidth{\frzwtmp}{\footnotesize \textcolor{worstcolor}{\itshape 59.2}}\ifdim\frzwtmp>\lgwAA\setlength{\lgwAA}{\frzwtmp}\fi
\settowidth{\frzwtmp}{\footnotesize {\color{basetext}$+$0.9}}\ifdim\frzwtmp>\lgwAA\setlength{\lgwAA}{\frzwtmp}\fi
\settowidth{\frzwtmp}{\footnotesize 61.0}\ifdim\frzwtmp>\lgwAA\setlength{\lgwAA}{\frzwtmp}\fi
\settowidth{\frzwtmp}{\footnotesize 59.3}\ifdim\frzwtmp>\lgwAA\setlength{\lgwAA}{\frzwtmp}\fi
\settowidth{\frzwtmp}{\footnotesize {\color{basetext}$-$3.4}}\ifdim\frzwtmp>\lgwAA\setlength{\lgwAA}{\frzwtmp}\fi
\settowidth{\frzwtmp}{\footnotesize \textbf{80.3}}\ifdim\frzwtmp>\lgwAA\setlength{\lgwAA}{\frzwtmp}\fi
\settowidth{\frzwtmp}{\footnotesize {\color{basetext}$+$3.4}}\ifdim\frzwtmp>\lgwAA\setlength{\lgwAA}{\frzwtmp}\fi
\settowidth{\frzwtmp}{\footnotesize 79.8}\ifdim\frzwtmp>\lgwAA\setlength{\lgwAA}{\frzwtmp}\fi
\settowidth{\frzwtmp}{\footnotesize {\color{basetext}$+$2.9}}\ifdim\frzwtmp>\lgwAA\setlength{\lgwAA}{\frzwtmp}\fi
\settowidth{\frzwtmp}{\footnotesize 71.8}\ifdim\frzwtmp>\lgwAA\setlength{\lgwAA}{\frzwtmp}\fi
\settowidth{\frzwtmp}{\footnotesize {\color{basetext}$+$3.9}}\ifdim\frzwtmp>\lgwAA\setlength{\lgwAA}{\frzwtmp}\fi
\settowidth{\frzwtmp}{\footnotesize 76.6}\ifdim\frzwtmp>\lgwAA\setlength{\lgwAA}{\frzwtmp}\fi
\settowidth{\frzwtmp}{\footnotesize {\color{basetext}$+$9.3}}\ifdim\frzwtmp>\lgwAA\setlength{\lgwAA}{\frzwtmp}\fi
\settowidth{\frzwtmp}{\footnotesize 77.3}\ifdim\frzwtmp>\lgwAA\setlength{\lgwAA}{\frzwtmp}\fi
\settowidth{\frzwtmp}{\footnotesize {\color{basetext}$+$9.7}}\ifdim\frzwtmp>\lgwAA\setlength{\lgwAA}{\frzwtmp}\fi
\settowidth{\frzwtmp}{\footnotesize 62.0}\ifdim\frzwtmp>\lgwAA\setlength{\lgwAA}{\frzwtmp}\fi
\settowidth{\frzwtmp}{\footnotesize {\color{basetext}$+$3.3}}\ifdim\frzwtmp>\lgwAA\setlength{\lgwAA}{\frzwtmp}\fi
\settowidth{\frzwtmp}{\footnotesize 63.6}\ifdim\frzwtmp>\lgwAA\setlength{\lgwAA}{\frzwtmp}\fi
\settowidth{\frzwtmp}{\footnotesize {\color{basetext}$+$2.3}}\ifdim\frzwtmp>\lgwAA\setlength{\lgwAA}{\frzwtmp}\fi
\settowidth{\frzwtmp}{\footnotesize \textcolor{worstcolor}{\itshape 63.8}}\ifdim\frzwtmp>\lgwAA\setlength{\lgwAA}{\frzwtmp}\fi
\settowidth{\frzwtmp}{\footnotesize {\color{basetext}$+$4.7}}\ifdim\frzwtmp>\lgwAA\setlength{\lgwAA}{\frzwtmp}\fi
\settowidth{\frzwtmp}{\footnotesize 62.2}\ifdim\frzwtmp>\lgwAA\setlength{\lgwAA}{\frzwtmp}\fi
\settowidth{\frzwtmp}{\footnotesize {\color{basetext}$-$3.8}}\ifdim\frzwtmp>\lgwAA\setlength{\lgwAA}{\frzwtmp}\fi
\settowidth{\frzwtmp}{\footnotesize 65.5}\ifdim\frzwtmp>\lgwAA\setlength{\lgwAA}{\frzwtmp}\fi
\settowidth{\frzwtmp}{\footnotesize {\color{basetext}$-$0.5}}\ifdim\frzwtmp>\lgwAA\setlength{\lgwAA}{\frzwtmp}\fi
\settowidth{\frzwtmp}{\footnotesize 77.0}\ifdim\frzwtmp>\lgwAA\setlength{\lgwAA}{\frzwtmp}\fi
\settowidth{\frzwtmp}{\footnotesize 76.5}\ifdim\frzwtmp>\lgwAA\setlength{\lgwAA}{\frzwtmp}\fi
\settowidth{\frzwtmp}{\footnotesize {\color{basetext}$-$1.9}}\ifdim\frzwtmp>\lgwAA\setlength{\lgwAA}{\frzwtmp}\fi
\settowidth{\frzwtmp}{\footnotesize 68.3}\ifdim\frzwtmp>\lgwAA\setlength{\lgwAA}{\frzwtmp}\fi
\settowidth{\frzwtmp}{\footnotesize {\color{basetext}$-$1.0}}\ifdim\frzwtmp>\lgwAA\setlength{\lgwAA}{\frzwtmp}\fi
\settowidth{\frzwtmp}{\footnotesize 81.5}\ifdim\frzwtmp>\lgwAA\setlength{\lgwAA}{\frzwtmp}\fi
\settowidth{\frzwtmp}{\footnotesize {\color{basetext}$+$11.8}}\ifdim\frzwtmp>\lgwAA\setlength{\lgwAA}{\frzwtmp}\fi
\settowidth{\frzwtmp}{\footnotesize \textbf{82.6}}\ifdim\frzwtmp>\lgwAA\setlength{\lgwAA}{\frzwtmp}\fi
\settowidth{\frzwtmp}{\footnotesize {\color{basetext}$+$12.3}}\ifdim\frzwtmp>\lgwAA\setlength{\lgwAA}{\frzwtmp}\fi
\settowidth{\frzwtmp}{\footnotesize 69.4}\ifdim\frzwtmp>\lgwAA\setlength{\lgwAA}{\frzwtmp}\fi
\settowidth{\frzwtmp}{\footnotesize {\color{basetext}$+$4.9}}\ifdim\frzwtmp>\lgwAA\setlength{\lgwAA}{\frzwtmp}\fi
\settowidth{\frzwtmp}{\footnotesize 69.7}\ifdim\frzwtmp>\lgwAA\setlength{\lgwAA}{\frzwtmp}\fi
\settowidth{\frzwtmp}{\footnotesize {\color{basetext}$+$3.8}}\ifdim\frzwtmp>\lgwAA\setlength{\lgwAA}{\frzwtmp}\fi
\settowidth{\frzwtmp}{\footnotesize 80.7}\ifdim\frzwtmp>\lgwAA\setlength{\lgwAA}{\frzwtmp}\fi
\settowidth{\frzwtmp}{\footnotesize {\color{basetext}$+$11.0}}\ifdim\frzwtmp>\lgwAA\setlength{\lgwAA}{\frzwtmp}\fi
\settowidth{\frzwtmp}{\footnotesize 65.2}\ifdim\frzwtmp>\lgwAA\setlength{\lgwAA}{\frzwtmp}\fi
\settowidth{\frzwtmp}{\footnotesize {\color{basetext}$-$11.5}}\ifdim\frzwtmp>\lgwAA\setlength{\lgwAA}{\frzwtmp}\fi
\settowidth{\frzwtmp}{\footnotesize 75.5}\ifdim\frzwtmp>\lgwAA\setlength{\lgwAA}{\frzwtmp}\fi
\settowidth{\frzwtmp}{\footnotesize {\color{basetext}$-$1.2}}\ifdim\frzwtmp>\lgwAA\setlength{\lgwAA}{\frzwtmp}\fi
\settowidth{\frzwtmp}{\footnotesize \textcolor{worstcolor}{\itshape 51.2}}\ifdim\frzwtmp>\lgwAA\setlength{\lgwAA}{\frzwtmp}\fi
\settowidth{\frzwtmp}{\footnotesize {\color{basetext}$-$33.4}}\ifdim\frzwtmp>\lgwAA\setlength{\lgwAA}{\frzwtmp}\fi
\settowidth{\frzwtmp}{\footnotesize 53.6}\ifdim\frzwtmp>\lgwAA\setlength{\lgwAA}{\frzwtmp}\fi
\settowidth{\frzwtmp}{\footnotesize {\color{basetext}$-$30.9}}\ifdim\frzwtmp>\lgwAA\setlength{\lgwAA}{\frzwtmp}\fi
\settowidth{\frzwtmp}{\footnotesize 66.7}\ifdim\frzwtmp>\lgwAA\setlength{\lgwAA}{\frzwtmp}\fi
\settowidth{\frzwtmp}{\footnotesize {\color{basetext}$-$4.8}}\ifdim\frzwtmp>\lgwAA\setlength{\lgwAA}{\frzwtmp}\fi
\settowidth{\frzwtmp}{\footnotesize \textbf{87.0}}\ifdim\frzwtmp>\lgwAA\setlength{\lgwAA}{\frzwtmp}\fi
\settowidth{\frzwtmp}{\footnotesize {\color{basetext}$+$14.7}}\ifdim\frzwtmp>\lgwAA\setlength{\lgwAA}{\frzwtmp}\fi
\settowidth{\frzwtmp}{\footnotesize 84.1}\ifdim\frzwtmp>\lgwAA\setlength{\lgwAA}{\frzwtmp}\fi
\settowidth{\frzwtmp}{\footnotesize {\color{basetext}$+$11.9}}\ifdim\frzwtmp>\lgwAA\setlength{\lgwAA}{\frzwtmp}\fi
\settowidth{\frzwtmp}{\footnotesize 82.4}\ifdim\frzwtmp>\lgwAA\setlength{\lgwAA}{\frzwtmp}\fi
\settowidth{\frzwtmp}{\footnotesize {\color{basetext}$+$7.5}}\ifdim\frzwtmp>\lgwAA\setlength{\lgwAA}{\frzwtmp}\fi
\settowidth{\frzwtmp}{\footnotesize 83.3}\ifdim\frzwtmp>\lgwAA\setlength{\lgwAA}{\frzwtmp}\fi
\settowidth{\frzwtmp}{\footnotesize {\color{basetext}$+$7.3}}\ifdim\frzwtmp>\lgwAA\setlength{\lgwAA}{\frzwtmp}\fi
\settowidth{\frzwtmp}{\footnotesize {\scriptsize \%}}\ifdim\frzwtmp>\lgwAA\setlength{\lgwAA}{\frzwtmp}\fi
\settowidth{\frzwtmp}{\footnotesize {\scriptsize\color{basetext} $\Delta$F1$^{*}$ (pp)}}\ifdim\frzwtmp>\lgwAA\setlength{\lgwAA}{\frzwtmp}\fi
\ifdefined\lgwAB\else\newlength{\lgwAB}\fi\setlength{\lgwAB}{0pt}
\settowidth{\frzwtmp}{\footnotesize $-$24.5}\ifdim\frzwtmp>\lgwAB\setlength{\lgwAB}{\frzwtmp}\fi
\settowidth{\frzwtmp}{\footnotesize {\color{basetext}0.76$\to$0.46}}\ifdim\frzwtmp>\lgwAB\setlength{\lgwAB}{\frzwtmp}\fi
\settowidth{\frzwtmp}{\footnotesize $-$4.0}\ifdim\frzwtmp>\lgwAB\setlength{\lgwAB}{\frzwtmp}\fi
\settowidth{\frzwtmp}{\footnotesize {\color{basetext}0.47$\to$0.42}}\ifdim\frzwtmp>\lgwAB\setlength{\lgwAB}{\frzwtmp}\fi
\settowidth{\frzwtmp}{\footnotesize $-$2.1}\ifdim\frzwtmp>\lgwAB\setlength{\lgwAB}{\frzwtmp}\fi
\settowidth{\frzwtmp}{\footnotesize {\color{basetext}0.55$\to$0.50}}\ifdim\frzwtmp>\lgwAB\setlength{\lgwAB}{\frzwtmp}\fi
\settowidth{\frzwtmp}{\footnotesize \textbf{$+$7.9}}\ifdim\frzwtmp>\lgwAB\setlength{\lgwAB}{\frzwtmp}\fi
\settowidth{\frzwtmp}{\footnotesize {\color{basetext}0.75$\to$0.85}}\ifdim\frzwtmp>\lgwAB\setlength{\lgwAB}{\frzwtmp}\fi
\settowidth{\frzwtmp}{\footnotesize $+$3.0}\ifdim\frzwtmp>\lgwAB\setlength{\lgwAB}{\frzwtmp}\fi
\settowidth{\frzwtmp}{\footnotesize {\color{basetext}0.83$\to$0.83}}\ifdim\frzwtmp>\lgwAB\setlength{\lgwAB}{\frzwtmp}\fi
\settowidth{\frzwtmp}{\footnotesize $+$4.9}\ifdim\frzwtmp>\lgwAB\setlength{\lgwAB}{\frzwtmp}\fi
\settowidth{\frzwtmp}{\footnotesize {\color{basetext}0.86$\to$0.84}}\ifdim\frzwtmp>\lgwAB\setlength{\lgwAB}{\frzwtmp}\fi
\settowidth{\frzwtmp}{\footnotesize $+$4.6}\ifdim\frzwtmp>\lgwAB\setlength{\lgwAB}{\frzwtmp}\fi
\settowidth{\frzwtmp}{\footnotesize {\color{basetext}0.87$\to$0.85}}\ifdim\frzwtmp>\lgwAB\setlength{\lgwAB}{\frzwtmp}\fi
\settowidth{\frzwtmp}{\footnotesize $-$22.3}\ifdim\frzwtmp>\lgwAB\setlength{\lgwAB}{\frzwtmp}\fi
\settowidth{\frzwtmp}{\footnotesize {\color{basetext}0.78$\to$0.52}}\ifdim\frzwtmp>\lgwAB\setlength{\lgwAB}{\frzwtmp}\fi
\settowidth{\frzwtmp}{\footnotesize \textcolor{worstcolor}{\itshape $-$25.4}}\ifdim\frzwtmp>\lgwAB\setlength{\lgwAB}{\frzwtmp}\fi
\settowidth{\frzwtmp}{\footnotesize {\color{basetext}0.77$\to$0.49}}\ifdim\frzwtmp>\lgwAB\setlength{\lgwAB}{\frzwtmp}\fi
\settowidth{\frzwtmp}{\footnotesize \textcolor{worstcolor}{\itshape $-$20.0}}\ifdim\frzwtmp>\lgwAB\setlength{\lgwAB}{\frzwtmp}\fi
\settowidth{\frzwtmp}{\footnotesize {\color{basetext}0.69$\to$0.43}}\ifdim\frzwtmp>\lgwAB\setlength{\lgwAB}{\frzwtmp}\fi
\settowidth{\frzwtmp}{\footnotesize $+$1.4}\ifdim\frzwtmp>\lgwAB\setlength{\lgwAB}{\frzwtmp}\fi
\settowidth{\frzwtmp}{\footnotesize {\color{basetext}0.50$\to$0.44}}\ifdim\frzwtmp>\lgwAB\setlength{\lgwAB}{\frzwtmp}\fi
\settowidth{\frzwtmp}{\footnotesize $+$0.6}\ifdim\frzwtmp>\lgwAB\setlength{\lgwAB}{\frzwtmp}\fi
\settowidth{\frzwtmp}{\footnotesize {\color{basetext}0.53$\to$0.52}}\ifdim\frzwtmp>\lgwAB\setlength{\lgwAB}{\frzwtmp}\fi
\settowidth{\frzwtmp}{\footnotesize \textbf{$+$4.6}}\ifdim\frzwtmp>\lgwAB\setlength{\lgwAB}{\frzwtmp}\fi
\settowidth{\frzwtmp}{\footnotesize {\color{basetext}0.78$\to$0.84}}\ifdim\frzwtmp>\lgwAB\setlength{\lgwAB}{\frzwtmp}\fi
\settowidth{\frzwtmp}{\footnotesize $+$4.2}\ifdim\frzwtmp>\lgwAB\setlength{\lgwAB}{\frzwtmp}\fi
\settowidth{\frzwtmp}{\footnotesize {\color{basetext}0.78$\to$0.83}}\ifdim\frzwtmp>\lgwAB\setlength{\lgwAB}{\frzwtmp}\fi
\settowidth{\frzwtmp}{\footnotesize $+$3.6}\ifdim\frzwtmp>\lgwAB\setlength{\lgwAB}{\frzwtmp}\fi
\settowidth{\frzwtmp}{\footnotesize {\color{basetext}0.85$\to$0.85}}\ifdim\frzwtmp>\lgwAB\setlength{\lgwAB}{\frzwtmp}\fi
\settowidth{\frzwtmp}{\footnotesize $+$4.5}\ifdim\frzwtmp>\lgwAB\setlength{\lgwAB}{\frzwtmp}\fi
\settowidth{\frzwtmp}{\footnotesize {\color{basetext}0.85$\to$0.86}}\ifdim\frzwtmp>\lgwAB\setlength{\lgwAB}{\frzwtmp}\fi
\settowidth{\frzwtmp}{\footnotesize {\color{basetext}0.86$\to$0.87}}\ifdim\frzwtmp>\lgwAB\setlength{\lgwAB}{\frzwtmp}\fi
\settowidth{\frzwtmp}{\footnotesize $-$19.9}\ifdim\frzwtmp>\lgwAB\setlength{\lgwAB}{\frzwtmp}\fi
\settowidth{\frzwtmp}{\footnotesize {\color{basetext}0.77$\to$0.52}}\ifdim\frzwtmp>\lgwAB\setlength{\lgwAB}{\frzwtmp}\fi
\settowidth{\frzwtmp}{\footnotesize $-$18.8}\ifdim\frzwtmp>\lgwAB\setlength{\lgwAB}{\frzwtmp}\fi
\settowidth{\frzwtmp}{\footnotesize {\color{basetext}0.75$\to$0.51}}\ifdim\frzwtmp>\lgwAB\setlength{\lgwAB}{\frzwtmp}\fi
\settowidth{\frzwtmp}{\footnotesize $-$11.0}\ifdim\frzwtmp>\lgwAB\setlength{\lgwAB}{\frzwtmp}\fi
\settowidth{\frzwtmp}{\footnotesize {\color{basetext}0.75$\to$0.52}}\ifdim\frzwtmp>\lgwAB\setlength{\lgwAB}{\frzwtmp}\fi
\settowidth{\frzwtmp}{\footnotesize $-$6.3}\ifdim\frzwtmp>\lgwAB\setlength{\lgwAB}{\frzwtmp}\fi
\settowidth{\frzwtmp}{\footnotesize {\color{basetext}0.55$\to$0.49}}\ifdim\frzwtmp>\lgwAB\setlength{\lgwAB}{\frzwtmp}\fi
\settowidth{\frzwtmp}{\footnotesize $-$0.7}\ifdim\frzwtmp>\lgwAB\setlength{\lgwAB}{\frzwtmp}\fi
\settowidth{\frzwtmp}{\footnotesize {\color{basetext}0.57$\to$0.56}}\ifdim\frzwtmp>\lgwAB\setlength{\lgwAB}{\frzwtmp}\fi
\settowidth{\frzwtmp}{\footnotesize $+$3.2}\ifdim\frzwtmp>\lgwAB\setlength{\lgwAB}{\frzwtmp}\fi
\settowidth{\frzwtmp}{\footnotesize {\color{basetext}0.78$\to$0.82}}\ifdim\frzwtmp>\lgwAB\setlength{\lgwAB}{\frzwtmp}\fi
\settowidth{\frzwtmp}{\footnotesize $+$2.0}\ifdim\frzwtmp>\lgwAB\setlength{\lgwAB}{\frzwtmp}\fi
\settowidth{\frzwtmp}{\footnotesize \textbf{$+$6.2}}\ifdim\frzwtmp>\lgwAB\setlength{\lgwAB}{\frzwtmp}\fi
\settowidth{\frzwtmp}{\footnotesize {\color{basetext}0.85$\to$0.89}}\ifdim\frzwtmp>\lgwAB\setlength{\lgwAB}{\frzwtmp}\fi
\settowidth{\frzwtmp}{\footnotesize $+$1.0}\ifdim\frzwtmp>\lgwAB\setlength{\lgwAB}{\frzwtmp}\fi
\settowidth{\frzwtmp}{\footnotesize {\color{basetext}0.88$\to$0.87}}\ifdim\frzwtmp>\lgwAB\setlength{\lgwAB}{\frzwtmp}\fi
\settowidth{\frzwtmp}{\footnotesize {\color{basetext}0.88$\to$0.88}}\ifdim\frzwtmp>\lgwAB\setlength{\lgwAB}{\frzwtmp}\fi
\settowidth{\frzwtmp}{\footnotesize \textcolor{worstcolor}{\itshape $-$11.5}}\ifdim\frzwtmp>\lgwAB\setlength{\lgwAB}{\frzwtmp}\fi
\settowidth{\frzwtmp}{\footnotesize {\color{basetext}0.79$\to$0.63}}\ifdim\frzwtmp>\lgwAB\setlength{\lgwAB}{\frzwtmp}\fi
\settowidth{\frzwtmp}{\footnotesize $-$7.5}\ifdim\frzwtmp>\lgwAB\setlength{\lgwAB}{\frzwtmp}\fi
\settowidth{\frzwtmp}{\footnotesize {\color{basetext}0.75$\to$0.64}}\ifdim\frzwtmp>\lgwAB\setlength{\lgwAB}{\frzwtmp}\fi
\settowidth{\frzwtmp}{\footnotesize {\color{basetext}0.78$\to$0.75}}\ifdim\frzwtmp>\lgwAB\setlength{\lgwAB}{\frzwtmp}\fi
\settowidth{\frzwtmp}{\footnotesize $-$5.5}\ifdim\frzwtmp>\lgwAB\setlength{\lgwAB}{\frzwtmp}\fi
\settowidth{\frzwtmp}{\footnotesize {\color{basetext}0.54$\to$0.48}}\ifdim\frzwtmp>\lgwAB\setlength{\lgwAB}{\frzwtmp}\fi
\settowidth{\frzwtmp}{\footnotesize $-$4.4}\ifdim\frzwtmp>\lgwAB\setlength{\lgwAB}{\frzwtmp}\fi
\settowidth{\frzwtmp}{\footnotesize {\color{basetext}0.57$\to$0.52}}\ifdim\frzwtmp>\lgwAB\setlength{\lgwAB}{\frzwtmp}\fi
\settowidth{\frzwtmp}{\footnotesize $+$5.3}\ifdim\frzwtmp>\lgwAB\setlength{\lgwAB}{\frzwtmp}\fi
\settowidth{\frzwtmp}{\footnotesize $+$1.9}\ifdim\frzwtmp>\lgwAB\setlength{\lgwAB}{\frzwtmp}\fi
\settowidth{\frzwtmp}{\footnotesize {\color{basetext}0.86$\to$0.89}}\ifdim\frzwtmp>\lgwAB\setlength{\lgwAB}{\frzwtmp}\fi
\settowidth{\frzwtmp}{\footnotesize \textcolor{worstcolor}{\itshape $+$0.2}}\ifdim\frzwtmp>\lgwAB\setlength{\lgwAB}{\frzwtmp}\fi
\settowidth{\frzwtmp}{\footnotesize {\color{basetext}0.87$\to$0.87}}\ifdim\frzwtmp>\lgwAB\setlength{\lgwAB}{\frzwtmp}\fi
\settowidth{\frzwtmp}{\footnotesize $+$0.5}\ifdim\frzwtmp>\lgwAB\setlength{\lgwAB}{\frzwtmp}\fi
\settowidth{\frzwtmp}{\footnotesize {\color{basetext}0.87$\to$0.88}}\ifdim\frzwtmp>\lgwAB\setlength{\lgwAB}{\frzwtmp}\fi
\settowidth{\frzwtmp}{\footnotesize \textbf{$+$6.9}}\ifdim\frzwtmp>\lgwAB\setlength{\lgwAB}{\frzwtmp}\fi
\settowidth{\frzwtmp}{\footnotesize {\color{basetext}0.77$\to$0.82}}\ifdim\frzwtmp>\lgwAB\setlength{\lgwAB}{\frzwtmp}\fi
\settowidth{\frzwtmp}{\footnotesize $+$5.6}\ifdim\frzwtmp>\lgwAB\setlength{\lgwAB}{\frzwtmp}\fi
\settowidth{\frzwtmp}{\footnotesize {\color{basetext}0.77$\to$0.83}}\ifdim\frzwtmp>\lgwAB\setlength{\lgwAB}{\frzwtmp}\fi
\settowidth{\frzwtmp}{\footnotesize {\scriptsize $10^{-3}$/month}}\ifdim\frzwtmp>\lgwAB\setlength{\lgwAB}{\frzwtmp}\fi
\settowidth{\frzwtmp}{\footnotesize {\scriptsize\color{basetext} first$\to$last}}\ifdim\frzwtmp>\lgwAB\setlength{\lgwAB}{\frzwtmp}\fi
\ifdefined\lgwAC\else\newlength{\lgwAC}\fi\setlength{\lgwAC}{0pt}
\settowidth{\frzwtmp}{\footnotesize $-$0.11}\ifdim\frzwtmp>\lgwAC\setlength{\lgwAC}{\frzwtmp}\fi
\settowidth{\frzwtmp}{\footnotesize {\color{basetext}3.0}}\ifdim\frzwtmp>\lgwAC\setlength{\lgwAC}{\frzwtmp}\fi
\settowidth{\frzwtmp}{\footnotesize $-$0.15\rlap{$^{\dagger}$}}\ifdim\frzwtmp>\lgwAC\setlength{\lgwAC}{\frzwtmp}\fi
\settowidth{\frzwtmp}{\footnotesize {\color{basetext}1.1}\rlap{$^{\dagger}$}}\ifdim\frzwtmp>\lgwAC\setlength{\lgwAC}{\frzwtmp}\fi
\settowidth{\frzwtmp}{\footnotesize $-$0.07}\ifdim\frzwtmp>\lgwAC\setlength{\lgwAC}{\frzwtmp}\fi
\settowidth{\frzwtmp}{\footnotesize {\color{basetext}4.6}}\ifdim\frzwtmp>\lgwAC\setlength{\lgwAC}{\frzwtmp}\fi
\settowidth{\frzwtmp}{\footnotesize $-$0.31}\ifdim\frzwtmp>\lgwAC\setlength{\lgwAC}{\frzwtmp}\fi
\settowidth{\frzwtmp}{\footnotesize {\color{basetext}2.1}}\ifdim\frzwtmp>\lgwAC\setlength{\lgwAC}{\frzwtmp}\fi
\settowidth{\frzwtmp}{\footnotesize $-$0.29}\ifdim\frzwtmp>\lgwAC\setlength{\lgwAC}{\frzwtmp}\fi
\settowidth{\frzwtmp}{\footnotesize $-$0.09}\ifdim\frzwtmp>\lgwAC\setlength{\lgwAC}{\frzwtmp}\fi
\settowidth{\frzwtmp}{\footnotesize {\color{basetext}2.8}}\ifdim\frzwtmp>\lgwAC\setlength{\lgwAC}{\frzwtmp}\fi
\settowidth{\frzwtmp}{\footnotesize $-$0.25\rlap{$^{\dagger}$}}\ifdim\frzwtmp>\lgwAC\setlength{\lgwAC}{\frzwtmp}\fi
\settowidth{\frzwtmp}{\footnotesize {\color{basetext}2.8}\rlap{$^{\dagger}$}}\ifdim\frzwtmp>\lgwAC\setlength{\lgwAC}{\frzwtmp}\fi
\settowidth{\frzwtmp}{\footnotesize $-$0.20}\ifdim\frzwtmp>\lgwAC\setlength{\lgwAC}{\frzwtmp}\fi
\settowidth{\frzwtmp}{\footnotesize {\color{basetext}2.7}}\ifdim\frzwtmp>\lgwAC\setlength{\lgwAC}{\frzwtmp}\fi
\settowidth{\frzwtmp}{\footnotesize $-$0.12}\ifdim\frzwtmp>\lgwAC\setlength{\lgwAC}{\frzwtmp}\fi
\settowidth{\frzwtmp}{\footnotesize {\color{basetext}2.9}}\ifdim\frzwtmp>\lgwAC\setlength{\lgwAC}{\frzwtmp}\fi
\settowidth{\frzwtmp}{\footnotesize $-$0.33}\ifdim\frzwtmp>\lgwAC\setlength{\lgwAC}{\frzwtmp}\fi
\settowidth{\frzwtmp}{\footnotesize {\color{basetext}5.8}}\ifdim\frzwtmp>\lgwAC\setlength{\lgwAC}{\frzwtmp}\fi
\settowidth{\frzwtmp}{\footnotesize $-$0.28}\ifdim\frzwtmp>\lgwAC\setlength{\lgwAC}{\frzwtmp}\fi
\settowidth{\frzwtmp}{\footnotesize {\color{basetext}3.3}}\ifdim\frzwtmp>\lgwAC\setlength{\lgwAC}{\frzwtmp}\fi
\settowidth{\frzwtmp}{\footnotesize $-$0.36}\ifdim\frzwtmp>\lgwAC\setlength{\lgwAC}{\frzwtmp}\fi
\settowidth{\frzwtmp}{\footnotesize {\color{basetext}7.1}}\ifdim\frzwtmp>\lgwAC\setlength{\lgwAC}{\frzwtmp}\fi
\settowidth{\frzwtmp}{\footnotesize $-$0.64}\ifdim\frzwtmp>\lgwAC\setlength{\lgwAC}{\frzwtmp}\fi
\settowidth{\frzwtmp}{\footnotesize {\color{basetext}3.6}}\ifdim\frzwtmp>\lgwAC\setlength{\lgwAC}{\frzwtmp}\fi
\settowidth{\frzwtmp}{\footnotesize $-$0.66}\ifdim\frzwtmp>\lgwAC\setlength{\lgwAC}{\frzwtmp}\fi
\settowidth{\frzwtmp}{\footnotesize {\color{basetext}3.7}}\ifdim\frzwtmp>\lgwAC\setlength{\lgwAC}{\frzwtmp}\fi
\settowidth{\frzwtmp}{\footnotesize $-$0.47}\ifdim\frzwtmp>\lgwAC\setlength{\lgwAC}{\frzwtmp}\fi
\settowidth{\frzwtmp}{\footnotesize {\color{basetext}5.4}}\ifdim\frzwtmp>\lgwAC\setlength{\lgwAC}{\frzwtmp}\fi
\settowidth{\frzwtmp}{\footnotesize $-$0.41}\ifdim\frzwtmp>\lgwAC\setlength{\lgwAC}{\frzwtmp}\fi
\settowidth{\frzwtmp}{\footnotesize {\color{basetext}5.1}}\ifdim\frzwtmp>\lgwAC\setlength{\lgwAC}{\frzwtmp}\fi
\settowidth{\frzwtmp}{\footnotesize $-$0.37}\ifdim\frzwtmp>\lgwAC\setlength{\lgwAC}{\frzwtmp}\fi
\settowidth{\frzwtmp}{\footnotesize $-$0.35}\ifdim\frzwtmp>\lgwAC\setlength{\lgwAC}{\frzwtmp}\fi
\settowidth{\frzwtmp}{\footnotesize $-$0.32}\ifdim\frzwtmp>\lgwAC\setlength{\lgwAC}{\frzwtmp}\fi
\settowidth{\frzwtmp}{\footnotesize {\color{basetext}5.2}}\ifdim\frzwtmp>\lgwAC\setlength{\lgwAC}{\frzwtmp}\fi
\settowidth{\frzwtmp}{\footnotesize $-$0.78}\ifdim\frzwtmp>\lgwAC\setlength{\lgwAC}{\frzwtmp}\fi
\settowidth{\frzwtmp}{\footnotesize {\color{basetext}10.5}}\ifdim\frzwtmp>\lgwAC\setlength{\lgwAC}{\frzwtmp}\fi
\settowidth{\frzwtmp}{\footnotesize $-$0.55}\ifdim\frzwtmp>\lgwAC\setlength{\lgwAC}{\frzwtmp}\fi
\settowidth{\frzwtmp}{\footnotesize {\color{basetext}8.2}}\ifdim\frzwtmp>\lgwAC\setlength{\lgwAC}{\frzwtmp}\fi
\settowidth{\frzwtmp}{\footnotesize {\color{basetext}11.0}}\ifdim\frzwtmp>\lgwAC\setlength{\lgwAC}{\frzwtmp}\fi
\settowidth{\frzwtmp}{\footnotesize $-$1.10}\ifdim\frzwtmp>\lgwAC\setlength{\lgwAC}{\frzwtmp}\fi
\settowidth{\frzwtmp}{\footnotesize {\color{basetext}7.8}}\ifdim\frzwtmp>\lgwAC\setlength{\lgwAC}{\frzwtmp}\fi
\settowidth{\frzwtmp}{\footnotesize $-$1.23}\ifdim\frzwtmp>\lgwAC\setlength{\lgwAC}{\frzwtmp}\fi
\settowidth{\frzwtmp}{\footnotesize {\color{basetext}8.0}}\ifdim\frzwtmp>\lgwAC\setlength{\lgwAC}{\frzwtmp}\fi
\settowidth{\frzwtmp}{\footnotesize $-$0.67}\ifdim\frzwtmp>\lgwAC\setlength{\lgwAC}{\frzwtmp}\fi
\settowidth{\frzwtmp}{\footnotesize {\color{basetext}10.7}}\ifdim\frzwtmp>\lgwAC\setlength{\lgwAC}{\frzwtmp}\fi
\settowidth{\frzwtmp}{\footnotesize $-$0.65}\ifdim\frzwtmp>\lgwAC\setlength{\lgwAC}{\frzwtmp}\fi
\settowidth{\frzwtmp}{\footnotesize {\color{basetext}10.1}}\ifdim\frzwtmp>\lgwAC\setlength{\lgwAC}{\frzwtmp}\fi
\settowidth{\frzwtmp}{\footnotesize $-$0.56}\ifdim\frzwtmp>\lgwAC\setlength{\lgwAC}{\frzwtmp}\fi
\settowidth{\frzwtmp}{\footnotesize $-$0.50}\ifdim\frzwtmp>\lgwAC\setlength{\lgwAC}{\frzwtmp}\fi
\settowidth{\frzwtmp}{\footnotesize {\color{basetext}10.6}}\ifdim\frzwtmp>\lgwAC\setlength{\lgwAC}{\frzwtmp}\fi
\settowidth{\frzwtmp}{\footnotesize $-$0.68}\ifdim\frzwtmp>\lgwAC\setlength{\lgwAC}{\frzwtmp}\fi
\settowidth{\frzwtmp}{\footnotesize {\color{basetext}10.3}}\ifdim\frzwtmp>\lgwAC\setlength{\lgwAC}{\frzwtmp}\fi
\settowidth{\frzwtmp}{\footnotesize $-$1.15}\ifdim\frzwtmp>\lgwAC\setlength{\lgwAC}{\frzwtmp}\fi
\settowidth{\frzwtmp}{\footnotesize {\color{basetext}21.1}}\ifdim\frzwtmp>\lgwAC\setlength{\lgwAC}{\frzwtmp}\fi
\settowidth{\frzwtmp}{\footnotesize {\color{basetext}12.9}}\ifdim\frzwtmp>\lgwAC\setlength{\lgwAC}{\frzwtmp}\fi
\settowidth{\frzwtmp}{\footnotesize $-$1.05}\ifdim\frzwtmp>\lgwAC\setlength{\lgwAC}{\frzwtmp}\fi
\settowidth{\frzwtmp}{\footnotesize {\color{basetext}26.1}}\ifdim\frzwtmp>\lgwAC\setlength{\lgwAC}{\frzwtmp}\fi
\settowidth{\frzwtmp}{\footnotesize $-$1.51}\ifdim\frzwtmp>\lgwAC\setlength{\lgwAC}{\frzwtmp}\fi
\settowidth{\frzwtmp}{\footnotesize {\color{basetext}17.1}}\ifdim\frzwtmp>\lgwAC\setlength{\lgwAC}{\frzwtmp}\fi
\settowidth{\frzwtmp}{\footnotesize $-$1.56}\ifdim\frzwtmp>\lgwAC\setlength{\lgwAC}{\frzwtmp}\fi
\settowidth{\frzwtmp}{\footnotesize $-$1.03}\ifdim\frzwtmp>\lgwAC\setlength{\lgwAC}{\frzwtmp}\fi
\settowidth{\frzwtmp}{\footnotesize {\color{basetext}20.7}}\ifdim\frzwtmp>\lgwAC\setlength{\lgwAC}{\frzwtmp}\fi
\settowidth{\frzwtmp}{\footnotesize $-$1.02}\ifdim\frzwtmp>\lgwAC\setlength{\lgwAC}{\frzwtmp}\fi
\settowidth{\frzwtmp}{\footnotesize {\color{basetext}19.8}}\ifdim\frzwtmp>\lgwAC\setlength{\lgwAC}{\frzwtmp}\fi
\settowidth{\frzwtmp}{\footnotesize $-$1.20}\ifdim\frzwtmp>\lgwAC\setlength{\lgwAC}{\frzwtmp}\fi
\settowidth{\frzwtmp}{\footnotesize {\color{basetext}19.7}}\ifdim\frzwtmp>\lgwAC\setlength{\lgwAC}{\frzwtmp}\fi
\settowidth{\frzwtmp}{\footnotesize {\color{basetext}20.6}}\ifdim\frzwtmp>\lgwAC\setlength{\lgwAC}{\frzwtmp}\fi
\settowidth{\frzwtmp}{\footnotesize {\color{basetext}20.4}}\ifdim\frzwtmp>\lgwAC\setlength{\lgwAC}{\frzwtmp}\fi
\settowidth{\frzwtmp}{\footnotesize {\scriptsize pp/month}}\ifdim\frzwtmp>\lgwAC\setlength{\lgwAC}{\frzwtmp}\fi
\settowidth{\frzwtmp}{\footnotesize {\scriptsize\color{basetext} $\bar{q}^{*}$ (\%)}}\ifdim\frzwtmp>\lgwAC\setlength{\lgwAC}{\frzwtmp}\fi
\begin{table*}[p]
  \centering
  \caption{\textbf{AndroZoo} (test Jan 2020--Nov 2021; full initial year) in the metric set, threshold protocol and layout of Table~\ref{tab:main-vtt2-b400}: the first six test months calibrate, the remaining 17 months evaluate, and the Cap batch size and the rejection quota $\rho$ equal $B_M$. Values are computed per month within seed, summarized within seed and averaged over seeds; rows with fewer seeds than the complete rows show their count and are not ranked. \textbf{Bold}=best, {\color{worstcolor}\itshape dark-red italics}=worst displayed value per ranked column and $B_M$ block.}
  \label{tab:legacy-azoo}
  \footnotesize
  \setlength{\tabcolsep}{4.2pt}
  \renewcommand{\arraystretch}{0.95}
  \begin{tabular}{@{}cl@{\hspace{7pt}}c@{\hspace{7pt}}ccw{c}{\lgwAA}w{c}{\lgwAA}@{\hspace{7pt}}cw{c}{\lgwAB}w{c}{\lgwAB}w{c}{\lgwAC}w{c}{\lgwAC}}
    \toprule
     &  & \multicolumn{1}{>{\columncolor{tabbaseband}[\tabcolsep][\tabcolsep]}c}{Baseline} & \multicolumn{4}{>{\columncolor{tabreliband}[\tabcolsep][\tabcolsep]}c}{Reliability} & \multicolumn{5}{>{\columncolor{tabstabband}[\tabcolsep][\tabcolsep]}c}{Stability} \\
    $B_M$ & Method & \cellcolor{tabbaseband}F1 $\uparrow$ & \cellcolor{tabreliband}E-AURC $\downarrow$ & \cellcolor{tabreliband}Cap$_{B_M}$ $\uparrow$ & \multicolumn{2}{>{\columncolor{tabreliband}[\tabcolsep][\tabcolsep]}c}{\makebox[0pt][c]{F1$^{*}$ $\uparrow$}\makebox[0pt][c]{\smash{\raisebox{-2.6pt}{\rule{\dimexpr 2\lgwAA+4\tabcolsep-5pt\relax}{0.4pt}}}}} & \cellcolor{tabstabband}Sen[F1] $\uparrow$ & \multicolumn{2}{>{\columncolor{tabstabband}[\tabcolsep][\tabcolsep]}c}{\makebox[0pt][c]{Sen[AUROC$_f$] $\uparrow$}\makebox[0pt][c]{\smash{\raisebox{-2.6pt}{\rule{\dimexpr 2\lgwAB+4\tabcolsep-5pt\relax}{0.4pt}}}}} & \multicolumn{2}{>{\columncolor{tabstabband}[\tabcolsep][\tabcolsep]}c}{\makebox[0pt][c]{Sen[$q^{*}$]}\makebox[0pt][c]{\smash{\raisebox{-2.6pt}{\rule{\dimexpr 2\lgwAC+4\tabcolsep-5pt\relax}{0.4pt}}}}} \\
     &  & \cellcolor{tabbaseband}{\scriptsize \%} & \cellcolor{tabreliband}{\scriptsize $\times$100} & \cellcolor{tabreliband}{\scriptsize \% of errors} & \cellcolor{tabreliband}{\scriptsize \%} & \cellcolor{tabreliband}{\scriptsize\color{basetext} $\Delta$F1$^{*}$ (pp)} & \cellcolor{tabstabband}{\scriptsize pp/month} & \cellcolor{tabstabband}{\scriptsize $10^{-3}$/month} & \cellcolor{tabstabband}{\scriptsize\color{basetext} first$\to$last} & \cellcolor{tabstabband}{\scriptsize pp/month} & \cellcolor{tabstabband}{\scriptsize\color{basetext} $\bar{q}^{*}$ (\%)} \\
    \midrule
    \multirow{11}{*}{50} & DeepDrebin & 54.2 & 5.34 & 14.3 & \textcolor{worstcolor}{\itshape 55.2} & {\color{basetext}$+$1.0} & \textbf{$+$1.81} & $-$24.5 & {\color{basetext}0.76$\to$0.46} & $-$0.11 & {\color{basetext}3.0} \\
     & CADE-OOD$^{c}$ & 62.4 & \textcolor{worstcolor}{\itshape 5.95} & \textcolor{worstcolor}{\itshape 4.1} & 62.4\rlap{$^{\dagger}$} & {\color{basetext}0.0}\rlap{$^{\dagger}$} & \textcolor{worstcolor}{\itshape $-$0.66} & $-$4.0 & {\color{basetext}0.47$\to$0.42} & $-$0.15\rlap{$^{\dagger}$} & {\color{basetext}1.1}\rlap{$^{\dagger}$} \\
     & CADE-MSP$^{c}$ & 62.4 & 5.43 & 11.1 & 60.7 & {\color{basetext}$-$1.7} & \textcolor{worstcolor}{\itshape $-$0.66} & $-$2.1 & {\color{basetext}0.55$\to$0.50} & $-$0.07 & {\color{basetext}4.6} \\
     & HCC-PL$^{a}$ & \textbf{75.1} & 1.27 & 26.8 & \textbf{76.5} & {\color{basetext}$+$1.3} & $+$0.79 & \textbf{$+$7.9} & {\color{basetext}0.75$\to$0.85} & $-$0.31 & {\color{basetext}2.1} \\
     & HCC-MSP$^{a}$ & \textbf{75.1} & 1.34 & \textbf{28.1} & 76.1 & {\color{basetext}$+$1.0} & $+$0.79 & \textbf{$+$7.9} & {\color{basetext}0.75$\to$0.85} & $-$0.29 & {\color{basetext}2.1} \\
     & Drebin & 62.6 & 1.46 & 20.8 & 61.3 & {\color{basetext}$-$1.3} & $+$0.97 & $+$3.0 & {\color{basetext}0.83$\to$0.83} & $-$0.09 & {\color{basetext}2.8} \\
    \addlinespace[2pt]
     & \multicolumn{11}{>{\columncolor{subheadcolor}[\tabcolsep][\tabcolsep]}l}{\itshape Transcendent: inductive (ICE) and cross-conformal (CCE)$^{b}$} \\
     & ICE (Drebin) & 63.2 & 1.33 & 25.2 & 68.6\rlap{$^{\dagger}$} & {\color{basetext}$+$5.5}\rlap{$^{\dagger}$} & $+$1.33 & $+$4.9 & {\color{basetext}0.86$\to$0.84} & $-$0.25\rlap{$^{\dagger}$} & {\color{basetext}2.8}\rlap{$^{\dagger}$} \\
     & CCE (Drebin) & 62.9 & \textbf{1.24} & 25.0 & 67.7 & {\color{basetext}$+$4.8} & $+$1.42 & $+$4.6 & {\color{basetext}0.87$\to$0.85} & $-$0.20 & {\color{basetext}2.7} \\
     & ICE (DeepDrebin) & 54.5 & 4.59 & 11.5 & 56.4 & {\color{basetext}$+$1.9} & $+$1.02 & $-$22.3 & {\color{basetext}0.78$\to$0.52} & $-$0.12 & {\color{basetext}2.8} \\
     & CCE (DeepDrebin) & \textcolor{worstcolor}{\itshape 53.9} & 4.35 & 12.4 & 55.6 & {\color{basetext}$+$1.6} & $+$1.39 & \textcolor{worstcolor}{\itshape $-$25.4} & {\color{basetext}0.77$\to$0.49} & $-$0.07 & {\color{basetext}2.9} \\
    \midrule
    \multirow{11}{*}{100} & DeepDrebin & \textcolor{worstcolor}{\itshape 58.2} & 5.73 & 19.9 & \textcolor{worstcolor}{\itshape 59.2} & {\color{basetext}$+$0.9} & $+$0.84 & \textcolor{worstcolor}{\itshape $-$20.0} & {\color{basetext}0.69$\to$0.43} & $-$0.33 & {\color{basetext}5.8} \\
     & CADE-OOD$^{c}$ & 62.7 & \textcolor{worstcolor}{\itshape 6.04} & \textcolor{worstcolor}{\itshape 8.6} & 61.0 & {\color{basetext}$-$1.7} & \textcolor{worstcolor}{\itshape $+$0.11} & $+$1.4 & {\color{basetext}0.50$\to$0.44} & $-$0.28 & {\color{basetext}3.3} \\
     & CADE-MSP$^{c}$ & 62.7 & 5.42 & 16.0 & 59.3 & {\color{basetext}$-$3.4} & \textcolor{worstcolor}{\itshape $+$0.11} & $+$0.6 & {\color{basetext}0.53$\to$0.52} & $-$0.36 & {\color{basetext}7.1} \\
     & HCC-PL$^{a}$ & \textbf{76.9} & 1.00 & \textbf{46.3} & \textbf{80.3} & {\color{basetext}$+$3.4} & $+$1.10 & \textbf{$+$4.6} & {\color{basetext}0.78$\to$0.84} & $-$0.64 & {\color{basetext}3.6} \\
     & HCC-MSP$^{a}$ & \textbf{76.9} & 1.01 & 45.8 & 79.8 & {\color{basetext}$+$2.9} & $+$1.10 & $+$4.2 & {\color{basetext}0.78$\to$0.83} & $-$0.66 & {\color{basetext}3.7} \\
     & Drebin & 67.9 & 1.01 & 34.8 & 71.8 & {\color{basetext}$+$3.9} & $+$1.08 & $+$3.6 & {\color{basetext}0.85$\to$0.85} & $-$0.47 & {\color{basetext}5.4} \\
    \addlinespace[2pt]
     & \multicolumn{11}{>{\columncolor{subheadcolor}[\tabcolsep][\tabcolsep]}l}{\itshape Transcendent: inductive (ICE) and cross-conformal (CCE)$^{b}$} \\
     & ICE (Drebin) & 67.3 & 1.02 & 37.8 & 76.6 & {\color{basetext}$+$9.3} & \textbf{$+$1.17} & $+$4.5 & {\color{basetext}0.85$\to$0.86} & $-$0.41 & {\color{basetext}5.1} \\
     & CCE (Drebin) & 67.6 & \textbf{0.93} & 37.4 & 77.3 & {\color{basetext}$+$9.7} & $+$1.07 & $+$4.5 & {\color{basetext}0.86$\to$0.87} & $-$0.37 & {\color{basetext}5.1} \\
     & ICE (DeepDrebin) & 58.7 & 4.22 & 19.6 & 62.0 & {\color{basetext}$+$3.3} & $+$0.46 & $-$19.9 & {\color{basetext}0.77$\to$0.52} & $-$0.35 & {\color{basetext}5.4} \\
     & CCE (DeepDrebin) & 61.3 & 3.70 & 17.9 & 63.6 & {\color{basetext}$+$2.3} & $+$0.47 & $-$18.8 & {\color{basetext}0.75$\to$0.51} & $-$0.32 & {\color{basetext}5.2} \\
    \midrule
    \multirow{11}{*}{200} & DeepDrebin & \textcolor{worstcolor}{\itshape 59.1} & \textcolor{worstcolor}{\itshape 4.89} & \textcolor{worstcolor}{\itshape 29.9} & \textcolor{worstcolor}{\itshape 63.8} & {\color{basetext}$+$4.7} & $+$1.06 & $-$11.0 & {\color{basetext}0.75$\to$0.52} & $-$0.78 & {\color{basetext}10.5} \\
     & CADE-OOD$^{c}$\,{\scriptsize(4/5)} & 66.1 & 4.79 & 20.8 & 62.2 & {\color{basetext}$-$3.8} & $+$0.65 & $-$6.3 & {\color{basetext}0.55$\to$0.49} & $-$0.55 & {\color{basetext}8.2} \\
     & CADE-MSP$^{c}$\,{\scriptsize(4/5)} & 66.1 & 4.58 & 29.9 & 65.5 & {\color{basetext}$-$0.5} & $+$0.65 & $-$0.7 & {\color{basetext}0.57$\to$0.56} & $-$0.66 & {\color{basetext}11.0} \\
     & HCC-PL$^{a}$ & \textbf{78.3} & 0.91 & 60.0 & 77.0 & {\color{basetext}$-$1.3} & $+$1.13 & $+$3.2 & {\color{basetext}0.78$\to$0.82} & $-$1.10 & {\color{basetext}7.8} \\
     & HCC-MSP$^{a}$ & \textbf{78.3} & 0.98 & 60.5 & 76.5 & {\color{basetext}$-$1.9} & $+$1.13 & $+$2.0 & {\color{basetext}0.78$\to$0.82} & $-$1.23 & {\color{basetext}8.0} \\
     & Drebin & 69.2 & 0.91 & 58.1 & 68.3 & {\color{basetext}$-$1.0} & $+$0.95 & \textbf{$+$6.2} & {\color{basetext}0.85$\to$0.89} & $-$0.67 & {\color{basetext}10.7} \\
    \addlinespace[2pt]
     & \multicolumn{11}{>{\columncolor{subheadcolor}[\tabcolsep][\tabcolsep]}l}{\itshape Transcendent: inductive (ICE) and cross-conformal (CCE)$^{b}$} \\
     & ICE (Drebin) & 69.7 & 0.84 & 61.3 & 81.5 & {\color{basetext}$+$11.8} & $+$1.29 & $+$1.0 & {\color{basetext}0.88$\to$0.87} & $-$0.65 & {\color{basetext}10.1} \\
     & CCE (Drebin) & 70.4 & \textbf{0.82} & \textbf{62.5} & \textbf{82.6} & {\color{basetext}$+$12.3} & $+$1.26 & $+$2.0 & {\color{basetext}0.88$\to$0.88} & $-$0.56 & {\color{basetext}10.5} \\
     & ICE (DeepDrebin) & 64.5 & 3.03 & 36.0 & 69.4 & {\color{basetext}$+$4.9} & \textbf{$+$1.56} & \textcolor{worstcolor}{\itshape $-$11.5} & {\color{basetext}0.79$\to$0.63} & $-$0.50 & {\color{basetext}10.6} \\
     & CCE (DeepDrebin) & 65.9 & 2.70 & 36.3 & 69.7 & {\color{basetext}$+$3.8} & \textcolor{worstcolor}{\itshape $+$0.94} & $-$7.5 & {\color{basetext}0.75$\to$0.64} & $-$0.68 & {\color{basetext}10.3} \\
    \midrule
    \multirow{11}{*}{400} & DeepDrebin & \textcolor{worstcolor}{\itshape 69.7} & \textcolor{worstcolor}{\itshape 2.42} & \textcolor{worstcolor}{\itshape 64.0} & 80.7 & {\color{basetext}$+$11.0} & \textbf{$+$1.87} & $+$1.0 & {\color{basetext}0.78$\to$0.75} & $-$1.15 & {\color{basetext}21.1} \\
     & CADE-OOD$^{c}$\,{\scriptsize(4/5)} & 76.7 & 3.90 & 40.0 & 65.2 & {\color{basetext}$-$11.5} & $+$0.77 & $-$5.5 & {\color{basetext}0.54$\to$0.48} & $-$1.15 & {\color{basetext}12.9} \\
     & CADE-MSP$^{c}$\,{\scriptsize(4/5)} & 76.7 & 3.66 & 41.4 & 75.5 & {\color{basetext}$-$1.2} & $+$0.77 & $-$4.4 & {\color{basetext}0.57$\to$0.52} & $-$1.05 & {\color{basetext}26.1} \\
     & HCC-PL$^{a}$ & \textbf{84.5} & 0.84 & 71.8 & \textcolor{worstcolor}{\itshape 51.2} & {\color{basetext}$-$33.4} & \textcolor{worstcolor}{\itshape $+$0.67} & $+$5.3 & {\color{basetext}0.78$\to$0.82} & $-$1.51 & {\color{basetext}17.1} \\
     & HCC-MSP$^{a}$ & \textbf{84.5} & 0.86 & 72.0 & 53.6 & {\color{basetext}$-$30.9} & \textcolor{worstcolor}{\itshape $+$0.67} & $+$4.5 & {\color{basetext}0.78$\to$0.82} & $-$1.56 & {\color{basetext}17.1} \\
     & Drebin & 71.5 & 0.71 & \textbf{84.9} & 66.7 & {\color{basetext}$-$4.8} & $+$1.21 & $+$1.9 & {\color{basetext}0.86$\to$0.89} & $-$1.03 & {\color{basetext}20.7} \\
    \addlinespace[2pt]
     & \multicolumn{11}{>{\columncolor{subheadcolor}[\tabcolsep][\tabcolsep]}l}{\itshape Transcendent: inductive (ICE) and cross-conformal (CCE)$^{b}$} \\
     & ICE (Drebin) & 72.3 & 0.66 & 83.1 & \textbf{87.0} & {\color{basetext}$+$14.7} & $+$1.34 & \textcolor{worstcolor}{\itshape $+$0.2} & {\color{basetext}0.87$\to$0.87} & $-$1.02 & {\color{basetext}19.8} \\
     & CCE (Drebin) & 72.2 & \textbf{0.63} & 81.7 & 84.1 & {\color{basetext}$+$11.9} & $+$1.39 & $+$0.5 & {\color{basetext}0.87$\to$0.88} & $-$1.20 & {\color{basetext}19.7} \\
     & ICE (DeepDrebin) & 74.9 & 1.60 & 68.0 & 82.4 & {\color{basetext}$+$7.5} & $+$1.51 & \textbf{$+$6.9} & {\color{basetext}0.77$\to$0.82} & $-$1.23 & {\color{basetext}20.6} \\
     & CCE (DeepDrebin) & 76.0 & 1.55 & 68.1 & 83.3 & {\color{basetext}$+$7.3} & $+$1.61 & $+$5.6 & {\color{basetext}0.77$\to$0.83} & $-$1.02 & {\color{basetext}20.4} \\
    \bottomrule
  \end{tabular}
  \par\smallskip
  {\scriptsize\raggedright Random Cap reference: 4.6\% at $B_M=50$; 9.2\% at $B_M=100$; 18.5\% at $B_M=200$; 37.0\% at $B_M=400$. Random Forest, SecSVM, DRMD, RAMDA, TIF and MalCertain were not run on this dataset. $^{a}$HCC with family-supervised training and warm updates; MSP and pseudo-loss (PL) score the same runs and predictions. $^{b}$Transcendent: folds re-drawn at each monthly cold refit; ICE fold 0, CCE five folds; both share CCE-credibility acquisition. $^{c}$CADE with a cold-refitted neural head; OOD and MSP score the same runs and predictions. $^{\dagger}$Tied scores make the rolling window miss its referral target by more than 25\% on average (CADE-OOD at $B_M=50$; ICE (Drebin) at $B_M=50$); quota-dependent cells are unranked.\par}
\end{table*}

% Generated by generate_satml_vtt2_budget_tables.py; do not edit numbers by hand.
% Layout, metrics and threshold protocol are those of Table I (generate_satml_vtt2_frozen_table.py).
\providecolor{tabbaseband}{RGB}{240,240,240}
\providecolor{tabreliband}{RGB}{230,245,220}
\providecolor{tabstabband}{RGB}{235,240,250}
\providecolor{worstcolor}{RGB}{176,32,32}
\providecolor{subheadcolor}{RGB}{246,246,246}
\providecolor{basetext}{RGB}{90,90,90}
\providecolor{relitext}{RGB}{46,125,50}
\providecolor{stabtext}{RGB}{40,53,147}
\providecommand{\codename}{\textsc{Aurora}}
\ifdefined\frzwtmp\else\newlength{\frzwtmp}\fi
\ifdefined\lgwBA\else\newlength{\lgwBA}\fi\setlength{\lgwBA}{0pt}
\settowidth{\frzwtmp}{\footnotesize \textbf{91.6}}\ifdim\frzwtmp>\lgwBA\setlength{\lgwBA}{\frzwtmp}\fi
\settowidth{\frzwtmp}{\footnotesize {\color{basetext}$+$1.8}}\ifdim\frzwtmp>\lgwBA\setlength{\lgwBA}{\frzwtmp}\fi
\settowidth{\frzwtmp}{\footnotesize \textcolor{worstcolor}{\itshape 84.8}}\ifdim\frzwtmp>\lgwBA\setlength{\lgwBA}{\frzwtmp}\fi
\settowidth{\frzwtmp}{\footnotesize {\color{basetext}$-$1.0}}\ifdim\frzwtmp>\lgwBA\setlength{\lgwBA}{\frzwtmp}\fi
\settowidth{\frzwtmp}{\footnotesize 86.4}\ifdim\frzwtmp>\lgwBA\setlength{\lgwBA}{\frzwtmp}\fi
\settowidth{\frzwtmp}{\footnotesize {\color{basetext}$+$0.6}}\ifdim\frzwtmp>\lgwBA\setlength{\lgwBA}{\frzwtmp}\fi
\settowidth{\frzwtmp}{\footnotesize 90.0}\ifdim\frzwtmp>\lgwBA\setlength{\lgwBA}{\frzwtmp}\fi
\settowidth{\frzwtmp}{\footnotesize {\color{basetext}$+$1.4}}\ifdim\frzwtmp>\lgwBA\setlength{\lgwBA}{\frzwtmp}\fi
\settowidth{\frzwtmp}{\footnotesize 90.3}\ifdim\frzwtmp>\lgwBA\setlength{\lgwBA}{\frzwtmp}\fi
\settowidth{\frzwtmp}{\footnotesize {\color{basetext}$+$1.7}}\ifdim\frzwtmp>\lgwBA\setlength{\lgwBA}{\frzwtmp}\fi
\settowidth{\frzwtmp}{\footnotesize 89.4}\ifdim\frzwtmp>\lgwBA\setlength{\lgwBA}{\frzwtmp}\fi
\settowidth{\frzwtmp}{\footnotesize 89.5}\ifdim\frzwtmp>\lgwBA\setlength{\lgwBA}{\frzwtmp}\fi
\settowidth{\frzwtmp}{\footnotesize {\color{basetext}$+$2.6}}\ifdim\frzwtmp>\lgwBA\setlength{\lgwBA}{\frzwtmp}\fi
\settowidth{\frzwtmp}{\footnotesize 89.7}\ifdim\frzwtmp>\lgwBA\setlength{\lgwBA}{\frzwtmp}\fi
\settowidth{\frzwtmp}{\footnotesize {\color{basetext}$+$2.5}}\ifdim\frzwtmp>\lgwBA\setlength{\lgwBA}{\frzwtmp}\fi
\settowidth{\frzwtmp}{\footnotesize {\color{basetext}$+$1.6}}\ifdim\frzwtmp>\lgwBA\setlength{\lgwBA}{\frzwtmp}\fi
\settowidth{\frzwtmp}{\footnotesize 90.5}\ifdim\frzwtmp>\lgwBA\setlength{\lgwBA}{\frzwtmp}\fi
\settowidth{\frzwtmp}{\footnotesize {\color{basetext}$+$1.5}}\ifdim\frzwtmp>\lgwBA\setlength{\lgwBA}{\frzwtmp}\fi
\settowidth{\frzwtmp}{\footnotesize \textbf{93.7}}\ifdim\frzwtmp>\lgwBA\setlength{\lgwBA}{\frzwtmp}\fi
\settowidth{\frzwtmp}{\footnotesize \textcolor{worstcolor}{\itshape 85.3}}\ifdim\frzwtmp>\lgwBA\setlength{\lgwBA}{\frzwtmp}\fi
\settowidth{\frzwtmp}{\footnotesize {\color{basetext}$-$1.9}}\ifdim\frzwtmp>\lgwBA\setlength{\lgwBA}{\frzwtmp}\fi
\settowidth{\frzwtmp}{\footnotesize 87.9}\ifdim\frzwtmp>\lgwBA\setlength{\lgwBA}{\frzwtmp}\fi
\settowidth{\frzwtmp}{\footnotesize 92.2}\ifdim\frzwtmp>\lgwBA\setlength{\lgwBA}{\frzwtmp}\fi
\settowidth{\frzwtmp}{\footnotesize {\color{basetext}$+$2.1}}\ifdim\frzwtmp>\lgwBA\setlength{\lgwBA}{\frzwtmp}\fi
\settowidth{\frzwtmp}{\footnotesize 91.4}\ifdim\frzwtmp>\lgwBA\setlength{\lgwBA}{\frzwtmp}\fi
\settowidth{\frzwtmp}{\footnotesize {\color{basetext}$+$3.1}}\ifdim\frzwtmp>\lgwBA\setlength{\lgwBA}{\frzwtmp}\fi
\settowidth{\frzwtmp}{\footnotesize 91.6}\ifdim\frzwtmp>\lgwBA\setlength{\lgwBA}{\frzwtmp}\fi
\settowidth{\frzwtmp}{\footnotesize {\color{basetext}$+$3.8}}\ifdim\frzwtmp>\lgwBA\setlength{\lgwBA}{\frzwtmp}\fi
\settowidth{\frzwtmp}{\footnotesize 91.8}\ifdim\frzwtmp>\lgwBA\setlength{\lgwBA}{\frzwtmp}\fi
\settowidth{\frzwtmp}{\footnotesize {\color{basetext}$+$3.7}}\ifdim\frzwtmp>\lgwBA\setlength{\lgwBA}{\frzwtmp}\fi
\settowidth{\frzwtmp}{\footnotesize {\color{basetext}$+$2.4}}\ifdim\frzwtmp>\lgwBA\setlength{\lgwBA}{\frzwtmp}\fi
\settowidth{\frzwtmp}{\footnotesize 92.8}\ifdim\frzwtmp>\lgwBA\setlength{\lgwBA}{\frzwtmp}\fi
\settowidth{\frzwtmp}{\footnotesize {\color{basetext}$+$2.2}}\ifdim\frzwtmp>\lgwBA\setlength{\lgwBA}{\frzwtmp}\fi
\settowidth{\frzwtmp}{\footnotesize \textbf{95.2}}\ifdim\frzwtmp>\lgwBA\setlength{\lgwBA}{\frzwtmp}\fi
\settowidth{\frzwtmp}{\footnotesize {\color{basetext}$+$3.0}}\ifdim\frzwtmp>\lgwBA\setlength{\lgwBA}{\frzwtmp}\fi
\settowidth{\frzwtmp}{\footnotesize \textcolor{worstcolor}{\itshape 82.7}}\ifdim\frzwtmp>\lgwBA\setlength{\lgwBA}{\frzwtmp}\fi
\settowidth{\frzwtmp}{\footnotesize {\color{basetext}$-$6.3}}\ifdim\frzwtmp>\lgwBA\setlength{\lgwBA}{\frzwtmp}\fi
\settowidth{\frzwtmp}{\footnotesize 94.0}\ifdim\frzwtmp>\lgwBA\setlength{\lgwBA}{\frzwtmp}\fi
\settowidth{\frzwtmp}{\footnotesize {\color{basetext}$+$2.7}}\ifdim\frzwtmp>\lgwBA\setlength{\lgwBA}{\frzwtmp}\fi
\settowidth{\frzwtmp}{\footnotesize 93.7\rlap{$^{\dagger}$}}\ifdim\frzwtmp>\lgwBA\setlength{\lgwBA}{\frzwtmp}\fi
\settowidth{\frzwtmp}{\footnotesize {\color{basetext}$+$2.4}\rlap{$^{\dagger}$}}\ifdim\frzwtmp>\lgwBA\setlength{\lgwBA}{\frzwtmp}\fi
\settowidth{\frzwtmp}{\footnotesize 93.1}\ifdim\frzwtmp>\lgwBA\setlength{\lgwBA}{\frzwtmp}\fi
\settowidth{\frzwtmp}{\footnotesize {\color{basetext}$+$4.2}}\ifdim\frzwtmp>\lgwBA\setlength{\lgwBA}{\frzwtmp}\fi
\settowidth{\frzwtmp}{\footnotesize 93.2}\ifdim\frzwtmp>\lgwBA\setlength{\lgwBA}{\frzwtmp}\fi
\settowidth{\frzwtmp}{\footnotesize {\color{basetext}$+$5.3}}\ifdim\frzwtmp>\lgwBA\setlength{\lgwBA}{\frzwtmp}\fi
\settowidth{\frzwtmp}{\footnotesize 93.3}\ifdim\frzwtmp>\lgwBA\setlength{\lgwBA}{\frzwtmp}\fi
\settowidth{\frzwtmp}{\footnotesize {\color{basetext}$+$5.2}}\ifdim\frzwtmp>\lgwBA\setlength{\lgwBA}{\frzwtmp}\fi
\settowidth{\frzwtmp}{\footnotesize 94.6}\ifdim\frzwtmp>\lgwBA\setlength{\lgwBA}{\frzwtmp}\fi
\settowidth{\frzwtmp}{\footnotesize {\color{basetext}$+$3.3}}\ifdim\frzwtmp>\lgwBA\setlength{\lgwBA}{\frzwtmp}\fi
\settowidth{\frzwtmp}{\footnotesize 95.0}\ifdim\frzwtmp>\lgwBA\setlength{\lgwBA}{\frzwtmp}\fi
\settowidth{\frzwtmp}{\footnotesize 96.1}\ifdim\frzwtmp>\lgwBA\setlength{\lgwBA}{\frzwtmp}\fi
\settowidth{\frzwtmp}{\footnotesize \textcolor{worstcolor}{\itshape 72.5}}\ifdim\frzwtmp>\lgwBA\setlength{\lgwBA}{\frzwtmp}\fi
\settowidth{\frzwtmp}{\footnotesize {\color{basetext}$-$16.9}}\ifdim\frzwtmp>\lgwBA\setlength{\lgwBA}{\frzwtmp}\fi
\settowidth{\frzwtmp}{\footnotesize {\color{basetext}$+$0.9}}\ifdim\frzwtmp>\lgwBA\setlength{\lgwBA}{\frzwtmp}\fi
\settowidth{\frzwtmp}{\footnotesize 88.1\rlap{$^{\dagger}$}}\ifdim\frzwtmp>\lgwBA\setlength{\lgwBA}{\frzwtmp}\fi
\settowidth{\frzwtmp}{\footnotesize {\color{basetext}$+$2.3}\rlap{$^{\dagger}$}}\ifdim\frzwtmp>\lgwBA\setlength{\lgwBA}{\frzwtmp}\fi
\settowidth{\frzwtmp}{\footnotesize 94.3}\ifdim\frzwtmp>\lgwBA\setlength{\lgwBA}{\frzwtmp}\fi
\settowidth{\frzwtmp}{\footnotesize {\color{basetext}$+$5.5}}\ifdim\frzwtmp>\lgwBA\setlength{\lgwBA}{\frzwtmp}\fi
\settowidth{\frzwtmp}{\footnotesize 94.7}\ifdim\frzwtmp>\lgwBA\setlength{\lgwBA}{\frzwtmp}\fi
\settowidth{\frzwtmp}{\footnotesize {\color{basetext}$+$6.3}}\ifdim\frzwtmp>\lgwBA\setlength{\lgwBA}{\frzwtmp}\fi
\settowidth{\frzwtmp}{\footnotesize {\color{basetext}$+$6.2}}\ifdim\frzwtmp>\lgwBA\setlength{\lgwBA}{\frzwtmp}\fi
\settowidth{\frzwtmp}{\footnotesize 96.5}\ifdim\frzwtmp>\lgwBA\setlength{\lgwBA}{\frzwtmp}\fi
\settowidth{\frzwtmp}{\footnotesize \textbf{96.7}}\ifdim\frzwtmp>\lgwBA\setlength{\lgwBA}{\frzwtmp}\fi
\settowidth{\frzwtmp}{\footnotesize {\scriptsize \%}}\ifdim\frzwtmp>\lgwBA\setlength{\lgwBA}{\frzwtmp}\fi
\settowidth{\frzwtmp}{\footnotesize {\scriptsize\color{basetext} $\Delta$F1$^{*}$ (pp)}}\ifdim\frzwtmp>\lgwBA\setlength{\lgwBA}{\frzwtmp}\fi
\ifdefined\lgwBB\else\newlength{\lgwBB}\fi\setlength{\lgwBB}{0pt}
\settowidth{\frzwtmp}{\footnotesize $+$1.4}\ifdim\frzwtmp>\lgwBB\setlength{\lgwBB}{\frzwtmp}\fi
\settowidth{\frzwtmp}{\footnotesize {\color{basetext}0.85$\to$0.85}}\ifdim\frzwtmp>\lgwBB\setlength{\lgwBB}{\frzwtmp}\fi
\settowidth{\frzwtmp}{\footnotesize $-$1.3}\ifdim\frzwtmp>\lgwBB\setlength{\lgwBB}{\frzwtmp}\fi
\settowidth{\frzwtmp}{\footnotesize {\color{basetext}0.69$\to$0.55}}\ifdim\frzwtmp>\lgwBB\setlength{\lgwBB}{\frzwtmp}\fi
\settowidth{\frzwtmp}{\footnotesize $+$0.1}\ifdim\frzwtmp>\lgwBB\setlength{\lgwBB}{\frzwtmp}\fi
\settowidth{\frzwtmp}{\footnotesize {\color{basetext}0.38$\to$0.43}}\ifdim\frzwtmp>\lgwBB\setlength{\lgwBB}{\frzwtmp}\fi
\settowidth{\frzwtmp}{\footnotesize \textcolor{worstcolor}{\itshape $-$1.9}}\ifdim\frzwtmp>\lgwBB\setlength{\lgwBB}{\frzwtmp}\fi
\settowidth{\frzwtmp}{\footnotesize {\color{basetext}0.78$\to$0.63}}\ifdim\frzwtmp>\lgwBB\setlength{\lgwBB}{\frzwtmp}\fi
\settowidth{\frzwtmp}{\footnotesize $-$1.5}\ifdim\frzwtmp>\lgwBB\setlength{\lgwBB}{\frzwtmp}\fi
\settowidth{\frzwtmp}{\footnotesize {\color{basetext}0.75$\to$0.59}}\ifdim\frzwtmp>\lgwBB\setlength{\lgwBB}{\frzwtmp}\fi
\settowidth{\frzwtmp}{\footnotesize $+$0.3}\ifdim\frzwtmp>\lgwBB\setlength{\lgwBB}{\frzwtmp}\fi
\settowidth{\frzwtmp}{\footnotesize {\color{basetext}0.91$\to$0.86}}\ifdim\frzwtmp>\lgwBB\setlength{\lgwBB}{\frzwtmp}\fi
\settowidth{\frzwtmp}{\footnotesize $+$1.1}\ifdim\frzwtmp>\lgwBB\setlength{\lgwBB}{\frzwtmp}\fi
\settowidth{\frzwtmp}{\footnotesize {\color{basetext}0.89$\to$0.90}}\ifdim\frzwtmp>\lgwBB\setlength{\lgwBB}{\frzwtmp}\fi
\settowidth{\frzwtmp}{\footnotesize $+$1.0}\ifdim\frzwtmp>\lgwBB\setlength{\lgwBB}{\frzwtmp}\fi
\settowidth{\frzwtmp}{\footnotesize {\color{basetext}0.90$\to$0.89}}\ifdim\frzwtmp>\lgwBB\setlength{\lgwBB}{\frzwtmp}\fi
\settowidth{\frzwtmp}{\footnotesize \textbf{$+$1.6}}\ifdim\frzwtmp>\lgwBB\setlength{\lgwBB}{\frzwtmp}\fi
\settowidth{\frzwtmp}{\footnotesize {\color{basetext}0.84$\to$0.84}}\ifdim\frzwtmp>\lgwBB\setlength{\lgwBB}{\frzwtmp}\fi
\settowidth{\frzwtmp}{\footnotesize {\color{basetext}0.83$\to$0.84}}\ifdim\frzwtmp>\lgwBB\setlength{\lgwBB}{\frzwtmp}\fi
\settowidth{\frzwtmp}{\footnotesize $+$2.1}\ifdim\frzwtmp>\lgwBB\setlength{\lgwBB}{\frzwtmp}\fi
\settowidth{\frzwtmp}{\footnotesize {\color{basetext}0.86$\to$0.88}}\ifdim\frzwtmp>\lgwBB\setlength{\lgwBB}{\frzwtmp}\fi
\settowidth{\frzwtmp}{\footnotesize $-$0.9}\ifdim\frzwtmp>\lgwBB\setlength{\lgwBB}{\frzwtmp}\fi
\settowidth{\frzwtmp}{\footnotesize {\color{basetext}0.65$\to$0.54}}\ifdim\frzwtmp>\lgwBB\setlength{\lgwBB}{\frzwtmp}\fi
\settowidth{\frzwtmp}{\footnotesize $+$0.4}\ifdim\frzwtmp>\lgwBB\setlength{\lgwBB}{\frzwtmp}\fi
\settowidth{\frzwtmp}{\footnotesize {\color{basetext}0.36$\to$0.45}}\ifdim\frzwtmp>\lgwBB\setlength{\lgwBB}{\frzwtmp}\fi
\settowidth{\frzwtmp}{\footnotesize {\color{basetext}0.78$\to$0.67}}\ifdim\frzwtmp>\lgwBB\setlength{\lgwBB}{\frzwtmp}\fi
\settowidth{\frzwtmp}{\footnotesize $-$1.2}\ifdim\frzwtmp>\lgwBB\setlength{\lgwBB}{\frzwtmp}\fi
\settowidth{\frzwtmp}{\footnotesize {\color{basetext}0.73$\to$0.61}}\ifdim\frzwtmp>\lgwBB\setlength{\lgwBB}{\frzwtmp}\fi
\settowidth{\frzwtmp}{\footnotesize $+$0.8}\ifdim\frzwtmp>\lgwBB\setlength{\lgwBB}{\frzwtmp}\fi
\settowidth{\frzwtmp}{\footnotesize {\color{basetext}0.91$\to$0.88}}\ifdim\frzwtmp>\lgwBB\setlength{\lgwBB}{\frzwtmp}\fi
\settowidth{\frzwtmp}{\footnotesize $+$0.9}\ifdim\frzwtmp>\lgwBB\setlength{\lgwBB}{\frzwtmp}\fi
\settowidth{\frzwtmp}{\footnotesize {\color{basetext}0.90$\to$0.88}}\ifdim\frzwtmp>\lgwBB\setlength{\lgwBB}{\frzwtmp}\fi
\settowidth{\frzwtmp}{\footnotesize {\color{basetext}0.83$\to$0.88}}\ifdim\frzwtmp>\lgwBB\setlength{\lgwBB}{\frzwtmp}\fi
\settowidth{\frzwtmp}{\footnotesize \textbf{$+$2.3}}\ifdim\frzwtmp>\lgwBB\setlength{\lgwBB}{\frzwtmp}\fi
\settowidth{\frzwtmp}{\footnotesize $-$0.7}\ifdim\frzwtmp>\lgwBB\setlength{\lgwBB}{\frzwtmp}\fi
\settowidth{\frzwtmp}{\footnotesize {\color{basetext}0.65$\to$0.55}}\ifdim\frzwtmp>\lgwBB\setlength{\lgwBB}{\frzwtmp}\fi
\settowidth{\frzwtmp}{\footnotesize $-$0.8}\ifdim\frzwtmp>\lgwBB\setlength{\lgwBB}{\frzwtmp}\fi
\settowidth{\frzwtmp}{\footnotesize {\color{basetext}0.37$\to$0.39}}\ifdim\frzwtmp>\lgwBB\setlength{\lgwBB}{\frzwtmp}\fi
\settowidth{\frzwtmp}{\footnotesize \textcolor{worstcolor}{\itshape $-$1.4}}\ifdim\frzwtmp>\lgwBB\setlength{\lgwBB}{\frzwtmp}\fi
\settowidth{\frzwtmp}{\footnotesize {\color{basetext}0.79$\to$0.69}}\ifdim\frzwtmp>\lgwBB\setlength{\lgwBB}{\frzwtmp}\fi
\settowidth{\frzwtmp}{\footnotesize {\color{basetext}0.75$\to$0.66}}\ifdim\frzwtmp>\lgwBB\setlength{\lgwBB}{\frzwtmp}\fi
\settowidth{\frzwtmp}{\footnotesize {\color{basetext}0.91$\to$0.89}}\ifdim\frzwtmp>\lgwBB\setlength{\lgwBB}{\frzwtmp}\fi
\settowidth{\frzwtmp}{\footnotesize {\color{basetext}0.91$\to$0.90}}\ifdim\frzwtmp>\lgwBB\setlength{\lgwBB}{\frzwtmp}\fi
\settowidth{\frzwtmp}{\footnotesize {\color{basetext}0.90$\to$0.90}}\ifdim\frzwtmp>\lgwBB\setlength{\lgwBB}{\frzwtmp}\fi
\settowidth{\frzwtmp}{\footnotesize $+$2.6}\ifdim\frzwtmp>\lgwBB\setlength{\lgwBB}{\frzwtmp}\fi
\settowidth{\frzwtmp}{\footnotesize {\color{basetext}0.83$\to$0.89}}\ifdim\frzwtmp>\lgwBB\setlength{\lgwBB}{\frzwtmp}\fi
\settowidth{\frzwtmp}{\footnotesize \textbf{$+$2.7}}\ifdim\frzwtmp>\lgwBB\setlength{\lgwBB}{\frzwtmp}\fi
\settowidth{\frzwtmp}{\footnotesize $+$2.0}\ifdim\frzwtmp>\lgwBB\setlength{\lgwBB}{\frzwtmp}\fi
\settowidth{\frzwtmp}{\footnotesize {\color{basetext}0.86$\to$0.90}}\ifdim\frzwtmp>\lgwBB\setlength{\lgwBB}{\frzwtmp}\fi
\settowidth{\frzwtmp}{\footnotesize $-$1.0}\ifdim\frzwtmp>\lgwBB\setlength{\lgwBB}{\frzwtmp}\fi
\settowidth{\frzwtmp}{\footnotesize {\color{basetext}0.31$\to$0.32}}\ifdim\frzwtmp>\lgwBB\setlength{\lgwBB}{\frzwtmp}\fi
\settowidth{\frzwtmp}{\footnotesize $-$0.1}\ifdim\frzwtmp>\lgwBB\setlength{\lgwBB}{\frzwtmp}\fi
\settowidth{\frzwtmp}{\footnotesize {\color{basetext}0.79$\to$0.82}}\ifdim\frzwtmp>\lgwBB\setlength{\lgwBB}{\frzwtmp}\fi
\settowidth{\frzwtmp}{\footnotesize \textcolor{worstcolor}{\itshape $-$1.7}}\ifdim\frzwtmp>\lgwBB\setlength{\lgwBB}{\frzwtmp}\fi
\settowidth{\frzwtmp}{\footnotesize {\color{basetext}0.76$\to$0.57}}\ifdim\frzwtmp>\lgwBB\setlength{\lgwBB}{\frzwtmp}\fi
\settowidth{\frzwtmp}{\footnotesize {\color{basetext}0.92$\to$0.88}}\ifdim\frzwtmp>\lgwBB\setlength{\lgwBB}{\frzwtmp}\fi
\settowidth{\frzwtmp}{\footnotesize $+$2.4}\ifdim\frzwtmp>\lgwBB\setlength{\lgwBB}{\frzwtmp}\fi
\settowidth{\frzwtmp}{\footnotesize \textbf{$+$2.6}}\ifdim\frzwtmp>\lgwBB\setlength{\lgwBB}{\frzwtmp}\fi
\settowidth{\frzwtmp}{\footnotesize {\color{basetext}0.85$\to$0.90}}\ifdim\frzwtmp>\lgwBB\setlength{\lgwBB}{\frzwtmp}\fi
\settowidth{\frzwtmp}{\footnotesize {\scriptsize $10^{-3}$/month}}\ifdim\frzwtmp>\lgwBB\setlength{\lgwBB}{\frzwtmp}\fi
\settowidth{\frzwtmp}{\footnotesize {\scriptsize\color{basetext} first$\to$last}}\ifdim\frzwtmp>\lgwBB\setlength{\lgwBB}{\frzwtmp}\fi
\ifdefined\lgwBC\else\newlength{\lgwBC}\fi\setlength{\lgwBC}{0pt}
\settowidth{\frzwtmp}{\footnotesize $-$0.01}\ifdim\frzwtmp>\lgwBC\setlength{\lgwBC}{\frzwtmp}\fi
\settowidth{\frzwtmp}{\footnotesize {\color{basetext}1.4}}\ifdim\frzwtmp>\lgwBC\setlength{\lgwBC}{\frzwtmp}\fi
\settowidth{\frzwtmp}{\footnotesize $+$0.01}\ifdim\frzwtmp>\lgwBC\setlength{\lgwBC}{\frzwtmp}\fi
\settowidth{\frzwtmp}{\footnotesize {\color{basetext}1.1}}\ifdim\frzwtmp>\lgwBC\setlength{\lgwBC}{\frzwtmp}\fi
\settowidth{\frzwtmp}{\footnotesize {\color{basetext}1.9}}\ifdim\frzwtmp>\lgwBC\setlength{\lgwBC}{\frzwtmp}\fi
\settowidth{\frzwtmp}{\footnotesize 0.00}\ifdim\frzwtmp>\lgwBC\setlength{\lgwBC}{\frzwtmp}\fi
\settowidth{\frzwtmp}{\footnotesize {\color{basetext}1.3}}\ifdim\frzwtmp>\lgwBC\setlength{\lgwBC}{\frzwtmp}\fi
\settowidth{\frzwtmp}{\footnotesize {\color{basetext}1.2}}\ifdim\frzwtmp>\lgwBC\setlength{\lgwBC}{\frzwtmp}\fi
\settowidth{\frzwtmp}{\footnotesize {\color{basetext}2.7}}\ifdim\frzwtmp>\lgwBC\setlength{\lgwBC}{\frzwtmp}\fi
\settowidth{\frzwtmp}{\footnotesize $+$0.02}\ifdim\frzwtmp>\lgwBC\setlength{\lgwBC}{\frzwtmp}\fi
\settowidth{\frzwtmp}{\footnotesize {\color{basetext}2.2}}\ifdim\frzwtmp>\lgwBC\setlength{\lgwBC}{\frzwtmp}\fi
\settowidth{\frzwtmp}{\footnotesize $-$0.03}\ifdim\frzwtmp>\lgwBC\setlength{\lgwBC}{\frzwtmp}\fi
\settowidth{\frzwtmp}{\footnotesize {\color{basetext}3.8}}\ifdim\frzwtmp>\lgwBC\setlength{\lgwBC}{\frzwtmp}\fi
\settowidth{\frzwtmp}{\footnotesize {\color{basetext}2.6}}\ifdim\frzwtmp>\lgwBC\setlength{\lgwBC}{\frzwtmp}\fi
\settowidth{\frzwtmp}{\footnotesize $-$0.02}\ifdim\frzwtmp>\lgwBC\setlength{\lgwBC}{\frzwtmp}\fi
\settowidth{\frzwtmp}{\footnotesize {\color{basetext}2.3}}\ifdim\frzwtmp>\lgwBC\setlength{\lgwBC}{\frzwtmp}\fi
\settowidth{\frzwtmp}{\footnotesize {\color{basetext}2.5}}\ifdim\frzwtmp>\lgwBC\setlength{\lgwBC}{\frzwtmp}\fi
\settowidth{\frzwtmp}{\footnotesize {\color{basetext}2.4}}\ifdim\frzwtmp>\lgwBC\setlength{\lgwBC}{\frzwtmp}\fi
\settowidth{\frzwtmp}{\footnotesize {\color{basetext}5.4}}\ifdim\frzwtmp>\lgwBC\setlength{\lgwBC}{\frzwtmp}\fi
\settowidth{\frzwtmp}{\footnotesize {\color{basetext}4.9}}\ifdim\frzwtmp>\lgwBC\setlength{\lgwBC}{\frzwtmp}\fi
\settowidth{\frzwtmp}{\footnotesize {\color{basetext}7.4}}\ifdim\frzwtmp>\lgwBC\setlength{\lgwBC}{\frzwtmp}\fi
\settowidth{\frzwtmp}{\footnotesize {\color{basetext}5.2}}\ifdim\frzwtmp>\lgwBC\setlength{\lgwBC}{\frzwtmp}\fi
\settowidth{\frzwtmp}{\footnotesize $-$0.07\rlap{$^{\dagger}$}}\ifdim\frzwtmp>\lgwBC\setlength{\lgwBC}{\frzwtmp}\fi
\settowidth{\frzwtmp}{\footnotesize {\color{basetext}3.2}\rlap{$^{\dagger}$}}\ifdim\frzwtmp>\lgwBC\setlength{\lgwBC}{\frzwtmp}\fi
\settowidth{\frzwtmp}{\footnotesize {\color{basetext}5.1}}\ifdim\frzwtmp>\lgwBC\setlength{\lgwBC}{\frzwtmp}\fi
\settowidth{\frzwtmp}{\footnotesize {\color{basetext}4.8}}\ifdim\frzwtmp>\lgwBC\setlength{\lgwBC}{\frzwtmp}\fi
\settowidth{\frzwtmp}{\footnotesize {\color{basetext}10.7}}\ifdim\frzwtmp>\lgwBC\setlength{\lgwBC}{\frzwtmp}\fi
\settowidth{\frzwtmp}{\footnotesize $-$0.07}\ifdim\frzwtmp>\lgwBC\setlength{\lgwBC}{\frzwtmp}\fi
\settowidth{\frzwtmp}{\footnotesize {\color{basetext}8.9}}\ifdim\frzwtmp>\lgwBC\setlength{\lgwBC}{\frzwtmp}\fi
\settowidth{\frzwtmp}{\footnotesize $-$0.12}\ifdim\frzwtmp>\lgwBC\setlength{\lgwBC}{\frzwtmp}\fi
\settowidth{\frzwtmp}{\footnotesize {\color{basetext}15.0}}\ifdim\frzwtmp>\lgwBC\setlength{\lgwBC}{\frzwtmp}\fi
\settowidth{\frzwtmp}{\footnotesize {\color{basetext}11.6}}\ifdim\frzwtmp>\lgwBC\setlength{\lgwBC}{\frzwtmp}\fi
\settowidth{\frzwtmp}{\footnotesize $-$0.18\rlap{$^{\dagger}$}}\ifdim\frzwtmp>\lgwBC\setlength{\lgwBC}{\frzwtmp}\fi
\settowidth{\frzwtmp}{\footnotesize {\color{basetext}4.1}\rlap{$^{\dagger}$}}\ifdim\frzwtmp>\lgwBC\setlength{\lgwBC}{\frzwtmp}\fi
\settowidth{\frzwtmp}{\footnotesize {\color{basetext}10.0}}\ifdim\frzwtmp>\lgwBC\setlength{\lgwBC}{\frzwtmp}\fi
\settowidth{\frzwtmp}{\footnotesize {\color{basetext}10.2}}\ifdim\frzwtmp>\lgwBC\setlength{\lgwBC}{\frzwtmp}\fi
\settowidth{\frzwtmp}{\footnotesize {\scriptsize pp/month}}\ifdim\frzwtmp>\lgwBC\setlength{\lgwBC}{\frzwtmp}\fi
\settowidth{\frzwtmp}{\footnotesize {\scriptsize\color{basetext} $\bar{q}^{*}$ (\%)}}\ifdim\frzwtmp>\lgwBC\setlength{\lgwBC}{\frzwtmp}\fi
\begin{table*}[p]
  \centering
  \caption{\textbf{APIGraph} (test Jan 2013--Nov 2018; full initial year) in the metric set, threshold protocol and layout of Table~\ref{tab:main-vtt2-b400}: the first six test months calibrate, the remaining 65 months evaluate, and the Cap batch size and the rejection quota $\rho$ equal $B_M$. Values are computed per month within seed, summarized within seed and averaged over seeds; rows with fewer seeds than the complete rows show their count and are not ranked. \textbf{Bold}=best, {\color{worstcolor}\itshape dark-red italics}=worst displayed value per ranked column and $B_M$ block.}
  \label{tab:legacy-apig}
  \footnotesize
  \setlength{\tabcolsep}{4.2pt}
  \renewcommand{\arraystretch}{0.95}
  \begin{tabular}{@{}cl@{\hspace{7pt}}c@{\hspace{7pt}}ccw{c}{\lgwBA}w{c}{\lgwBA}@{\hspace{7pt}}cw{c}{\lgwBB}w{c}{\lgwBB}w{c}{\lgwBC}w{c}{\lgwBC}}
    \toprule
     &  & \multicolumn{1}{>{\columncolor{tabbaseband}[\tabcolsep][\tabcolsep]}c}{Baseline} & \multicolumn{4}{>{\columncolor{tabreliband}[\tabcolsep][\tabcolsep]}c}{Reliability} & \multicolumn{5}{>{\columncolor{tabstabband}[\tabcolsep][\tabcolsep]}c}{Stability} \\
    $B_M$ & Method & \cellcolor{tabbaseband}F1 $\uparrow$ & \cellcolor{tabreliband}E-AURC $\downarrow$ & \cellcolor{tabreliband}Cap$_{B_M}$ $\uparrow$ & \multicolumn{2}{>{\columncolor{tabreliband}[\tabcolsep][\tabcolsep]}c}{\makebox[0pt][c]{F1$^{*}$ $\uparrow$}\makebox[0pt][c]{\smash{\raisebox{-2.6pt}{\rule{\dimexpr 2\lgwBA+4\tabcolsep-5pt\relax}{0.4pt}}}}} & \cellcolor{tabstabband}Sen[F1] $\uparrow$ & \multicolumn{2}{>{\columncolor{tabstabband}[\tabcolsep][\tabcolsep]}c}{\makebox[0pt][c]{Sen[AUROC$_f$] $\uparrow$}\makebox[0pt][c]{\smash{\raisebox{-2.6pt}{\rule{\dimexpr 2\lgwBB+4\tabcolsep-5pt\relax}{0.4pt}}}}} & \multicolumn{2}{>{\columncolor{tabstabband}[\tabcolsep][\tabcolsep]}c}{\makebox[0pt][c]{Sen[$q^{*}$]}\makebox[0pt][c]{\smash{\raisebox{-2.6pt}{\rule{\dimexpr 2\lgwBC+4\tabcolsep-5pt\relax}{0.4pt}}}}} \\
     &  & \cellcolor{tabbaseband}{\scriptsize \%} & \cellcolor{tabreliband}{\scriptsize $\times$100} & \cellcolor{tabreliband}{\scriptsize \% of errors} & \cellcolor{tabreliband}{\scriptsize \%} & \cellcolor{tabreliband}{\scriptsize\color{basetext} $\Delta$F1$^{*}$ (pp)} & \cellcolor{tabstabband}{\scriptsize pp/month} & \cellcolor{tabstabband}{\scriptsize $10^{-3}$/month} & \cellcolor{tabstabband}{\scriptsize\color{basetext} first$\to$last} & \cellcolor{tabstabband}{\scriptsize pp/month} & \cellcolor{tabstabband}{\scriptsize\color{basetext} $\bar{q}^{*}$ (\%)} \\
    \midrule
    \multirow{11}{*}{50} & DeepDrebin & \textbf{89.7} & \textbf{0.61} & \textbf{28.3} & \textbf{91.6} & {\color{basetext}$+$1.8} & \textbf{$+$0.12} & $+$1.4 & {\color{basetext}0.85$\to$0.85} & $-$0.01 & {\color{basetext}1.4} \\
     & CADE-OOD$^{c}$ & \textcolor{worstcolor}{\itshape 85.8} & 1.94 & \textcolor{worstcolor}{\itshape 4.3} & \textcolor{worstcolor}{\itshape 84.8} & {\color{basetext}$-$1.0} & \textcolor{worstcolor}{\itshape $+$0.03} & $-$1.3 & {\color{basetext}0.69$\to$0.55} & $+$0.01 & {\color{basetext}1.1} \\
     & CADE-MSP$^{c}$ & \textcolor{worstcolor}{\itshape 85.8} & \textcolor{worstcolor}{\itshape 4.15} & 5.4 & 86.4 & {\color{basetext}$+$0.6} & \textcolor{worstcolor}{\itshape $+$0.03} & $+$0.1 & {\color{basetext}0.38$\to$0.43} & $-$0.01 & {\color{basetext}1.9} \\
     & HCC-PL$^{a}$ & 88.6 & 1.78 & 19.6 & 90.0 & {\color{basetext}$+$1.4} & $+$0.06 & \textcolor{worstcolor}{\itshape $-$1.9} & {\color{basetext}0.78$\to$0.63} & 0.00 & {\color{basetext}1.4} \\
     & HCC-MSP$^{a}$ & 88.6 & 1.39 & 20.6 & 90.3 & {\color{basetext}$+$1.7} & $+$0.06 & $-$1.5 & {\color{basetext}0.75$\to$0.59} & $-$0.01 & {\color{basetext}1.3} \\
     & Drebin & 87.7 & 0.62 & 23.3 & 89.4 & {\color{basetext}$+$1.7} & $+$0.08 & $+$0.3 & {\color{basetext}0.91$\to$0.86} & $-$0.01 & {\color{basetext}1.4} \\
    \addlinespace[2pt]
     & \multicolumn{11}{>{\columncolor{subheadcolor}[\tabcolsep][\tabcolsep]}l}{\itshape Transcendent: inductive (ICE) and cross-conformal (CCE)$^{b}$} \\
     & ICE (Drebin) & 86.9 & 0.71 & 24.4 & 89.5 & {\color{basetext}$+$2.6} & $+$0.07 & $+$1.1 & {\color{basetext}0.89$\to$0.90} & $-$0.01 & {\color{basetext}1.2} \\
     & CCE (Drebin) & 87.2 & 0.70 & 23.6 & 89.7 & {\color{basetext}$+$2.5} & $+$0.07 & $+$1.0 & {\color{basetext}0.90$\to$0.89} & $-$0.01 & {\color{basetext}1.2} \\
     & ICE (DeepDrebin) & 88.2 & 0.93 & 15.9 & 89.7 & {\color{basetext}$+$1.6} & $+$0.06 & \textbf{$+$1.6} & {\color{basetext}0.84$\to$0.84} & 0.00 & {\color{basetext}1.2} \\
     & CCE (DeepDrebin) & 89.0 & 0.91 & 16.4 & 90.5 & {\color{basetext}$+$1.5} & $+$0.05 & \textbf{$+$1.6} & {\color{basetext}0.83$\to$0.84} & 0.00 & {\color{basetext}1.3} \\
    \midrule
    \multirow{11}{*}{100} & DeepDrebin & \textbf{91.2} & \textbf{0.54} & \textbf{42.6} & \textbf{93.7} & {\color{basetext}$+$2.5} & \textbf{$+$0.10} & $+$2.1 & {\color{basetext}0.86$\to$0.88} & $-$0.01 & {\color{basetext}2.7} \\
     & CADE-OOD$^{c}$ & \textcolor{worstcolor}{\itshape 87.3} & 1.84 & 8.3 & \textcolor{worstcolor}{\itshape 85.3} & {\color{basetext}$-$1.9} & $+$0.04 & $-$0.9 & {\color{basetext}0.65$\to$0.54} & $+$0.02 & {\color{basetext}2.2} \\
     & CADE-MSP$^{c}$ & \textcolor{worstcolor}{\itshape 87.3} & \textcolor{worstcolor}{\itshape 3.57} & \textcolor{worstcolor}{\itshape 5.1} & 87.9 & {\color{basetext}$+$0.6} & $+$0.04 & $+$0.4 & {\color{basetext}0.36$\to$0.45} & $-$0.03 & {\color{basetext}3.8} \\
     & HCC-PL$^{a}$ & 90.0 & 1.44 & 30.6 & 92.2 & {\color{basetext}$+$2.1} & \textcolor{worstcolor}{\itshape $+$0.03} & \textcolor{worstcolor}{\itshape $-$1.9} & {\color{basetext}0.78$\to$0.67} & $-$0.01 & {\color{basetext}2.6} \\
     & HCC-MSP$^{a}$ & 90.0 & 1.20 & 29.7 & 92.2 & {\color{basetext}$+$2.1} & \textcolor{worstcolor}{\itshape $+$0.03} & $-$1.2 & {\color{basetext}0.73$\to$0.61} & $-$0.02 & {\color{basetext}2.3} \\
     & Drebin & 88.3 & 0.64 & 38.6 & 91.4 & {\color{basetext}$+$3.1} & $+$0.05 & $+$0.8 & {\color{basetext}0.91$\to$0.88} & 0.00 & {\color{basetext}2.7} \\
    \addlinespace[2pt]
     & \multicolumn{11}{>{\columncolor{subheadcolor}[\tabcolsep][\tabcolsep]}l}{\itshape Transcendent: inductive (ICE) and cross-conformal (CCE)$^{b}$} \\
     & ICE (Drebin) & 87.8 & 0.71 & 36.1 & 91.6 & {\color{basetext}$+$3.8} & $+$0.05 & $+$0.9 & {\color{basetext}0.90$\to$0.89} & 0.00 & {\color{basetext}2.6} \\
     & CCE (Drebin) & 88.0 & 0.71 & 36.1 & 91.8 & {\color{basetext}$+$3.7} & $+$0.05 & $+$0.8 & {\color{basetext}0.90$\to$0.88} & 0.00 & {\color{basetext}2.5} \\
     & ICE (DeepDrebin) & 89.8 & 0.80 & 30.7 & 92.2 & {\color{basetext}$+$2.4} & $+$0.09 & $+$2.1 & {\color{basetext}0.83$\to$0.88} & $-$0.01 & {\color{basetext}2.4} \\
     & CCE (DeepDrebin) & 90.6 & 0.77 & 31.0 & 92.8 & {\color{basetext}$+$2.2} & $+$0.08 & \textbf{$+$2.3} & {\color{basetext}0.83$\to$0.88} & 0.00 & {\color{basetext}2.4} \\
    \midrule
    \multirow{11}{*}{200} & DeepDrebin & \textbf{92.2} & \textbf{0.46} & \textbf{56.5} & \textbf{95.2} & {\color{basetext}$+$3.0} & \textbf{$+$0.08} & $+$2.1 & {\color{basetext}0.86$\to$0.88} & 0.00 & {\color{basetext}5.4} \\
     & CADE-OOD$^{c}$ & 89.0 & 1.64 & 15.3 & \textcolor{worstcolor}{\itshape 82.7} & {\color{basetext}$-$6.3} & \textcolor{worstcolor}{\itshape $+$0.03} & $-$0.7 & {\color{basetext}0.65$\to$0.55} & $+$0.02 & {\color{basetext}4.9} \\
     & CADE-MSP$^{c}$ & 89.0 & \textcolor{worstcolor}{\itshape 3.25} & \textcolor{worstcolor}{\itshape 5.9} & 89.7 & {\color{basetext}$+$0.6} & \textcolor{worstcolor}{\itshape $+$0.03} & $-$0.8 & {\color{basetext}0.37$\to$0.39} & $-$0.02 & {\color{basetext}7.4} \\
     & HCC-PL$^{a}$ & 91.3 & 1.29 & 41.9 & 94.0 & {\color{basetext}$+$2.7} & $+$0.04 & \textcolor{worstcolor}{\itshape $-$1.4} & {\color{basetext}0.79$\to$0.69} & $+$0.01 & {\color{basetext}5.2} \\
     & HCC-MSP$^{a}$ & 91.3 & 1.02 & 38.7 & 93.7\rlap{$^{\dagger}$} & {\color{basetext}$+$2.4}\rlap{$^{\dagger}$} & $+$0.04 & $-$0.7 & {\color{basetext}0.75$\to$0.66} & $-$0.07\rlap{$^{\dagger}$} & {\color{basetext}3.2}\rlap{$^{\dagger}$} \\
     & Drebin & 88.9 & 0.67 & 56.0 & 93.1 & {\color{basetext}$+$4.2} & $+$0.04 & $+$0.8 & {\color{basetext}0.91$\to$0.89} & $-$0.01 & {\color{basetext}5.4} \\
    \addlinespace[2pt]
     & \multicolumn{11}{>{\columncolor{subheadcolor}[\tabcolsep][\tabcolsep]}l}{\itshape Transcendent: inductive (ICE) and cross-conformal (CCE)$^{b}$} \\
     & ICE (Drebin) & \textcolor{worstcolor}{\itshape 87.9} & 0.67 & 47.8 & 93.2 & {\color{basetext}$+$5.3} & $+$0.04 & $+$0.8 & {\color{basetext}0.91$\to$0.90} & 0.00 & {\color{basetext}5.1} \\
     & CCE (Drebin) & 88.1 & 0.66 & 47.5 & 93.3 & {\color{basetext}$+$5.2} & $+$0.04 & $+$0.9 & {\color{basetext}0.90$\to$0.90} & 0.00 & {\color{basetext}5.1} \\
     & ICE (DeepDrebin) & 91.3 & 0.64 & 50.4 & 94.6 & {\color{basetext}$+$3.3} & \textbf{$+$0.08} & $+$2.6 & {\color{basetext}0.83$\to$0.89} & 0.00 & {\color{basetext}4.8} \\
     & CCE (DeepDrebin) & 91.9 & 0.62 & 49.9 & 95.0 & {\color{basetext}$+$3.1} & \textbf{$+$0.08} & \textbf{$+$2.7} & {\color{basetext}0.83$\to$0.89} & 0.00 & {\color{basetext}4.8} \\
    \midrule
    \multirow{11}{*}{400} & DeepDrebin & 92.8 & \textbf{0.40} & \textbf{70.6} & 96.1 & {\color{basetext}$+$3.3} & \textbf{$+$0.08} & $+$2.0 & {\color{basetext}0.86$\to$0.90} & $-$0.01 & {\color{basetext}10.7} \\
     & CADE-OOD$^{c}$ & 89.4 & 1.73 & 23.1 & \textcolor{worstcolor}{\itshape 72.5} & {\color{basetext}$-$16.9} & $+$0.01 & $-$1.0 & {\color{basetext}0.65$\to$0.55} & $-$0.07 & {\color{basetext}8.9} \\
     & CADE-MSP$^{c}$ & 89.4 & \textcolor{worstcolor}{\itshape 3.81} & \textcolor{worstcolor}{\itshape 8.0} & 90.3 & {\color{basetext}$+$0.9} & $+$0.01 & $-$1.5 & {\color{basetext}0.31$\to$0.32} & $-$0.12 & {\color{basetext}15.0} \\
     & HCC-PL$^{a}$ & \textcolor{worstcolor}{\itshape 85.7} & 1.19 & 53.4 & 89.7 & {\color{basetext}$+$3.7} & \textcolor{worstcolor}{\itshape $-$0.01} & $-$0.1 & {\color{basetext}0.79$\to$0.82} & $-$0.03 & {\color{basetext}11.6} \\
     & HCC-MSP$^{a}$ & \textcolor{worstcolor}{\itshape 85.7} & 1.55 & 44.7 & 88.1\rlap{$^{\dagger}$} & {\color{basetext}$+$2.3}\rlap{$^{\dagger}$} & \textcolor{worstcolor}{\itshape $-$0.01} & \textcolor{worstcolor}{\itshape $-$1.7} & {\color{basetext}0.76$\to$0.57} & $-$0.18\rlap{$^{\dagger}$} & {\color{basetext}4.1}\rlap{$^{\dagger}$} \\
     & Drebin & 88.7 & 0.62 & 69.9 & 94.3 & {\color{basetext}$+$5.5} & $+$0.04 & $+$0.8 & {\color{basetext}0.92$\to$0.88} & $-$0.02 & {\color{basetext}10.7} \\
    \addlinespace[2pt]
     & \multicolumn{11}{>{\columncolor{subheadcolor}[\tabcolsep][\tabcolsep]}l}{\itshape Transcendent: inductive (ICE) and cross-conformal (CCE)$^{b}$} \\
     & ICE (Drebin) & 88.4 & 0.64 & 63.4 & 94.7 & {\color{basetext}$+$6.3} & $+$0.04 & $+$0.9 & {\color{basetext}0.91$\to$0.89} & $-$0.01 & {\color{basetext}10.0} \\
     & CCE (Drebin) & 88.6 & 0.64 & 63.1 & 94.7 & {\color{basetext}$+$6.2} & $+$0.04 & $+$0.9 & {\color{basetext}0.90$\to$0.89} & 0.00 & {\color{basetext}10.0} \\
     & ICE (DeepDrebin) & 92.3 & 0.51 & 69.1 & 96.5 & {\color{basetext}$+$4.2} & $+$0.07 & $+$2.4 & {\color{basetext}0.86$\to$0.90} & $+$0.01 & {\color{basetext}10.2} \\
     & CCE (DeepDrebin) & \textbf{92.9} & 0.50 & 68.8 & \textbf{96.7} & {\color{basetext}$+$3.8} & $+$0.06 & \textbf{$+$2.6} & {\color{basetext}0.85$\to$0.90} & $+$0.01 & {\color{basetext}10.0} \\
    \bottomrule
  \end{tabular}
  \par\smallskip
  {\scriptsize\raggedright Random Cap reference: 1.5\% at $B_M=50$; 2.9\% at $B_M=100$; 5.9\% at $B_M=200$; 11.8\% at $B_M=400$. Random Forest, SecSVM, DRMD, RAMDA, TIF and MalCertain were not run on this dataset. $^{a}$HCC with family-supervised training and warm updates; MSP and pseudo-loss (PL) score the same runs and predictions. $^{b}$Transcendent: folds re-drawn at each monthly cold refit; ICE fold 0, CCE five folds; both share CCE-credibility acquisition. $^{c}$CADE with a cold-refitted neural head; OOD and MSP score the same runs and predictions. $^{\dagger}$Tied scores make the rolling window miss its referral target by more than 25\% on average (HCC-MSP at $B_M=200$; HCC-MSP at $B_M=400$); quota-dependent cells are unranked. Months with fewer than 20 accepted samples have no F1$^{*}$ (HCC-PL at $B_M=400$: 1 of 325 seed-months).\par}
\end{table*}

% Generated by generate_satml_vtt2_budget_tables.py; do not edit numbers by hand.
% Layout, metrics and threshold protocol are those of Table I (generate_satml_vtt2_frozen_table.py).
\providecolor{tabbaseband}{RGB}{240,240,240}
\providecolor{tabreliband}{RGB}{230,245,220}
\providecolor{tabstabband}{RGB}{235,240,250}
\providecolor{worstcolor}{RGB}{176,32,32}
\providecolor{subheadcolor}{RGB}{246,246,246}
\providecolor{basetext}{RGB}{90,90,90}
\providecolor{relitext}{RGB}{46,125,50}
\providecolor{stabtext}{RGB}{40,53,147}
\providecommand{\codename}{\textsc{Aurora}}
\ifdefined\frzwtmp\else\newlength{\frzwtmp}\fi
\ifdefined\lgwCA\else\newlength{\lgwCA}\fi\setlength{\lgwCA}{0pt}
\settowidth{\frzwtmp}{\footnotesize \textbf{84.4}}\ifdim\frzwtmp>\lgwCA\setlength{\lgwCA}{\frzwtmp}\fi
\settowidth{\frzwtmp}{\footnotesize {\color{basetext}$+$1.9}}\ifdim\frzwtmp>\lgwCA\setlength{\lgwCA}{\frzwtmp}\fi
\settowidth{\frzwtmp}{\footnotesize 66.0}\ifdim\frzwtmp>\lgwCA\setlength{\lgwCA}{\frzwtmp}\fi
\settowidth{\frzwtmp}{\footnotesize {\color{basetext}$-$2.8}}\ifdim\frzwtmp>\lgwCA\setlength{\lgwCA}{\frzwtmp}\fi
\settowidth{\frzwtmp}{\footnotesize \textcolor{worstcolor}{\itshape 65.4}}\ifdim\frzwtmp>\lgwCA\setlength{\lgwCA}{\frzwtmp}\fi
\settowidth{\frzwtmp}{\footnotesize {\color{basetext}$-$3.3}}\ifdim\frzwtmp>\lgwCA\setlength{\lgwCA}{\frzwtmp}\fi
\settowidth{\frzwtmp}{\footnotesize 83.8}\ifdim\frzwtmp>\lgwCA\setlength{\lgwCA}{\frzwtmp}\fi
\settowidth{\frzwtmp}{\footnotesize {\color{basetext}$+$1.1}}\ifdim\frzwtmp>\lgwCA\setlength{\lgwCA}{\frzwtmp}\fi
\settowidth{\frzwtmp}{\footnotesize 83.9}\ifdim\frzwtmp>\lgwCA\setlength{\lgwCA}{\frzwtmp}\fi
\settowidth{\frzwtmp}{\footnotesize {\color{basetext}$+$1.2}}\ifdim\frzwtmp>\lgwCA\setlength{\lgwCA}{\frzwtmp}\fi
\settowidth{\frzwtmp}{\footnotesize 70.9}\ifdim\frzwtmp>\lgwCA\setlength{\lgwCA}{\frzwtmp}\fi
\settowidth{\frzwtmp}{\footnotesize {\color{basetext}0.0}}\ifdim\frzwtmp>\lgwCA\setlength{\lgwCA}{\frzwtmp}\fi
\settowidth{\frzwtmp}{\footnotesize 72.1}\ifdim\frzwtmp>\lgwCA\setlength{\lgwCA}{\frzwtmp}\fi
\settowidth{\frzwtmp}{\footnotesize {\color{basetext}$+$1.6}}\ifdim\frzwtmp>\lgwCA\setlength{\lgwCA}{\frzwtmp}\fi
\settowidth{\frzwtmp}{\footnotesize 72.4}\ifdim\frzwtmp>\lgwCA\setlength{\lgwCA}{\frzwtmp}\fi
\settowidth{\frzwtmp}{\footnotesize 83.2}\ifdim\frzwtmp>\lgwCA\setlength{\lgwCA}{\frzwtmp}\fi
\settowidth{\frzwtmp}{\footnotesize {\color{basetext}$+$1.5}}\ifdim\frzwtmp>\lgwCA\setlength{\lgwCA}{\frzwtmp}\fi
\settowidth{\frzwtmp}{\footnotesize 84.3}\ifdim\frzwtmp>\lgwCA\setlength{\lgwCA}{\frzwtmp}\fi
\settowidth{\frzwtmp}{\footnotesize \textbf{87.9}}\ifdim\frzwtmp>\lgwCA\setlength{\lgwCA}{\frzwtmp}\fi
\settowidth{\frzwtmp}{\footnotesize {\color{basetext}$+$2.8}}\ifdim\frzwtmp>\lgwCA\setlength{\lgwCA}{\frzwtmp}\fi
\settowidth{\frzwtmp}{\footnotesize 67.2\rlap{$^{\dagger}$}}\ifdim\frzwtmp>\lgwCA\setlength{\lgwCA}{\frzwtmp}\fi
\settowidth{\frzwtmp}{\footnotesize {\color{basetext}$-$3.4}\rlap{$^{\dagger}$}}\ifdim\frzwtmp>\lgwCA\setlength{\lgwCA}{\frzwtmp}\fi
\settowidth{\frzwtmp}{\footnotesize \textcolor{worstcolor}{\itshape 61.4}}\ifdim\frzwtmp>\lgwCA\setlength{\lgwCA}{\frzwtmp}\fi
\settowidth{\frzwtmp}{\footnotesize {\color{basetext}$-$9.3}}\ifdim\frzwtmp>\lgwCA\setlength{\lgwCA}{\frzwtmp}\fi
\settowidth{\frzwtmp}{\footnotesize 85.1}\ifdim\frzwtmp>\lgwCA\setlength{\lgwCA}{\frzwtmp}\fi
\settowidth{\frzwtmp}{\footnotesize {\color{basetext}$+$0.7}}\ifdim\frzwtmp>\lgwCA\setlength{\lgwCA}{\frzwtmp}\fi
\settowidth{\frzwtmp}{\footnotesize 78.9}\ifdim\frzwtmp>\lgwCA\setlength{\lgwCA}{\frzwtmp}\fi
\settowidth{\frzwtmp}{\footnotesize 81.5}\ifdim\frzwtmp>\lgwCA\setlength{\lgwCA}{\frzwtmp}\fi
\settowidth{\frzwtmp}{\footnotesize {\color{basetext}$+$3.1}}\ifdim\frzwtmp>\lgwCA\setlength{\lgwCA}{\frzwtmp}\fi
\settowidth{\frzwtmp}{\footnotesize 81.9}\ifdim\frzwtmp>\lgwCA\setlength{\lgwCA}{\frzwtmp}\fi
\settowidth{\frzwtmp}{\footnotesize 86.6}\ifdim\frzwtmp>\lgwCA\setlength{\lgwCA}{\frzwtmp}\fi
\settowidth{\frzwtmp}{\footnotesize {\color{basetext}$+$1.7}}\ifdim\frzwtmp>\lgwCA\setlength{\lgwCA}{\frzwtmp}\fi
\settowidth{\frzwtmp}{\footnotesize 87.3}\ifdim\frzwtmp>\lgwCA\setlength{\lgwCA}{\frzwtmp}\fi
\settowidth{\frzwtmp}{\footnotesize \textbf{91.0}}\ifdim\frzwtmp>\lgwCA\setlength{\lgwCA}{\frzwtmp}\fi
\settowidth{\frzwtmp}{\footnotesize {\color{basetext}$+$3.7}}\ifdim\frzwtmp>\lgwCA\setlength{\lgwCA}{\frzwtmp}\fi
\settowidth{\frzwtmp}{\footnotesize 70.3\rlap{$^{\dagger}$}}\ifdim\frzwtmp>\lgwCA\setlength{\lgwCA}{\frzwtmp}\fi
\settowidth{\frzwtmp}{\footnotesize {\color{basetext}$-$3.8}\rlap{$^{\dagger}$}}\ifdim\frzwtmp>\lgwCA\setlength{\lgwCA}{\frzwtmp}\fi
\settowidth{\frzwtmp}{\footnotesize \textcolor{worstcolor}{\itshape 59.0}}\ifdim\frzwtmp>\lgwCA\setlength{\lgwCA}{\frzwtmp}\fi
\settowidth{\frzwtmp}{\footnotesize {\color{basetext}$-$15.1}}\ifdim\frzwtmp>\lgwCA\setlength{\lgwCA}{\frzwtmp}\fi
\settowidth{\frzwtmp}{\footnotesize 86.7}\ifdim\frzwtmp>\lgwCA\setlength{\lgwCA}{\frzwtmp}\fi
\settowidth{\frzwtmp}{\footnotesize {\color{basetext}$+$0.3}}\ifdim\frzwtmp>\lgwCA\setlength{\lgwCA}{\frzwtmp}\fi
\settowidth{\frzwtmp}{\footnotesize 86.5}\ifdim\frzwtmp>\lgwCA\setlength{\lgwCA}{\frzwtmp}\fi
\settowidth{\frzwtmp}{\footnotesize {\color{basetext}$+$0.1}}\ifdim\frzwtmp>\lgwCA\setlength{\lgwCA}{\frzwtmp}\fi
\settowidth{\frzwtmp}{\footnotesize 83.0}\ifdim\frzwtmp>\lgwCA\setlength{\lgwCA}{\frzwtmp}\fi
\settowidth{\frzwtmp}{\footnotesize {\color{basetext}$+$2.0}}\ifdim\frzwtmp>\lgwCA\setlength{\lgwCA}{\frzwtmp}\fi
\settowidth{\frzwtmp}{\footnotesize 84.9}\ifdim\frzwtmp>\lgwCA\setlength{\lgwCA}{\frzwtmp}\fi
\settowidth{\frzwtmp}{\footnotesize {\color{basetext}$+$4.2}}\ifdim\frzwtmp>\lgwCA\setlength{\lgwCA}{\frzwtmp}\fi
\settowidth{\frzwtmp}{\footnotesize 85.2}\ifdim\frzwtmp>\lgwCA\setlength{\lgwCA}{\frzwtmp}\fi
\settowidth{\frzwtmp}{\footnotesize 89.3}\ifdim\frzwtmp>\lgwCA\setlength{\lgwCA}{\frzwtmp}\fi
\settowidth{\frzwtmp}{\footnotesize {\color{basetext}$+$2.3}}\ifdim\frzwtmp>\lgwCA\setlength{\lgwCA}{\frzwtmp}\fi
\settowidth{\frzwtmp}{\footnotesize 89.9}\ifdim\frzwtmp>\lgwCA\setlength{\lgwCA}{\frzwtmp}\fi
\settowidth{\frzwtmp}{\footnotesize {\color{basetext}$+$2.5}}\ifdim\frzwtmp>\lgwCA\setlength{\lgwCA}{\frzwtmp}\fi
\settowidth{\frzwtmp}{\footnotesize \textbf{94.4}}\ifdim\frzwtmp>\lgwCA\setlength{\lgwCA}{\frzwtmp}\fi
\settowidth{\frzwtmp}{\footnotesize {\color{basetext}$+$4.5}}\ifdim\frzwtmp>\lgwCA\setlength{\lgwCA}{\frzwtmp}\fi
\settowidth{\frzwtmp}{\footnotesize 72.5\rlap{$^{\dagger}$}}\ifdim\frzwtmp>\lgwCA\setlength{\lgwCA}{\frzwtmp}\fi
\settowidth{\frzwtmp}{\footnotesize {\color{basetext}$-$4.3}\rlap{$^{\dagger}$}}\ifdim\frzwtmp>\lgwCA\setlength{\lgwCA}{\frzwtmp}\fi
\settowidth{\frzwtmp}{\footnotesize \textcolor{worstcolor}{\itshape 58.9}}\ifdim\frzwtmp>\lgwCA\setlength{\lgwCA}{\frzwtmp}\fi
\settowidth{\frzwtmp}{\footnotesize {\color{basetext}$-$18.0}}\ifdim\frzwtmp>\lgwCA\setlength{\lgwCA}{\frzwtmp}\fi
\settowidth{\frzwtmp}{\footnotesize 89.5}\ifdim\frzwtmp>\lgwCA\setlength{\lgwCA}{\frzwtmp}\fi
\settowidth{\frzwtmp}{\footnotesize 88.6\rlap{$^{\dagger}$}}\ifdim\frzwtmp>\lgwCA\setlength{\lgwCA}{\frzwtmp}\fi
\settowidth{\frzwtmp}{\footnotesize {\color{basetext}$+$0.4}\rlap{$^{\dagger}$}}\ifdim\frzwtmp>\lgwCA\setlength{\lgwCA}{\frzwtmp}\fi
\settowidth{\frzwtmp}{\footnotesize 85.5}\ifdim\frzwtmp>\lgwCA\setlength{\lgwCA}{\frzwtmp}\fi
\settowidth{\frzwtmp}{\footnotesize 88.4}\ifdim\frzwtmp>\lgwCA\setlength{\lgwCA}{\frzwtmp}\fi
\settowidth{\frzwtmp}{\footnotesize {\color{basetext}$+$6.0}}\ifdim\frzwtmp>\lgwCA\setlength{\lgwCA}{\frzwtmp}\fi
\settowidth{\frzwtmp}{\footnotesize 88.7}\ifdim\frzwtmp>\lgwCA\setlength{\lgwCA}{\frzwtmp}\fi
\settowidth{\frzwtmp}{\footnotesize {\color{basetext}$+$6.1}}\ifdim\frzwtmp>\lgwCA\setlength{\lgwCA}{\frzwtmp}\fi
\settowidth{\frzwtmp}{\footnotesize 92.9}\ifdim\frzwtmp>\lgwCA\setlength{\lgwCA}{\frzwtmp}\fi
\settowidth{\frzwtmp}{\footnotesize {\color{basetext}$+$4.0}}\ifdim\frzwtmp>\lgwCA\setlength{\lgwCA}{\frzwtmp}\fi
\settowidth{\frzwtmp}{\footnotesize 93.3}\ifdim\frzwtmp>\lgwCA\setlength{\lgwCA}{\frzwtmp}\fi
\settowidth{\frzwtmp}{\footnotesize {\scriptsize \%}}\ifdim\frzwtmp>\lgwCA\setlength{\lgwCA}{\frzwtmp}\fi
\settowidth{\frzwtmp}{\footnotesize {\scriptsize\color{basetext} $\Delta$F1$^{*}$ (pp)}}\ifdim\frzwtmp>\lgwCA\setlength{\lgwCA}{\frzwtmp}\fi
\ifdefined\lgwCB\else\newlength{\lgwCB}\fi\setlength{\lgwCB}{0pt}
\settowidth{\frzwtmp}{\footnotesize $-$3.2}\ifdim\frzwtmp>\lgwCB\setlength{\lgwCB}{\frzwtmp}\fi
\settowidth{\frzwtmp}{\footnotesize {\color{basetext}0.83$\to$0.74}}\ifdim\frzwtmp>\lgwCB\setlength{\lgwCB}{\frzwtmp}\fi
\settowidth{\frzwtmp}{\footnotesize $-$0.9}\ifdim\frzwtmp>\lgwCB\setlength{\lgwCB}{\frzwtmp}\fi
\settowidth{\frzwtmp}{\footnotesize {\color{basetext}0.55$\to$0.52}}\ifdim\frzwtmp>\lgwCB\setlength{\lgwCB}{\frzwtmp}\fi
\settowidth{\frzwtmp}{\footnotesize \textbf{$+$1.2}}\ifdim\frzwtmp>\lgwCB\setlength{\lgwCB}{\frzwtmp}\fi
\settowidth{\frzwtmp}{\footnotesize {\color{basetext}0.48$\to$0.52}}\ifdim\frzwtmp>\lgwCB\setlength{\lgwCB}{\frzwtmp}\fi
\settowidth{\frzwtmp}{\footnotesize $-$3.0}\ifdim\frzwtmp>\lgwCB\setlength{\lgwCB}{\frzwtmp}\fi
\settowidth{\frzwtmp}{\footnotesize {\color{basetext}0.68$\to$0.61}}\ifdim\frzwtmp>\lgwCB\setlength{\lgwCB}{\frzwtmp}\fi
\settowidth{\frzwtmp}{\footnotesize $-$2.1}\ifdim\frzwtmp>\lgwCB\setlength{\lgwCB}{\frzwtmp}\fi
\settowidth{\frzwtmp}{\footnotesize {\color{basetext}0.67$\to$0.62}}\ifdim\frzwtmp>\lgwCB\setlength{\lgwCB}{\frzwtmp}\fi
\settowidth{\frzwtmp}{\footnotesize $-$1.6}\ifdim\frzwtmp>\lgwCB\setlength{\lgwCB}{\frzwtmp}\fi
\settowidth{\frzwtmp}{\footnotesize {\color{basetext}0.82$\to$0.73}}\ifdim\frzwtmp>\lgwCB\setlength{\lgwCB}{\frzwtmp}\fi
\settowidth{\frzwtmp}{\footnotesize $-$1.4}\ifdim\frzwtmp>\lgwCB\setlength{\lgwCB}{\frzwtmp}\fi
\settowidth{\frzwtmp}{\footnotesize {\color{basetext}0.83$\to$0.76}}\ifdim\frzwtmp>\lgwCB\setlength{\lgwCB}{\frzwtmp}\fi
\settowidth{\frzwtmp}{\footnotesize $-$1.7}\ifdim\frzwtmp>\lgwCB\setlength{\lgwCB}{\frzwtmp}\fi
\settowidth{\frzwtmp}{\footnotesize {\color{basetext}0.84$\to$0.77}}\ifdim\frzwtmp>\lgwCB\setlength{\lgwCB}{\frzwtmp}\fi
\settowidth{\frzwtmp}{\footnotesize {\color{basetext}0.83$\to$0.72}}\ifdim\frzwtmp>\lgwCB\setlength{\lgwCB}{\frzwtmp}\fi
\settowidth{\frzwtmp}{\footnotesize \textcolor{worstcolor}{\itshape $-$3.4}}\ifdim\frzwtmp>\lgwCB\setlength{\lgwCB}{\frzwtmp}\fi
\settowidth{\frzwtmp}{\footnotesize $-$3.7}\ifdim\frzwtmp>\lgwCB\setlength{\lgwCB}{\frzwtmp}\fi
\settowidth{\frzwtmp}{\footnotesize {\color{basetext}0.84$\to$0.70}}\ifdim\frzwtmp>\lgwCB\setlength{\lgwCB}{\frzwtmp}\fi
\settowidth{\frzwtmp}{\footnotesize $-$1.3}\ifdim\frzwtmp>\lgwCB\setlength{\lgwCB}{\frzwtmp}\fi
\settowidth{\frzwtmp}{\footnotesize {\color{basetext}0.56$\to$0.51}}\ifdim\frzwtmp>\lgwCB\setlength{\lgwCB}{\frzwtmp}\fi
\settowidth{\frzwtmp}{\footnotesize \textbf{$+$2.4}}\ifdim\frzwtmp>\lgwCB\setlength{\lgwCB}{\frzwtmp}\fi
\settowidth{\frzwtmp}{\footnotesize {\color{basetext}0.49$\to$0.53}}\ifdim\frzwtmp>\lgwCB\setlength{\lgwCB}{\frzwtmp}\fi
\settowidth{\frzwtmp}{\footnotesize $-$3.9}\ifdim\frzwtmp>\lgwCB\setlength{\lgwCB}{\frzwtmp}\fi
\settowidth{\frzwtmp}{\footnotesize {\color{basetext}0.68$\to$0.59}}\ifdim\frzwtmp>\lgwCB\setlength{\lgwCB}{\frzwtmp}\fi
\settowidth{\frzwtmp}{\footnotesize $-$3.4}\ifdim\frzwtmp>\lgwCB\setlength{\lgwCB}{\frzwtmp}\fi
\settowidth{\frzwtmp}{\footnotesize {\color{basetext}0.68$\to$0.60}}\ifdim\frzwtmp>\lgwCB\setlength{\lgwCB}{\frzwtmp}\fi
\settowidth{\frzwtmp}{\footnotesize $-$1.9}\ifdim\frzwtmp>\lgwCB\setlength{\lgwCB}{\frzwtmp}\fi
\settowidth{\frzwtmp}{\footnotesize {\color{basetext}0.80$\to$0.71}}\ifdim\frzwtmp>\lgwCB\setlength{\lgwCB}{\frzwtmp}\fi
\settowidth{\frzwtmp}{\footnotesize $-$3.5}\ifdim\frzwtmp>\lgwCB\setlength{\lgwCB}{\frzwtmp}\fi
\settowidth{\frzwtmp}{\footnotesize {\color{basetext}0.84$\to$0.74}}\ifdim\frzwtmp>\lgwCB\setlength{\lgwCB}{\frzwtmp}\fi
\settowidth{\frzwtmp}{\footnotesize {\color{basetext}0.84$\to$0.73}}\ifdim\frzwtmp>\lgwCB\setlength{\lgwCB}{\frzwtmp}\fi
\settowidth{\frzwtmp}{\footnotesize \textcolor{worstcolor}{\itshape $-$4.0}}\ifdim\frzwtmp>\lgwCB\setlength{\lgwCB}{\frzwtmp}\fi
\settowidth{\frzwtmp}{\footnotesize {\color{basetext}0.84$\to$0.71}}\ifdim\frzwtmp>\lgwCB\setlength{\lgwCB}{\frzwtmp}\fi
\settowidth{\frzwtmp}{\footnotesize {\color{basetext}0.85$\to$0.71}}\ifdim\frzwtmp>\lgwCB\setlength{\lgwCB}{\frzwtmp}\fi
\settowidth{\frzwtmp}{\footnotesize $-$1.1}\ifdim\frzwtmp>\lgwCB\setlength{\lgwCB}{\frzwtmp}\fi
\settowidth{\frzwtmp}{\footnotesize {\color{basetext}0.54$\to$0.49}}\ifdim\frzwtmp>\lgwCB\setlength{\lgwCB}{\frzwtmp}\fi
\settowidth{\frzwtmp}{\footnotesize \textbf{$+$2.5}}\ifdim\frzwtmp>\lgwCB\setlength{\lgwCB}{\frzwtmp}\fi
\settowidth{\frzwtmp}{\footnotesize {\color{basetext}0.48$\to$0.53}}\ifdim\frzwtmp>\lgwCB\setlength{\lgwCB}{\frzwtmp}\fi
\settowidth{\frzwtmp}{\footnotesize $-$1.5}\ifdim\frzwtmp>\lgwCB\setlength{\lgwCB}{\frzwtmp}\fi
\settowidth{\frzwtmp}{\footnotesize {\color{basetext}0.69$\to$0.63}}\ifdim\frzwtmp>\lgwCB\setlength{\lgwCB}{\frzwtmp}\fi
\settowidth{\frzwtmp}{\footnotesize $-$2.0}\ifdim\frzwtmp>\lgwCB\setlength{\lgwCB}{\frzwtmp}\fi
\settowidth{\frzwtmp}{\footnotesize {\color{basetext}0.81$\to$0.74}}\ifdim\frzwtmp>\lgwCB\setlength{\lgwCB}{\frzwtmp}\fi
\settowidth{\frzwtmp}{\footnotesize {\color{basetext}0.83$\to$0.75}}\ifdim\frzwtmp>\lgwCB\setlength{\lgwCB}{\frzwtmp}\fi
\settowidth{\frzwtmp}{\footnotesize $-$2.9}\ifdim\frzwtmp>\lgwCB\setlength{\lgwCB}{\frzwtmp}\fi
\settowidth{\frzwtmp}{\footnotesize $-$4.1}\ifdim\frzwtmp>\lgwCB\setlength{\lgwCB}{\frzwtmp}\fi
\settowidth{\frzwtmp}{\footnotesize \textcolor{worstcolor}{\itshape $-$4.6}}\ifdim\frzwtmp>\lgwCB\setlength{\lgwCB}{\frzwtmp}\fi
\settowidth{\frzwtmp}{\footnotesize $-$4.5}\ifdim\frzwtmp>\lgwCB\setlength{\lgwCB}{\frzwtmp}\fi
\settowidth{\frzwtmp}{\footnotesize {\color{basetext}0.88$\to$0.72}}\ifdim\frzwtmp>\lgwCB\setlength{\lgwCB}{\frzwtmp}\fi
\settowidth{\frzwtmp}{\footnotesize $-$1.2}\ifdim\frzwtmp>\lgwCB\setlength{\lgwCB}{\frzwtmp}\fi
\settowidth{\frzwtmp}{\footnotesize {\color{basetext}0.54$\to$0.51}}\ifdim\frzwtmp>\lgwCB\setlength{\lgwCB}{\frzwtmp}\fi
\settowidth{\frzwtmp}{\footnotesize \textbf{$+$4.0}}\ifdim\frzwtmp>\lgwCB\setlength{\lgwCB}{\frzwtmp}\fi
\settowidth{\frzwtmp}{\footnotesize {\color{basetext}0.48$\to$0.57}}\ifdim\frzwtmp>\lgwCB\setlength{\lgwCB}{\frzwtmp}\fi
\settowidth{\frzwtmp}{\footnotesize \textcolor{worstcolor}{\itshape $-$5.9}}\ifdim\frzwtmp>\lgwCB\setlength{\lgwCB}{\frzwtmp}\fi
\settowidth{\frzwtmp}{\footnotesize {\color{basetext}0.71$\to$0.59}}\ifdim\frzwtmp>\lgwCB\setlength{\lgwCB}{\frzwtmp}\fi
\settowidth{\frzwtmp}{\footnotesize $-$2.2}\ifdim\frzwtmp>\lgwCB\setlength{\lgwCB}{\frzwtmp}\fi
\settowidth{\frzwtmp}{\footnotesize {\color{basetext}0.71$\to$0.66}}\ifdim\frzwtmp>\lgwCB\setlength{\lgwCB}{\frzwtmp}\fi
\settowidth{\frzwtmp}{\footnotesize {\color{basetext}0.80$\to$0.72}}\ifdim\frzwtmp>\lgwCB\setlength{\lgwCB}{\frzwtmp}\fi
\settowidth{\frzwtmp}{\footnotesize {\color{basetext}0.82$\to$0.74}}\ifdim\frzwtmp>\lgwCB\setlength{\lgwCB}{\frzwtmp}\fi
\settowidth{\frzwtmp}{\footnotesize $-$4.8}\ifdim\frzwtmp>\lgwCB\setlength{\lgwCB}{\frzwtmp}\fi
\settowidth{\frzwtmp}{\footnotesize {\color{basetext}0.86$\to$0.73}}\ifdim\frzwtmp>\lgwCB\setlength{\lgwCB}{\frzwtmp}\fi
\settowidth{\frzwtmp}{\footnotesize $-$5.0}\ifdim\frzwtmp>\lgwCB\setlength{\lgwCB}{\frzwtmp}\fi
\settowidth{\frzwtmp}{\footnotesize {\scriptsize $10^{-3}$/month}}\ifdim\frzwtmp>\lgwCB\setlength{\lgwCB}{\frzwtmp}\fi
\settowidth{\frzwtmp}{\footnotesize {\scriptsize\color{basetext} first$\to$last}}\ifdim\frzwtmp>\lgwCB\setlength{\lgwCB}{\frzwtmp}\fi
\ifdefined\lgwCC\else\newlength{\lgwCC}\fi\setlength{\lgwCC}{0pt}
\settowidth{\frzwtmp}{\footnotesize 0.00}\ifdim\frzwtmp>\lgwCC\setlength{\lgwCC}{\frzwtmp}\fi
\settowidth{\frzwtmp}{\footnotesize {\color{basetext}1.3}}\ifdim\frzwtmp>\lgwCC\setlength{\lgwCC}{\frzwtmp}\fi
\settowidth{\frzwtmp}{\footnotesize {\color{basetext}1.2}}\ifdim\frzwtmp>\lgwCC\setlength{\lgwCC}{\frzwtmp}\fi
\settowidth{\frzwtmp}{\footnotesize {\color{basetext}2.7}}\ifdim\frzwtmp>\lgwCC\setlength{\lgwCC}{\frzwtmp}\fi
\settowidth{\frzwtmp}{\footnotesize $-$0.01}\ifdim\frzwtmp>\lgwCC\setlength{\lgwCC}{\frzwtmp}\fi
\settowidth{\frzwtmp}{\footnotesize {\color{basetext}1.4}}\ifdim\frzwtmp>\lgwCC\setlength{\lgwCC}{\frzwtmp}\fi
\settowidth{\frzwtmp}{\footnotesize {\color{basetext}1.1}}\ifdim\frzwtmp>\lgwCC\setlength{\lgwCC}{\frzwtmp}\fi
\settowidth{\frzwtmp}{\footnotesize $-$0.02}\ifdim\frzwtmp>\lgwCC\setlength{\lgwCC}{\frzwtmp}\fi
\settowidth{\frzwtmp}{\footnotesize {\color{basetext}2.9}}\ifdim\frzwtmp>\lgwCC\setlength{\lgwCC}{\frzwtmp}\fi
\settowidth{\frzwtmp}{\footnotesize $-$0.02\rlap{$^{\dagger}$}}\ifdim\frzwtmp>\lgwCC\setlength{\lgwCC}{\frzwtmp}\fi
\settowidth{\frzwtmp}{\footnotesize {\color{basetext}2.4}\rlap{$^{\dagger}$}}\ifdim\frzwtmp>\lgwCC\setlength{\lgwCC}{\frzwtmp}\fi
\settowidth{\frzwtmp}{\footnotesize $+$0.03}\ifdim\frzwtmp>\lgwCC\setlength{\lgwCC}{\frzwtmp}\fi
\settowidth{\frzwtmp}{\footnotesize {\color{basetext}4.0}}\ifdim\frzwtmp>\lgwCC\setlength{\lgwCC}{\frzwtmp}\fi
\settowidth{\frzwtmp}{\footnotesize {\color{basetext}3.2}}\ifdim\frzwtmp>\lgwCC\setlength{\lgwCC}{\frzwtmp}\fi
\settowidth{\frzwtmp}{\footnotesize {\color{basetext}3.0}}\ifdim\frzwtmp>\lgwCC\setlength{\lgwCC}{\frzwtmp}\fi
\settowidth{\frzwtmp}{\footnotesize {\color{basetext}2.6}}\ifdim\frzwtmp>\lgwCC\setlength{\lgwCC}{\frzwtmp}\fi
\settowidth{\frzwtmp}{\footnotesize {\color{basetext}2.4}}\ifdim\frzwtmp>\lgwCC\setlength{\lgwCC}{\frzwtmp}\fi
\settowidth{\frzwtmp}{\footnotesize $-$0.03}\ifdim\frzwtmp>\lgwCC\setlength{\lgwCC}{\frzwtmp}\fi
\settowidth{\frzwtmp}{\footnotesize $-$0.04}\ifdim\frzwtmp>\lgwCC\setlength{\lgwCC}{\frzwtmp}\fi
\settowidth{\frzwtmp}{\footnotesize {\color{basetext}2.5}}\ifdim\frzwtmp>\lgwCC\setlength{\lgwCC}{\frzwtmp}\fi
\settowidth{\frzwtmp}{\footnotesize $-$0.06}\ifdim\frzwtmp>\lgwCC\setlength{\lgwCC}{\frzwtmp}\fi
\settowidth{\frzwtmp}{\footnotesize {\color{basetext}6.1}}\ifdim\frzwtmp>\lgwCC\setlength{\lgwCC}{\frzwtmp}\fi
\settowidth{\frzwtmp}{\footnotesize $+$0.04\rlap{$^{\dagger}$}}\ifdim\frzwtmp>\lgwCC\setlength{\lgwCC}{\frzwtmp}\fi
\settowidth{\frzwtmp}{\footnotesize {\color{basetext}5.0}\rlap{$^{\dagger}$}}\ifdim\frzwtmp>\lgwCC\setlength{\lgwCC}{\frzwtmp}\fi
\settowidth{\frzwtmp}{\footnotesize $+$0.12}\ifdim\frzwtmp>\lgwCC\setlength{\lgwCC}{\frzwtmp}\fi
\settowidth{\frzwtmp}{\footnotesize {\color{basetext}8.9}}\ifdim\frzwtmp>\lgwCC\setlength{\lgwCC}{\frzwtmp}\fi
\settowidth{\frzwtmp}{\footnotesize {\color{basetext}5.7}}\ifdim\frzwtmp>\lgwCC\setlength{\lgwCC}{\frzwtmp}\fi
\settowidth{\frzwtmp}{\footnotesize {\color{basetext}4.8}}\ifdim\frzwtmp>\lgwCC\setlength{\lgwCC}{\frzwtmp}\fi
\settowidth{\frzwtmp}{\footnotesize {\color{basetext}5.2}}\ifdim\frzwtmp>\lgwCC\setlength{\lgwCC}{\frzwtmp}\fi
\settowidth{\frzwtmp}{\footnotesize {\color{basetext}4.9}}\ifdim\frzwtmp>\lgwCC\setlength{\lgwCC}{\frzwtmp}\fi
\settowidth{\frzwtmp}{\footnotesize $-$0.05}\ifdim\frzwtmp>\lgwCC\setlength{\lgwCC}{\frzwtmp}\fi
\settowidth{\frzwtmp}{\footnotesize {\color{basetext}5.1}}\ifdim\frzwtmp>\lgwCC\setlength{\lgwCC}{\frzwtmp}\fi
\settowidth{\frzwtmp}{\footnotesize $-$0.12}\ifdim\frzwtmp>\lgwCC\setlength{\lgwCC}{\frzwtmp}\fi
\settowidth{\frzwtmp}{\footnotesize {\color{basetext}11.8}}\ifdim\frzwtmp>\lgwCC\setlength{\lgwCC}{\frzwtmp}\fi
\settowidth{\frzwtmp}{\footnotesize $+$0.14\rlap{$^{\dagger}$}}\ifdim\frzwtmp>\lgwCC\setlength{\lgwCC}{\frzwtmp}\fi
\settowidth{\frzwtmp}{\footnotesize {\color{basetext}4.8}\rlap{$^{\dagger}$}}\ifdim\frzwtmp>\lgwCC\setlength{\lgwCC}{\frzwtmp}\fi
\settowidth{\frzwtmp}{\footnotesize {\color{basetext}16.5}}\ifdim\frzwtmp>\lgwCC\setlength{\lgwCC}{\frzwtmp}\fi
\settowidth{\frzwtmp}{\footnotesize {\color{basetext}10.5}}\ifdim\frzwtmp>\lgwCC\setlength{\lgwCC}{\frzwtmp}\fi
\settowidth{\frzwtmp}{\footnotesize {\color{basetext}5.8}\rlap{$^{\dagger}$}}\ifdim\frzwtmp>\lgwCC\setlength{\lgwCC}{\frzwtmp}\fi
\settowidth{\frzwtmp}{\footnotesize {\color{basetext}10.8}}\ifdim\frzwtmp>\lgwCC\setlength{\lgwCC}{\frzwtmp}\fi
\settowidth{\frzwtmp}{\footnotesize $-$0.07}\ifdim\frzwtmp>\lgwCC\setlength{\lgwCC}{\frzwtmp}\fi
\settowidth{\frzwtmp}{\footnotesize {\color{basetext}10.0}}\ifdim\frzwtmp>\lgwCC\setlength{\lgwCC}{\frzwtmp}\fi
\settowidth{\frzwtmp}{\footnotesize $-$0.08}\ifdim\frzwtmp>\lgwCC\setlength{\lgwCC}{\frzwtmp}\fi
\settowidth{\frzwtmp}{\footnotesize {\color{basetext}10.7}}\ifdim\frzwtmp>\lgwCC\setlength{\lgwCC}{\frzwtmp}\fi
\settowidth{\frzwtmp}{\footnotesize {\scriptsize pp/month}}\ifdim\frzwtmp>\lgwCC\setlength{\lgwCC}{\frzwtmp}\fi
\settowidth{\frzwtmp}{\footnotesize {\scriptsize\color{basetext} $\bar{q}^{*}$ (\%)}}\ifdim\frzwtmp>\lgwCC\setlength{\lgwCC}{\frzwtmp}\fi
\begin{table*}[p]
  \centering
  \caption{\textbf{Transcendent} (41 aligned test months, 2015--2018; full initial year) in the metric set, threshold protocol and layout of Table~\ref{tab:main-vtt2-b400}: the first six test months calibrate, the remaining 35 months evaluate, and the Cap batch size and the rejection quota $\rho$ equal $B_M$. Values are computed per month within seed, summarized within seed and averaged over seeds; rows with fewer seeds than the complete rows show their count and are not ranked. \textbf{Bold}=best, {\color{worstcolor}\itshape dark-red italics}=worst displayed value per ranked column and $B_M$ block.}
  \label{tab:legacy-trans}
  \footnotesize
  \setlength{\tabcolsep}{4.2pt}
  \renewcommand{\arraystretch}{0.95}
  \begin{tabular}{@{}cl@{\hspace{7pt}}c@{\hspace{7pt}}ccw{c}{\lgwCA}w{c}{\lgwCA}@{\hspace{7pt}}cw{c}{\lgwCB}w{c}{\lgwCB}w{c}{\lgwCC}w{c}{\lgwCC}}
    \toprule
     &  & \multicolumn{1}{>{\columncolor{tabbaseband}[\tabcolsep][\tabcolsep]}c}{Baseline} & \multicolumn{4}{>{\columncolor{tabreliband}[\tabcolsep][\tabcolsep]}c}{Reliability} & \multicolumn{5}{>{\columncolor{tabstabband}[\tabcolsep][\tabcolsep]}c}{Stability} \\
    $B_M$ & Method & \cellcolor{tabbaseband}F1 $\uparrow$ & \cellcolor{tabreliband}E-AURC $\downarrow$ & \cellcolor{tabreliband}Cap$_{B_M}$ $\uparrow$ & \multicolumn{2}{>{\columncolor{tabreliband}[\tabcolsep][\tabcolsep]}c}{\makebox[0pt][c]{F1$^{*}$ $\uparrow$}\makebox[0pt][c]{\smash{\raisebox{-2.6pt}{\rule{\dimexpr 2\lgwCA+4\tabcolsep-5pt\relax}{0.4pt}}}}} & \cellcolor{tabstabband}Sen[F1] $\uparrow$ & \multicolumn{2}{>{\columncolor{tabstabband}[\tabcolsep][\tabcolsep]}c}{\makebox[0pt][c]{Sen[AUROC$_f$] $\uparrow$}\makebox[0pt][c]{\smash{\raisebox{-2.6pt}{\rule{\dimexpr 2\lgwCB+4\tabcolsep-5pt\relax}{0.4pt}}}}} & \multicolumn{2}{>{\columncolor{tabstabband}[\tabcolsep][\tabcolsep]}c}{\makebox[0pt][c]{Sen[$q^{*}$]}\makebox[0pt][c]{\smash{\raisebox{-2.6pt}{\rule{\dimexpr 2\lgwCC+4\tabcolsep-5pt\relax}{0.4pt}}}}} \\
     &  & \cellcolor{tabbaseband}{\scriptsize \%} & \cellcolor{tabreliband}{\scriptsize $\times$100} & \cellcolor{tabreliband}{\scriptsize \% of errors} & \cellcolor{tabreliband}{\scriptsize \%} & \cellcolor{tabreliband}{\scriptsize\color{basetext} $\Delta$F1$^{*}$ (pp)} & \cellcolor{tabstabband}{\scriptsize pp/month} & \cellcolor{tabstabband}{\scriptsize $10^{-3}$/month} & \cellcolor{tabstabband}{\scriptsize\color{basetext} first$\to$last} & \cellcolor{tabstabband}{\scriptsize pp/month} & \cellcolor{tabstabband}{\scriptsize\color{basetext} $\bar{q}^{*}$ (\%)} \\
    \midrule
    \multirow{11}{*}{50} & DeepDrebin & 82.5 & 1.42 & \textbf{18.7} & \textbf{84.4} & {\color{basetext}$+$1.9} & $-$0.02 & $-$3.2 & {\color{basetext}0.83$\to$0.74} & 0.00 & {\color{basetext}1.3} \\
     & CADE-OOD$^{c}$ & \textcolor{worstcolor}{\itshape 68.7} & 4.65 & 5.8 & 66.0 & {\color{basetext}$-$2.8} & \textcolor{worstcolor}{\itshape $-$0.46} & $-$0.9 & {\color{basetext}0.55$\to$0.52} & 0.00 & {\color{basetext}1.2} \\
     & CADE-MSP$^{c}$ & \textcolor{worstcolor}{\itshape 68.7} & \textcolor{worstcolor}{\itshape 6.11} & \textcolor{worstcolor}{\itshape 5.2} & \textcolor{worstcolor}{\itshape 65.4} & {\color{basetext}$-$3.3} & \textcolor{worstcolor}{\itshape $-$0.46} & \textbf{$+$1.2} & {\color{basetext}0.48$\to$0.52} & 0.00 & {\color{basetext}2.7} \\
     & HCC-PL$^{a}$ & 82.7 & 4.47 & 15.6 & 83.8 & {\color{basetext}$+$1.1} & $-$0.05 & $-$3.0 & {\color{basetext}0.68$\to$0.61} & $-$0.01 & {\color{basetext}1.4} \\
     & HCC-MSP$^{a}$ & 82.7 & 2.73 & 15.3 & 83.9 & {\color{basetext}$+$1.2} & $-$0.05 & $-$2.1 & {\color{basetext}0.67$\to$0.62} & $-$0.01 & {\color{basetext}1.4} \\
     & Drebin & 70.8 & 1.65 & 13.4 & 70.9 & {\color{basetext}0.0} & $-$0.36 & $-$1.6 & {\color{basetext}0.82$\to$0.73} & 0.00 & {\color{basetext}1.1} \\
    \addlinespace[2pt]
     & \multicolumn{11}{>{\columncolor{subheadcolor}[\tabcolsep][\tabcolsep]}l}{\itshape Transcendent: inductive (ICE) and cross-conformal (CCE)$^{b}$} \\
     & ICE (Drebin) & 70.4 & 1.63 & 12.7 & 72.1 & {\color{basetext}$+$1.6} & $-$0.13 & $-$1.4 & {\color{basetext}0.83$\to$0.76} & 0.00 & {\color{basetext}1.2} \\
     & CCE (Drebin) & 70.8 & 1.58 & 13.3 & 72.4 & {\color{basetext}$+$1.6} & $-$0.14 & $-$1.7 & {\color{basetext}0.84$\to$0.77} & 0.00 & {\color{basetext}1.2} \\
     & ICE (DeepDrebin) & 81.7 & 1.19 & 13.8 & 83.2 & {\color{basetext}$+$1.5} & \textbf{$+$0.07} & $-$3.2 & {\color{basetext}0.83$\to$0.72} & $-$0.01 & {\color{basetext}1.3} \\
     & CCE (DeepDrebin) & \textbf{83.2} & \textbf{1.13} & 12.9 & 84.3 & {\color{basetext}$+$1.1} & $+$0.06 & \textcolor{worstcolor}{\itshape $-$3.4} & {\color{basetext}0.83$\to$0.72} & $-$0.02 & {\color{basetext}1.2} \\
    \midrule
    \multirow{11}{*}{100} & DeepDrebin & 85.1 & 1.17 & \textbf{28.2} & \textbf{87.9} & {\color{basetext}$+$2.8} & $+$0.03 & $-$3.7 & {\color{basetext}0.84$\to$0.70} & $-$0.02 & {\color{basetext}2.9} \\
     & CADE-OOD$^{c}$ & \textcolor{worstcolor}{\itshape 70.6} & 4.36 & 10.6 & 67.2\rlap{$^{\dagger}$} & {\color{basetext}$-$3.4}\rlap{$^{\dagger}$} & \textcolor{worstcolor}{\itshape $-$0.55} & $-$1.3 & {\color{basetext}0.56$\to$0.51} & $-$0.02\rlap{$^{\dagger}$} & {\color{basetext}2.4}\rlap{$^{\dagger}$} \\
     & CADE-MSP$^{c}$ & \textcolor{worstcolor}{\itshape 70.6} & \textcolor{worstcolor}{\itshape 4.85} & \textcolor{worstcolor}{\itshape 9.0} & \textcolor{worstcolor}{\itshape 61.4} & {\color{basetext}$-$9.3} & \textcolor{worstcolor}{\itshape $-$0.55} & \textbf{$+$2.4} & {\color{basetext}0.49$\to$0.53} & $+$0.03 & {\color{basetext}4.0} \\
     & HCC-PL$^{a}$ & 84.4 & 4.41 & 20.5 & 85.1 & {\color{basetext}$+$0.7} & $-$0.07 & $-$3.9 & {\color{basetext}0.68$\to$0.59} & 0.00 & {\color{basetext}3.2} \\
     & HCC-MSP$^{a}$ & 84.4 & 2.58 & 20.7 & 85.1 & {\color{basetext}$+$0.7} & $-$0.07 & $-$3.4 & {\color{basetext}0.68$\to$0.60} & 0.00 & {\color{basetext}3.0} \\
     & Drebin & 78.2 & 1.31 & 22.6 & 78.9 & {\color{basetext}$+$0.7} & $-$0.14 & $-$1.9 & {\color{basetext}0.80$\to$0.71} & $+$0.03 & {\color{basetext}2.6} \\
    \addlinespace[2pt]
     & \multicolumn{11}{>{\columncolor{subheadcolor}[\tabcolsep][\tabcolsep]}l}{\itshape Transcendent: inductive (ICE) and cross-conformal (CCE)$^{b}$} \\
     & ICE (Drebin) & 78.4 & 1.46 & 21.4 & 81.5 & {\color{basetext}$+$3.1} & $+$0.05 & $-$3.5 & {\color{basetext}0.84$\to$0.74} & 0.00 & {\color{basetext}2.4} \\
     & CCE (Drebin) & 78.8 & 1.41 & 21.3 & 81.9 & {\color{basetext}$+$3.1} & $+$0.05 & $-$3.7 & {\color{basetext}0.84$\to$0.74} & 0.00 & {\color{basetext}2.4} \\
     & ICE (DeepDrebin) & 84.9 & 1.09 & 19.2 & 86.6 & {\color{basetext}$+$1.7} & \textbf{$+$0.14} & $-$3.7 & {\color{basetext}0.84$\to$0.73} & $-$0.03 & {\color{basetext}2.6} \\
     & CCE (DeepDrebin) & \textbf{85.8} & \textbf{1.05} & 18.8 & 87.3 & {\color{basetext}$+$1.6} & $+$0.13 & \textcolor{worstcolor}{\itshape $-$4.0} & {\color{basetext}0.84$\to$0.71} & $-$0.04 & {\color{basetext}2.5} \\
    \midrule
    \multirow{11}{*}{200} & DeepDrebin & 87.3 & \textbf{0.91} & \textbf{41.8} & \textbf{91.0} & {\color{basetext}$+$3.7} & $+$0.06 & $-$3.5 & {\color{basetext}0.85$\to$0.71} & $-$0.06 & {\color{basetext}6.1} \\
     & CADE-OOD$^{c}$ & \textcolor{worstcolor}{\itshape 74.2} & 4.06 & 16.6 & 70.3\rlap{$^{\dagger}$} & {\color{basetext}$-$3.8}\rlap{$^{\dagger}$} & \textcolor{worstcolor}{\itshape $-$0.34} & $-$1.1 & {\color{basetext}0.54$\to$0.49} & $+$0.04\rlap{$^{\dagger}$} & {\color{basetext}5.0}\rlap{$^{\dagger}$} \\
     & CADE-MSP$^{c}$ & \textcolor{worstcolor}{\itshape 74.2} & \textcolor{worstcolor}{\itshape 4.56} & \textcolor{worstcolor}{\itshape 13.8} & \textcolor{worstcolor}{\itshape 59.0} & {\color{basetext}$-$15.1} & \textcolor{worstcolor}{\itshape $-$0.34} & \textbf{$+$2.5} & {\color{basetext}0.48$\to$0.53} & $+$0.12 & {\color{basetext}8.9} \\
     & HCC-PL$^{a}$ & 86.4 & 3.44 & 33.1 & 86.7 & {\color{basetext}$+$0.3} & $-$0.04 & $-$3.7 & {\color{basetext}0.68$\to$0.59} & 0.00 & {\color{basetext}5.7} \\
     & HCC-MSP$^{a}$ & 86.4 & 1.98 & 32.0 & 86.5 & {\color{basetext}$+$0.1} & $-$0.04 & $-$1.5 & {\color{basetext}0.69$\to$0.63} & 0.00 & {\color{basetext}4.8} \\
     & Drebin & 81.0 & 1.14 & 34.7 & 83.0 & {\color{basetext}$+$2.0} & $-$0.05 & $-$2.0 & {\color{basetext}0.81$\to$0.74} & $-$0.03 & {\color{basetext}5.2} \\
    \addlinespace[2pt]
     & \multicolumn{11}{>{\columncolor{subheadcolor}[\tabcolsep][\tabcolsep]}l}{\itshape Transcendent: inductive (ICE) and cross-conformal (CCE)$^{b}$} \\
     & ICE (Drebin) & 80.7 & 1.39 & 30.0 & 84.9 & {\color{basetext}$+$4.2} & $+$0.02 & $-$3.0 & {\color{basetext}0.83$\to$0.75} & $-$0.03 & {\color{basetext}4.9} \\
     & CCE (Drebin) & 81.0 & 1.37 & 30.2 & 85.2 & {\color{basetext}$+$4.2} & 0.00 & $-$2.9 & {\color{basetext}0.83$\to$0.75} & $-$0.03 & {\color{basetext}4.8} \\
     & ICE (DeepDrebin) & 87.0 & 1.08 & 26.5 & 89.3 & {\color{basetext}$+$2.3} & \textbf{$+$0.11} & $-$4.1 & {\color{basetext}0.84$\to$0.74} & $-$0.05 & {\color{basetext}5.2} \\
     & CCE (DeepDrebin) & \textbf{87.5} & 1.02 & 28.0 & 89.9 & {\color{basetext}$+$2.5} & $+$0.09 & \textcolor{worstcolor}{\itshape $-$4.6} & {\color{basetext}0.84$\to$0.73} & $-$0.06 & {\color{basetext}5.1} \\
    \midrule
    \multirow{11}{*}{400} & DeepDrebin & \textbf{89.9} & \textbf{0.66} & \textbf{60.5} & \textbf{94.4} & {\color{basetext}$+$4.5} & 0.00 & $-$4.5 & {\color{basetext}0.88$\to$0.72} & $-$0.12 & {\color{basetext}11.8} \\
     & CADE-OOD$^{c}$ & \textcolor{worstcolor}{\itshape 76.8} & 3.77 & 23.6 & 72.5\rlap{$^{\dagger}$} & {\color{basetext}$-$4.3}\rlap{$^{\dagger}$} & \textcolor{worstcolor}{\itshape $-$0.12} & $-$1.2 & {\color{basetext}0.54$\to$0.51} & $+$0.14\rlap{$^{\dagger}$} & {\color{basetext}4.8}\rlap{$^{\dagger}$} \\
     & CADE-MSP$^{c}$ & \textcolor{worstcolor}{\itshape 76.8} & \textcolor{worstcolor}{\itshape 4.11} & \textcolor{worstcolor}{\itshape 21.7} & \textcolor{worstcolor}{\itshape 58.9} & {\color{basetext}$-$18.0} & \textcolor{worstcolor}{\itshape $-$0.12} & \textbf{$+$4.0} & {\color{basetext}0.48$\to$0.57} & $-$0.02 & {\color{basetext}16.5} \\
     & HCC-PL$^{a}$ & 88.3 & 3.55 & 44.5 & 89.5 & {\color{basetext}$+$1.2} & 0.00 & \textcolor{worstcolor}{\itshape $-$5.9} & {\color{basetext}0.71$\to$0.59} & $-$0.04 & {\color{basetext}10.5} \\
     & HCC-MSP$^{a}$ & 88.3 & 1.80 & 42.8 & 88.6\rlap{$^{\dagger}$} & {\color{basetext}$+$0.4}\rlap{$^{\dagger}$} & 0.00 & $-$2.2 & {\color{basetext}0.71$\to$0.66} & $+$0.04\rlap{$^{\dagger}$} & {\color{basetext}5.8}\rlap{$^{\dagger}$} \\
     & Drebin & 83.2 & 1.44 & 52.1 & 85.5 & {\color{basetext}$+$2.3} & $-$0.01 & $-$3.7 & {\color{basetext}0.80$\to$0.72} & $-$0.01 & {\color{basetext}10.8} \\
    \addlinespace[2pt]
     & \multicolumn{11}{>{\columncolor{subheadcolor}[\tabcolsep][\tabcolsep]}l}{\itshape Transcendent: inductive (ICE) and cross-conformal (CCE)$^{b}$} \\
     & ICE (Drebin) & 82.4 & 1.26 & 46.3 & 88.4 & {\color{basetext}$+$6.0} & $-$0.02 & $-$3.5 & {\color{basetext}0.83$\to$0.74} & $-$0.07 & {\color{basetext}10.0} \\
     & CCE (Drebin) & 82.6 & 1.23 & 46.1 & 88.7 & {\color{basetext}$+$6.1} & $-$0.02 & $-$3.5 & {\color{basetext}0.82$\to$0.74} & $-$0.08 & {\color{basetext}10.0} \\
     & ICE (DeepDrebin) & 88.9 & 0.89 & 48.2 & 92.9 & {\color{basetext}$+$4.0} & \textbf{$+$0.01} & $-$4.8 & {\color{basetext}0.86$\to$0.73} & $-$0.07 & {\color{basetext}10.8} \\
     & CCE (DeepDrebin) & 89.6 & 0.85 & 48.6 & 93.3 & {\color{basetext}$+$3.7} & 0.00 & $-$5.0 & {\color{basetext}0.86$\to$0.73} & $-$0.08 & {\color{basetext}10.7} \\
    \bottomrule
  \end{tabular}
  \par\smallskip
  {\scriptsize\raggedright Random Cap reference: 4.0\% at $B_M=50$; 6.7\% at $B_M=100$; 10.6\% at $B_M=200$; 15.4\% at $B_M=400$. Random Forest, SecSVM, DRMD, RAMDA, TIF and MalCertain were not run on this dataset. $^{a}$HCC with family-supervised training and warm updates; MSP and pseudo-loss (PL) score the same runs and predictions. $^{b}$Transcendent: folds re-drawn at each monthly cold refit; ICE fold 0, CCE five folds; both share CCE-credibility acquisition. $^{c}$CADE with a cold-refitted neural head; OOD and MSP score the same runs and predictions. $^{\dagger}$Tied scores make the rolling window miss its referral target by more than 25\% on average (CADE-OOD at $B_M=100$; CADE-OOD at $B_M=200$; CADE-OOD at $B_M=400$; HCC-MSP at $B_M=400$); quota-dependent cells are unranked. Months with fewer than 20 accepted samples have no F1$^{*}$ (CADE-MSP at $B_M=400$: 1 of 175 seed-months).\par}
\end{table*}

\begin{figure*}[!t]
\centering
\includegraphics[width=0.95\textwidth]{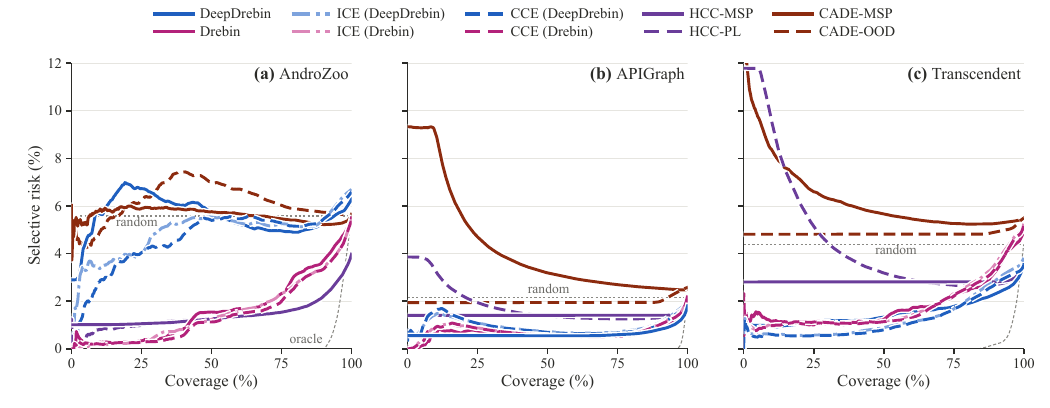}
\caption{Month-averaged risk--coverage curves at $\budgetmonthlymath=50$ on AndroZoo, APIGraph and Transcendent, one curve per row of Tables~\ref{tab:legacy-azoo}--\ref{tab:legacy-trans}, with the colors of Fig.~\ref{fig:eaurc_vtt2}(a). The axis is truncated at 12\%; on Transcendent, CADE-MSP and HCC-PL start above it at coverages below 2\%.}
\label{fig:risk_coverage_tau50}
\end{figure*}

\section{Deployment Simulation Algorithm}

Algorithm~\ref{alg:posthoc_rej_routine} is the deployment simulation used for Table~\ref{tab:main-vtt2-b400} and the appendix Tables~\ref{tab:vtt2-lower-budgets}--\ref{tab:legacy-trans}. The growing-pool variant replaces the window by all preceding months and the quota $W\rho$ by $(i-1)\rho$; Appendix~\ref{appendix:threshold_protocol} compares both with a frozen threshold.

\label{appendix:posthoc_rej_simulation}

% NOTE (page budget, restructure 2026-06-12): surrounding prose condensed
% ~50%; algorithm floats kept; original in .backup-pre-restructure-0900.
Algorithmic definitions of the \textit{deployment simulation} (\S\ref{subsec:operational_assessment}); practitioners only need the stored scores $S$ for each test month $M_i$ in $\mathcal{D}_{\text{test}}$.
Figure~\ref{fig:overview_rejection} applies the rolling-window variant with the fixed quota $\cutoffrejmath=400$ at $\budgetmonthlymath=50$.

\begin{algorithm}
\caption{Deployment simulation with rolling threshold re-calibration (Protocol~\ref{prot:posthoc_abstention})}
\label{alg:posthoc_rej_routine}
\begin{algorithmic}[1]
\Require monthly uncertainty scores $S_{M_1},\dots,S_{M_N}$ (larger $=$ less confident), quota $\rho$, window $W$ ($W=6$)
\For{$i=W+1$ \textbf{to} $N$}
  \State $A \gets S_{M_{i-W}}\cup\dots\cup S_{M_{i-1}}$ \Comment{unlabeled scores of the $W$ preceding months}
  \State $t_i \gets$ the $(W\rho)$-th largest value in $A$
  \State reject every $j\in M_i$ with $S_{M_i}[j]>t_i$
  \State record the referrals $q_i$ and F1$^{*}$ on the accepted samples \Comment{undefined if fewer than 20 accepted}
\EndFor
\end{algorithmic}
\end{algorithm}

\section{Choice of the Threshold-Calibration Protocol}
\label{appendix:threshold_protocol}

The threshold-based metrics of Table~\ref{tab:main-vtt2-b400} (F1$^{*}$, $\Delta$F1$^{*}$, $q^{*}$) depend on how the rejection threshold is set over the stream. We compared three label-free protocols on the same predictions and scores, all targeting $\rho=\budgetmonthlymath=400$ referrals per month: \emph{frozen}, one threshold per seed calibrated on January--June 2022 and kept fixed; \emph{growing}, Algorithm~\ref{alg:posthoc_rej_routine} with all preceding months as the pool, which re-calibrates each month on the pool of all preceding months; and \emph{rolling}, which re-calibrates each month on the six preceding months only. Table~\ref{tab:threshold-protocols} reports, for every row, the accepted-set F1, its change against F1, the realized referrals per month and their root mean square deviation from the quota.

Adherence to the quota decides the choice. A frozen threshold fixes a referral \emph{rate}; as monthly arrivals fall from 5.6k to 1.7k over the window, its count drifts away from the quota with no change in the scores (RMSE of the monthly count against the quota: mean 340, median 257 referrals per month across rows). The growing pool corrects this only slowly, because the early months dominate the pool (mean 313, median 205). The rolling window follows the drift and keeps the realized count closest to the quota (mean 305, median 148; 417 referrals per month on average against the target of 400), so it is the protocol of Table~\ref{tab:main-vtt2-b400}; its window length equals the calibration window. The means are pulled up by the two CADE rows and HCC-PL, whose tied or drifting scores make the quota unattainable under any protocol. Tracking the quota costs accepted-set F1 on average: mean F1$^{*}$ falls from 34.1 under the frozen threshold to 31.5 under rolling re-calibration and mean $\Delta$F1$^{*}$ from $-2.3$ to $-4.8$\,pp, because most rows refer far fewer samples than the quota under the frozen threshold, and referring more removes correct predictions along with errors. The choice matters for one headline reading. Under the frozen threshold, HCC-PL's accepted-set gain is obtained by referring 38\% of the stream (1,159 referrals per month); with monthly re-calibration its referrals fall to 570 and its gain from $+14.6$ to $+4.1$\,pp, and CCE~(DeepDrebin) has the highest accepted-set F1. Rank agreement with the frozen protocol across the sixteen rows is Spearman $r_s=0.91$ for F1$^{*}$ and $0.88$ for $\Delta$F1$^{*}$, and seven rows have a positive $\Delta$F1$^{*}$ under every protocol.

\begin{table*}[!t]
  \centering
  \caption{Threshold-calibration protocols on the sixteen rows of Table~\ref{tab:main-vtt2-b400} (Hypercube VTT2, $B_M=400$, five seeds), all targeting $\rho=400$ referrals per month on the same predictions and uncertainty scores. F1$^{*}$: accepted-set F1 (\%); $\Delta$F1$^{*}$: F1$^{*}-$F1 (pp); Ref./mo: mean realized referrals per month; RMSE$_{400}$: root mean square deviation of the monthly referral count from the quota. Frozen: one threshold per seed calibrated on January--June 2022 and kept fixed. Growing pool: re-calibrated each month on all preceding months. Rolling: re-calibrated each month on the six preceding months, the protocol of Table~\ref{tab:main-vtt2-b400}. Every value is a mean over months within seed, then over seeds. \textbf{Bold} = best per column (highest F1$^{*}$ and $\Delta$F1$^{*}$, lowest RMSE$_{400}$; Ref./mo is not ranked)}.
  \label{tab:threshold-protocols}
  \footnotesize
  \setlength{\tabcolsep}{4pt}
  \begin{tabular}{@{}lcccccccccccc@{}}
    \toprule
    Method & \multicolumn{4}{c}{Frozen (Jan--Jun 2022)} & \multicolumn{4}{c}{Growing pool} & \multicolumn{4}{c}{Rolling six months (Alg.~\ref{alg:posthoc_rej_routine}, Table~\ref{tab:main-vtt2-b400})} \\
    \cmidrule(lr){2-5} \cmidrule(lr){6-9} \cmidrule(lr){10-13}
    & F1$^{*}$ & $\Delta$F1$^{*}$ & Ref./mo & RMSE$_{400}$ & F1$^{*}$ & $\Delta$F1$^{*}$ & Ref./mo & RMSE$_{400}$ & F1$^{*}$ & $\Delta$F1$^{*}$ & Ref./mo & RMSE$_{400}$ \\
    \midrule
    DeepDrebin & 36.0 & $-$2.4 & 104 & 301 & 34.5 & $-$4.0 & 163 & 245 & 31.1 & $-$7.4 & 333 & 163 \\
    CADE-OOD & 39.7 & $+$1.4 & 196 & 459 & 40.4 & $+$2.1 & 276 & 536 & 42.5 & $+$4.2 & 462 & 830 \\
    CADE-MSP & 6.8 & $-$31.9 & 775 & 1135 & 3.1 & $-$35.6 & 843 & 1511 & 2.7 & $-$35.5 & 993 & 1877 \\
    HCC-PL & \textbf{53.7} & \textbf{$+$14.6} & 1159 & 1021 & \textbf{46.8} & \textbf{$+$7.7} & 867 & 647 & 43.2 & $+$4.1 & 570 & 288 \\
    HCC-MSP & 36.5 & $-$2.6 & 145 & 263 & 33.2 & $-$5.9 & 226 & 194 & 25.8 & $-$13.3 & 352 & 170 \\
    Drebin & 26.3 & $-$4.1 & 184 & 227 & 24.9 & $-$5.5 & 244 & 171 & 23.1 & $-$7.3 & 351 & 124 \\
    Random Forest & 24.5 & $-$13.3 & 415 & \textbf{94} & 24.9 & $-$12.9 & 381 & \textbf{89} & 24.7 & $-$13.0 & 396 & \textbf{90} \\
    SecSVM & 24.2 & $-$2.6 & 147 & 269 & 23.6 & $-$3.2 & 197 & 222 & 22.5 & $-$4.3 & 333 & 156 \\
    DRMD & 44.4 & $+$0.9 & 126 & 277 & 44.1 & $+$0.5 & 180 & 228 & 43.6 & $+$0.1 & 365 & 119 \\
    RAMDA & 36.3 & $-$5.2 & 242 & 185 & 35.9 & $-$5.5 & 274 & 163 & 34.2 & $-$7.3 & 384 & 291 \\
    TIF & 31.4 & $-$6.9 & 159 & 251 & 28.8 & $-$9.6 & 182 & 230 & 20.2 & $-$18.1 & 316 & 163 \\
    MalCertain & 26.2 & $-$3.5 & 148 & 275 & 25.3 & $-$4.4 & 214 & 216 & 24.7 & $-$5.0 & 325 & 140 \\
    \addlinespace[2pt]
    \multicolumn{13}{@{}l}{\itshape Transcendent: inductive (ICE) and cross-conformal (CCE)} \\
    ICE (Drebin) & 35.8 & $+$4.3 & 224 & 201 & 36.4 & $+$4.8 & 260 & 164 & 37.5 & $+$6.0 & 350 & 116 \\
    CCE (Drebin) & 36.5 & $+$4.4 & 225 & 200 & 37.1 & $+$4.9 & 260 & 164 & 38.2 & $+$6.1 & 352 & 112 \\
    ICE (DeepDrebin) & 43.0 & $+$5.3 & 284 & 147 & 43.7 & $+$5.9 & 343 & 120 & 44.2 & $+$6.4 & 392 & 123 \\
    CCE (DeepDrebin) & 44.1 & $+$5.8 & 288 & 140 & 44.8 & $+$6.5 & 349 & 113 & \textbf{45.4} & \textbf{$+$7.0} & 397 & 116 \\
    \midrule
    {\itshape Mean (16 rows)} & 34.1 & $-$2.3 & 301 & 340 & 33.0 & $-$3.4 & 329 & 313 & 31.5 & $-$4.8 & 417 & 305 \\
    {\itshape Median (16 rows)} & 36.2 & $-$2.5 & 210 & 257 & 35.2 & $-$3.6 & 260 & 205 & 32.6 & $-$4.7 & 358 & 148 \\
    \bottomrule
  \end{tabular}
\end{table*}

\section{Hyperparameter Configurations}
\label{hyperparameters_configurations}

\subsection{Hypercube VTT2 comparison}
\label{appendix:hp_vtt2}

All rows of Tables~\ref{tab:main-vtt2-b400} and~\ref{tab:vtt2-lower-budgets} use the 10,000 Drebin features with the largest absolute weights of a linear SVM ($C=1$) fitted on the 2021 training year, and fit their initial model on all 67,953 labeled 2021 applications. Each month, a model predicts the month's applications and then acquires the $\budgetmonthlymath$ applications of that month with the lowest confidence under its acquisition score. Neural models keep their weights across months and update on the most recent 5,000 labeled applications with a fresh optimizer; unless stated otherwise, scikit-learn models are refitted from scratch on the cumulative labeled pool. Following the time-consistent validation of~\cite{HCC}, hyperparameters are selected on the 2021 training year and the six months after it (January--June 2022), which enter no reported metric; every reported number covers July 2022--December 2024. The configurations are identical at every budget: the lower budgets share each method's initial model per seed with the $\budgetmonthlymath=400$ run of their own campaign (Appendix~\ref{appendix:vtt2_budgets}), not with the runs of Table~\ref{tab:main-vtt2-b400}, and change only the number of acquired labels. Monthly refits of Drebin and Random Forest use a fixed random state, so the seed acts through the initial model and the acquired samples. The rows of Table~\ref{tab:main-vtt2-b400} were trained in successive runs of the same training pipeline. The base models of the Transcendent ICE/CCE rows were trained by a later revision of the pipeline than the Drebin and DeepDrebin rows, with the configurations stated below. DRMD seed~1 was trained before, and seeds 2--5 after, a correction that passes the run seed to the agent; before the correction, every run initialized the agent with seed~1. A later run of the same configurations reproduces the HCC, TIF and MalCertain trajectories exactly but not those of Drebin, DeepDrebin, Random Forest, SecSVM, DRMD and RAMDA, whose training code revision was not preserved.

\paragraph{\textbf{Drebin}} Linear SVM (squared hinge loss, $\ell_2$ penalty, $C=1$, no class weights, at most 50,000 iterations). Acquisition and confidence use the absolute decision margin.

\paragraph{\textbf{SecSVM}} Linear SVM with every weight bounded to $[-K,K]$ with $K=2$~\cite{demontis2019secureml}, trained by projected subgradient descent with line search ($C=1$, step 0.5, at most 10,000 iterations, tolerance $10^{-4}$). Margin confidence as for Drebin. The bound was chosen by a sweep over $K\in\{0.25,0.5,1,2,4,8\}$ with $C=1$ and $K\in\{1,4\}$ with $C=10$, each configuration fitted once on the 2021 training year without updates and scored by the area under its monthly F1 curve. The initial sweep scored all 36 months of the stream, including the evaluation months. Following the time-consistent validation of~\cite{HCC}, which selects hyperparameters on the six months after the training year and tests on the remaining months, the sweep was repeated on the six calibration months (January--June 2022), which enter no reported metric. It places $K=2$, 4 and 8 on a plateau within 0.002 of each other (0.705--0.707, against at most 0.598 for $K\le1$ and 0.585 for $C=10$); Table~\ref{tab:main-vtt2-b400} uses the tightest bound on this plateau, $K=2$.

\paragraph{\textbf{Random Forest}} 101 trees of depth at most 64 (Gini impurity, $\sqrt{d}$ of the $d$ features per split, no class weights). The confidence is the forest's class probability, the mean over trees of the leaf class fractions.

\paragraph{\textbf{DeepDrebin}} Multilayer perceptron 10,000--200--200--2 with ReLU and dropout 0.5, cross-entropy loss, SGD at learning rate 0.05 with batch size 64 for 10 epochs, trained on a 66\% split of its training set; used for the initial fit and for each monthly update. These values equal those of the DeepDrebin baseline in the official code release of DRMD~\cite{mcfadden2026drmd}; the batch size and learning rate were selected on the 2021 training year and calibrated on the six months after it, following~\cite{HCC}; no evaluation month was used.

\paragraph{\textbf{HCC}} Benchmark adaptation of HCC~\cite{HCC}: encoder 512--384--256--128 with ReLU and a classifier head 100--100--2 with dropout 0.2, trained on 100 times the binary cross-entropy (on one-hot targets) plus a contrastive loss with margin 10 on mean squared embedding distances, using binary labels only (no malware families). Adam (learning rate $10^{-3}$), batches of 1,024 formed as in Chen et al.~\cite{HCC} from 512 random samples and, for each of them, a random sample of the same class; 40 epochs for the initial fit and for each update. HCC-MSP acquires and scores by the logit margin. HCC-PL rescores the same checkpoints with the pseudo-loss of Chen et al.~\cite{HCC}: 1,023 nearest neighbors among the normalized embeddings of the pre-month labeled pool, margin 10, cross-entropy weight 100, evaluation mode.

\paragraph{\textbf{CADE}} Contrastive autoencoder with encoder 512--384--256--128 (mirrored decoder), Adam at 0.003 for 150 epochs with step decay 0.95 every 10 epochs, triplet loss with weight 0.1 and margin 10, batch size 1,536; groups are benign, each labeled malware family, and one group for malware without a family label. A separate classifier 100--100--2 with dropout 0.2 is trained with Adam at 0.001 for 50 epochs (batch size 32). Both are refitted from scratch on the cumulative pool each month. Acquisition uses the MAD-normalized distance to the nearest class centroid (CADE-OOD); CADE-MSP scores the same predictions by one minus the maximum softmax probability.

\paragraph{\textbf{CADE and HCC uncertainty scores}}\label{appendix:ood_scores}
Let $z$ be the encoder embedding of a sample. CADE computes, for every training group $c$ (the groups above), the centroid $\mu_c$ of the training embeddings, the median $m_c$ of their distances $\lVert z_j-\mu_c\rVert_2$ to it, and the scaled median absolute deviation $\mathrm{MAD}_c=1.4826\cdot\mathrm{median}_j\,\bigl|\lVert z_j-\mu_c\rVert_2-m_c\bigr|$~\cite{CADE}. The OOD score is the smallest normalized deviation over groups,
\begin{equation*}
  U_{\mathrm{OOD}}(x)=\min_c \frac{\bigl|\lVert z-\mu_c\rVert_2-m_c\bigr|}{\mathrm{MAD}_c}.
\end{equation*}
HCC's pseudo-loss~\cite{HCC} assigns the predicted class $\hat y$ as pseudo-label and compares $z$ with its $1{,}023$ nearest neighbors $\mathcal{N}$ in the labeled pool available before the month, split into neighbors of the same label $\mathcal{S}$ and of the other label $\mathcal{D}$. With squared distances $D_j=\lVert z-z_j\rVert_2^2$, margin $m=10$, the predicted malware probability $p$ and $\lambda=100$,
\begin{align*}
  U_{\mathrm{PL}}(x)={}&\frac{\bigl[\sum_{j\in\mathcal{S}}D_j-|\mathcal{S}|\,m\bigr]_{+}}{|\mathcal{S}|}\\
  &+\frac{\bigl[2m\,|\mathcal{D}|-\sum_{j\in\mathcal{D}}D_j\bigr]_{+}}{|\mathcal{D}|}+\lambda\,\mathrm{BCE}(p,\hat y).
\end{align*}
Larger values of both scores denote greater uncertainty.

\paragraph{\textbf{DRMD}} PPO bandit with separate actor and critic networks, each with three hidden layers of 512 LeakyReLU units and dropout 0.5; Adam at $2.5\times10^{-4}$ with annealing, minibatches of 250, five training epochs, reward weight 10 on the minority class. The reject action, temporal rewards and DRMD's own active-learning mode are disabled. DRMD keeps its weights and retrains each month on the full cumulative labeled pool. Acquisition and confidence use the actor's logit margin.

\paragraph{\textbf{RAMDA}} Variational autoencoder (600 hidden units, 80 latent dimensions, dropout 0.1) with a detector 160--40--40--40--2 on the latent mean and scale; loss of 10 times the reconstruction error, the KL term and 10 times a disentanglement term with margin 60. Adam at $10^{-3}$, batch size 512, 50 epochs for the autoencoder, then 50 for the detector with the autoencoder frozen. An application is predicted malicious if its reconstruction error exceeds 30 or the detector predicts malware; the confidence in the tables is the detector's probability.

\paragraph{\textbf{TIF}} Network 512--256 (batch normalization, ReLU, dropout 0.5) to a 200-unit layer-normalized embedding and a binary head. Stage one trains for 30 epochs (batch size 512) with a multi-proxy contrastive loss (three proxies per class, temperature 0.2); stage two trains for 20 epochs (batch size 1,024, weight decay $10^{-3}$) with contrastive weight 0.1 and an invariance penalty of weight 1.0 over monthly environments, with a 10-epoch warm-up and early stopping (patience 5) on a 20\% validation split. Adam at $10^{-4}$ in both stages; each update repeats both stages on the most recent 5,000 labeled applications.

\paragraph{\textbf{MalCertain}} Base network 200--200--1 (ReLU, dropout 0.4, binary cross-entropy, Adam at $10^{-4}$, batch size 8) whose epoch-30 state serves as the base model; uncertainty features come from Bayesian, Monte Carlo dropout, epoch-snapshot and ensemble members (45 features). A class-balanced linear SVM ($C=1000$) on these features may only flip predicted-benign applications to malware. The most recent 20\% of the labeled pool is held out to train this correction, and all components are refitted from scratch on the cumulative pool each month. The tables pair the corrected predictions with the base network's confidence.

\paragraph{\textbf{Transcendent ICE/CCE}} Five stratified folds, fixed per seed; newly acquired applications are assigned to folds with class balance and never move. Nonconformity is the signed decision score of the base classifier. ICE fits on four folds and calibrates on the first; CCE uses all five fold models, with majority-vote predictions and median class p-values. The Drebin base is refitted on the cumulative proper-training folds; the DeepDrebin base (as above) updates on its latest 5,000 proper-training applications. All four rows acquire by one minus CCE credibility.

\subsection{AndroZoo, APIGraph and Transcendent}
The configurations below describe the experiments of Appendix~\ref{appendix:prior_hypercube}.

\paragraph{\textbf{Drebin}} Linear SVM (squared hinge loss, $\ell_2$ penalty), refitted from scratch each month. On AndroZoo and APIGraph (at most 1,000 iterations), $C\in\{0.01,0.1,1\}$ was run for every combination of dataset and $\budgetmonthlymath$, and the tables use the value with the highest F1 pooled over the seeds and the first six test months, which calibrate the rejection threshold and are not evaluated; this follows the time-consistent validation of~\cite{HCC}. The selected value is $C=0.1$ on APIGraph and $C=0.01$ on AndroZoo. On Transcendent, a single configuration was run ($C=1$, at most 10,000 iterations).

\paragraph{\textbf{DeepDrebin}} Two hidden layers of 200 ReLU units with dropout 0.5~\cite{DeepDrebin}, Adam at $10^{-3}$, batch size 512, refitted from scratch each month. On AndroZoo and APIGraph (two-unit softmax output, cross-entropy loss), 30 and 50 epochs were run for every combination of dataset and $\budgetmonthlymath$ and selected as for Drebin; 50 epochs are selected in 4 of the 8 combinations. On Transcendent (one output unit, binary cross-entropy weighted by the class ratio), a single configuration with 30 epochs was run.

% NOTE (page budget, restructure 2026-06-12): phrasing condensed, all factual
% settings preserved; the stale "single run / truncated budgets" claim for
% HCC-on-Hypercube was updated to match the completed reruns reflected in
% Table~\ref{tab:main-hc} (full window, 4-5 seeds per budget). Original in
% .backup-pre-restructure-0900.

The HCC and CADE runs on these datasets follow the official implementations~\cite{HCC,CADE} with the settings below, selected on the training year and the six months after it; only the subsequent months are reported.

\paragraph{\textbf{HCC}} The model follows Chen et al.~\cite{HCC}: a four-layer MLP encoder ($512$–$384$–$256$–$128$, each layer ReLU-activated) yields a $128$-dimensional embedding; the classifier adds two hidden layers with $100$ ReLU units apiece and a binary softmax; mini-batch size is $1\,024$. \emph{Cold-start} hyperparameter selection explores the Cartesian product of two optimizers (SGD, Adam), learning rates $\{0.001,\,0.003,\,0.005,\,0.007\}$, three schedulers (cosine annealing without restart; step decay $0.95/10$ epochs; step decay $0.5/10$), and $\{100,\,150,\,200,\,250\}$ epochs; the \emph{warm-start} phase further evaluates learning-rate scales $\{1\,\%,\,5\,\%\}$ of the initial value and $50$ or $100$ additional epochs, again with both optimizers. Best settings -- APIGraph: cold SGD, lr $0.003$, decay $0.95/10$, $250$ epochs; warm Adam at $1.5\times10^{-4}$ ($5\,\%$) for $100$ epochs after each monthly update. AndroZoo: cold SGD, lr $0.001$, decay $0.5/10$, $200$ epochs; warm Adam at $1\times10^{-5}$ ($1\,\%$) for $50$ epochs. Transcendent: cold SGD, lr $0.003$, decay $0.95/10$, $250$ epochs; warm Adam at $3\times10^{-5}$ ($1\,\%$) for $100$ epochs.

\paragraph{\textbf{CADE}} The CADE runs replicate the CADE setup of Chen et al.~\cite{HCC}: the OOD selector is paired with an MLP (cold and warm start) that mirrors the encoder dimensions in its decoder and shares the same classifier sub-network; batch size is 1\,536, the MLP is trained for 50 epochs at lr 0.001, and we grid-search the autoencoder hyperparameters (the original CADE study did not).
\emph{Cold start}: APIGraph -- Adam, lr 0.001, step decay 0.95/10, 150 epochs; AndroZoo -- Adam, lr 0.001, step decay 0.5/10, 100 epochs; Transcendent -- Adam, lr 0.003, cosine decay, 100 epochs.
\emph{Warm start}: APIGraph -- Adam, AE lr 0.001 with cosine annealing for 250 pre-training epochs, then lr reduced to 5\,\% for both AE and MLP over 50 active-learning epochs; AndroZoo -- Adam, AE lr 0.001 with cosine annealing for 100 epochs, lr scaled to 1\,\% during 50 active-learning epochs; Transcendent -- Adam, AE lr 0.007 with step decay 0.95 over 200 epochs, then lr 0.00035 for 50 epochs.

\paragraph{\textbf{Transcendent ICE/CCE}} Each month, five stratified folds are drawn afresh over the cumulative labeled pool and five fold models are refitted from scratch; fold 0 serves as the ICE calibration split, and CCE aggregates all five as in Appendix~\ref{appendix:hp_vtt2}. The Drebin base uses $C=0.01$ (AndroZoo), 0.1 (APIGraph) and 1 (Transcendent dataset); the DeepDrebin base is a 200--200--1 network (ReLU, dropout 0.5, Adam at $10^{-3}$, batch size 512, 30 epochs). Acquisition by CCE credibility; seeds 0--4.

% NOTE: a commented-out rejection-simulation subfigure block was removed here;
% it was the only user of the subcaption package (now removed from the preamble).
% See the backup if it needs to be restored (figs/rejection_simulation/rej_sim_{apigraph,transcendent}.pdf).

%\section{Additional Results}
%\label{appendix:additional_results}

% NOTE: the old commented-out metrics tables (tab:combined-metrics_old,
% tab:appendix-metrics_old, with retired CV/MAPD/MD metrics) were removed;
% they are preserved in the backup copy of this directory.

\section{Candidate and Retired Metrics}
\label{appendix:augrc}
\label{appendix:secondary_metrics}
\label{appendix:retired_metrics}

Table~\ref{tab:retired-metrics} records the candidate metrics that were considered for Table~\ref{tab:main-vtt2-b400} but are not headline metrics, with the literature that supports each, the reason it was not adopted, and the evidence on the VTT2 cohort. It replaces per-metric appendix sections.

\providecolor{tabreliband}{RGB}{230,245,220}
\providecolor{tabstabband}{RGB}{235,240,250}
\begin{table*}[!t]
\caption{Candidate metrics not adopted as headline metrics in Table~\ref{tab:main-vtt2-b400}. $r_s$: Spearman rank correlation across the sixteen seed-mean rows of Table~\ref{tab:main-vtt2-b400} (Hypercube VTT2, $\budgetmonthlymath=400$). Cells marked ``13 rows'' exclude HCC-PL and the two CADE rows. AUGRC uses Traub et al.'s closed form from monthly failure AUROC (AUROC$_f$) and accuracy.}
\label{tab:retired-metrics}
\scriptsize
\setlength{\tabcolsep}{3.5pt}
\renewcommand{\arraystretch}{1.0}
\begin{tabular}{@{}>{\raggedright\arraybackslash}p{1.9cm}>{\raggedright\arraybackslash}p{3.6cm}>{\raggedright\arraybackslash}p{3.4cm}>{\raggedright\arraybackslash}p{4.3cm}>{\raggedright\arraybackslash}p{3.3cm}@{}}
\toprule
\textbf{Metric} & \textbf{Definition} & \textbf{Literature} & \textbf{Exclusion rationale} & \textbf{VTT2 evidence} \\
\midrule
\rowcolor{tabreliband}\multicolumn{5}{@{}l@{}}{\textbf{Reliability}} \\
Mean failure AUROC (AUROC$_f$) & Probability that an error receives higher uncertainty than a correct prediction; threshold-free ranking. & Error-detection baseline~\cite{hendrycks2016baseline}; AUROC~\cite{hanley1982auroc}; ranking quality~\cite{galil2023can}. & Omits error frequency and high-confidence error concentration~\cite{jaeger2023failureevaluation,ding2020revisiting}; order-preserving score transforms tie. Retained for Sen[AUROC$_f$] and its first$\to$last companion. & $r_s=-0.71$ with E-AURC. \\
\addlinespace[1.5pt]
AUGRC & Area under generalized risk (probability of an accepted error) across coverage. & Traub et al.~\cite{traub2024overcoming}. & Weights selective risk by coverage, downweighting the high-confidence region targeted by \codename{}. & $r_s=0.91$ with E-AURC; DRMD--DeepDrebin gap: 0.87 E-AURC points versus 0.26 AUGRC points. \\
\addlinespace[1.5pt]
NAURC & Excess area normalized by its ceiling: 0 for oracle ranking, 1 for an uninformative score; removes error-rate dependence. & Proposed by Cattelan and Silva~\cite{cattelan2024naurc}; used by Heng and Soh~\cite{heng2026knowwhen}. & Closely tracks failure AUROC on this cohort; E-AURC retains error weighting and is interpreted alongside F1. & $r_s=-0.90$ with mean failure AUROC, $0.83$ with E-AURC. \\
\addlinespace[1.5pt]
SR$_{10}$ & Error rate among the 10\% most confident monthly predictions. & Fixed-coverage risk~\cite{geifman2017selective,kamath2020selective}; partial AURC~\cite{liang2024selectiveshift}. & Covers the region E-AURC weights most; reported with E-AURC in the results text. & $r_s=0.90$ with E-AURC (13 rows); range 3.4--7.0\% versus about 8\% overall error. \\
\addlinespace[1.5pt]
AURC[F1]$^{*}$ & Area under $1-$accepted-set F1 over target coverages 5--100\%; thresholds fixed after calibration on the preceding window. & No prior precedent; Transcend reports accepted-set F1 at one rate~\cite{transcend,barbero2022transcendent}. & F1 is non-additive over coverage; drift separates target from realized coverage. Excluding months with fewer than 20 accepted samples gives methods different support. Replaced by F1$^{*}$ at the single calibrated operating point. & Largest $|r_s|$ with a headline metric: 0.68 ($\Delta$F1$^{*}$, per-seed values); distinct but not interpretable. \\
\addlinespace[1.5pt]
Pooled-month failure AUROC and E-AURC & Errors ranked against correct predictions over the full window, with one hindsight threshold. & AUROC pair decomposition~\cite{hanley1982auroc,borkan2019nuanced}; pooling bias~\cite{airola2011pooled,forman2010apples}. & Dominated by between-month pairs; reflects score drift across months more than actionable within-month ranking. & Within-month pairs: 3.7\% of pooled pairs. Pooling reverses DeepDrebin and CCE~(DeepDrebin) (13 rows). \\
\addlinespace[1.5pt]
Missed malware Miss$^{*}$, Miss$_{400}$; deployment-threshold capture and yield & Shares of arriving malware accepted and missed, errors referred, and referrals that are errors. & Non-rejected accuracy~\cite{condessa2017rejectionmeasures}; review efficiency~\cite{kivlichan2021moderator}; CADE inspection precision--recall~\cite{CADE}. & Redundant with deployment F1$^{*}$/$\Delta$F1$^{*}$ and $\bar q^{*}$, and month-end batch Cap$_{400}$. & Protocol~\ref{prot:posthoc_abstention} capture: $r_s=0.57$ with Cap$_{400}$ and 0.55 with $\bar q^{*}$ (13 rows); combines ranking and workload. \\
\addlinespace[1.5pt]
\rowcolor{tabstabband}\multicolumn{5}{@{}l@{}}{\textbf{Stability}} \\
$\sigma$[F1] & Monthly F1 standard deviation (pp). & No temporal-evaluation precedent. & Restates decline in a bounded, declining series. & $r_s=-0.96$ with Sen[F1], $-0.71$ with F1; seed-pooling change up to 0.4\,pp. \\
\addlinespace[1.5pt]
CV[F1] & Monthly F1 standard deviation divided by mean F1. & Coefficient of variation~\cite{abdi2010coefficient}; used for run-to-run reproducibility, not drift. & Depends on the mean F1 used for normalization. & $r_s=-0.92$ with Sen[F1], $-0.83$ with F1. \\
\addlinespace[1.5pt]
Min F1; worst-three mean & Minimum monthly F1; mean F1 of the three worst months. & Worst-time performance~\cite{yao2022wildtime}; worst-group accuracy~\cite{koh2021wilds,sagawa2020groupdro}; selective worst-group accuracy~\cite{jones2021selective}; minimum alongside mean in malware continual learning~\cite{rahman2022limitations}; failures hidden by averages~\cite{demsar2006statistical}. & Minima occur in the same 2024 months for all methods, restating level and trend. & $r_s=0.92$ / $0.90$ with Sen[F1], $0.83$ / $0.88$ with F1; minima in Oct--Nov 2024 for 48 of 80 seed trajectories. \\
\addlinespace[1.5pt]
Maximum drawdown & Largest monthly F1 decline from peak to trough. & Chan et al.~\cite{chan2020reliability}; Chaudhry et al.~\cite{chaudhry2018riemannian}. & Depends on the evaluation horizon; collinear with CV[F1]. & $r_s=0.93$ with CV[F1], $-0.86$ with Sen[F1]. \\
\addlinespace[1.5pt]
Mann--Kendall $\tau$ & Consistency of signs across all monthly pairwise changes. & Mann~\cite{mann1945nonparametric}. & Omits magnitude; the Sen slope~\cite{sen1968slope} gives a similar ranking with interpretable units. & $r_s=0.93$ with Sen[F1]; negative for every row. \\
\addlinespace[1.5pt]
RMSSD[F1] & Root mean square of successive monthly differences; order-sensitive volatility independent of level. & Instability measure~\cite{jahng2008instability}; differencing precedent~\cite{chan2020reliability}. & Ranking stability and workload took priority. Least seed-stable F1 statistic: spread across rows 3.1\,pp versus mean seed SD 0.7\,pp (Sen[F1]: 0.80 versus 0.08\,pp/month). & Distinct: largest $|r_s|$ is 0.61 ($\bar q^{*}$), $0.40$ with F1. Accepted-stream RMSSD adds little ($r_s=0.87$, 13 rows). \\
\addlinespace[1.5pt]
$\Delta$Rej$^{*}$, \sigrej, $Q^{*}$ & Mean referral deviation from quota $\rho$ (400/month), referral SD, and RMSE relative to quota. & Referral-rate overshoot under drift~\cite{barbero2022transcendent}; quarantine cost~\cite{pendlebury2019tesseract}. & Arrivals fall from 5.6k to 1.7k per month; a constant referral rate implies count bias of $-98$ to $-109$ referrals/month without score drift. Replaced by rates: Sen[$q^{*}$], $\bar q^{*}$. & $\Delta$Rej$^{*}$: $r_s=0.98$ with $\bar q^{*}$; $Q^{*}$ range 23--469\% of quota $\rho$ (Table~\ref{tab:threshold-protocols}). \\
\addlinespace[1.5pt]
BF$^{*}$ & Share of abstention-induced F1 change that is beneficial, weighted by magnitude. & No prior precedent; specific to \codename{}. & $\Delta$F1$^{*}$ gives an interpretable effect size; BF$^{*}$ is undefined when total movement is zero. & Included in released results. \\
\bottomrule
\end{tabular}
\end{table*}

% Reviews of the previous submission and how they were addressed (SaTML CfP).
% Reviews of the previous submission (IEEE S&P 2027, cycle 1), reproduced unedited as required by the SaTML CfP.
% Generated from reviews_S&P.txt; the authors' rebuttal is omitted (it names an author). Do not add the rebuttal.
\clearpage
\section{Changes Since the Previous Submission}
\label{appendix:review_changes}

An earlier version of this paper was submitted to the IEEE Symposium on Security and Privacy 2027 (first cycle). Its reviews are reproduced unedited in Appendix~\ref{appendix:prior_reviews}. The revision addresses them as follows.
\begin{itemize}
\item \textbf{Novelty (Review D, meta-review).} \codename{} does not propose a new uncertainty estimator. It evaluates, over a longitudinal stream with a monthly labeling budget, whether confidence ranking, threshold transfer and analyst workload hold up under drift. Prior work evaluates uncertainty under shift~\cite{ovadia2019can} and selective classification~\cite{traub2024overcoming,jaeger2023failureevaluation} on fixed test sets, not over months of drift with monthly retraining. MalCertain~\cite{li2024malcertain} is discussed in Section~\ref{sec:background} and evaluated as a baseline in Table~\ref{tab:main-vtt2-b400}.
\item \textbf{Threshold protocol (Reviews A, D).} The previous text, Algorithm~\ref{alg:posthoc_rej_routine} and the released code described different protocols. Protocol~\ref{prot:posthoc_abstention} and Algorithm~\ref{alg:posthoc_rej_routine} now state the one used for every table, a rolling six-month re-calibration on the unlabeled score-history, and all threshold-based metrics are recomputed under it. Appendix~\ref{appendix:threshold_protocol} compares it with a growing pool and a frozen threshold.
\item \textbf{Ranking versus threshold transfer (Review C).} Within-month ranking is measured at a fixed quota by Cap$_{400}$, which also serves as the per-month quantile oracle. Threshold transfer is measured by F1$^{*}$ and $\Delta$F1$^{*}$, and workload by $\bar q^{*}$ and Sen[$q^{*}$]. $\Delta$Rej$^{*}$ and $\sigma$[Rej]$^{*}$ are retired (Table~\ref{tab:retired-metrics}).
\item \textbf{Temporal decay (Review C).} The Mann--Kendall $\tau$ is replaced by the Theil--Sen slope Sen[F1], an effect size in percentage points per month. The decline holds in every seed: all per-seed slopes are negative at every budget. Results are descriptive, and the decline is reported as specific to Hypercube VTT2.
\item \textbf{Comparison set (Reviews C, D, meta-review).} Table~\ref{tab:main-vtt2-b400} covers ten classifiers, including MalCertain, DRMD, RAMDA, TIF, Random Forest and SecSVM, and four Transcendent configurations with Drebin and DeepDrebin as base classifiers. The research questions are scoped to recent Android malware classifiers.
\item \textbf{Number of metrics (Review B).} Table~\ref{tab:main-vtt2-b400} reports F1 and three metrics per dimension. The remaining candidates, including AURC[F1]$^{*}$, are listed with their redundancy in Table~\ref{tab:retired-metrics}.
\item \textbf{Interpretation (Review B).} Section~\ref{subsec:reliability} defines risk--coverage curves before the results. Section~\ref{subsec:no_single_metric} states which metrics to check for which deployment need, and Section~\ref{sec:results} explains over-referral as poor threshold transfer. Risk--coverage curves on the additional datasets are reported per dataset (Appendix~\ref{appendix:prior_hypercube}) without a single-cause explanation.
\item \textbf{Terminology and active-learning scope (Review C).} Confidence quality is defined in the ranking sense, not as ECE-style calibration, and the introduction and background use this sense. The only AUROC reported is the failure AUROC (AUROC$_f$), which scores errors against correct predictions. Rejection is applied post hoc and does not affect the active-learning loop, so the reliability metrics evaluate rejection, not the value of acquired labels.
\item \textbf{Presentation (Review B).} Table marking is regenerated from the data: bold marks the best and dark-red italics the worst displayed value per ranked column. The Windows malware statistics are replaced by Android-specific figures.
\item \textbf{Not addressed.} Calibration-aware training (Review A); feature spaces other than Drebin and settings beyond Android (Reviews A, D; stated as a limitation); robustness to adversarial samples (Review A, meta-review; the threat model is natural drift); a random-acquisition control and an explicit cost-sensitive objective (Review C); further generic uncertainty and OOD scores beyond MSP and the evaluated methods (Review C).
\end{itemize}

\section{Reviews of a Previous Submission}
\label{appendix:prior_reviews}
{\small
\subsection*{Review \#1389A}

\paragraph*{Paper Summary}

The paper proposes AURORA, a framework to evaluate selective Android malware classifiers under distribution drift. The framework explores two crucial dimensions: reliability of the confidence function of SOTA malware classifiers under distribution shift, and temporal stability of SOTA malware detectors with respect to performance and rejection behavior over time. The paper evaluates four representative detectors based on Drebin's feature space across numerous metrics based on four select datasets (AndroZoo, APIGraph, Transcendent, Hypercube) and across different labelling budgets. The paper's analysis surfaces numerous valuable insights and tensions between typical evaluation metrics and the paper's own metrics, providing valuable directions for future malware detection research.

\paragraph*{Technical Correctness}

1. No Apparent Flaws

\paragraph*{Technical Correctness Comments}

The paper appears to be technically correct

\paragraph*{Scientific Contribution}

3. Creates a New Tool to Enable Future Science

4. Addresses a Long-Known Issue

6. Provides a Valuable Step Forward in an Established Field

\paragraph*{Scientific Contribution Comments}

The paper and framework provide numerous contributions. The framework provides a standardized testing setting for future papers to evaluate against. The paper provides numerous insights into the shortcomings of current SOTA Android malware detectors, offering future works valuable directions to explore. The analysis identifies numerous failure modes for existing approaches, in particular in relation to stability  and rejection quotas.

\paragraph*{Presentation}

1. No Flaws in Presentation

\paragraph*{Presentation Comments}

The presentation of the paper is clear and well-structured.

\paragraph*{Comments to Authors}

Thank you for submitting to S\&P. I find the paper to be a solid systems/methodology contribution to the field, with only minor concerns

- The paper identifies a valid gap in the malware literature. While the general disconnect between performance and the reliability of the confidence signal of a classifier is a known issue, the paper highlights its significance to the selective classification malware detection space.

- The performance metric coverage and depth of experimental analysis are thorough. The proposed metrics provide a seemingly complete characterization of the reliability and stability behaviors of the detectors under drift. The note about AURC[F1]* and the silent degrading of the precision-recall balance is a good example of this.

- The paper shows a deep methodological self-awareness. The experimental setup is carefully adapted to avoid the many technical correctness pitfalls that exist. I particularly appreciated the paper explicitly separating Hypercube from the three legacy benchmarks and keeping the latter for comparability.

- While the main takeaway (no silver bullet) is somewhat expected, the practical method selection guideline provides useful, actionable directions and is, in my opinion, the correct way to look at the problem.

I have only some very minor notes/questions:

- There are several ML papers on confidence/calibration-aware training objectives. It would have been interesting if the paper had included some exploratory analysis of how such approaches can be integrated into sota detectors and whether it would have addressed some of the failures identified here.

- The paper mentions that thresholds are calibrated using a rolling window. However, Algorithm 1 shows a monotonically increasing pool of unlabeled scores. Which one of the two is the correct reading? This has repercussions on how the rejection threshold is set and on the rejection metrics. If early months' scores are kept indefinitely, they dilute the weight of more recent scores.

- While it is reasonable to use a shared feature space to control variables, this leaves as an open question whether the insights regarding OOD/HCC pseudo-loss generalize to more complex/different input spaces.

- While I understand that the focus of the paper is on drift, since we are considering an adversarial setting (malware detection), it could have been interesting to include adversarial considerations in the discussion/threat model on how adversaries can actively exploit the confidence/ood functions.

\paragraph*{Recommended Decision}

1. Accept with Meta Review

\paragraph*{Reviewer Confidence}

3. Fairly Confident

\paragraph*{Should this submission be reviewed by the Research Ethics Committee?}

1. No

\subsection*{Review \#1389B}

\paragraph*{Paper Summary}

The paper discusses (and tries to capture with new metrics) several inter-connected problems that affect modern ML-based malware classifiers. These includes i) the problem of (a lack of) calibrated confidence, ii) the problem of degraded classification under distribution drift, and iii) the problem of selecting representative samples to label for continual learning (under a limited budget). In particular, the authors observe that SOTA approaches based on selective classification relies on the assumption that mis-classifications occur prevalently when the model is less confident, and set their goal to measure this assumption by proposing new metrics to capture reliability and temporal stability.

\paragraph*{Technical Correctness}

1. No Apparent Flaws

\paragraph*{Technical Correctness Comments}

None

\paragraph*{Scientific Contribution}

3. Creates a New Tool to Enable Future Science

6. Provides a Valuable Step Forward in an Established Field

\paragraph*{Scientific Contribution Comments}

I like the direction of this paper: new metrics can help to better understand the pros/cons of different classifiers and better capture some aspects (such as the reliability) that so far received little attention.

\paragraph*{Presentation}

2. Minor Flaws in Presentation

\paragraph*{Presentation Comments}

Excellent introduction and motivation for the research. Well, written and clear presentation of the concept and results. But still very dense and I wish there was more space for examples, more detailed explanations of the metrics and better discussion of the results.

\paragraph*{Comments to Authors}

I found that the metrics proposed in this paper (not novel per se, but AFAIK new to the malware classification domain) provide an interesting and informative view of the performance of a classifier. However, since I (and I guess the average reader) might  not be familiar with the Risk-Coverage curve, and the associated AURC and E-AURC, I wish the paper presented these metrics better in Section 5. An example figure and few lines on how to interpret the curve and its values would make the paper much easier to read (now one has to wait for the final results presented five pages later to finally see how this look like). The AURC is better for small values and is more difficult to intuitively assess. For instance, all ROC curves have roughly the same shape, but RC curves might vary.. so the same area value might correspond to very different behaviors (high errors only on low-confidence samples, or flat line). This makes absolute values more difficult to interpret at first.

This is complicated by the fact that the paper proposes many metrics (7 reported in table 1, but few more discussed in the paper) and the authors correctly point out that none of them is sufficient in isolation and all are useful in their own way. Thus, assessing the performance of a model requires to understand the combined effect of two baseline metrics, two reliability metrics, and five stability metrics. I cannot decide whether I absolutely love it or hate it!

As this is a very dense paper with lots of non-intuitive numbers, the interpretation of the results offered by the authors is of paramount importance. In this aspect, the authors do a fairly good job at capturing the main results. However, some findings remain a bit obscure (for instance, Point 4 in the RQ1 takeaways, which seems to come a bit out of nowhere), some are hard to understand (for instance, why the RC plots differ so much between Hypercube and Androzoo? Not jut in terms of raw values but even macro trends) and some are more a summary of the table than an interpretation (e.g., CADE overshoot its quote on a given dataset more than Drebin. Yes, that's what the numbers say. But why? What does this mean in relationship with other metrics?)

In summary it is difficult to draw any general conclusion. Some classifiers are better at low budget on some metrics, but worse in others - and viceversa. Some are reliable on one dataset but not on another, or more stable in one than in the others. The authors capture this randomness in the well-named section ''No Universal Winner''. However, I do not buy some of their conclusions. For instance, the authors say that the absence of absolute best means that ''deployment decisions should instead weigh the individual dimensions''. However, the same section also shows that many choices (e.g., subsampling) can improve results in some dataset and degrade them in others. So, what should an expert do in a real deployment if everything is data-set specific?

At the end, adding new metrics and suggesting a multi-dimension assessment does not make things clearer. It makes them less so. But I still think this paper is important because of what the authors say at the end: even if methods that pair high stationary performance with reliable selective prediction under drift remain an open challenge, this framework makes at least this aspect *measurable*. Yes, I agree.

Small comments:

In the last column of Table 1, some of the red-emphasized values are wrong (e.g., 8 is marked as good, 18 good, but 17 bad)

In the motivation section the authors describe Android classifiers but then justify the cost of retraining by quoting the huge number of new samples reported by AvTest. However, the reported numbers are not for Android. In fact, out of the 3M+ new malware detected by AVTest in the past two weeks, only 2\% are android apps. I am not saying that the problem is not real, but the authors should be clear in the motivation and avoid using Windows malware figures in an Android setting.

\paragraph*{Recommended Decision}

1. Accept with Meta Review

\paragraph*{Reviewer Confidence}

2. Highly Confident

\paragraph*{Questions for Text Rebuttal}

My main issues with the paper are related to the presentation and discussion of results (I wanted more). No specific question for the rebuttal.

\paragraph*{Should this submission be reviewed by the Research Ethics Committee?}

1. No

\subsection*{Review \#1389C}

\paragraph*{Paper Summary}

This paper argues that temporal malware-classification evaluation should assess not only predictive performance but also whether the confidence or OOD scores used by selective classification and active learning can remain reliable under drift. The authors propose AURORA, an evaluation framework built around three groups of measurements: baseline classification performance, confidence-ranking reliability, and temporal stability. The framework combines risk-coverage analysis and E-AURC with an F1-based operational coverage summary and monthly stability metrics such as F1 volatility, Mann-Kendall trend, selective-classification benefit, and rejection-count bias/variance. The paper evaluates Drebin, DeepDrebin, CADE, and HCC on four temporally ordered malware datasets. The main empirical observation is that stronger headline F1 does not necessarily imply better error ranking or more useful rejection behavior, and that the relatively simple DeepDrebin+MSP baseline is often competitive or stronger than the more complex CADE/HCC configurations.

\paragraph*{Technical Correctness}

2. Minor Issues

\paragraph*{Technical Correctness Comments}

E-AURC directly evaluates whether confidence ranks current prediction errors correctly, which is well aligned with selective classification. However, the paper sometimes extends this result to active learning by arguing that poor confidence ranking misdirects annotation budget toward non-informative samples. That implication is not established by E-AURC alone: an actively useful sample can be currently classified correctly but still have high expected learning value, while many currently misclassified samples can be redundant. Although the evaluation does run a full monthly active-learning loop, so downstream F1 trajectories partially reflect annotation utility, the design does not isolate the contribution of the selection strategy itself. This is a limitation the authors themselves acknowledge in section 6. The paper should therefore separate ''error-ranking quality for rejection'' from ''future model improvement per queried label,'' or add a selection-strategy ablation that attributes performance to the query rule.

The proposed AURC[F1]* is useful as an operational descriptive summary, but it should not be treated as equivalent to standard selective risk. Standard selective risk is an average per-sample loss, whereas F1 is a nonlinear function of the accepted set. In addition, F1 does not encode the stated asymmetric cost that false negatives are more expensive than false positives: in $F1 = 2TP/(2TP+FP+FN)$, $FP$ and $FN$ enter symmetrically. A robustness comparison against an explicitly cost-sensitive objective or class-conditional selective risks would make the operational interpretation substantially stronger.

The temporal-decay conclusion needs stronger statistical support. The paper interprets negative Mann-Kendall point estimates as evidence of degradation and later describes ''universal decay'' on Hypercube, even though some reported values are close to zero ($\tau=-0.04$). While the paper does conduct uncertainty analyses elsewhere (CI-based tie marking in the tables, and a seed-resampling/block-bootstrap audit of the Pareto frontier in section 5.3), no such analysis is applied to the trend statistic itself: the submission reports neither significance information nor effect size slopes for $\tau$ and the monthly sequence is naturally dependent because models are updated over time.

The starred deployment metrics characterize the full score-thresholding pipeline, not confidence ranking alone. Large $\Delta\mathrm{Rej}^{*}$ or $\sigma[\mathrm{Rej}]^{*}$ may result from cross-month score-distribution shift or historical-threshold transfer even when within-month ranking remains useful. Since the paper later interprets large deviations as an unstable “rejection mechanism,” it would be useful to separate these effects for example with a per-month quantile oracle or an alternative online threshold-calibration baseline.

\paragraph*{Scientific Contribution}

3. Creates a New Tool to Enable Future Science

6. Provides a Valuable Step Forward in an Established Field

\paragraph*{Scientific Contribution Comments}

The paper makes contribution in the evaluation perspective: it makes a convincing case that F1/FNR alone are insufficient for systems whose deployment decisions depend on confidence, uncertainty, or OOD scores. Applying risk-coverage analysis and operational stability measurements to temporal malware detection is useful and likely to influence how future systems are evaluated.

The empirical finding that DeepDrebin+MSP can match or outperform more expensive CADE/HCC configurations is valuable because it challenges the assumption that more sophisticated representation learning automatically produces better confidence behavior. The compute and label-complexity comparison strengthens this result. This result raises questions regarding what kinds of models we should choose to use in real world situations.

However the contribution is weakened by the limited comparison set for a paper that makes broad claims about SOTA selective/drift-adaptive malware classification. Transcendent is directly related to classification with rejection yet is excluded from the evaluation, and the paper does not compare against stronger general uncertainty/OOD baselines or more recent malware uncertainty/adaptation approaches. This makes the conclusion best read as applying to the four evaluated frameworks rather than to SOTA selective malware methods in general.

The paper's most novel operational metrics would be more compelling if they were connected to explicit deployment objectives. For example, a cost-sensitive risk combining false negatives, false positives, and analyst deferral cost would provide a clearer decision-theoretic interpretation than relying mainly on accepted-set F1 and rejection counts.

\paragraph*{Presentation}

2. Minor Flaws in Presentation

\paragraph*{Presentation Comments}

The paper sometimes mixes probability calibration, confidence-ranking quality, OOD detection, and operational rejection stability. The authors clarify in section 3 that ''calibrated'' is used in a broad ranking-fidelity sense rather than the ECE sense, but this terminology is not used consistently: the abstract and earlier discussion still read naturally as claims about probability calibration, while section 3.2.1 explicitly distinguishes calibration from ranking quality. Consistently distinct names for these related but different properties would make the framework easier to follow.

The AUROC discussion should distinguish classification AUROC (malware vs. benign) from failure-detection AUROC (incorrect vs. correct predictions). The current presentation uses closely related notation in different contexts, which makes the comparison with E-AURC harder to interpret.

The deployment-threshold description would benefit from one concrete worked example showing exactly what score is stored for MSP-based methods, how the historical threshold is selected, and how the same threshold is applied to the next month. This would remove ambiguity without changing the method.

\paragraph*{Comments to Authors}

The central claim of the paper is useful. A classifier that achieves good F1 but cannot rank its own errors is not sufficient for selective classification, and the paper demonstrates this convincingly with risk-coverage analysis. The use of a recent, better-curated Hypercube dataset together with legacy benchmarks is also a strength, as is the explicit comparison of compute and label requirements.

My main concern is that the paper currently treats one confidence score as serving two somewhat different roles: deciding which predictions to reject now and deciding which samples are most useful to label for future retraining. The former is directly evaluated by E-AURC; the latter is only indirectly reflected in the end-to-end active-learning results, which do not include a random or diversity-based selection control to attribute gains to the query rule. I recommend narrowing the active-learning claim or adding evidence such as future $F1$/FNR improvement per queried label relative to random/diversity-aware selection.

The custom operational metrics are potentially useful, but their interpretation should be tightened. AURC[F1]* measures accepted-set F1 over deployment coverages, not a standard expected risk, and F1 itself does not express an asymmetric malware \& false alarm cost. Likewise, rejection bias/variance measure end-to-end threshold-transfer stability, which is not the same property as within-month confidence ranking. These distinctions should be explicit in both the metric definitions and the conclusions.

The evaluation breadth is not yet strong enough for the broadest SOTA claims. In particular, excluding Transcendent is difficult to justify for a paper centered on selective classification, even if Transcendent is not an active-learning adaptation framework. At minimum, the claims should be scoped to the evaluated methods; stronger support would come from including a direct reject-option baseline and modern uncertainty/OOD baselines. The Hypercube degradation result is interesting, but ''universal decay'' should be supported by extending the paper's existing uncertainty analyses (CI ties, frontier bootstrap) to the trend statistic itself like effect-size slopes with dependence-aware uncertainty rather than the sign of $\tau$ alone.

\paragraph*{Recommended Decision}

2. Accept with Noteworthy Concerns in Meta Review

\paragraph*{Reviewer Confidence}

2. Highly Confident

\paragraph*{Questions for Text Rebuttal}

Can the authors clarify the scope of the active-learning claim? Specifically, what evidence shows that poor E-AURC corresponds to low annotation value rather than only poor current-error ranking, and can the authors report any downstream learning-utility comparison like future F1/FNR gain per queried label if it is already available?

Can the authors justify AURC[F1]* as the chosen operational objective and clarify what it is intended to measure? In particular, how should it be interpreted given that F1 is not a per-sample selective risk and does not encode a larger cost for $FN$ than $FP$? Any existing robustness result using an explicitly cost-sensitive or class-conditional selective objective would be helpful.

Can the authors provide stronger support for the Hypercube ''universal decay'' conclusion and clarify the breadth of the SOTA claim? In particular, please explain whether trend uncertainty/effect-size analyses were performed and why Transcendent or other direct uncertainty/rejection baselines were not included in the empirical comparison.

\paragraph*{Should this submission be reviewed by the Research Ethics Committee?}

1. No

\subsection*{Review \#1389D}

\paragraph*{Paper Summary}

This paper examines the reliability of malware classifiers under distribution shift. It argues that standard metrics such as F1, FPR, and FNR do not show whether a model's confidence or OOD score can identify samples it is likely to misclassify. The proposed AURORA framework evaluates both confidence-ranking reliability and temporal stability in selective-classification and active-learning settings.

The authors evaluate Drebin, DeepDrebin, CADE, and HCC on four timestamped Android malware datasets using E-AURC, risk-coverage curves, F1 variation and trends, and rejection-count bias and variability. The results suggest that more complex drift-adaptation methods do not necessarily provide more reliable confidence rankings: DeepDrebin with MSP matches or outperforms CADE and HCC in some settings, and strong offline performance does not consistently translate into stable identification of high-risk samples during long-term deployment.

\paragraph*{Technical Correctness}

4. Substantial Major Issues

\paragraph*{Technical Correctness Comments}

The released Figure 3 and main-table code calibrate once on the first six validation months and freezes the threshold over all test months. The generated table files match the paper's reported values, so the released evidence indicates that the results use the fixed-threshold protocol rather than the rolling protocol described in the paper.

\paragraph*{Scientific Contribution}

1. Independent Confirmation of Important Results with Limited Prior Research

\paragraph*{Scientific Contribution Comments}

The paper mainly confirms findings already established in selective classification and uncertainty evaluation in malware classification. Can You Trust Your Model's Uncertainty? Evaluating Predictive Uncertainty under Dataset Shift (NeurIPS 2019) showed that distribution shift can substantially degrade uncertainty estimates. A Call to Reflect on Evaluation Practices for Failure Detection in Image Classification (ICLR 2023) found that sophisticated confidence methods do not necessarily outperform a simple softmax baseline. More recent work, including Overcoming Common Flaws in the Evaluation of Selective Classification Systems (NeurIPS 2024) and What Does It Take to Build a Performant Selective Classifier? (NeurIPS 2025), further studies selective-classification metrics and confidence ranking. AURORA's main contribution is to demonstrate these issues in longitudinal Android malware evaluation and show their practical effects on methods such as CADE and HCC, rather than to introduce a fundamentally new uncertainty method.

\paragraph*{Presentation}

1. No Flaws in Presentation

\paragraph*{Comments to Authors}

This paper re-evaluates malware classification methods from the perspectives of confidence reliability and long-term stability. The problem is practically relevant, but the current submission has major issues in novelty, experimental protocol, and baseline coverage.

First, the novelty and research objective are not sufficiently clear. The observations that strong classification performance does not imply reliable confidence estimates, that distribution shift degrades uncertainty estimates, and that sophisticated uncertainty methods may not outperform simple softmax baselines have already been studied in many related works, such as Can You Trust Your Model's Uncertainty? (NIPS'19), A Call to Reflect on Evaluation Practices for Failure Detection in Image Classification (ICLR'23), Overcoming Common Flaws in the Evaluation of Selective Classification Systems (NIPS'24), and MalCertain: Enhancing Deep Neural Network Based Android Malware Detection by Tackling Prediction Uncertainty (ICSE '24). In particular, the MalCertain paper studies the same problem in Android malware classification and presents uncertainty estimation methods to help correct wrong predictions for addressing concept drift problems. AURORA primarily re-examines these findings in Android malware classification. Furthermore, studying OOD in Android malware classification has been extensively studied. The key question is whether a more fundamental approach can be proposed for all types of malware analysis, beyond just Android and classification; otherwise, the research findings are limited.

Second, the threshold protocol is internally inconsistent. Section 3.1 describes monthly recalibration from all preceding scores. However, the released code instead calibrates once on six validation months and freezes the threshold over the test period. The authors should state the actual protocol, correct Algorithm 1, and verify whether the conclusions hold under the claimed rolling recalibration.

Finally, the baseline coverage is too limited for broad claims about malware selective methods. The paper omits Transcendent precisely because it is a purely selective-classification method, even though selective classification is one of AURORA's central settings and the experiments use the Transcendent dataset.

\paragraph*{Recommended Decision}

3. Weak Reject (Can be Convinced by a Champion)

\paragraph*{Reviewer Confidence}

2. Highly Confident

\paragraph*{Questions for Text Rebuttal}

The code and paper are inconsistent regarding the threshold protocol. Which protocol was actually used in the experiments?

\paragraph*{Should this submission be reviewed by the Research Ethics Committee?}

1. No

\subsection*{Comment @A1 by Reviewer D}

We thank the authors for submitting their paper to IEEE S\&P 2027. The paper and the authors’ response have been carefully reviewed, and we agree that the paper addresses an important security problem, namely concept drift in Android malware detection, and that the preliminary results are promising. However, the reviewers identified major concerns regarding the novelty of the proposed approach and its comparison with closely related work, particularly MalCertain [ICSE’24]. In particular, the paper should more clearly explain how the proposed metrics can identify concept drift issues that are distinct from those already investigated in prior work. The evaluation should also assess the robustness of the proposed approach against diverse adversarial Android malware samples and provide a more comprehensive comparison with existing approaches to better demonstrate its improvements over the state of the art. We hope that these comments will help the authors further strengthen the work and better position its contributions in a future submission.
}

\end{document}